\pdfoutput=1
\makeatletter
\let\my@xfloat\@xfloat
\makeatother

\documentclass[12pt,a4paper] {book}

\usepackage[dvips]{graphicx}
\usepackage{caption}
\usepackage{setspace}
\DeclareCaptionLabelSeparator{vline}{ $\mathbf{|}$ }
\usepackage{mathrsfs}
\usepackage{fancyhdr}
\usepackage{verbatim}
\usepackage{hyperref}
\usepackage{sectsty}
\usepackage{pslatex}
\usepackage{amssymb}
\usepackage{amsmath}
\usepackage{pifont}
\newcommand{\cmark}{\ding{51}}
\newcommand{\xmark}{\ding{55}}
\usepackage[normalem]{ulem}
\usepackage{titlesec}
\usepackage[numbers,sort&compress]{natbib}
\usepackage{fourier-orns}
\usepackage{bm}
\DeclareMathAlphabet\mathbfcal{OMS}{cmsy}{b}{n}
\newcommand{\figrule}{\vspace{-5mm}\flushleft\noindent\rule[0.5ex]
{\linewidth/3}{0.5pt}\vspace{-3mm}}
\renewcommand{\chaptername}{}{\Large}

\newcommand{\ssubsection}[1]{\subsection*{\centering #1}}
\usepackage[final]{pdfpages}

\usepackage{xcolor}
\definecolor{cover}{cmyk}{0.2,0.2,0.0,0.3}
\definecolor{darko}{cmyk}{0.0,0.6,1.0,0.1}
\definecolor{darkblue}{cmyk}{1,1,0,0.1}
\definecolor{darkgreen}{rgb}{0.2,0.6,0.3}
\definecolor{grey30}{gray}{0.3}
\definecolor{grey50}{gray}{0.5}
\definecolor{grey75}{gray}{0.75}
\definecolor{grey95}{gray}{0.95}
\definecolor{numcol}{cmyk}{0.2,0.1,0.01,0.01}
\definecolor{boxcol}{cmyk}{0.1,0.02,0.01,0.01}
\definecolor{boxcold}{cmyk}{0.4,0.08,0.04,0.01}
\definecolor{pb}{cmyk}{0.8,0.16,0.08,0.02}
\definecolor{pg}{cmyk}{0.65,0.0,0.38,0.02}
\newcommand{\cb}{\color{blue}}
\newcommand{\cdb}{\color{darkblue}}
\newcommand{\cbl}{\color{black}}
\usepackage[framemethod=tikz]{mdframed}

\makeatletter
\def\@xfloat#1[#2]{
        \my@xfloat#1[#2]%
        \def\baselinestretch{1}%
        \@normalsize \normalsize
}
\makeatother

\usepackage{tocloft}

\usepackage{float}

\fancypagestyle{abstract}{
\fancyhead{} 
 
}

\fancypagestyle{contents}{
\fancyhead[L]{\leftmark}
 
}

\fancypagestyle{chapter}{
\fancyhead{} 
 
\fancyhead[R]{\rightmark \ \ \ \bfseries \thepage}
}

\fancypagestyle{normal}{
 
\fancyhead[L]{\nouppercase\leftmark}
\fancyhead[R]{Hall\quad \thepage}
\fancyfoot{} 
\addtolength{\headheight}{3.0pt} 
}

\begin{document}

\raggedbottom

\pagestyle{empty}
\begin{titlepage} 
\begin{center}

\vspace{\stretch{1}}
\includegraphics[height=130pt]{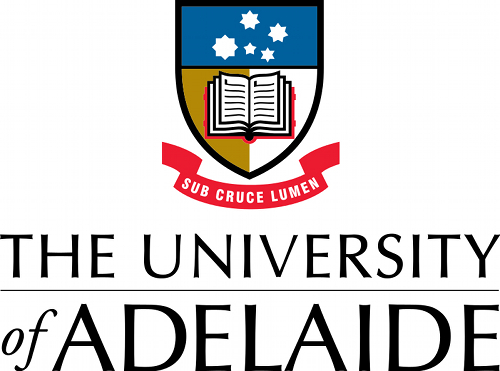} \\

\vspace{13mm}

{\LARGE \textbf{\cdb Biological Cell Resonators}} \\

\vspace{1cm}

\hspace{-5cm}
\begin{tikzpicture}
\shade[left color=cover,right color=grey95] (-10,0) rectangle (3.5,5);
\node (image) at (-0.7,2.5) {
\includegraphics[height=140pt]{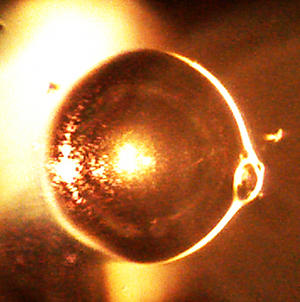}};
\end{tikzpicture}

\vspace{-4mm}
\hspace{-5cm}
\begin{tikzpicture}
\shade[left color=cover,right color=grey95] (-10,0) rectangle (3.5,0.2);
\end{tikzpicture}

\vspace{2mm}
{\large by}\\

\vspace{\stretch{1}}

{\large \cdb Dr. Jonathan Michael MacGillivray Hall, FRSA} \\

\vspace{\stretch{1}}

\begin{singlespace}
\textit{Supervisors:} \\
Prof. Tanya M. Monro \\
Assoc. Prof. Shahraam Afshar V. \\
Dr. Alexandre Fran\c{c}ois\\
\end{singlespace}

\vspace{\stretch{2}}

{
A thesis submitted to the degree of Doctor of Philosophy}

\vspace{\stretch{2}}

\begin{singlespace}
{
in the \\
\cdb Faculty of Sciences \\
School of Physical Sciences
}
\end{singlespace}

\vspace{\stretch{6}}
{%\large  
August 2017} \\

\pagebreak

\end{center} 
\end{titlepage}

\null
\par\vspace*{\fill}
\noindent Legal disclaimer: A digital copy of this thesis 
retaining copyright protection
has been placed on the Cornell 
University arXiv, under the non-excusive distribution licence. 

\vspace{3mm}
%\noindent \boxed{\quad\text{When citing this work, use: \href{http://arxiv.org}
%{\cb http://arxiv.org/abs/????.?????\cbl}}\quad}
%\vspace{3mm}\\
\noindent Academic website: 
\href{http://drjonathanmmhallfrsa.wordpress.com}
{\cb http://drjonathanmmhallfrsa.wordpress.com\cbl}
\cleardoublepage
\clearpage

\pagenumbering{roman}
\setcounter{page}{1}

\pagestyle{abstract}

\ssubsection{\cdb Declaration of Authorship}
\addcontentsline{toc}{chapter}{Declaration of Authorship}

I, Jonathan Michael MacGillivray Hall, certify that this work contains no 
material which has been accepted for the award
of any other degree or diploma in any university or other tertiary
institution and, to the best of my knowledge and belief, 
contains no material previously published or written by another person,
 except where due reference has been made in the text. 

I give consent to this copy of my thesis, when deposited in the University
Library, being available for loan and photocopying, subject to the provisions 
of the Copyright Act 1968. 

I also give permission for the digital version of my thesis to be made 
available on the web, via the University's digital research repository, 
the Library catalogue and also through web search engines, 
unless permission has been granted by the University to restrict 
access for a period of time. 

I acknowledge the support I have received for my research through the provision
of an Australian Government Research Training Program Scholarship. \\

\vspace{5mm}

\begin{singlespace}
\noindent Signed:\\
\rule{0.5\textwidth}{0.5pt}\\

\vspace{5mm}

\noindent Date:\\
\rule{0.5\textwidth}{0.5pt}
\end{singlespace}
\newpage
\thispagestyle{empty}

\cleardoublepage

\begin{center}
{\cdb THE UNIVERSITY OF ADELAIDE}
\end{center}
\ssubsection{\cdb Abstract}
\addcontentsline{toc}{chapter}{Abstract}
\begin{center}
\vspace{4mm}
\begin{singlespace}
{\cdb Faculty of Sciences \\
School of Physical Sciences}
\end{singlespace}
\vspace{6mm}
Doctor of Philosophy \\
\vspace{7mm}
by \cdb Jonathan M. M. Hall
\end{center}
\vspace{10mm}

Modern sensing technologies developed within the field of photonics 
incorporate a number of optical and acoustic phenomena. 
One such effect that has become a focal point in biosensing is 
\textit{whispering gallery modes}. These modes occur within 
optical cavities that exhibit a degree of symmetry, and are thus able to 
support resonating waves. This thesis develops the theory of resonances, 
exploring under what conditions a micro or nanoscale device 
can sustain these resonances, and for which physical criteria the resonance 
conditions deteriorate. The study is then extended to consider the 
biological cell. 
The discovery of a biological cell resonator, in which modes are definitively 
sustained without artificial assistance, represents the culmination 
of this thesis. 

The properties of resonators and their emitted 
energy spectra are studied within the general framework of the 
Finite Difference Time Domain method, requiring supercomputing resources to 
probe the transient behaviour and interactions among the electromagnetic 
fields. The formal theory of Mie scattering is extended to develop a 
cutting-edge, computationally efficient model for general, multilayer 
microspheres, which represents a valuable achievement for the 
scientific community in its own right. 
The model unifies the approaches in the field of mathematical modelling 
to express the energy spectrum in a single encompassing equation, 
which is then applied in a range of contexts. 

The gulf between modelling and biological resonators is bridged 
by an in-depth study of the physical characteristics of a range of 
biological cells, and the selection criteria for viable resonator 
candidates are developed through a number of detailed 
feasibility studies. The bovine embryo is consequently selected as the 
optimal choice. 

The scientific advancements contained within each chapter, 
including the improved models, 
the selection criteria and the experimental techniques developed, 
are integrated 
together to perform the principal measurements of the spectra within 
a biological cell. Evidence is established for the ability 
of a bovine embryo to sustain whispering gallery modes. 
This is a significant finding covering extensive research ground, 
since it is the first such measurement world-wide.  
The ability of a cell to sustain modes on its own  
represents a conceptually elegant paradigm for 
new technologies involving on-site cell interrogation and reporting of 
the status and health of a biological cell in the future. 
The methodological and technological  
developments contained within this interdisciplinary thesis thus 
become a vital asset for the future realisation of autonomous biological cell 
sensors. 

\newpage
\thispagestyle{empty}
\cleardoublepage

\newpage
\thispagestyle{abstract}
\ssubsection{\cdb Acknowledgments}
\addcontentsline{toc}{chapter}{Acknowledgments}

\vspace{0.5cm}

In this thesis lies a significant {\oe}uvre of work across multiple 
disciplines, including all the inherent logistical challenges of 
such a task. 
I would like to formally acknowledge my supervisors for their 
respective roles 
in this project: Prof. Tanya Monro, 
through the Australian Research Council Georgina Sweet Laureate Fellowship 
which supported the Laureate Scholarship, 
Assoc. Prof. Shahraam Afshar Vahid, and Dr. Alexandre Fran\c{c}ois.  

The unending positivity of Prof. Mark Hutchinson must be mentioned, along with 
the dedicated professional staff of the Centre for Nanoscale BioPhotonics, 
especially Ms. Melodee Trebilcock, both of whom I was able to turn to 
when the management psychology of such diverse interdisciplinary research 
led inevitably to conflicting procedural expectations. 

My friends and colleagues, Dr. Tess Reynolds and Dr. Matthew Henderson, 
were also always supportive. 
I wish to thank Mr. Steven Amos for his help during all the hours 
I spent over at the School of Chemical Engineering, 
and for teaching me the chemistry and 
practices required to mix my own media, and also Dr. Nicolas Riesen for 
conversations about measurement apparatus. 
My thanks also go to Dr. Wenle Weng, who assisted me in the clear 
measurement and identification of whispering gallery modes early 
in the experimental portion of the project. 

For all my friends in the OSA and SPIE Adelaide University Chapters, 
the IONS-KOALA 2014 International Conference organisation committee members, 
and the wonderful experiences we shared together to pull off the best conference
 in the series. 

Interstate, the moral support I received for the supercomputing portion 
of the thesis from Prof. Andrew Greentree and my friend and 
colleague Dr. Daniel Drumm (RMIT) should not be understated. 
In addition, the directional insights of Prof. Ewa Goldys (Macquarie University)
have helped keep my research priorities focused. 

The resources from eResearch SA, and The 
National Computational Infrastructure (NCI) 
Facility (ANU) were vital in the completion of the early modelling 
investigations.  

I wish to thank The University of Adelaide, Adelaide Enterprise, and 
The University of South Australia, as many people came forward to assist 
me at different phases of the project. 
I thank the Robinson Research Institute in reproductive health for their 
time and resources, including the assistance and understanding of my friend Mr. 
Avishkar Saini.  
 
\vspace{1cm}
On a personal level, I thank my friends for their support and understanding, 
especially Ian Kennedy, for all the conversations we had. I also thank the 
Burnside Symphony Orchestra and Haydn Chamber Orchestra for doing without me 
by the end of the project. 
The ongoing interest I keep to heart with Prof. Derek Leinweber, 
Elder Prof. Anthony Thomas and the CSSM, and the perpetual  
Visiting Research Associate status they granted me early in the project 
to continue integrating my research skills across multiple sub-fields was 
greatly valued. 
Finally, and most importantly, I acknowledge 
my family for unending support through all the complex 
phases of the project for whom any thanks I could bring to bear 
would be inadequate.

\clearpage
\thispagestyle{empty}
\cleardoublepage

\begin{singlespace} 

\thispagestyle{contents}
\tableofcontents

\clearpage
\thispagestyle{empty}
\cleardoublepage

\clearpage
\addcontentsline{toc}{chapter}{List of Figures}
\listoffigures

\clearpage
\thispagestyle{empty}
\cleardoublepage

\clearpage
\addcontentsline{toc}{chapter}{List of Tables}
\listoftables

\clearpage
\thispagestyle{empty}
\cleardoublepage

\end{singlespace} 
\clearpage

\pagestyle{normal}
\pagenumbering{arabic}
\setcounter{page}{1}

\renewcommand{\topfraction}{0.85}
\renewcommand{\textfraction}{0.1}
\renewcommand{\floatpagefraction}{0.75}

%Chapter heading formatting commands:
\titleformat{\chapter}[display]
  {\vspace{5cm}\flushright\fontseries{b}\fontsize{80}{100}\selectfont}
  {\fontseries{b}\fontsize{100}{130}\selectfont\textcolor{numcol}\thechapter}
  {0pt}
  {\Huge\bfseries}[]
\chapter*{\cdb \decofourright\,\, Prologue}
\label{chpt:pro}

\addcontentsline{toc}{chapter}{Prologue}
\renewcommand{\chaptermark}[1]{\markboth{\textbf{\chaptername #1}}{}}
\chaptermark{\cdb \decofourright\,\, Prologue}
\renewcommand{\chaptermark}[1]{\markboth{\textbf{\chaptername\ 
\thechapter. \,\, #1}}{}}

\section*{\cdb Overview}
\addcontentsline{toc}{section}{\cdb Overview}

\textit{``Interdisciplinary research (IDR) is a mode of research by teams or 
individuals that integrates information, data, techniques, tools, 
perspectives, concepts, and/or theories from two or more disciplines or bodies 
of specialized knowledge to advance fundamental understanding or to solve 
problems whose solutions are beyond the scope of a single discipline or area of 
research practice.''} \vspace{5mm}\\
\indent
National Academy of Engineering, National Academy of Sciences, Policy and 
Global Affairs, Institute of Medicine, Committee on Science, Engineering, and \\
Public Policy, Committee on Facilitating Interdisciplinary Research, 
\\\textit{Facilitating Interdisciplinary Research}, p.~39 (2005) 
\cite{engineering2005facilitating}.\\

The synthesis of physics, chemistry and biology involves the development of a 
wide arc of complementary techniques, and the exploration of new 
methodologies for addressing key mysteries in nature.  
In combining disparate fields, a cogent way forward is to apply the knowledge 
from each discipline in order to pose questions that are not within reach 
of any one field alone. In this thesis, such a synthesis is attained through 
the use of fundamental physics principles applied to biological cells. 

While a range of scientific developments developed herein, both technological 
and methodological, may be viewed as standalone achievements providing value 
to their respective fields in their own right, in the broader vision of the 
project, one may integrate the experimental methodologies, the models, the 
findings, the analyses and the conclusions in order to encapsulate 
the discoveries into a coherent narrative. 

I begin this narrative by noting a simple, compelling observation: that 
\emph{symmetry} can lead to \emph{resonance}. In the context of photonics, 
introducing light into a microscopic symmetrical object under specific  
conditions will cause it to resonate in a way that is highly sensitive to its 
immediate environment, as well as to its own geometry. In this sense, a 
\emph{resonator} can act as a \emph{sensor}. 
The resonances, and their behaviour under a variety of conditions,  
yield insights into both the structure and the electromagnetic influences 
acting on the resonator, and many instances where such effects represent 
a new frontier in scientific understanding are in the field of biology. 

To have a 
cell act as a resonator is to have a new window into its internal structure 
and its immediate surroundings. 
The fundamental symmetries extant in the building blocks of living matter 
present a compelling new direction in biosensing, where cells and their 
environments can be interrogated directly, producing or trapping photons at 
the site of inquiry in order to provide crucial information on their status 
and their features. 

At the outset of the journey, one must take stock of the specific aims and 
tools at one's disposal for exploring the research question of the present 
study. But first, I outline the vision. 

\newpage
\section*{\cdb Vision}
\addcontentsline{toc}{section}{\cdb Vision}

To begin, I pose the question of whether a biological cell can sustain 
resonances, with the ultimate aim being the generation of 
such resonances within a living cell.
I explore the physical requirements for achieving this goal, and the 
parameters that might serve to hamper its realisation, 
making particular use of \emph{whispering gallery modes} -- a phenomenon 
indicative of the resonant behaviour of waves within a symmetrical object. 

\begin{figure}
\begin{center}
\includegraphics[width=1.0\hsize]{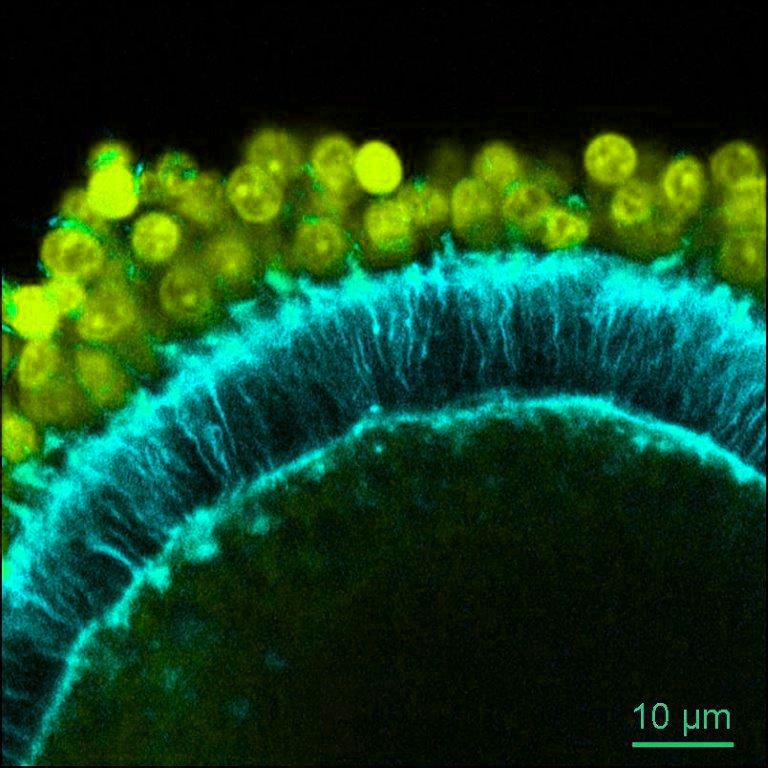}
\end{center}
\vspace{-5mm}\caption[Prologue: In the search for a biological resonator, the 
bovine oocyte is presented as the most viable candidate.]{
In this thesis, the cell 
determined as the most compelling candidate for a biological resonator is the 
\emph{bovine oocyte}. This scale image shows the cumulus 
cell nuclei (\color{darkgreen} \textit{green}\color{black}) surrounding the 
\textit{zona pellucida} region (\color{cyan} 
\textit{cyan}\color{black}). The cyan-coloured `strings' are the very fine 
filaments that connect the cumulus cells to the oocyte, 
allowing communication between the two cell types (\emph{transzonal} 
processes). 
\textit{Image:} courtesy of Assoc. Prof. J. G. Thompson, The University of 
Adelaide.}
\label{fig:cum}
\figrule
\end{figure}

First, the underlying principles of resonators are investigated and developed 
through a number of feasibility studies, with the focus on 
resonators that either resemble cells or include features that are useful for 
understanding the methodological requirements 
 for cell resonators. In pursuing such an investigation, new computational 
methods and analytic models are developed,  
and as a result, new aspects of resonator behaviour are uncovered that 
extend beyond the known work in the field. 
Upon presenting the necessary tools for a systematic inquiry into the 
possibility of a cell resonator, the cells that  
may provide an adequate testing ground for this 
purpose are enumerated, and their necessary physical requirements are 
explained. 
The most promising cell candidate is then identified -- the bovine embryo,
which is then subjected to a detailed 
analysis of its properties with respect to these physical requirements. 
The results of the studies are verified through the development of 
sophisticated modelling capabilities, each of which 
has been carefully designed to address 
the key challenges that present themselves at each milestone of the project. 
For example, bovine oocytes comprise a spherical outer shell, 
known as the \textit{zona pellucida}. Unfertilised oocytes also 
exhibit special outer cells attached to the \textit{zona pellucida} 
known as \textit{cumulus cells}, as shown in Fig.~\ref{fig:cum}, 
which disengage and dissipate upon fertilisation. 
It will become clear through the course of this thesis 
that for such a cell to be realised as a viable 
resonator, these cells need to be 
removed before achieving a resonance, either by selecting the right 
stage of embryo development after fertilisation, or otherwise rendering 
the embryo free, or \textit{denuded}, of these extra cells. 

The details involved in carrying out this process are then explained, which 
involve complex logistical demands, and new developments in 
interdisciplinary scientific methodology. 
Finally, the results are summarised and interpreted in the context of the main 
research question, reaching their denouement in Section~\ref{sec:fin} of the 
thesis. 

With the overall vision in mind, this study represents a new step in 
characterising the ability of biological cells to act as resonators, and in 
outlining and explaining the methodological challenges faced by researchers 
embarking upon a similar path in the future.

\section*{\cdb Aims}
\addcontentsline{toc}{section}{\cdb Aims}

In this thesis, the main aim is to develop the methodology required for 
examining the key research question, namely, whether a biological cell can act 
as a resonator,  making use of whispering gallery mode resonances. In order to 
achieve this aim, I will explore: \\

\noindent$\bullet\,$ The development and tailoring of models designed to mimic 
spherical, near-spherical and imperfect resonators;\\
$\bullet\,$ The properties and use of hollow cavity resonators, such as 
microbubbles, which share geometrical features with certain 
biological cells;\\
$\bullet\,$ A non-invasive method for extracting the geometric parameters from 
well studied microresonator structures;\\
$\bullet\,$ Multiple-layer resonator models that can account for a variety of 
mode excitation strategies, including active fluorescent-layer coatings such 
as those that have proved expedient in biosensing developments in the 
literature;\\
$\bullet\,$ The physical criteria required for a biological cell to act as a 
resonator;\\
$\bullet\,$ A number of candidate cell types that are most likely to be 
feasible;\\
\noindent$\bullet\,$ A variety of interrogation techniques, both passive and 
active, for the detection of whispering gallery modes, including 
the prism coupler, the fibre taper and doping with fluorescent materials;\\
$\bullet\,$ A simple proof-of-concept using a similar-sized artificial 
analogue of a cell in a sample of media; and \\
$\bullet\,$ A critical study of the most promising cell candidate, using 
the techniques listed above.

\section*{\cdb Roadmap of the thesis}
\addcontentsline{toc}{section}{\cdb Roadmap of the thesis}

While the narrative of the thesis involves the advancement of knowledge 
in a range of contrasting research areas, material pertaining to 
separate disciplines has been grouped together in chapters, where possible. 
Chapter~\ref{chpt:intro} provides a summary of the field of biosensing, 
specifically relating to whispering gallery mode resonators, and how 
the project is to be formulated in the context of biosensing. 
Those from a physics and engineering background will find particular value 
in Chapters~\ref{chpt:sph} and \ref{chpt:bub}, which explores the 
properties and the behaviour of power spectra obtained from
 microsphere and microbubble resonators using computational and analytic 
tools to yield new insights in facilitating resonator design. 
The heart of the mathematical 
modelling work presented herein is Chapter~\ref{chpt:mod}, 
in which a unified description of general, multilayer resonators is 
presented. 
Experts in the field of biology will find Chapter~\ref{chpt:cel} provides 
a comprehensive description of the challenges encountered in measuring 
physical attributes of cells using a variety of methods. 
The criteria developed in this chapter are applied experimentally 
in Chapter~\ref{chpt:wgm}, in which the apparatus, measurements and 
final analysis are presented, drawing together the multiple strands of the 
thesis. This chapter contains a combination of experimental optical physics, 
chemistry and biology. 

In addition, it is intended that an exploration of these interdisciplinary 
themes will yield new insights 
into the methods of characterising resonances, the models used to investigate 
them, and the physical parameters required of a resonator, 
as key developments in themselves. 

Therefore, as much as possible, each critical development explored in this 
thesis is explained in terms of its wider application 
to the fields of photonics and biosensing, as well as keeping with the 
directed narrative of the interdisciplinary vision. \\

\textit{``Interdisciplinary thinking is rapidly becoming an integral feature of 
research as a result of four powerful ``drivers'': the inherent complexity of 
nature and society, the desire to explore problems and questions that are not 
confined to a single discipline, the need to solve societal problems, and the 
power of new technologies.''}\vspace{5mm}\\
\indent
National Academy of Engineering, National Academy of Sciences, Policy and 
Global Affairs, Institute of Medicine, Committee on Science, Engineering, and \\
Public Policy, Committee on Facilitating Interdisciplinary Research, 
\\\textit{Facilitating Interdisciplinary Research}, p.~40 (2005) 
\cite{engineering2005facilitating}.

\titleformat{\chapter}[display]
  {\vspace{-1.5cm}\flushright\fontseries{b}\fontsize{80}{100}\selectfont}
  {\fontseries{b}\fontsize{100}{130}\selectfont\textcolor{numcol}\thechapter}
  {0pt}
  {\Huge\bfseries}[]
\chapter{\cdb Introduction}
\label{chpt:intro}

This thesis primarily focuses on the investigation into the optical phenomenon 
known as \emph{whispering gallery modes} (WGMs), and its   
application to biosensing and biological cells. 

The use of WGMs for biosensing has risen to a prominent place in the literature 
\cite{Boyd:01,Arnold:03,Ksendzov:05,Quan:2005a,Zhu2006,Ren:07,Weller,
pmid18036809,Francois:09,Huang2011,pmid21983134,Vollmer:2012a,Baaske2014}. 
Part of the reason for this is that the applications extend well beyond 
physics, in the detection of biological matter such as macromolecules. 
It has been demonstrated that devices known as \emph{microresonators} -- 
microscopic objects exhibiting a degree of symmetry which can support WGMs 
(described more precisely in Section~\ref{sec:arch}) --  
can be sensitive to the presence  
of virions, animal cells and bacteria \cite{Vollmer:08}, as well as proteins 
\cite{Vollmer:02a,Boyd:01,Arnold:03,Ksendzov:05} and DNA \cite{pmid18036809,
Vollmer:2012a}. In some cases this can be achieved to the level of a single 
molecule \cite{Vollmer30122008,Baaske2014}. 
Consequently, WGMs can be used to circumvent the requirement 
of fluorophore markers for the labelling of proteins, antibodies, amino acids 
and peptides, which are restricted in their use to specific 
biological targets \cite{Vollmer:08,Vollmer:2012a,Armani:2007a,Nguyen:13}. 
Such \emph{label-free} detection technologies also include 
Raman scattering microscopy \cite{doi:10.1167/iovs.03-0628,Freudiger1857}, 
autofluorescence \cite{ELPS:ELPS200406070}, intracavity spectroscopy 
\cite{4014191}, microphotonic sensors \cite{Kita:11} and the use of 
surface plasmon resonances \cite{doi:10.1021/ac0513459,BIOT:BIOT200800316}, to 
name a few, and have been widely used in the field of biosensing for some time. 

Parallel to the development of biosensing applications, WGMs have also been 
used for a range of other fundamental scientific studies, such as 
high-efficiency optical frequency combs \cite{Liang:11, Li:13,Riesen:15}, 
nonlinear optics 
\cite{Schliesser2010207} and quantum electrodynamics (QED) 
\cite{1367-2630-3-1-314,PhysRevA.72.031801,4675187,PhysRevA.83.063847}. 
The construction of chains or arrays of WGM resonators has also been reported 
in the literature \cite{Astratov:04,Astratov:14}, 
potentially leading to the development of perfect absorbers 
for use in solar panels \cite{He:13,Qu:16}, focusing microscopes, 
laser scalpels 
and polarisation filters \cite{Astratov:14}, and for the investigation of 
metamaterials \cite{6062374,He:13}.

The specific topics that form the basis for this thesis are 
the techniques for the generation of 
whispering gallery modes within a cavity, the biosensing applications of WGMs, 
the development of mathematical models of microresonator behaviour, 
including both analytic and computational methods used in computing the 
resonances, and cell biology. 
The goal will be to investigate the possibility of generating whispering 
gallery modes within a biological cell, as explained in Section~\ref{sec:form}. 
A significant amount of physics will be 
involved in interpreting both the computational results and the 
experimental outcomes.

\section{\cdb Whispering gallery modes}
\label{sec:nut}

Whispering gallery modes can be produced in microresonators, 
which exhibit a geometry that includes at least one axis of revolution, 
so that electromagnetic waves travel along the 
interface between the two materials. These waves largely remain bound due to 
\textit{total internal reflection} (TIR). 
Resonances occur when the round trip of a wave is an integer multiple of the 
wavelength. 

\begin{figure}[t]
\centering
\includegraphics[height=120pt]{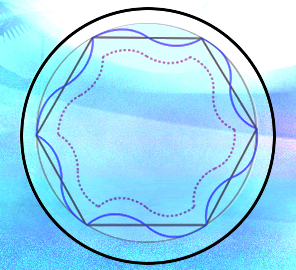}
\includegraphics[height=120pt]{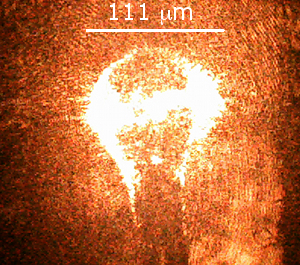}
\includegraphics[height=120pt]{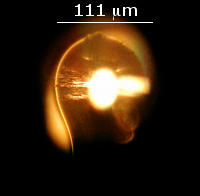}\\
\hspace{2mm}(a)\hspace{4.2cm}(b)\hspace{4.2cm}(c)\\
\vspace{-3mm}
\caption[Whispering gallery modes are illustrated in a microsphere resonator.]
{Whispering gallery modes (a) illustrated schematically in a circular 
cross section of a microsphere resonator. The two concentric oscillating 
lines indicate two configurations of the electric or magnetic field 
of the same azimuthal mode number, but different radial mode numbers. 
The azimuthal mode number is related to the number of nodes in the 
oscillating wave. The \color{blue} \textit{blue} \cbl line near the 
surface corresponds to a smaller radial mode number, while  
the \color{purple} \textit{purple} \cbl line corresponds to a larger 
radial mode number with modes extending deeper into 
the centre of the microsphere. 
(b) A silica microsphere is shown exhibiting WGMs with scale bar shown. 
(c) The microsphere boundary is shown under a white light source. 
}
\label{fig:wgmex}
\figrule
\end{figure}

Consider, for example, a spherical object constructed from a material with a 
refractive index higher than its surrounding environment, which acts as a 
resonator. 
Figure~\ref{fig:wgmex}(a) depicts a circular cross section of such a resonator. 
A microscope image of this effect in a silica 
microsphere (diameter $111$ $\mu$m) in the laboratory is shown in 
Fig.~\ref{fig:wgmex}(b), with 
the same microsphere shown under a white light source in 
Fig.~\ref{fig:wgmex}(c). The sphere is connected to a silica glass rod. 
The central bright spot corresponds to the reflection of the white 
light source into the objective. The WGMs are apparent around the 
equator of the microsphere. 
The wave pattern near the outer 
boundary of the resonator illustrates the minima and maxima of, for example, 
the electric field. In this scenario, 
the wave returns to its original position in phase, which will only occur for 
those specific wavelength values that 
correspond to the WGM resonance positions. As a result, the modes are 
\emph{quantised}, in that only certain 
values are admissible as resonances for a given resonator and its environment. 
A detailed discussion on this quantisation condition, and the method 
for the determination of the mode positions, is presented in 
Section~\ref{sec:an}, and developed as part of the multilayer 
model in Chapter~\ref{chpt:mod}.

\begin{figure}
\centering
\includegraphics[height=160pt]{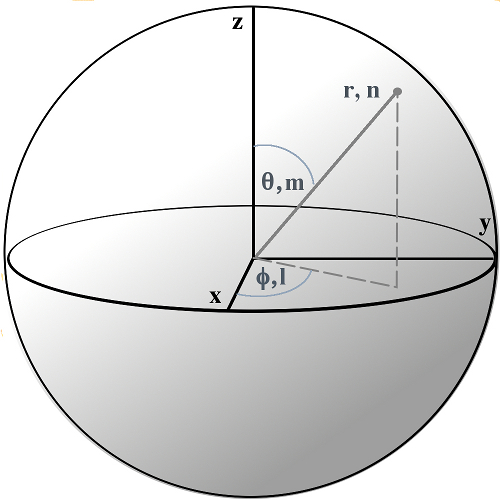}
\vspace{-3mm}
\caption[The spherical polar coordinates and their mode numbers are 
illustrated.]{Spherical polar coordinates, with corresponding 
spherical harmonic mode numbers, as illustrated on a microsphere.
}
\label{fig:coords}
\figrule
\end{figure}

As will be seen in the discussion of microsphere sensors in 
Chapter~\ref{chpt:sph}, 
the mathematical description of these modes requires 
the three eigenvalues of the quantisation 
condition. These \emph{quantum numbers} correspond to the 
azimuthal ($l$), polar ($m$) and radial directions ($n$) in spherical polar 
coordinates, as illustrated in Fig~\ref{fig:coords}. 
For an example circular cross section in the $x-y$ plane, shown in 
Fig.~\ref{fig:wgmex}(a), the number of surface nodes required to bring the 
wave back to its initial position in phase is related to the \emph{azimuthal} 
quantum number, whereas modes that extend further into the centre of the 
microsphere are related to the \emph{radial} mode number. 
The modes of the microsphere also fall into one of two categories, describing 
their polarisation -- \emph{transverse electric} (TE) or \emph{transverse 
magnetic} (TM) modes. Simply put, 
this notation states that the orientation of the three-vector 
electric fields ($\mathbf{E}$) or magnetic fields ($\mathbf{H}$) 
have no component in the 
radial direction of the coordinates of the sphere, shown in 
Fig.~\ref{fig:coords}. A more complete discussion of the conventions 
that exist for the definition of the polarisations is given in 
Section~\ref{sec:mod} and summarised in Appendix~\ref{chpt:app1}. 
Although more complicated mode patterns exist in 
general, the resonances of a microsphere can always be decomposed into a 
superposition of modes corresponding to these polarisations \cite{Hall:17a}. 

WGMs include a set of radiation modes and bound modes, which are 
discussed in the context of computational modelling in Section~\ref{sec:FDTD}, 
and investigated within a unified framework in Section~\ref{sec:mod}. 
They   
produce an \emph{evanescent field}, which is the portion of the 
electromagnetic field that extends beyond the 
surface of the resonator into the surrounding medium. As a consequence, the 
wavelength values that correspond to the 
resonances are highly sensitive to the geometric characteristics of the 
resonator, such as its diameter, its asphericity, and the refractive index 
contrast between the resonator and its surrounding medium \cite{Teraoka:06a}. 
The index contrast determines the amount that the evanescent field extends 
into the surrounding medium, thus affecting the refractive 
index \emph{sensitivity} of the modes within the resonator \cite{Arnold:03}. 
It is the mode sensitivity in particular that makes the WGM resonator a 
compelling candidate for biosensing applications.

\section{\cdb Microresonator architecture}
\label{sec:arch}

Whispering gallery modes can be generated in a wide range of different 
geometries, including those with a circular cross section outlined in 
Section~\ref{sec:nut}, as well as more exotic shapes, as follows. 

The literature on microresonators 
includes a wealth of varied examples, such as 
micropillars \cite{Vahala2003,pmid20941155}, 
microbottles \cite{Sumetsky:04,Lu:15,Wang:16}, toroids \cite{Armani2003}, 
disks \cite{:/content/aip/journal/apl/60/3/10.1063/1.106688,PhysRevA.84.063828,
Kuo2009,Ku:11}, rings \cite{Armani:2007a,so85559}, 
photonic crystals \cite{Ryu:04,doi:10.1117/12.532297,
:/content/aip/journal/apl/88/20/10.1063/1.2203943,4181729,
ValdiviaValero20111726,Liu} 
and other resonator designs that exhibit some degree of symmetry in at least 
one axis of revolution 
\cite{:/content/aip/journal/apl/82/4/10.1063/1.1540242,
:/content/aip/journal/apl/82/18/10.1063/1.1572966,1347-4065-42-6B-L652,
Huang:08,Tang:15}, which can even include the cross sections of 
waveguides or extruded slabs of dielectric material \cite{Tong2003,AfsharV:14}. 
The materials used in constructing artificial resonators include amorphous 
\cite{Vollmer:02a,Baaske2014,Ruan:14,Wildgen:15} and 
crystalline \cite{doi:10.1021/nn5061207,PhysRevA.91.023812,
PhysRevLett.92.043903,Grudinin200633,Grudinin:07,PhysRevA.83.063847,
PhysRevLett.114.093902} inorganic materials and polymers 
\cite{:/content/aip/journal/apl/83/8/10.1063/1.1605261,Francois:15b,
C5LC00670H}. 

In this work, the primary focus will be on resonators in spherical or 
microbubble form. The reason for this is that the biological cell candidates 
considered in Chapter~\ref{chpt:cel}  
are spherical at first approximation, and can contain multiple protein layers. 
Spheres represent optimally symmetric objects, 
and are often found in natural settings where 
the external force of surface tension applied evenly to the outermost 
layer of the cell encourages such shapes to form. 
Although cells 
exist that are closely related to other symmetrical shapes (such as the 
\emph{erythrocyte} -- the red blood cell, which resembles a biconcave disk 
\cite{ery}), the dependence of the quality of the modes 
on the diameter and the refractive index of a resonator, described in 
Sections~\ref{sec:bubbehav} and \ref{sec:selec},  
indicates that cells of much larger diameter represent the 
most viable candidates for sustaining WGMs. The ensuing investigation will 
also guide us towards those that are predominantly spherical in shape.

\subsection{\cdb Passive resonators}
\label{subsec:passr}

Although the wavelength positions of resonances are highly dependent on the 
geometric parameters of the resonator, the measurement of these resonances 
cannot occur in practice unless electromagnetic radiation is introduced. 
In such circumstances the electromagnetic field is said to \emph{couple} 
to the resonances, and these resonances are then said to be \emph{excited} 
by the radiation. 

\begin{figure}
\begin{center}
\includegraphics[width=0.6\hsize]{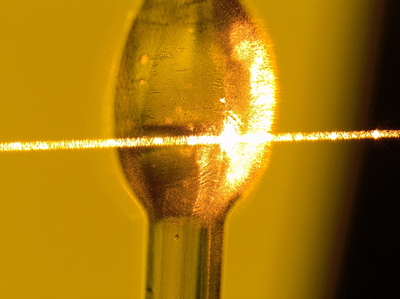}
\end{center}
\vspace{-5mm}\caption[An example glass microbubble is excited via a fibre 
taper.]{An image of a glass microbubble resonator, with WGMs 
excited via an SMF$28$ optical fibre taper placed 
less than a micron away from the outer shell of the microbubble. This 
represents a \emph{passive} interrogation method, whereby radiation is 
introduced into the resonator through phase-matched coupling. 
\textit{Image:} 
courtesy of Dr. Nicolas Riesen, The University of Adelaide.}
\label{fig:tapbubex}
\figrule
\end{figure}

A wide range of strategies has been investigated in the literature 
for the excitation of WGMs in microresonators. 
\emph{Passive} interrogation describes the case where 
external radiation is 
introduced into the resonator through the coupling of light at the 
material interface, rather than 
\emph{active} interrogation, where the excitation occurs from 
within the resonator itself, such as via a fluorescent medium.  
These methods will be explored more thoroughly in 
Chapters~\ref{chpt:cel} and \ref{chpt:wgm}, where the advantages and 
disadvantages of each method are investigated in depth. 
The methods of passive interrogation 
include the coupling of the light through a prism via frustrated 
TIR \cite{Gorodetsky:99,:/content/aip/journal/apl/82/4/10.1063/1.1540242,
BRAGINSKY1989393}. 
Waveguides \cite{Astratov:04,0022-3727-39-24-006} and fibre tapers 
\cite{Armani2003,1674-1056-17-3-047,6525394} have also 
been used to achieve mode coupling. 
In the case of the prism, the incidence angle must be tuned to fulfill the 
\emph{phase-matching} condition, describing the equalisation of the propagation 
constant 
of the incident beam with the WGMs of the resonator \cite{Knight:97}. 
Phase-matching represents an important topic in the generation of WGMs, 
and will be discussed in detail in Section~\ref{sec:an}, 
and used extensively in Chapter~\ref{chpt:wgm}, particularly for the 
experimental interrogation methods reported in Section~\ref{subsec:pass}. 
In the case of optical fibres, the taper waist must be 
fabricated to fulfill this same condition 
\cite{:/content/aip/journal/apl/82/4/10.1063/1.1540242,BRAGINSKY1989393} in 
order to allow coupling to the WGMs, as shown in Fig.~\ref{fig:tapbubex}. 

The requirement of an external evanescent coupling 
configuration, which typically must be tuned or 
calibrated carefully in order to achieve mode coupling, can 
render applications outside a laboratory setting problematic. This is because 
the physical space and time constraints involved in 
achieving on-site measurement within a biological environment are often 
limited, especially if living matter is involved. 
An alternative approach, which is able to ameliorate some of these limitations, 
is that of \emph{active} resonators. 

\subsection{\cdb Active resonators}
\label{subsec:act}

It is possible for the excitation of WGMs to occur from within the resonator 
itself, using fluorescent nanocrystals \cite{Kaufman2012,Ilchenko:13,
Tao21062016}, quantum dots \cite{C1JM10531K,6062374,1063-7818-44-3-189} or 
organic dyes \cite{Gupta201515,C5OB00299K,doi:10.1021/acs.joc.5b00077,
Francois:15b,doi:10.1021/acsami.5b10710,ADOM:ADOM201500776}.  
This allows for free-space excitation of the WGMs \cite{Francois:15a,
Francois:15b,LPOR:LPOR201600265}, in which the fluorescent medium 
is excited by a laser through a microscope objective in the absence 
of a radiation coupling device, such as those explored in the case 
of passive interrogation. 
This excitation strategy is worth exploring in more detail, 
because it presents a range of compelling features and advantages that 
passive techniques are not able to match at the present time. 
The active interrogation method
will be a focus for the development of the models described in 
Chapters~\ref{chpt:bub} and \ref{chpt:mod}. 

The first report of an active WGM resonator in the literature occurred in  
1961, shortly after the construction of the first laser by T. Maiman 
\cite{PhysRev.124.1807}.  
In this case, samarium-doped CaF$_2$ microspheres were excited using a 
flash-lamp. It was found that a major advantage of active resonators is that 
they allow for remote excitation and collection of radiation, ultimately 
allowing the resonators to probe biological entities, or to sense 
macromolecules \cite{Vollmer:08,Vollmer:02a,Boyd:01,Arnold:03,Ksendzov:05,
pmid18036809,Vollmer:2012a,Vollmer30122008,Baaske2014}. 

Active resonators contain a \emph{gain medium}, described below, and upon 
excitation, the emitted fluorescence intensity is modulated by the resonance 
frequency. This is a direct result of the \emph{Purcell effect} 
\cite{PhysRev.69.37}, which, in brief, describes the reduction of 
the lifetime of the emission, thus increasing fluorescent intensity at 
resonant wavelength values \cite{benisty1999confined,1263971,Zhu2014}. 
The Purcell effect will be explained in more detail in 
Section~\ref{sec:an}. 
The most important aspect of fluorescence-based resonators, besides the 
selection of the resonator geometry, is this gain medium, 
which provides the required fluorescence emission. 

Fluorescent dyes are the most common gain medium used for active 
microresonators, as they typically provide a broad range of 
emission wavelengths, stretching from the ultraviolet to the visible 
\cite{Gupta201515}, and in some cases, through to the near infrared 
\cite{C5OB00299K,doi:10.1021/acs.joc.5b00077}.  
The literature on the chemistry of various organic dyes in polymer matrices is 
extensive, and includes dyes for which the solubility and 
reactivity can be classified on the basis of the presence of \emph{azo} 
(--N=N--) groups, \emph{quinone} groups 
(\emph{anthraquinone} dyes), or \emph{phthalocyanines} \cite{WHO:2010}. 
These organic fluorophores are commonly mixed with a polymer, 
such as PMMA \cite{LPOR:LPOR201200074,C5LC00670H,Linslal:16}, 
SU$8$ \cite{Vanga2015209,Chandrahalim2015}, 
PDMS \cite{Saito:08} or polystyrene \cite{Himmelhaus2009418,
OPPH:OPPH201600006,LPOR:LPOR201600265},  
and subsequently fabricated directly into a chosen resonator 
design. Alternatively, fluorescent dyes can be introduced into already-formed 
polymer microresonators. One common method, especially for 
polystyrene microspheres, is to use a two-phase liquid system. In this case, 
the resonators are suspended in an aqueous solution while the 
fluorophore is dissolved in an organic dye that is not miscible with water 
\cite{Francois:15b}. 
As an alternative, resonators can simply be 
coated with fluorescent dye molecules using chemical moieties on both the dye 
and the resonator surface \cite{doi:10.1021/acsami.5b10710,ADOM:ADOM201500776}. 
It is this latter technique, in particular, which is followed in 
Chapter~\ref{chpt:bub} to examine the properties 
of dye-coated silica microbubbles \cite{Hall:17}, with a view toward 
simulating the fluorescent signal obtained from biological cells. 

Exactly which of these methods, both passive and active, 
 is most suitable for achieving the realisation of a functional, biological 
cell resonator, and what methodological alterations need to be made to 
accommodate this unique scenario, will be the principal subject of 
Chapter~\ref{chpt:wgm}. It is clear, however, that it is important to 
understand how each method relates to biosensing.

\section{\cdb Biosensing}
\label{sec:bio}

One particularly engaging aspect of modern research developments into resonator 
technology is the application to the sensing 
or measurement of important quantities in biology or medicine. 
WGMs represent a key optical phenomenon for sensing due to the 
sensitivity of their evanescent field to nearby entities, such as 
biomolecules, which break the symmetry of the electromagnetic field, thereby 
causing a measurable shift in either the resonance positions or intensities 
\cite{doi:10.1021/nl401633y}. 

In biosensing, the detection of single particles has become well established 
using passive resonators, with multiple demonstrations being documented 
\cite{doi:10.1021/nl401633y,:/content/aip/journal/apl/98/24/10.1063/1.3599584,
CPHC:CPHC201100757}. 
However, the free-space excitation and collection platform provided by active 
resonators 
enables novel applications that cannot be achieved with passive resonators. 
For example, free-floating resonators can be inserted into living cells for 
sensing \cite{Himmelhaus2009418}, or tagging and tracking purposes, using the 
specific spectral fingerprint of each individual WGM resonator 
\cite{doi:10.1021/acs.nanolett.5b02491,Humar2015}. 
The development of WGM measurement techniques in the context of biological 
cells is a key emphasis of this thesis, and will be discussed in more detail 
as the project formulation is presented in Section~\ref{sec:form}. 

Fluorescent-based interrogation methods have also been used in conjunction 
with smaller resonators, 
with extents of the order of $10$ microns ($\mu$m) 
\cite{Francois:15b,Reynolds:15,Francois:16a}. 
These resonators exhibit a number of 
advantages over their larger counterparts. Firstly, they are more localised, 
providing greater scope for placement of the resonator at the site 
required for sensing. The size of a resonator, in addition to the index 
contrast, also affects the refractive index sensitivity of the modes, 
the combination of which has recently been explored through the use of a 
figure of merit \cite{Reynolds:15}. Smaller resonators constructed from 
a given material with a certain refractive index, and in a given medium, 
exhibit an evanescent field that 
extends further into the medium than is the case for larger diameter resonators 
with the same index contrast. In the 
context of biosensing, this means that less material is required for observing 
a shift in the WGM positions, 
allowing scope for the detection of single particles 
\cite{Zhu2010,:/content/aip/journal/apl/98/24/10.1063/1.3599584,
CPHC:CPHC201100757}, and in some cases, single molecules \cite{Vollmer30122008,
Baaske2014}. 

On the other hand, smaller resonators are more susceptible 
to loss of confinement and more sensitive to geometric deformations or 
increased ellipticity, as a consequence of their extended evanescent field 
\cite{Reynolds:15,Riesen:15a}. This has the effect 
of reducing the clear definition of the modes, which is most commonly 
characterised by the \emph{quality factor}, or $Q$-factor -- a 
quantity that will be a particular focus in this thesis, and will 
be discussed in detail in Chapter~\ref{chpt:sph}. It is worth noting that 
active resonators typically exhibit lower $Q$-factors than their passive 
resonator counterparts, owing to their reliance on excitation methods that 
typically have no preferred electromagnetic alignment, which offsets the 
practical advantages afforded by the convenience of free-space interrogation. 

It has been shown that fluorescent resonators can be  
turned into microscopic laser sources using a lasing gain medium and a 
suitable pump source, enabling higher $Q$-factors to be 
realised upon reaching the lasing threshold 
\cite{Francois:15b,Francois:16,C5LC00670H,Karow:16,C6LC00070C,Ding2016,
Yakunin2015,Fan2015}
and lowering the detection limits for sensing applications 
\cite{Francois:15a,Francois:16}.
The determination of the lasing threshold can be achieved 
numerically \cite{doi:10.1080/09500340.2014.887799}  
through use of the lasing eigenvalue problem (LEP) formulation, which 
introduces the gain through an 
imaginary part of the refractive index  \cite{Smotrova2004,Chang2015}. 
However, exactly how the $Q$-factor and the effective mode volume 
of the resonator, $V_{\text{eff}}$, both influence the lasing threshold and the 
mode behaviour above the lasing threshold remains an unresolved issue.  
Early work by Sandoghdar \textit{et al.}  indicates that the lasing threshold 
of neodymium doped silica microspheres 
has a linear dependence on $Q^{-1}$ \cite{PhysRevA.54.R1777}, whereas Spillane 
\textit{et al.} suggested a dependence on $V_{\text{eff}}/Q^2$ 
\cite{Spillane2002}. More recently, Gargas \textit{et al.} 
established a dependence on $Q^{-1}$ for the lasing 
threshold of a ZnO microdisk \cite{doi:10.1021/nn9018174}, among other factors 
such as the diameter of the disk, further indicating that the behaviour of the 
lasing WGMs and the determination of the lasing threshold are not fully 
understood. 
 
Single cells have been reported to be 
transformed into WGM resonators by injecting a mixture of fluorescent dye and 
high refractive index oil, thus providing the required laser gain medium to 
generate WGMs, as well as some degree of confinement due to the refractive 
index contrast between the oil droplet and its surrounding 
environment \cite{Humar2015}. Furthermore, by combining fluorescent 
microspheres with flow cytometry, automated high-throughput 
sensing has been achieved, using a robust data analysis 
algorithm to extract real-time information about resonator properties from 
their WGM spectra \cite{:/content/aapt/journal/ajp/82/5/10.1119/1.4870185}. 
Fluorescent cholesteric liquid crystal core shell structures 
have also shown potential as magnetically transportable light sources for 
in-channel illumination applications \cite{C6LC00070C}. 

With such a wide variety of research avenues in the field of biosensing, it is 
worthwhile reminding the reader 
how the research question is to be formulated in the context of the 
techniques and methods developed in this thesis.

\section{\cdb Formulation of the project}
\label{sec:form}

\begin{figure}[t]
\begin{center}
\includegraphics[height=0.3\hsize]{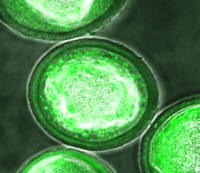}
\includegraphics[height=0.3\hsize]{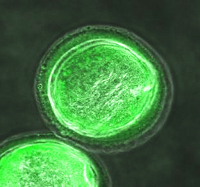}
\includegraphics[height=0.3\hsize]{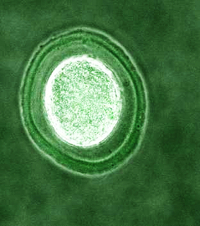}\\
\end{center}
\vspace{-5mm}\caption[Murine embryos exhibiting autofluorescence 
under confocal laser microscopy are shown.]
{Examples of murine embryos exhibiting autofluorescence under confocal laser 
microscopy are shown. 
Both these and related types of cell are considered extensively in 
Chapter~\ref{chpt:cel}.}
\label{fig:fluoro}
\figrule
\end{figure}

As a complementary approach to the myriad techniques that have been developed 
in the literature, the concept of a biological cell acting as a resonator 
\emph{itself} will be explored herein. 
This research direction is conceptually quite separate from the integration of 
existing microresonator sensors with biological cells, 
involving the introduction of microspheres onto or within 
the cell membrane \cite{Himmelhaus2009418}. 
In effect, what is being proposed is to investigate whether a cell is able to 
support resonances on its own, that is, without the use of an 
artificial cavity known to be able to sustain modes. 
By generating photons at the site of inquiry through the 
aforementioned methods of active interrogation,
the measurement of resonances would, in a sense, be 
reporting on the immediate internal and external environment of the cell 
itself. 
If successful, this would represent an elegant conceptual development in the 
field of biosensing. 
It is also possible that the natural autofluorescence of cells themselves 
could potentially be harnessed to generate modes, 
for example, in mammalian oocytes (such as those shown in 
Fig.~\ref{fig:fluoro}, and described in more detail in 
Chapter~\ref{chpt:cel}). 
The consequences of being able to measure modes sustained 
naturally by a living cell could potentially lead to new insights into the 
complex biological environment inhabited by living matter.

As a result, the microresonators that will be considered in detail will be 
confined to 
those that act as a direct analogue for a particular class of biological cell. 
The models developed in this thesis become progressively more 
sophisticated as the understanding of the mode behaviour and the criteria for 
sustaining WGMs are refined. The class of biological cell identified in 
Chapter~\ref{chpt:cel} is then explored for its potential use 
as a biological resonator. 

A detailed summary of the structure of the thesis is as follows.

\section{\cdb Structure}

Like the structure of a cell, the thesis is multi-layered. 
While the roadmap presented in the Prologue indicates the respective 
disciplines and areas of interest treated within each chapter, a 
technical summary of the thesis structure is as follows. 

In Chapter~\ref{chpt:sph}, the technology underpinning WGM sensors is 
explained, and a computational toolkit developed, which is able to handle 
deformations and inhomogeneities 
in microspheres. The model is tested against known and 
verified analytic models from the 
literature. Chapter~\ref{chpt:bub} demonstrates the customisability of the 
toolkit, incorporating a single-layer `bubble' structure, as well as a uniform 
layer of dipole sources, representing a fluorescent active medium. 
A non-invasive 
spectrum-based method for the extraction of key geometric 
data from such resonators is presented. 
In order to counteract the resource limitations of these complex computational 
models, especially for large scale (diameter $> 100$ $\mu$m) cells, an 
efficient, multilayer microsphere model 
is developed in Chapter~\ref{chpt:mod}, and harmonised with the existing 
models in the literature. The unique feature of 
this model is its ability to incorporate a number of different excitation 
strategies, including any number of dipole sources or active fluorescent 
layers, into a unified formula. 

Throughout the initial chapters, the vision underpins each summary discussion, 
in order to maintain the scope of the topic on 
cell resonators. This theme becomes the prominent discussion topic in 
Chapter~\ref{chpt:cel}, where the search for a candidate cell, 
its required physical criteria, and the feasibility of the most suitable 
choice are explored. The technologies developed thus far are then applied to 
the best candidate cell, and the experimental setup and methodology 
are expounded in Chapter~\ref{chpt:wgm}, which represents the denouement of 
the thesis. With the experiment and analysis concluded, the future impacts of 
the research are discussed in Chapter~\ref{chpt:fdi}. 
The final evaluation of the work presented herein, and a summary of the main 
contribution of this thesis, forms Chapter~\ref{chpt:con}. After briefly 
exploring the status of this work in the broader context of 
the vision, I append proofs, checks and ancillary information ill-suited to 
the main body of the work, and supporting publications for the thesis. 

\chapter{\cdb Spherical Sensors}
\label{chpt:sph}

\section{\cdb Microsphere sensor technology}
\label{sec:micsens}

Modern sensing technologies form a vital part of photonics research 
applications. Microspheres in particular have attracted 
much interest in different fields
of research, such as remote atmospheric sensing \cite{PhysRevA.13.396,Moroz05}, 
 photonic band-gap devices \cite{0022-3727-39-24-006,Imakita:13},
fluorescence spectroscopy \cite{chance:78,Chew:88,Kolmakov10}, nonlinear 
optics \cite{Li:13,Farnesi:15,Riesen:15,LPOR:LPOR201000036,
PhysRevLett.112.093901,Liang:15}, superscattering 
\cite{PhysRevLett.105.013901}, and metamaterial absorbers \cite{He:13,Qu:16}. 
There are also numerous theoretical works on light scattering from microspheres 
\cite{PhysRevA.13.396,Chew:88,Chew:87a,Xu:03,Xu:04,Liang:04,Moroz05,Teraoka:06,
Teraoka:06a,Teraoka:07}, 
commonly based on the fundamental principles of \emph{Mie scattering} 
\cite{BohrenHuffman}, which even the most sophisticated 
models must be able to reproduce. 

It is of special interest in this thesis to explore aspects of 
microspheres that can be adapted for biosensing applications 
\cite{Vollmer:02a,Arnold:03,Quan:2005a,Fan:2009a,
Saetchnikov2010714,Huang2011,Vollmer:2012a,Francois:13,Francois:15b}. 
One should note that the microsphere is 
the simplest three-dimensional test case for a more extensive biological cell 
analogue, the development of which will occupy Chapters~\ref{chpt:bub} and 
\ref{chpt:mod} of this thesis. It is therefore worthwhile to understand the 
key aspects of microsphere sensors, their WGM behaviour, the models that 
describe them, and their expected similarities to and differences from 
biological cell resonators. The development of a customisable toolkit, based 
on the Finite-Difference Time-Domain (FDTD) method, represents 
an important part of this exploratory study. 
FDTD is particularly suitable in this context, because of its ability 
to incorporate a range of deformations and novel structures, while 
at the same time achieving a slice-by-slice record of the behaviour 
of the electromagnetic fields. This wealth of simulated data 
can be analysed, aggregated into 
a power spectrum, defined as the power emitted from the 
resonator over a given wavelength range, or used to access transient or 
emergent optical phenomena, as will be explained in Section~\ref{sec:FDTD}. 

In FDTD, once the geometry of the 
microsphere resonator and flux collection methods have been 
established, the WGM spectrum is introduced, and a range of useful 
properties is analysed in a variety of scenarios. 
The results of the simulations are connected to the well established physical 
phenomenon of Mie scattering, and finally, the distribution of the emitted 
radiation in the vicinity of a microsphere is examined. 

Recent technological advances have led to new
possibilities for using microspheres with several dielectric layers, 
coatings of active materials, or doped with fluorescence nanoparticles, as 
described in Chapter~\ref{chpt:intro}. As a result, there is now a necessity 
to develop more advanced models, as well as efficient numerical simulation 
tools, for describing these resonators and their excitation schemes accurately. 
Currently in the literature, a number of models exist that 
individually describe one of the following: 
the wavelength values of microsphere 
resonances \cite{Johnson:93,Teraoka:06a}, the emitted spectra anticipated from 
embedded dipole sources \cite{Chew:76,Chew:87a} or 
active layer coatings \cite{Chew:88}, and also single \cite{Teraoka:06,
Teraoka:07} and multiple layer \cite{Chew:76,Xu:03,Xu:04,Liang:04,Moroz05} 
microsphere scenarios that do not include the ability to calculate 
WGM spectra. 
These models are not all written using the same notation, making the 
combined use in their present forms impractical. 
Some computational methods 
are numerically unstable in certain parameter regions, such as resonators with 
very thin layers, introducing large artefacts 
that adversely affect the simulated spectrum \cite{Yang:03}.
Some approaches require a 
multitude of separate equations \cite{Moroz05}, thus 
making a unified analysis that 
incorporates all previously developed features an attractive prospect. 
This unified model is developed in Chapter~\ref{chpt:mod}, and 
represents a chief accomplishment of this thesis. 

The simulation techniques in the aforementioned works represent an important 
class of tools for the interpretation of WGM spectra obtained from optical 
resonators. They may be used to identify the mode orders in a given WGM
spectrum \cite{Francois:13,Preston:15}, and facilitate resonator design 
\cite{Hall:15,Reynolds:15,doi:10.1117/12.2078526}. 
Furthermore, the ability to calculate the underlying geometric parameters of a 
given resonator based solely on its spectrum \cite{Hall:17} 
makes the development of robust models for resonators of high importance.

\section{\cdb Analytic models for spherical resonators}
\label{sec:an}

In this section, the literature on the mathematical modelling of 
microsphere resonators is summarised, as a prelude to development 
of a new, unified model presented in Chapter~\ref{chpt:mod}, 
which allows the features of these previous models to be combined 
using a consistent notation. A comparison of the different methods 
in the literature with the fully developed model of Chapter~\ref{chpt:mod} 
can be found in Appendix~\ref{chpt:app3}.

To begin, I consider that analytic models describing WGMs in the literature 
can be split into two categories: models that provide only the positions of 
the TE and TM modes \cite{Johnson:93,Teraoka:06a}, which are 
unable to predict the 
profile of the output spectrum, and models that consider the full 
behaviour of the electric and magnetic fields both inside and outside 
the resonator \cite{PhysRevA.13.396,chance1978molecular,Gersten:80,Gersten:81,
Chew:87a,Chew:88,Ruppin:82,Schmidt:12,pmid24921827}. 
Although the simpler analytic models used in the literature are comparatively 
efficient to calculate, the TE and TM modes obtained   
through geometrical arguments \cite{Johnson:93,Teraoka:06a} do not include the 
ability to generate spectra, nor do they take into account radiation losses.

Both these two types of analytic model rely on the assumptions 
that a microsphere resonator is perfectly spherical with no surface roughness, 
that it consists of a medium that is homogeneous in refractive index, 
 and that the total radiated power is collected without directional bias. 
The formulae for the resonance positions and spectra must therefore be 
rederived separately for scenarios that consider different 
resonator shapes or inhomogeneities, and this can become cumbersome.

\subsection{\cdb Johnson model for WGM mode positions in microspheres}

The simplest analytic model that provides the positions of the TE and TM modes 
is particularly useful for identifying the mode numbers of the WGM peaks 
that occur in the spectrum of a dielectric microsphere 
resonator \cite{Johnson:93,Teraoka:06a}. For brevity, this 
model will be denoted the Johnson model \cite{Johnson:93}. 
The resonance positions identified by this 
model exactly match the results of Mie scattering theory 
 \cite{BohrenHuffman}, and this will provide a benchmark of comparison that 
will be used in Section~\ref{sec:mie} for the investigation of alternative 
excitation strategies. 

According to the Johnson model \cite{Johnson:93,Teraoka:06a}, 
the derivation of the WGM 
resonance condition for a dielectric sphere is now shown. 
Starting from a general construction, 
a continuous electric field, oscillating with frequency $\omega$, takes the 
form $\mathbf{E}(\mathbf{r}) = 
\mathbf{E}_0(\mathbf{r})\,\mathrm{exp}({\mathrm{i}\,\omega\,t})$, 
for a vector position 
$\mathbf{r}$ with respect to spherical coordinates, with the origin 
placed at the centre of the microsphere. This satisfies the 
Helmholtz wave equation
\begin{equation}
\label{eq:w}
\bm{\nabla}\times\bm{\nabla}
\times\mathbf{E}_0(\mathbf{r}) 
- k^2 \mathrm{n}^2(\mathbf{r}) 
\mathbf{E}_0(\mathbf{r}) = 0,  
\end{equation}
for a
wave-number $k\equiv\omega/c$, where $c$ is the speed of light in a vacuum. 
The refractive index function $\mathrm{n}(\mathbf{r})$ 
changes value across the boundary of the resonator 
and its surrounding medium, and in the simplest case of a microsphere 
of radius $R$ constructed from a dielectric medium,  
without the inclusion of absorption loss, takes the form
\begin{align}
\mathrm{n}(\mathbf{r}) &= \mathrm{n}_1, \quad r\equiv |\mathbf{r}| \leq R,\\
\mathrm{n}(\mathbf{r}) &= \mathrm{n}_2, \quad r > R.
\end{align}

The TE mode positions can be extracted from this equation using separation of 
variables in spherical polar coordinates 
\begin{align}
\label{eq:S}
\mathbf{E}_0 &= 
\mathrm{S}_m(r) \frac{\mathrm{exp}({\mathrm{i}\,m\,\phi})}{k\,r}\,
\mathbf{X}_{lm}(\theta),\\
\mathbf{X}_{lm}(\theta) &=\frac{\mathrm{i}\,m}{\sin\theta}
\mathrm{P}_l^m(\cos\theta)\hat{\mathbf{e}}_\theta - 
\frac{\partial}{\partial\theta}\mathrm{P}_l^m(\cos\theta)\hat{\mathbf{e}}_\phi, 
\end{align}
where the angular vector functions $\mathbf{X}_{lm}(\theta)$ are 
defined in terms of the associated Legendre polynomials $\mathrm{P}_l^m$.
These functions $\mathbf{X}_{lm}(\theta)$ \cite{Liang:04,Jackson89} 
occur in all models derived from the wave equation of Eq.~(\ref{eq:w}), 
and a consistent notation must be established. A summary of the 
vector spherical harmonics (VSH), harmonising the notations in the literature, 
can be found in Appendix~\ref{chpt:app1}. 
In the formalism of the Johnson model, the VSH contain 
contributions in the directions $\hat{\mathbf{e}}_\theta$ and 
$\hat{\mathbf{e}}_\phi$. 
The function $\mathrm{S}_m(r)$ in Eq.~(\ref{eq:S}) 
satisfies the second order differential equation
\begin{equation}
\label{eq:de}
\frac{\mathrm{d}\mathrm{S}_m(r)}{\mathrm{d} r^2} + \Big(k^2\,\mathrm{n}^2(r) - 
\frac{m(m+1)}{r^2}\Big)\mathrm{S}_m(r)=0. 
\end{equation}
From the differential equation in Eq.~(\ref{eq:de}), 
the resonance condition for the TE modes is obtained 
directly from an application of the continuity condition across the boundary,  
in the transverse component of the electric field
\begin{align}
\mathbf{E}^{\text{trans}}(\mathrm{n}_1kr)  & =\mathbf{E}^{\text{trans}}
(\mathrm{n}_2kr), \\
\text{or}\quad \hat{\mathbf{r}}\times\mathbf{E}_1 &= 
\hat{\mathbf{r}}\times\mathbf{E}_2, 
\label{eq:Econt}
\end{align}
evaluated at $r=R$, 
with a similar continuity equation for the magnetic field $\mathbf{H}$ 
applying for the TM modes. 

The radial function $\mathrm{S}_m(r)$ takes a Riccati-Bessel form, 
which can be expressed in terms of the spherical Bessel functions 
of the first kind, $j_m$, for the incoming waves, 
and the spherical Hankel function, $h_m^{(1)}$, for the outgoing waves
\begin{equation}
\mathrm{S}_m(r) = \Big\{
\begin{array}{ll} 
 z^r_1 \,\,j_m(z^r_1) \qquad & r< R \\
 A_m \, z^r_2 \,\,h_m^{(1)}(z^r_2) 
\qquad & r> R.
\end{array}
\end{equation}
The use of the outgoing spherical Hankel function, $h_m^{(1)}(z)$ 
 takes into account leaky WGMs, which radiate energy outwards, and are 
described by normalisable quasi-modes \cite{PhysRevA.41.5187}. 
Here, the arguments $z^r_{1,2} \equiv  \mathrm{n}_{1,2} \,k\, r$ are 
defined as a useful shorthand called the \emph{size parameters}, and the 
coefficients $A_m$ can be solved for using the continuity condition above. 
The resonance condition for the TE modes of the sphere thus takes the 
following form, for a radial coordinate equal to the boundary radius 
of the sphere, $r=R$ \cite{Teraoka:06a} 
\begin{equation}
\label{eq:TE}
 \frac{\mathrm{n}_1}{z^R_1} \frac{(m+1) j_m(z^R_1) - z^R_1\, j_{m+1}(z^R_1)}
{j_m(z^R_1)} = 
\frac{\mathrm{n}_2}{z^R_2} \frac{(m+1) h^{(1)}_m(z^R_2) -  z^R_2\, 
h^{(1)}_{m+1}(z^R_2)}
{h^{(1)}_m(z^R_2)}.
\end{equation}

For the TM modes, take note that 
the form of $\mathbf{E}_0$ contains both angular and radial 
vector components (for $\hat{\mathbf{e}}_r$ as the unit vector in the radial 
direction)
\begin{align}
\mathbf{E}_0 &= \frac{\mathrm{exp}({\mathrm{i} m \phi})}
{k^2\,\mathrm{n}^2(r)}\left[
\frac{1}{r}\frac{\partial \mathrm{T}_m(r) }{\partial r}
\mathbf{Y}_{lm}(\theta) + \frac{1}{r^2}\mathrm{T}_m(r)\mathbf{Z}_{lm}(\theta)
\right], \\
\mathbf{Y}_{lm}(\theta) &\equiv \hat{\mathbf{e}}_r\times\mathbf{X}_{lm}(\theta),\\
\mathbf{Z}_{lm}(\theta) &\equiv l(l+1)\mathrm{P}_l^m(\cos\theta)\hat{\mathbf{e}}_r, 
\end{align}
where the function $\mathrm{T}_m(r)$ obeys the equation
\begin{equation}
\label{eqn:Teqn}
\frac{\mathrm{d}\mathrm{T}_m(r)}{\mathrm{d} r^2}
-\frac{2}{\mathrm{n}(r)}\frac{\mathrm{d} \mathrm{n}(r)}{\mathrm{d}r}
\frac{\mathrm{d}\mathrm{T}_m(r)}{\mathrm{d} r} 
+ \Big(k^2\,\mathrm{n}^2(r) - \frac{m(m+1)}{r^2}\Big)\mathrm{T}_m(r)=0. 
\end{equation}
The solutions to Eq.~(\ref{eqn:Teqn}) for $\mathrm{T}_m(r)$ take a similar 
form to those of $\mathrm{S}_m(r)$ above
\begin{equation}
\mathrm{T}_m(r) = \Big\{
\begin{array}{ll} 
 z^r_1 \,\,j_m(z^r_1) \qquad & r< R \\
 B_m \, z^r_2 \,\,h_m^{(1)}(z^r_2) 
\qquad & r> R, 
\end{array}
\end{equation}
where the coefficients $B_m$ may be defined through enforcing continuity at the 
boundary in the same way as $A_m$, thus leading to the resonance condition 
\begin{equation}
\label{eq:TM}
 \frac{1}{\mathrm{n}_1 z^R_1} \frac{(m+1) j_m(z^R_1) - z_1\, j_{m+1}(z^R_1)}
{j_m(z^R_1)} = 
\frac{1}{\mathrm{n}_2 z^R_2} \frac{(m+1) h^{(1)}_m(z^R_2) -  z_2\, h^{(1)}_{m+1}
(z^R_2)}
{h^{(1)}_m(z^R_2)}.
\end{equation}
Thus, the resonance conditions of Eqs.~(\ref{eq:TE}) and (\ref{eq:TM}) 
may be used to specify the positions of the modes as a function 
of wavenumber $k$, or wavelength, $\lambda = 2\pi/k$.

\subsection{\cdb Chew model for the power emitted from active microspheres}

A more general type of analytic model, such as that 
developed by Chance, Prock and Silbey using dyadic 
Green's functions 
\cite{chance1978molecular} and separately by Chew \textit{et al.} 
\cite{PhysRevA.13.396,Chew:87a,Chew:88}, 
is able to simulate the full electric and magnetic field behaviour both 
inside and outside the resonator. 
These models deal particularly with the case of a dipole source placed on the 
surface of the resonator. Both of these models were shown to be equivalent 
to each other as formulations of spherical resonators excited by a dipole 
source \cite{Chew:87a}, and were also shown to obtain the same resonance 
positions as Mie scattering \cite{BohrenHuffman}. The resonance conditions 
derived from the Johnson model can thus serve as an excellent benchmark 
comparison for 
spectra generated from the variety of models developed throughout this work, 
in order to check that each model correctly reproduces the established 
Mie coefficients for known resonance positions of a given radial 
and azimuthal mode order. This checking process will be carried out 
explicitly in Section~\ref{sec:mie}. 

A thorough discussion of the well known Chew model of $1976$ \cite{Chew:76},  
and its relation to other models, including the unified 
multilayer model developed in Chapter~\ref{chpt:mod}, 
is left until Appendix~\ref{chpt:app3}. 
The Chew model is derived from the expansion of the vector electric 
and magnetic fields of a radiating dipole in terms of the VSH 
\cite{PhysRevA.13.396,
chance1978molecular,Gersten:80,Gersten:81,Chew:87a,Chew:88,Ruppin:82,
Schmidt:12,pmid24921827}
\begin{align}
\mathbf{E}_d &= \sum_{n,m}\big(\frac{\mathrm{i}c}{\mathrm{n}_2 \omega} 
a^d_E(n,m) \bm{\nabla}\times [h^{(1)}_n(k_nr)\mathbf{Y}_{nnm}(\hat{\mathbf{r}})]
+a^d_M(n,m)h^{(1)}_n(k_2r)\mathbf{Y}_{nnm}(\hat{\mathbf{r}})\big),\\
\mathbf{B}_d &= \sum_{n,m}\big(
a^d_E(n,m) h^{(1)}_n(k_nr)\mathbf{Y}_{nnm}(\hat{\mathbf{r}})
-\frac{\mathrm{i}c}{\omega} a^d_M(n,m)\bm{\nabla}\times 
[h^{(1)}_n(k_2r)\mathbf{Y}_{nnm}(\hat{\mathbf{r}})]\big),
\end{align}
where the convention used for the VSH, $\mathbf{Y}_{nnm}$, 
defined in Appendix~\ref{chpt:app1}, 
will be reformed when the unified model is presented, in Section~\ref{sec:mod}. 
The quantities 
$a^d_E$ and $a^d_M$ are the electric and magnetic dipole coefficients, 
respectively, defined in terms of the vector dipole moment $\mathbf{P}$, 
the permittivity ($\epsilon$) and 
permeability ($\mu$) of the surrounding medium, and specified 
at the radial position of the dipole $r'$, 
with the gradient operator $\bm{\nabla}'$ acting on this coordinate
\begin{align}
a^d_E(n,m) &= 4\pi k_2 \sqrt{\frac{\mu_2}{\epsilon_2}} \mathbf{P}\cdot
\{\bm{\nabla}'\times[j_n(k_2 r')\mathbf{Y}^{*}_{nnm}(\hat{\mathbf{r}})]\}, \\
a^d_M(n,m) &= 4\pi\mathrm{i} \frac{k_2^3}{\epsilon_2}j_n(k_2 r')\mathbf{P}
\cdot\mathbf{Y}^{*}_{nnm}(\hat{\mathbf{r}}). 
\end{align}
For $r < r'$, the spherical Bessel functions $j_n$ must be interchanged with 
$h_n^{(1)}$. 

In the Chew model, the scattered electric field in the outer region takes a 
similar form to that obtained from a radiating dipole
\begin{equation}
\mathbf{E}_2 = \sum_{n,m}\big(\frac{\mathrm{i}c}{\mathrm{n}_2 \omega} 
c_E(n,m) \bm{\nabla}\times [h^{(1)}_n(k_nr)\mathbf{Y}_{nnm}(\hat{\mathbf{r}})]
+c_M(n,m)h^{(1)}_n(k_2r)\mathbf{Y}_{nnm}(\hat{\mathbf{r}})\big),
\end{equation}
where the coefficients $c_E$ and $c_M$ are related to the Mie scattering 
coefficients of Ref.~\cite{BohrenHuffman}, and may be expressed in terms 
of the dipole coefficients $a^d_E$ and $a^d_M$, making use of the 
continuity condition of Eq.~(\ref{eq:Econt}) and its magnetic field 
variant \cite{Chew:87a}. 

The total emitted power can be obtained by integrating the Poynting 
vector $\mathbf{S}$ with respect to the total solid angle $\Omega$
\begin{equation}
P = r^2\int\!\mathbf{S}\cdot\hat{\mathbf{r}}\,\mathrm{d}\Omega 
= \frac{c}{8\pi}\frac{\epsilon_2}{\mu_2}r^2\int\!|\mathbf{E}_d + \mathbf{E}_s|^2
\mathrm{d}\Omega. 
\end{equation}
In the specific case of a dipole source oriented in the radial (normal) or 
tangential (transverse) direction placed on the surface of the sphere, 
the components of the emitted power take the following form when
 normalised to the power of a dipole in 
the surrounding medium of index $\mathrm{n}_2$ \cite{Chew:87a}
\begin{align}
\label{eq:chewex}
P_{\bot}/P_{\bot}^0 &= \frac{3 \epsilon^{3/2}_1 \mathrm{n}_1}{2\, {z^R_1}^2}\left(
\frac{\epsilon_2}{\mu_2}\right)^{1/2}\sum_{m=1}^\infty m(m+1)(2m+1)
\frac{j_m\,^2(z^R_1)}{{z^R_1}^2\,|D_m|^2}, \\
\label{eq:chewey}
P_{\Vert}/P_{\Vert}^0 &= \frac{3 \epsilon^{3/2}_1 \mathrm{n}_1}{4\, {z^R_1}^2}\left(
\frac{\epsilon_2}{\mu_2}\right)^{1/2}\sum_{m=1}^\infty(2m+1)\left[
\Big|\frac{[z^R_1\,\,j_m(z^R_1)]'}{z^R_1\,D_m}\Big|^2 
+ \frac{\mu_1 \mu_2}{\epsilon_1 \epsilon_2}\frac{j_m\,^2(z^R_1)}
{|\tilde{D}_m|^2}\right], \\
%\end{align}
%
%\begin{align}
\mbox{for}\quad D_m &= \epsilon_1 j_m(z^R_1)[z^R_2\,h^{(1)}_m(z^R_2)]' - 
\epsilon_2 h^{(1)}_m(z^R_2)[z^R_1\,j_m(z^R_1)]',\\
\tilde{D}_m &= D_m(\epsilon_{1,2}\rightarrow\mu_{1,2}). 
\end{align}
Note that the radial oscillations contain a single term, which encodes 
the contributions from the TM modes only, as a consequence of spherical 
symmetry, which will be discussed more thoroughly in Chapter~\ref{chpt:mod}. 
In the case of the tangential oscillations, contributions from both 
TE and TM modes appear, as shown in Eq.~(\ref{eq:chewey}). 
This observation will become important for the interpretation of the 
results obtained from FDTD simulations, described in Section~\ref{sec:spec}. 

A range of models has been reported in the literature 
for the incorporation of multiple dielectric layers. These include a 
version of the Chew model for $N$ layers, 
in which the continuity equations for the 
quantities $c_E$ and $c_M$ for both TE and TM modes are expanded 
to a large matrix equation to calculate all $4N$ unknown parameters
\cite{Chew:76}. The calculation of the resonance conditions 
for a microsphere with a single layer coating has also 
been solved explicitly by Teraoka and Arnold \cite{Teraoka:06,Teraoka:07}, 
presented in detail in Appendix~\ref{sec:TA}. 
In addition, there are several publications that solve the continuity 
condition of an $N$-layer microsphere system using a method known 
as the \emph{transfer matrix approach}. These include 
the study of Bragg `onion' resonators by Liang, Xu, Huang and Yariv
\cite{Xu:03,Xu:04,Liang:04} and their 
scattering cross sections in the far field region, and the WGM spectrum 
for dipoles placed on the surface, as in the Chew model, by Moroz 
\cite{Moroz05}. Note that while the development of a model that incorporates 
the ability to obtain the WGM spectrum, as well as to simulate the 
inclusion of one or more dipole sources, or layers of dipole sources, 
will be left until Chapter~\ref{chpt:mod}, 
it will become necessary, as described in Section~\ref{sec:micsens}, 
to unify the disparate conventions and frameworks so that the behaviour 
of the WGMs in fluorescent, multilayer resonators can be simulated and 
compared to experiment.

\subsection{\cdb Quality factor and loss}
\label{subsect:gq}

One particular performance characteristic of a resonator 
is the quality factor, or $Q$-factor, which expresses the ability of a 
resonator to store 
energy. A resonator with a high $Q$-factor is able to contain the 
electromagnetic fields that circulate around the surface for longer periods of 
time compared to resonators with a low $Q$-factor. For each round trip taken 
by the field, an amount of energy \emph{leaks} from the resonator, and 
the time taken for a sufficient energy loss to transpire is known as the 
cavity \emph{ring-down} time. This relationship between $Q$-factor and 
ring-down time is a consequence of the definition of the $Q$-factor 
in terms of the exponential decay constant, as derived from signal analysis 
\cite{ResSens}, and will be described below. 

The resultant light scattered from the resonator 
during the process of leaking radiation 
can be represented by the value of the emitted power over 
a range of 
wavelengths or frequencies, known as the spectrum. The usefulness of the 
spectrum in analysing the behaviour of WGMs is introduced in 
Section~\ref{sec:spec}, and will comprise a significant portion of this thesis. 
 
One particular contribution to the $Q$-factor, the \emph{geometric} 
portion, $Q_g$, is the easiest to 
estimate theoretically, and is often 
considered to be a reliable indicator of the performance of a resonator 
\cite{Little:99}. The total $Q$-factor, as measured in experiment, includes 
a range of other contributions to the loss of energy, such as material 
absorption, surface scattering, topological deformations such as asphericity, 
and the radiation loss, derived from the coupling condition to the underlying 
geometric modes of the resonator \cite{doi:10.1117/12.349244,Reynolds:15}. 
These contributions must also be considered, and are discussed later 
in this section. 

Consider $Q_g$, which has its origin in the signal analysis 
of inductor-resistor-capacitor (LRC) circuits, where a resonant frequency 
$\omega$ of an oscillating wave, $E(t)$, 
can be expanded to first order \cite{Prkna2004}
\begin{align}
\label{eq:Eteq}
E(t) &= E_0 \mathrm{exp}(-{\mathrm{i}\,\omega\,t}) 
= E_0 \mathrm{exp}(-t/\tau), \\
\omega &\approx \omega_0 \Big(1 - \frac{\mathrm{i}}{2 Q_g}\Big), 
\label{eq:omQ}
\end{align}
where $E_0$ is the wave amplitude and $\tau$ is the exponential time constant. 
Substituting Eq.~(\ref{eq:omQ}) back into Eq.~(\ref{eq:Eteq}), 
the distribution of energy in the cavity is 
proportional to the Lorentzian function
\begin{equation}
|E(\omega)|^2 \propto \frac{1}{(\omega-\omega_0)^2 + (\omega_0/2Q_g)^2}. 
\end{equation}
Essentially, $Q_g$ describes the imaginary part of the eigenfrequency 
$\omega_0$, 
which is related to the time constant $\tau$ via \cite{doi:10.1063/1.4920922}
\begin{equation}
Q_g = \frac{\tau\,\omega_0}{2}.
\end{equation}
This relationship between the $Q$-factor and the decay constant is an important 
point in the definition of the Purcell factor, a quantity 
associated with the Purcell effect introduced in Section~\ref{subsec:act}. 

If the wave is simply propagating along a waveguide, or within a prism 
of uniform refractive index $\mathrm{n}$, as will be used  
in Section~\ref{subsec:pris}, 
the propagation constant takes the 
form for a plane wave $\beta = k_0 \mathrm{n}$ \cite{Prkna2004,Little:99}. 
For a wave travelling along a circular cross section of a 
resonator of radius $R$, the propagation constant can be obtained 
from the mathematical form of the near (evanescent) field, or the bound portion 
of the field in the vicinity of the outer boundary of the resonator. 
Using separation of variables, this field takes the form
\begin{equation}
\Psi_{lmn}(r, \theta, \phi) = N_1 \,\psi_r(r)\psi_\theta(\theta)\psi_\phi(\phi),
\end{equation}
with component-wise contributions as follows \cite{Little:99}
\begin{align}
\psi_r(r) &= \begin{cases}{j_l(z_1),\qquad r\leq R}\\
{j_l(z_1)\mathrm{exp}(-\alpha_1(r-R)),\quad r> R},\end{cases}\\
\psi_\theta(\theta) &= \mathrm{exp}(-m\theta^2/2)H_N(\sqrt{m}\theta), 
\quad m \gg 1 \gg \theta, \\
\psi_\phi(\phi) &= \mathrm{exp}(\pm j m \phi). 
\end{align}
The coefficients in these formulae are defined as
\begin{align}
N_1 &= \Big\{\sqrt{\frac{\pi}{m}} 2^{l-m-1}(l-m)!\,R^2\Big[\Big(
1+\frac{1}{\alpha_1 R}\Big)j_l^2(z_1^R) - j_{l-1}(z_1^R)j_{l+1}(z_1^R)\Big]
\Big\}^{-1/2},\\
\alpha_1 &= \sqrt{\beta_l^2 - k^2 \mathrm{n}_2^2},\qquad 
\beta_l = \frac{\sqrt{l(l+1)}}{R}.
\end{align}
Note that the radial component, $\psi(r)$, takes the form 
of a spherical Bessel function within the resonator. However, 
immediately outside the resonator the fields decay exponentially 
away from the boundary in the radial direction, with a decay constant 
of $\alpha_1$. The propagation constant $\beta_l$, projected 
onto the equator of the sphere such that $l=m$, takes the form
\begin{equation}
\beta_m = \frac{m}{R}. 
\end{equation}
As a result, the phase-matching condition for microsphere resonators, 
introduced 
in Section~\ref{subsec:passr}, can be specified explicitly, by setting 
the propagation constants of a plane wave within a coupler to that of 
the equatorial modes
\begin{equation}
\label{eq:kwgm}
\beta_{\mathrm{WGM}}= \frac{m}{R} = k_0\mathrm{n}_1. 
\end{equation}
Thus, noting that $k_0 = m/(\mathrm{n}_1 R)$, and recalling 
that the complex wavenumber is related to the eigenfrequency, 
$k_0\equiv\omega_0/c$, as stated below Eq.~(\ref{eq:w}), $Q_g$ takes 
the form
\begin{equation}
\label{eqn:qg}
Q_g = \frac{\mathrm{Re}\,k_0}{-2\, \mathrm{Im}\, k_0}, 
\end{equation}
for the wave number $k_0 = 2\pi/\lambda_0$. 

The scaling behaviour of this $Q$-factor for large resonators of 
refractive index $\mathrm{n}_1$ can be estimated using a 
model derived from an asymptotic analysis of the resonance 
conditions of Eqs.~(\ref{eq:TE}) and (\ref{eq:TM}), as described 
in Refs~\cite{Datsyuk1992,Lam:92}
\begin{align}
Q_a &= \frac{1}{2}\nu\, \mathrm{n}_1^{1-2 b} \sqrt{\mathrm{n}_1^2 - 1} \,
\mathrm{exp}({2 \mathscr{T}_l}),\\
\mathscr{T}_l &\equiv \nu\, (\eta_l - \mathrm{tanh}\, \eta_l),\\
\eta_l &\equiv \mathrm{arccosh}\,\left\{\mathrm{n}_1 \Bigg[1 - \frac{1}{\nu}
\Bigg(t_p^0 \, (\nu/2)^{1/3} + \frac{\mathrm{n}_1^{1-2b}}
{\sqrt{\mathrm{n}_1^2-1}}\Bigg)\Bigg]^{-1}\right\}. 
\end{align}
Here, $\nu\equiv l+1/2$, and the factor $b$ is equal to $1$ for the TM modes, 
and equal to $0$ for the TE modes. The exterior refractive index 
is assumed to be $1$ in this example. 
In this asymptotic form, one may choose the order, $p$, to which to calculate 
the ordered zeroes of the Airy function, $t_p^0$, 
which appears in the asymptotic expansion of the characteristic equation. 
The formula above is only valid for a large azimuthal mode number ($l$) 
approximation, and assumes that only the \emph{fundamental} mode 
is observed, for which the quantum numbers 
introduced in Section~\ref{sec:nut} and Fig.~\ref{fig:coords} 
take the values $l=m$, and $n=0$. 
The value taken by $l$ in this case can be related simply to the number 
of round trips taken by the resonating waves, as shown in 
Fig.~\ref{fig:wgmex}(a), which depends upon the radius of the microsphere, 
$R$, and takes the form
\begin{equation}
\label{eq:l}
l_{\mathrm{fund}} = \frac{2 \pi \,\mathrm{n}_1\,R}{\lambda_0}. 
\end{equation}
This approach is not generally suitable 
for predicting the absolute magnitude of the measured $Q$-factor, especially 
for small resonators capable of small azimuthal mode numbers at their 
resonance condition. Of course, to gain a more realistic value of the total 
$Q$-factor, one must estimate the effect of other sources of loss. 

While the geometric $Q$-factor expresses one portion 
of the loss, there are other sources of loss that need to be accounted 
for before one can attempt an accurate comparison with the $Q$-factors 
measured in experiment \cite{doi:10.1117/12.349244,Reynolds:15}. 
These sources include material absorption, scattering from surface 
roughness, and radiation loss, that is, loss incurred from selecting 
a particular mode excitation strategy, with a given coupling efficiency. 
The losses can be modelled, as in the case of an LRC series circuit, where the total 
$Q$-factor is inversely proportional to the loss over a resistance $R_C$, $Q\propto 1/R_C$ 
\cite{di2000networks}. 
In the context of optical resonators, the total loss derived from the imaginary part 
of the eigenfrequency $\omega_0$, from Eq.~(\ref{eq:omQ}), can thus be added in parallel
\begin{equation}
\label{eqn:Q}
\frac{1}{Q} = \frac{1}{Q_m} + \frac{1}{Q_s} + \frac{1}{Q_c}.  
\end{equation}
Here, the quantity $Q_m$ is the $Q$-factor derived from 
material loss such as absorption, 
$Q_s$ is due to scattering from surface roughness, and $Q_c$ is the cavity 
tunnelling loss, which can derived from one of several analytic models 
described in Section~\ref{sec:an}. 
The cavity tunnelling loss itself is composed of the geometric loss, $Q_g$, 
and the radiation loss, which depends intimately on the method of excitation 
\cite{pmid24921827}. 
Note that experimental measurement of the geometric portion of the $Q$-factor 
cannot be conducted directly, as the measurement 
can only take place in the presence of radiation through which the modes are 
excited. Therefore, analysis of the spectrum represents 
a more realistic assessment of the $Q$-factor, as the selection of a specific 
excitation strategy is implicit in the calculation of any spectrum. 
In the case of microspheres, it is also possible to 
separate the asphericity component from the cavity loss in the same way
 \cite{Riesen:15a}, by introducing an extra term into Eq.~(\ref{eqn:Q}). 

The cavity $Q$-factor, $Q_c$, can be obtained directly from the spectra, for a 
peak centred on a wavelength $\lambda_0$
\begin{equation}
\label{eqn:qc}
Q_c = \frac{\lambda_0}{\delta \lambda}. 
\end{equation}
WGM resonators can exhibit high performance in terms of their $Q$-factors, 
which in some cases exceed $10^{11}$ \cite{Gorodetskii2015} experimentally. 
In most cases, however, the limiting factor on $Q$ is the material absorption 
or the surface scattering. 
The limit of the $Q$-factor due to material losses can be determined by 
considering the absorption of light by both the resonator 
and the surrounding medium using an absorption decay constant $\alpha_m$
\begin{equation}
Q_m = \frac{2\pi \mathrm{n}_1}{\lambda_0 \alpha_m}. 
\end{equation}
The magnitude of the absorption decay constant depends on how tightly the 
light is confined within the resonator, the wavelength range, 
and the surrounding refractive index.
Scattering contributions can be calculated by modelling the surface roughness 
as a changing dielectric constant, and this has been used to 
determine upper limits on the $Q$-factors of small silica spheres. 
One such expression for the limit of the $Q$-factor 
due to surface scattering can be derived by considering the scattered 
power $P_s$ in the case of an inhomogeneity in the dielectric constant, 
$\delta\epsilon = (\epsilon_0 - 1)f(x,y)\delta(z)$, for a 
Dirac delta function $\delta(z)$ \cite{Gorodetsky:00}
\begin{equation}
Q_s \approx \frac{3 \lambda_0^3 R}{8 \mathrm{n}_1 \pi^2 \mathrm{B}^2 \sigma^2}, 
\end{equation}
where $\mathrm{B}$ is the correlation length and 
$\sigma = (\langle f(x,y)^2\rangle)^{1/2}$ 
is the variance of 
the surface roughness \cite{Ilchenko:13}. 
Each of these quantities can be determined by measuring the extent to which 
a fluctuating beam exhibits coherence upon being transmitted through the 
surface \cite{1347-4065-12-11-1693}. 
Using high-quality polishing techniques, surface roughnesses of $20$ nm can be 
achieved for some resonators. According to one study \cite{Ilchenko:13}, 
assuming that $\sigma$ and $B$ are roughly equal in magnitude at $20$ nm, 
the scattering limitation of the $Q$-factor can be estimated as 
$Q_s = 7\times 10^8$ for a diamond resonator with a radius of $1$ mm, at a 
wavelength of $1.55$ $\mu$m. For a polystyrene microsphere of radius 
$15$ $\mu$m in water, at a wavelength of $770$ nm, a 
limiting factor is reported closer to $Q_s = 4\times 10^6$ -- significantly 
less than the geometric $Q$-factor expected from Mie scattering theory, 
$Q_g = 1.5\times 10^8$
\cite{:/content/aip/journal/apl/91/14/10.1063/1.2795332}. 
It is important to note that it is only when the roughness approaches the 
order of $50$ nanometres (nm), that $Q_s$ becomes a limiting factor for the 
overall $Q$-factor, as estimated for the specific case of polystyrene 
microspheres with radii of approximately $10$ $\mu$m 
\cite{:/content/aip/journal/apl/91/14/10.1063/1.2795332,doi:10.1021/la047584t,
Yakubov2001}. 
While the biological cells considered as viable resonator candidates in 
Chapter~\ref{chpt:cel} are typically much larger in diameter than $10$ $\mu$m, 
it will be discovered that they can exhibit significant surface roughness, 
and thus one must keep in mind these circumstances in which scattering loss can 
dominate. 

In the case of microspheres, it is sometimes convenient to define a separate 
$Q$-factor corresponding to the asphericity, 
which can also limit the value of the total $Q$-factor. This is useful when 
considering active microspheres \cite{Riesen:15a}, 
particularly the variety that are coated with a fluorescent dye, which tend 
to emit without a direction preference. Such active resonators are thus 
susceptible to degradation of the $Q$-factor, even for relatively 
small deviations in their sphericity -- down to the nanometre level 
\cite{Reynolds:15}. 
This is a consequence of the presence of multiple microsphere diameters 
experienced by rays propagating inside the resonator. 
For example, an elliptical resonator is capable of sustaining modes 
along both its minor and major axes, as well as skew rays, which propagate 
along a spiral path, and thus experience a different effective diameter 
for each round trip. The relatively 
uniform emission of radiation from active resonators thus includes 
contributions from multiple paths, leading to mode-splitting and 
reduced  $Q$-factors \cite{Reynolds:15,Riesen:15a,LPOR:LPOR201600265}. 

It is well known that the geometric portion of the $Q$-factor can overestimate 
the measured $Q$-factor by many orders of magnitude, as seen by comparing 
analytic models in the absence of radiation \cite{Little:99,Talebi} with 
methods that include the missing radiation losses \cite{Kaplan:13,Shirazi:13,
Yang:14,Hall:15,Reynolds:15}. 
The contribution to the total $Q$-factor from Eq.~(\ref{eqn:qc}) expresses 
both the intrinsic geometric loss and the loss of radiation through the 
excitation strategy that was used to generate the spectrum, effectively 
renormalising the width of the resonance. 
Equation~(\ref{eqn:qc}) describes a general way of extracting the $Q$-factor 
of a cavity from a spectrum, irrespective of the specific excitation method. 
However, in the case of active resonators, which include an embedded emitter 
or fluorescent layer, additional spectral features may be observed, such as 
the Purcell effect. The magnitude of this effect is typically ascertained by 
calculation of a quantity known as the Purcell factor
\cite{PhysRev.69.37, PhysRev.69.681}, which is defined in terms of the 
decay rate of the emitter, $\Gamma^{\mathrm{em}}$. 

It has long been known that the local environment is able to alter the behaviour 
of an emitter \cite{PhysRev.69.681,Lukosz:77,BROKMANN200591,PhysRevLett.96.113002}. 
In the case of a resonant cavity, the presence of a refractive index contrast 
between the resonator boundary and its surrounding medium can modify 
the local density of states available for the emitted energy to be stored, 
affecting both the emission rate and the emitted power 
\cite{PhysRev.69.681,PhysRevA.62.013820,PhysRevA.64.033812,PhysRevB.69.035316}. 
Since the loss of radiation through emission into free space is related to the 
solid angle into which it radiates (which is handled explicitly when 
calculating the emitted power in Appendix~\ref{chpt:app2}), 
it is useful to define the effective mode volume, $V_{\text{eff}}$, which can be 
used to describe the distribution of power of the modes within a cavity. For 
an electric field $\mathbf{E}$, permittivity $\epsilon$ and a radial coordinate 
$\mathbf{r}$, one may define \cite{PhysRev.69.37,762379}
\begin{equation}
V_{\text{eff}} = \frac{\int\! \epsilon(\mathbf{r}) |\mathbf{E}(\mathbf{r})|^2 
\mathrm{d}^3\mathbf{r}}
{\epsilon(\mathbf{r}) |\mathbf{E}(\mathbf{r})|^2_{\text{max}}},
\end{equation}
where the integral is taken over all space. 

If an emitter with a dipole moment $\mathbf{P}$ and transition frequency 
$\omega$ has an 
excitation rate $\Gamma^{\mathrm{ex}}$  near its 
emission frequency, then a single-frequency approximation of the emission 
rate can be calculated in terms of the radiative ($\Gamma^{\mathrm{r}}$) 
and non-radiative ($\Gamma^{\mathrm{nr}}$) decay rates, 
respectively \cite{PhysRevLett.96.113002}
\begin{equation}
\label{eq:Gamma}
\Gamma^{\mathrm{em}} = \Gamma^{\mathrm{ex}}(\Gamma^{\mathrm{r}}/\Gamma) 
= \Gamma^{\mathrm{ex}}(1-\Gamma^{\mathrm{nr}}/\Gamma).
\end{equation}
Since the excitation rate depends on local electric field   
$\mathbf{E}=\mathbf{E}(\mathrm{r},\omega)$ 
such that $\Gamma^{\mathrm{ex}}\propto|\mathbf{P}\cdot\mathbf{E}|^2$, 
and the total spontaneous decay rate can be calculated via 
Fermi's golden rule
\begin{equation}
\Gamma = \frac{2 \omega}{3 \hbar \epsilon_0}
|\mathbf{P}|^2\rho(\mathbf{r},\omega), 
\end{equation}
where $\hbar$ is Planck's constant and $\rho$ is the density of states 
available to the emitter, then the emission behaviour of the system 
can be entirely determined by the dyadic Green's function 
of $\mathbf{E}$, as follows. 
Solutions to the Helmholtz equation of Eq.~(\ref{eq:w}) can be 
described by integrating a Green's function 
$\stackrel{\leftrightarrow}{\mathbf{G}}$, together with  
the electric current $\mathbf{J}$, throughout a volume $V$ 
that entirely encompasses 
the resonator \cite{NARAYANASWAMY20101877}
\begin{equation}
\mathbf{E}(\mathbf{r}) = 
\int_V\!\stackrel{\leftrightarrow}{\mathbf{G}}(\mathbf{r}, 
\mathbf{r}')\cdot \mathbf{J}(\mathbf{r}')\mathrm{d}^3r'. 
\end{equation}
The local density of states, $\rho(\mathbf{r}, \omega)$, 
can then be expressed in terms of this function 
\cite{PhysRevA.64.033812,PhysRevB.68.245405}
\begin{equation}
\rho(\mathbf{r}, \omega) = \frac{6 \omega}{\pi c^2}
\left(\hat{\mathbf{P}}\cdot\stackrel{\leftrightarrow}{\mathbf{G}} 
\cdot\hat{\mathbf{P}}\right), 
\end{equation}
for a unit vector $\hat{\mathbf{P}}$. To take a simple example, in the case 
of an emitter in free space, the density and decay rates simplify 
to $\rho = \omega^2/(\pi^2 c^2)$, and 
$\Gamma = \Gamma^0 = \omega^3|\mathbf{P}|^2/(3 \pi \epsilon_0\hbar c^3)$. 
In the case of a dipole emitter with a high quantum yield, $\Gamma^{\mathrm{r}}/\Gamma$, 
corresponding to an emitted electric field 
$\mathbf{E}_d(\mathbf{r}) = 
(k^2/\epsilon_0)\stackrel{\leftrightarrow}{\mathbf{G}}(\mathbf{r},\mathbf{r}_d)\cdot\mathbf{P}$, 
the non-radiative rate takes the form \cite{chance1978molecular}
\begin{align}
\label{eq:Gnr}
\frac{\Gamma^{\mathrm{nr}}}{\Gamma} &= \frac{1}{P^0}\frac{1}{2}\int\!\mathrm{Re}
\{\mathbf{J}(\mathbf{r})\cdot\mathbf{E}_{d}^*(\mathbf{r})\}\mathrm{d}^3r, \\
P^0 &= \frac{\omega^4|\mathbf{P}|^2}{12\pi\epsilon_0c^3},
\end{align}
in terms of the emitted power of a dipole in free space, $P^0$. 
The normalised excitation rate from an incident radiation 
field $\mathbf{E}_0$ that interacts with the emitter takes the form
\begin{equation}
\label{eq:PP0}
\frac{\Gamma^{\mathrm{ex}}}{\Gamma^0} = 
\left|\frac{\hat{\mathbf{P}}\cdot\mathbf{E}(\mathbf{r})}
{\hat{\mathbf{P}}\cdot\mathbf{E}_0(\mathbf{r})}\right|^2. 
\end{equation}
The Purcell factor for a resonance near a wavelength of $\lambda_0$ 
is then defined as the ratio 
\cite{PhysRev.69.37, PhysRev.69.681}
\begin{equation}
\Gamma^{\mathrm{em}}/\Gamma^0 = \frac{3\,\lambda_0^3\,Q}
{4\pi^2\,\mathrm{n}_1^3\,V_{\text{eff}}}. 
\end{equation}
By substituting the non-radiative 
and excitation decay rates into the formula of Eq.~(\ref{eq:Gamma}), 
it can be seen that there is a relationship between 
the normalised emitted power, $P/P^0$, and the emission ratio \cite{PhysRevA.64.033812}. 
In some cases this can result in an emission of more power within an environment 
than would have been the case in free space, $P/P^0 > 1$. This particular 
effect will be explored further in Section~\ref{sec:spec}.

\subsection{\cdb Free spectral range}

WGM spectra exhibit relatively sharp peaks corresponding to the WGM resonance 
positions, and can be well spaced, particularly for small 
resonators. One can calculate the peak-to-peak wavelength range, or 
\textit{free spectral range} (FSR), and the relative linewidth, 
$\delta\lambda$, in the following manner. 
The FSR of a microsphere resonator of radius $R$ and effective 
refractive index contrast 
of $\mathrm{n}_1$, for two consecutive polar mode numbers $m$ and $m+1$, 
can be approximated from \cite{Bacardit2011}
\begin{align}
m\,\lambda_m &= \mathrm{n}_{1} 2\pi R,\\
(m+1)\,\lambda_{m+1} &= \Big(\mathrm{n}_{1} - 
\frac{\partial\mathrm{n}_1}{\partial\lambda}(\lambda_m - \lambda_{m+1})\Big)2\pi R, 
\end{align}
such that the difference in the modes takes the following formula, 
\begin{equation}
\mathrm{FSR} = \lambda_m - \lambda_{m+1} \approx \frac{\tilde{\lambda}^2}
{2 \pi\,\mathrm{n}_1^g \,R}. 
\label{eqn:FSR} 
\end{equation}
Here, $\tilde{\lambda}^2 \approx \lambda_m.\lambda_{m+1}$, and the group 
refractive index $\mathrm{n}_1^g$ takes the form $\mathrm{n}_1 - 
\frac{\partial\mathrm{n}_1}{\partial\lambda}\lambda_m$. 

Alternatively, the FSR can be estimated from the wavelength 
dependence of the propagation constant 
$\beta$ \cite{Rabus2007}
\begin{equation}
\frac{\partial\beta}{\partial\lambda} = 
-\frac{\beta}{\lambda} + k \frac{\partial\mathrm{n}_1}{\partial\lambda}, 
\end{equation}
such that the formula of Eq.~(\ref{eqn:FSR}) can be reproduced from
\begin{equation}
\mathrm{FSR} = \Delta\lambda = 
-\frac{2\pi}{R}\left(\frac{\partial\beta}{\partial\lambda}\right)^{-1}
\approx = \frac{\tilde{\lambda}^2}
{2 \pi\,\mathrm{n}_1^g \,R}. 
\end{equation}
In Eq.~(\ref{eqn:FSR}), it can be seen that, as a general rule, the FSR 
decreases as the radius increases. This will be an important 
consideration in the analysis of the behaviour of the FSR in the 
remainder of this chapter, as well as in Chapters~\ref{chpt:bub} 
through \ref{chpt:cel}.

\section{\cdb A customisable FDTD-based toolkit}
\label{sec:FDTD}

The FDTD technique simulates the evolution of electromagnetic fields by 
discretising a volume into a three-dimensional spatial lattice 
\cite{taflove1995computational}. Maxwell's equations are then solved for 
\emph{every point} on the lattice, for a finite number of time increments. 
FDTD represents a critical tool in the simulation of novel resonator 
architectures, specifically for its ability to incorporate material 
inhomogeneities and a wide variety of resonator shapes, 
such as shells, ellipsoids or deformations to a microsphere. 

The use of two-dimensional FDTD simulations to describe the resonance 
peak positions has been used previously in the case of microdisks 
\cite{Kuo2009}.  However, the accurate prediction of the $Q$-factor of the 
resonances represents a principal challenge  
due to the significant dependence of the $Q$-factor 
on minute characteristics 
of the resonator, such as the surface roughness, material inhomogeneities and 
microscopic deformations in the shape of the resonator
\cite{Min:2009a} described in the previous section. 
As a result, theoretical $Q$-factors evaluated through analytic 
models can be difficult to match experimentally  
\cite{Min:2009a,Vahala:2004}. 

The key flexibility of the FDTD method, however, is the ability to address 
this issue by incorporating geometric, material and refractive index features 
in a way that is both intuitive and easy to implement. 
In addition, specific modes or wavelength regions 
can be reliably identified, and tailored for specific tasks 
by altering the initial configuration of the resonator. 
Furthermore, the ability to scan over a wide range of parameters 
may lead to valuable design solutions for biosensing 
that would not otherwise have been found.  

The FDTD method can lead to the possibility of accessing transient or emergent 
optical effects due to the fact that an emitter source that covers a wide 
range of wavelength values is used (in this case a Gaussian source), and each 
time step is evaluated separately, allowing scope for tracking the intermediate 
configurations of the fields both inside and outside the resonator. 
However, FDTD is computationally expensive. For example, time steps 
totalling a few hundred wave periods frequently require up to $100$ hours of 
simulation time, for approximately $30$ CPUs on modern supercomputers. 
Table~\ref{tab:comp} summarises typical computing resources for a range of 
spatial and temporal resolutions. 
Nevertheless, for each simulation run, a large amount of data can be extracted 
by accessing the electric, magnetic and Poynting 
vector fields at each grid point. 
One particular feature of this approach is that one is able to map out the 
angular distribution of the energy density for a 
typical flux plane, which will be discussed further in Section~\ref{sec:ang}.

FDTD is also able to consider any desired position or alignment 
of dipole sources, including inhomogeneous distributions of dipoles 
placed on the surface or throughout the medium. 
The freedom to define a specific flux collection region at any point in space, 
or for any length of time,  
is an important feature for ongoing investigations into 
a range of coupling scenarios. Measuring the flux from a particular direction 
and aperture automatically 
biases the shape of the collected power 
spectrum. FDTD is therefore especially suitable for simulating 
the effect of collecting radiation through a restricted region. 

\begin{table*}[tp]
  \caption[The computational requirements of FDTD simulations are summarised.]
{Computing resources required for a three-dimensional 
FDTD simulation of a $6$ $\mu$m diameter sphere excited by a dipole source 
with a central wavelength of $0.6$ $\mu$m. The Tizard 
machine at eResearchSA (\href{http://www.ersa.edu.au/tizard}
{\cb http://www.ersa.edu.au/tizard\cbl}) is used in these simulations, 
which uses AMD 6238, 2.6 GHz CPUs. The number of CPUs, the memory (RAM), 
virtual memory (VM) and the amount of computing time (wall-time) for the 
simulations are listed for a variety of FDTD grid resolution values, 
$\Delta x$. The resolution in the frequency domain is quoted after being 
converted to wavelength, $\Delta \lambda$. The wall-time increases linearly 
with the flux collection time, which is held fixed at $0.6$ picosecond (ps) in 
this table. 
}
\vspace{-6pt}
  \newcommand\T{\rule{0pt}{2.8ex}}
  \newcommand\B{\rule[-1.4ex]{0pt}{0pt}}
  \begin{center}
    \begin{tabular}{llllll}
      \hline
      \hline
      \T\B            
       $\Delta x$ (nm) & $\Delta \lambda$ (nm) & CPUs & RAM (GB) & VM (GB) & 
Wall-time (hrs : mins) \\
      \hline     
      \quad$33$ &\,\, $0.62$ &\,\, $24$ &\,\, $28.45$ &\,\, $34.82$ & 
\qquad\,\, $15:08$ \\
      \quad$30$ &\,\, $0.62$ &\,\, $24$ &\,\, $36.73$ &\,\, $43.00$ & 
\qquad\,\, $26:12$ \\
      \quad$29$ &\,\, $0.62$ &\,\, $24$ &\,\, $41.44$ &\,\, $47.75$ & 
\qquad\,\, $27:19$  \\
      \quad$27$ &\,\, $0.62$ &\,\, $24$ &\,\, $46.57$ &\,\, $52.77$ & 
\qquad\,\, $30:09$ \\ 
      \quad$26$ &\,\, $0.62$ &\,\, $24$ &\,\, $53.33$ &\,\, $59.61$ & 
\qquad\,\, $37:31$ \\
      \quad$25$ &\,\, $0.62$ &\,\, $24$ &\,\, $59.31$ &\,\, $65.62$ & 
\qquad\,\, $43:12$ \\
      \quad$22$ &\,\, $0.31$ &\,\, $32$ &\,\, $100.19$ &\,\, $108.02$ & 
\qquad\,\, $90:37$ \\ 
      \hline
    \end{tabular}    
  \end{center}
  \label{tab:comp}
\end{table*}

\begin{figure}[t]
\centering
\includegraphics[height=160pt]{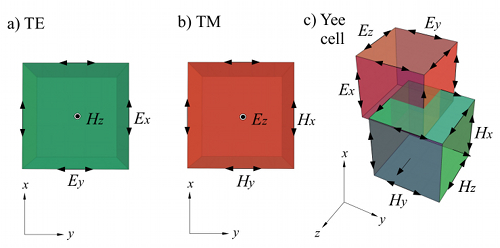}
\vspace{-3mm}
\caption[The \emph{Yee cell} of the Finite-Difference Time-Domain method is 
shown.]{Diagram of the Yee cell for the electric and magnetic 
fields in coordinate space. The orientation of the fields along each boundary 
of the cell is also shown.}
\label{fig:yee}
\figrule
\end{figure}

The principal drawbacks in the FDTD method are the computational load, 
and systematic effects due to the discretisation of space. 
The accuracy of each FDTD calculation is limited by the size of the \emph{Yee 
cell}, which determines the grid resolution of the simulated volume. An 
example Yee cell is illustrated in Fig.~\ref{fig:yee}, showing the orientation 
of the electric and magnetic field components, 
as well as the contributions from both the TE and TM components. 
The Yee cell contains an electric field `sub-cell' and a magnetic field 
sub-cell, 
offset from one another spatially by one half cell length in all three 
directions, so that the cells intersect one another as shown. 
The components of the electric and magnetic fields are simulated on 
the edges of each sub-cell. 
In the following simulations, the cell width is $20$ to $30$ nm. 
A decrease in the cell size increases the computing time requirement 
to the fourth power (three spatial dimensions, and time) \cite{OskooiRo10}. 
As a result, the available hardware resources typically limit the 
improvement in accuracy of the FDTD simulations. 
This has an additional impact on the behaviour of a microresonator,
 as the surface of the 
resonator gains an `effective roughness', due to the finite cell size. 
This effect 
can be incorporated into the systematic uncertainty of the calculation.   
An unanticipated benefit of this effective roughness is a more realistic 
prediction of the $Q$-factor of the resonator. 
Note also that the simulated roughness is comparable in size to  
that of the typical surface roughness of a physical polystyrene 
microsphere, which is approximately $15$ nm or 
greater \cite{doi:10.1021/la047584t}. 

The three-dimensional FDTD method is simulated using the free software 
package, MEEP \cite{OskooiRo10}. 
This work is presented in the publication, Ref.~\cite{Hall:15}, which is 
listed as \textbf{Item 1} in Section~\ref{sec:jou}. 
For reference, code and scripts associated with these calculations have been 
placed online for public availability --
\footnote{\href{https://sourceforge.net/projects/npps/files/FDTD_WGM_Simulator}
{\cb https://sourceforge.net/projects/npps/files/FDTD\_WGM\_Simulator\cbl}} 
see Appendix~\ref{sec:code}.
Since the simulation includes the complete set of radiation and bound modes 
occurring in the system, the only simulation artefacts in an 
FDTD calculation involve discretisation and finite-volume effects, and any 
assumptions of ideal material properties one might make. 
By altering the grid size, the convergence of the simulation can be estimated, 
and this can be quantified in the form of a systematic 
uncertainty in the positions of the WGMs. The capability of including and 
handling this effect is discussed in more detail in Section~{\ref{sec:spec}}.

\subsection{\cdb Establishing the resonator geometry}

The FDTD method is used to consider a variety of flux collection scenarios, 
source distribution, and resonator properties. 
A specific example scenario of a polystyrene microsphere in a surrounding 
medium of air or water is considered, and the $Q$-factors obtained from the 
simulation will then be compared to those of the Chew model \cite{Chew:88}. 

The geometry of the resonator is defined for a dielectric medium, and 
placed in the discretised space, with the edges of the total 
three-dimensional FDTD volume simulated as though 
padded with a field-absorbing \textit{perfectly-matched layer}. 
To generate an electromagnetic current to excite the WGMs, one or several 
electric dipole sources are placed in the vicinity of the sphere. 
These sources can be used to simulate active resonators, such as those 
that incorporate fluorescent nanocrystals \cite{Kaufman2012,Ilchenko:13,
Tao21062016} or quantum dots \cite{C1JM10531K,6062374,1063-7818-44-3-189}. 
The exploration of the electromagnetic effects of 
sources is of contemporary interest due to the recent advancement of the 
detection of single particles by WGM resonators 
\cite{Zhu2010,:/content/aip/journal/apl/98/24/10.1063/1.3599584,
CPHC:CPHC201100757,Vollmer30122008,Baaske2014}. 
In addition, a distribution of a large number of dipole sources may be used to 
simulate a fluorescent layer, analogous to active resonators that use organic 
dyes, or to natural autofluorescence \cite{Gupta201515,C5OB00299K,
doi:10.1021/acs.joc.5b00077,Francois:15b,doi:10.1021/acsami.5b10710,
ADOM:ADOM201500776}. 
However, since the focus of this chapter is to explore and verify the FDTD 
approach by examining the behaviour of the TE and TM modes with respect to the 
dipole orientation, and the spatial distribution of the energy, this latter 
option will be dealt with chiefly in Chapters~\ref{chpt:bub} and \ref{chpt:mod}.

\begin{figure}
\centering
\includegraphics[height=200pt]{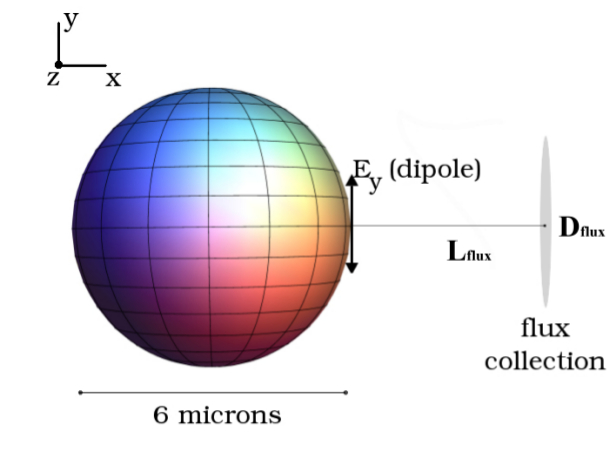}
\vspace{-3mm}
\caption[The flux collection from a simulated microsphere resonator is shown.]
{A circular flux collection region, offset from a 
$6$ $\mu$m diameter microsphere in the $x$-direction, is placed so that its 
normal is aligned radially outward along the same axis. 
In this illustration, the WGMs are excited from a tangentially-oriented 
electric dipole source. 
}
\label{fig:Sphere}
\figrule
\end{figure}

In the first example, a 
microsphere with a diameter of $6$ microns, as illustrated in 
Fig.~\ref{fig:Sphere}, is placed in the centre of the FDTD volume  
described at the beginning of Section~\ref{sec:FDTD}. An 
electric dipole is then placed on the surface of the sphere and oriented 
tangentially to the surface, while a circular flux collection region is 
defined a distance $L_{\mathrm{flux}}$, normal to the sphere, with diameter 
$D_{\mathrm{flux}}$. This region aggregates the flux in the frequency domain. 
One may vary the orientation and position of both source and collection region 
to build up a map of the coupling efficiencies to different WGMs.

\subsection{\cdb Introducing the spectrum}
\label{sec:spec}

The power output spectrum represents an important quantity for assessing the 
$Q$-factors and the wavelength positions of the excited WGMs. 
The total output power ($P$) in terms of wavelength 
($\lambda$) is obtained by integrating the projection of the Poynting vector 
($\mathbf{S} \equiv \mathbf{E} \times \mathbf{H}^*$) onto a  
flux region of area $\mathscr{A}$: 
\begin{equation}
\label{eq:PFDTD}
P(\lambda) = \int\!\mathbf{S}\cdot \hat{\mathbf{n}}\,\, \mathrm{d}\mathscr{A},  
\end{equation}
for a unit vector $\hat{\mathbf{n}}$ normal to the collection region. 
Aggregating the power flux at each time slice, as opposed to time-averaging, 
ensures that the full electromagnetic behaviour is incorporated into the 
spectrum, including all transient states. 

The profile of the power spectrum, plotted as a function of wavelength 
$\lambda$, will show sharp peaks that correspond to the positions of TE and TM 
WGMs. The relative heights of the peaks indicate the coupling efficiencies to 
different modes, which are highly dependent on the geometry of the dielectric 
medium, the refractive index contrast, and also the method of mode excitation, 
such as dipole sources of different alignments, or other coupling scenarios. 
The power spectrum is then normalised to the power output of the sources 
in the absence of the dielectric sphere. This power, $P^0$, is calculated 
in the same manner as $P$ in Eq.~\ref{eq:PFDTD}, with the dipole sources 
within the FDTD volume emitted. 

An FDTD simulation of a polystyrene microsphere, $\mathrm{n}_1 = 1.59$, 
with a diameter of $6$ $\mu$m, is carried out for a tangentially-oriented 
electric dipole source emitting a Gaussian pulse with a central wavelength of 
$600$ nm, and a width of $5$ femtosecond (fs), or a spectral 
full-width at half-maximum of $75$ nm. 
Note that the pulse width is 
narrower than the decay transition rate expected from a typical
 fluorescent source, such as Rhodamine dye, which is roughly $1$ to $3$ 
nanosecond (ns) \cite{Barnes:92,pmid19950338}. The pulse is taken to be 
effectively instantaneous with respect to the phenomena being measured.   
The source is introduced in the form of a vector current that is separable 
in the space and time components, $\mathbf{A}$ and $f$, respectively
\begin{equation}
\label{eqn:src}
\mathbf{J}(\mathbf{x}, t) = A\,\times\,\mathbf{A}(\mathbf{x})f(t). 
\end{equation}
$A$ is a complex number that represents the total amplitude of the source, 
which in this chapter will be set to $1$. 
A circular flux collection region with a diameter of $2.58$ $\mu$m is placed a 
distance of $240$ nm from the surface of the sphere in the 
$x$-direction, with its normal aligned along the same axis.  Spectral 
information is then collected for the wavelength range encompassed by 
the Gaussian pulse. 
The finite grid resolution entails an asymmetry in the Gaussian 
peak, which diminishes as the resolution increases. 

\begin{figure}[H]
\begin{center}
\includegraphics[width=0.8\hsize,angle=0]{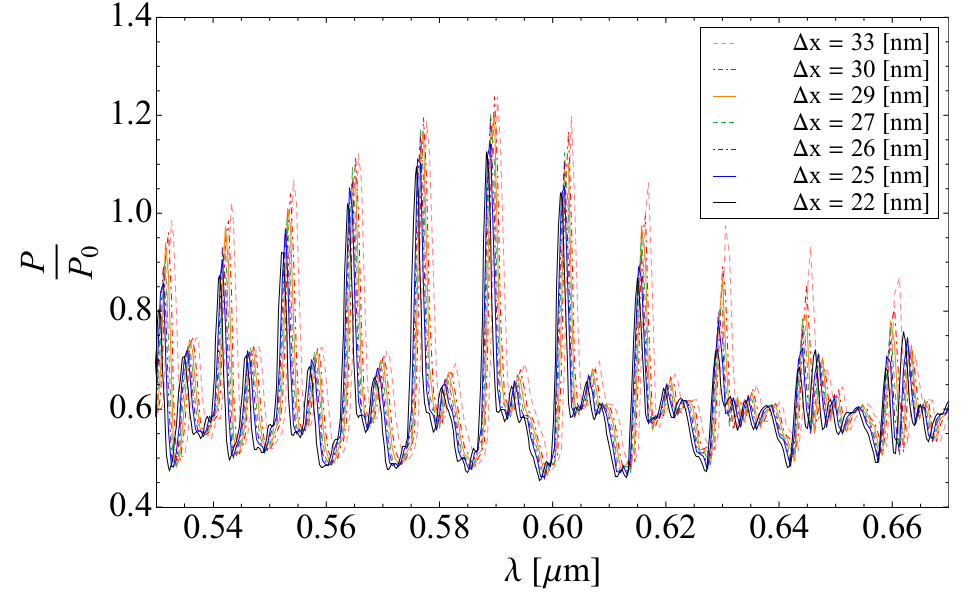}\\
\vspace{-5mm} 
\hspace{14mm}\mbox{(a)}\\
\includegraphics[width=0.8\hsize,angle=0]{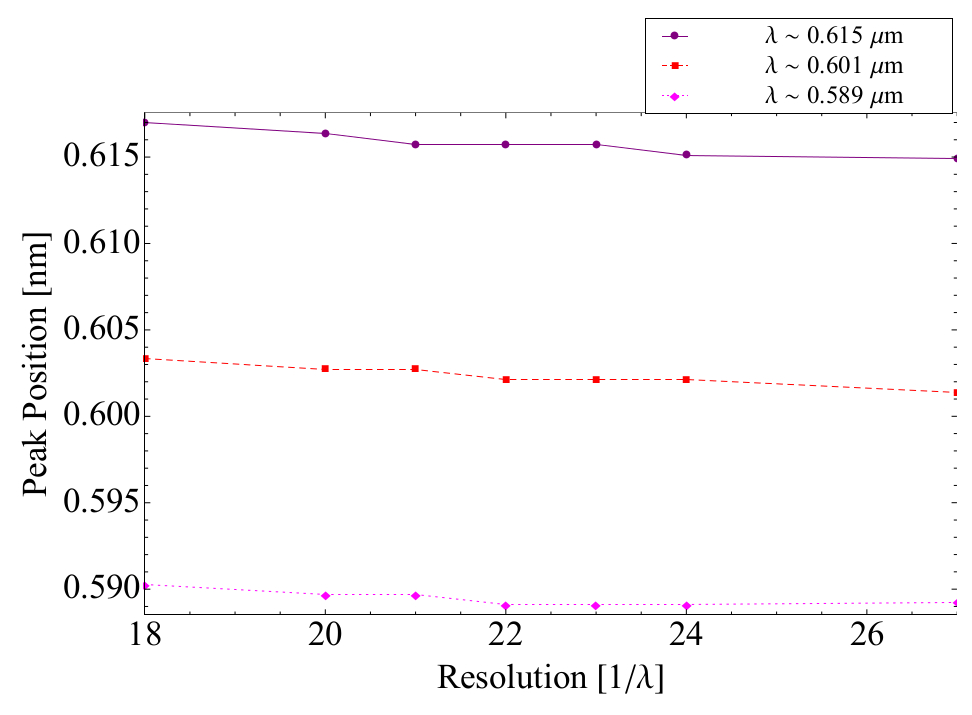}\\
\vspace{-5mm}
\hspace{11mm}\mbox{(b)}\\
\end{center}
\vspace{-6mm}
\caption[The convergence of the power spectrum in FDTD is explored.]{(a) 
A comparison of the power spectra of $6$ $\mu$m diameter 
microspheres for different grid resolutions. (b) The convergence of the 
position of three central peaks is shown as a function of 
resolution. Excitation occurs from a tangentially oriented dipole source 
placed on the surface, with a central wavelength of 
$\lambda = 0.6$ $\mu$m. The flux collection time for each simulation is 
$0.6$ ps.}
\label{fig:res}
\figrule
\end{figure}

A comparison of a variety of grid resolutions is shown in  
Fig.~\ref{fig:res}(a), using a total flux collection time of $0.6$ picosecond 
(ps) -- $120$ times the width of the pulse. 
The flux collection begins after the Gaussian pulse has decayed, that is, 
after $5$ fs. For resolutions in each spatial direction of $22$ to $33$ nm, 
one finds that as $\Delta x$ decreases, that is, at higher resolution, 
the profile features of the 
spectrum do not alter significantly; there is only a small offset in the 
positions of the peaks, and the peak heights. The positions of the WGMs are 
determined from collection of the frequencies, which are then converted to 
wavelengths. The temporal resolution is then interpolated to yield a value 
$<1$ nm. 

The systematic uncertainty in the resonance positions due to resolution 
can be quantified by tracking the positions of the WGM peaks. 
By examining the wavelength positions of the 
most prominent peaks, one can obtain a converged result to a chosen tolerance, 
as shown in Fig.~\ref{fig:res}(b).  An error estimate in the peak positions 
can be obtained by comparing the results for different grid resolutions. 
For example, using the lowest simulated resolution of 
$\Delta x = 33$ nm, the most-central peak has a wavelength position of 
$603.32$ nm. For the highest simulated 
resolution of $\Delta x = 22$ nm, the position is $601.35$ nm.  This yields an 
offset in the position of the central peak of $1.97$ nm. 
For comparison, note that a spectral resolution of approximately $0.03$ nm
has been shown to be necessary for the detection of nearby macromolecules by a 
microsphere \cite{PMID:22219711}. 
This offset, however, is uniform over a wide wavelength range, and provides 
a reference spectrum from which any shifts in the surrounding refractive index 
can be detected. In this investigation, it is nevertheless important to 
emphasise that the FDTD-based method is developed principally as a screening 
tool for assessing the feasibility of new designs of biosensors at the 
proof-of-concept stage. 

\begin{figure}[H]
\begin{center}
\includegraphics[width=0.8\hsize,angle=0]{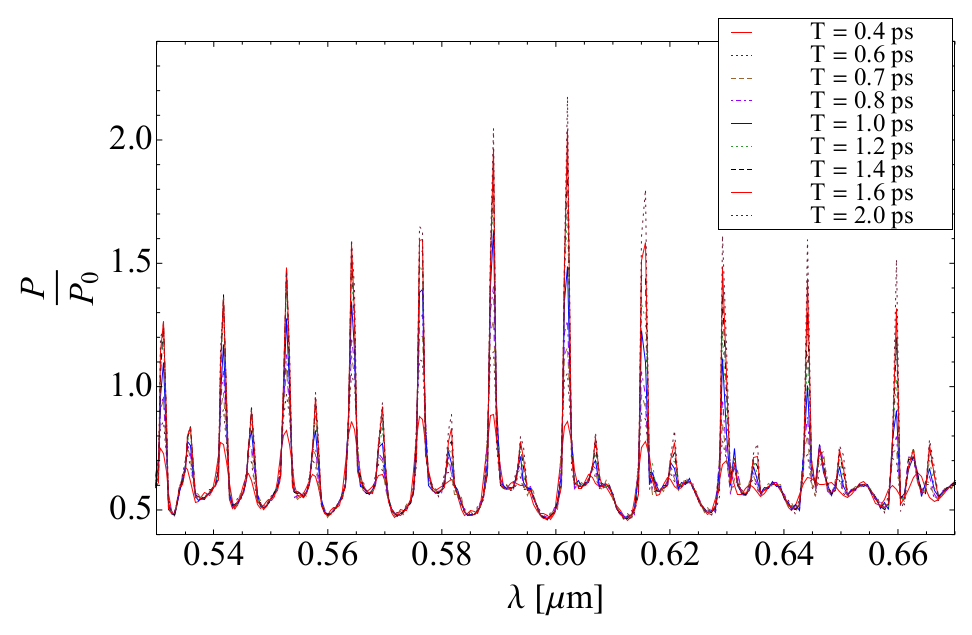}\\
\vspace{-5mm} 
\hspace{15mm}\mbox{(a)}\\
\includegraphics[width=0.8\hsize,angle=0]{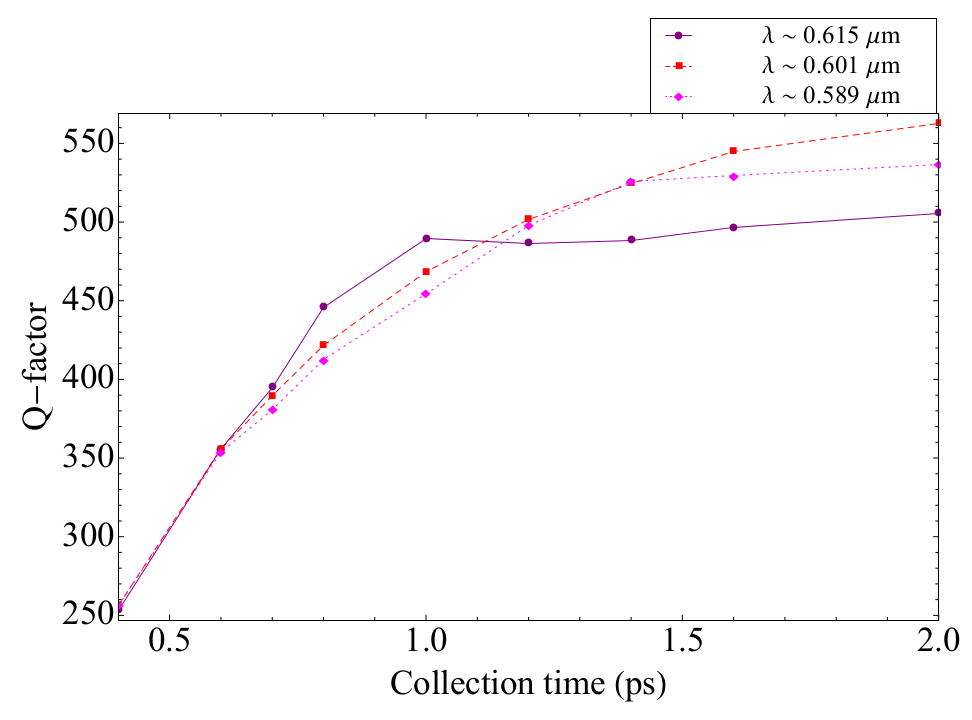}\\
\vspace{-5mm}
\hspace{14mm}\mbox{(b)}\\
\end{center}
\vspace{-6mm}
\caption[The impact of the flux collection time on the spectrum is examined.]
{(a) A comparison of the power spectra of $6$ $\mu$m 
diameter microspheres with a tangential source ($\lambda = 0.6$ $\mu$m), for 
different flux collection times. (b)  The behaviour of the 
$Q$-factors of three central peaks is shown as a function of collection time 
(ps).  The output power is normalised to the dipole emission 
rate in an infinite bulk medium of the same refractive index as the 
surrounding medium. }
\label{fig:WGMtime}
\figrule
\end{figure}

By changing the length of time allowed for flux collection, one can also 
obtain important insights into the distribution of modes in the power 
spectrum. Fig.~\ref{fig:WGMtime}(a) shows the resultant normalised power 
spectrum for a variety of flux collection times. A spatial resolution of 
$\Delta x=25$ nm is chosen, and also a density of sampled wavelengths 
corresponding to $\Delta \lambda = 0.62$ nm. As the collection time increases, 
the WGM 

\newpage
\noindent peaks become more prominent. Furthermore, double-peak structures seen 
at small collection times are no longer apparent for larger 
collection times, as the sampling of the full mode structure of the radiating 
cavity does not have sufficient time to be measured by the flux region. 
For the longer collection times, an enhancement in the normalised dipole power 
is observed ($P/P^0 > 1$), at wavelengths coinciding with the dominant WGMs. 
This indicates that the power output of the source is enhanced at these 
wavelengths by the presence of the dielectric microsphere. 
This effect is due to the difference in time scales 
between the cavity ring-down time and the time evolution in the spectrum. 
The resonances with high $Q$-factors 
originate from field configurations traversing paths around a 
highly symmetrical resonator as established in Section~\ref{sec:nut}, however, 
these resonances also have longer ring-down times, as 
discussed in Section~\ref{sec:an}. 
As a result, a certain amount 
of flux collection time is required to resolve the total underlying mode 
structure. This collection time can be calculated by analysing the 
structure of the spectrum as a function of collection time as follows. 

The behaviour of the $Q$-factor as a function of the flux collection time 
is shown in Fig.~\ref{fig:WGMtime}(b) for three central 
WGM peaks. For the peak at $\lambda \sim 0.615$ $\mu$m, convergence is achieved 
beyond a collection time of $1.0$ ps, with a variation of $3.4\%$. 
For the other two peaks, $\lambda \sim 0.601$ $\mu$m and $0.589$ 
$\mu$m, variations of $16.9\%$ and $15.3\%$ are measured, respectively. 
Note that, while peaks with lower $Q$-factors incur less variation 
with respect to collection time, peaks with higher $Q$-factors are more 
affected. 
However, an increase in collection time becomes 
increasingly computationally intensive, due to the asymptotic behaviour 
in the $Q$-factor 
with respect to collection time \cite{marki2007design}. Nevertheless, 
the uncertainty in the $Q$-factor due to the collection time can be 
quantified, and thus incorporated into the systematic uncertainty. 

\begin{figure}[H]
\begin{center}
\hspace{-5.5mm}
\includegraphics[width=0.8\hsize]{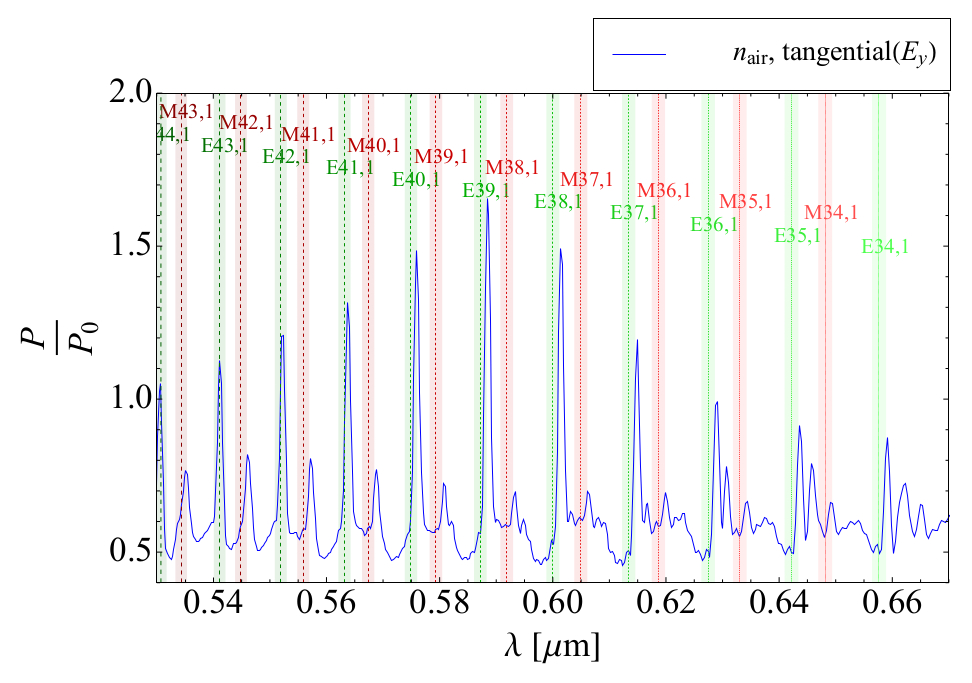}\\
\vspace{-5mm}
\hspace{11mm}\mbox{(a)}\\
\includegraphics[width=0.8\hsize]{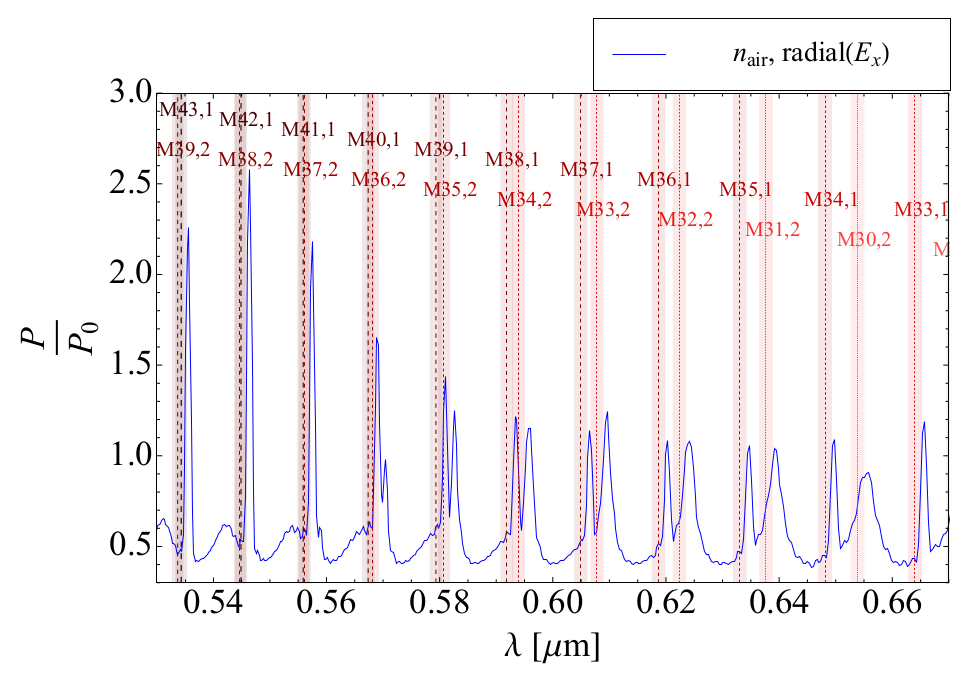}\\
\vspace{-5mm}
\hspace{14mm}\mbox{(b)}\\
\end{center}
\vspace{-6mm}
\caption[The dipole orientation is shown to affect energy coupling to the 
modes.]{FDTD simulation of the normalised power spectrum of a 
polystyrene microsphere, $6$ $\mu$m in diameter, with a surrounding medium of 
air. (a) Whispering gallery modes are excited from a 
tangential or (b) radial electric dipole source with a central wavelength of 
$0.6$ $\mu$m. Vertical lines indicate predictions of the TE$_{m,n}$ 
(\color{darkgreen}\textit{green}\color{black}) and TM$_{m,n}$ (\color{red}
\textit{red}\color{black}) modes derived from the 
Johnson model, for polar and radial mode numbers $m$ and $n$, respectively. 
The width of the bands indicates the systematic uncertainty 
in the positions due to the finite grid size of FDTD.}
\label{fig:WGM}
\figrule
\end{figure}

\begin{figure}[H]
\begin{center}
\includegraphics[width=0.8\hsize]{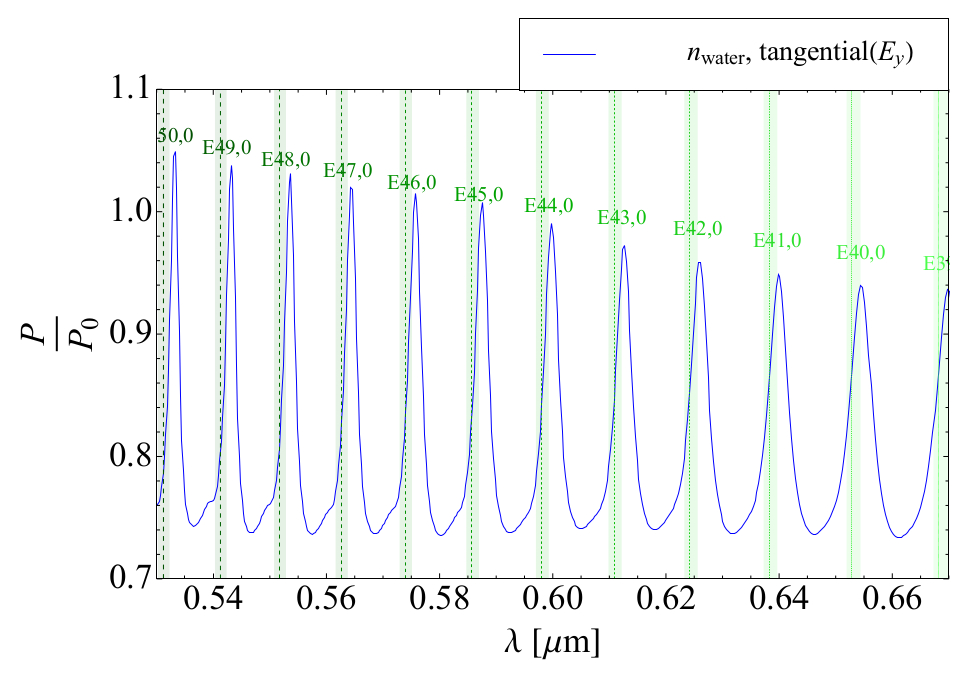}\\
\vspace{-5mm}
\hspace{13mm}\mbox{(a)}\\
\includegraphics[width=0.8\hsize]{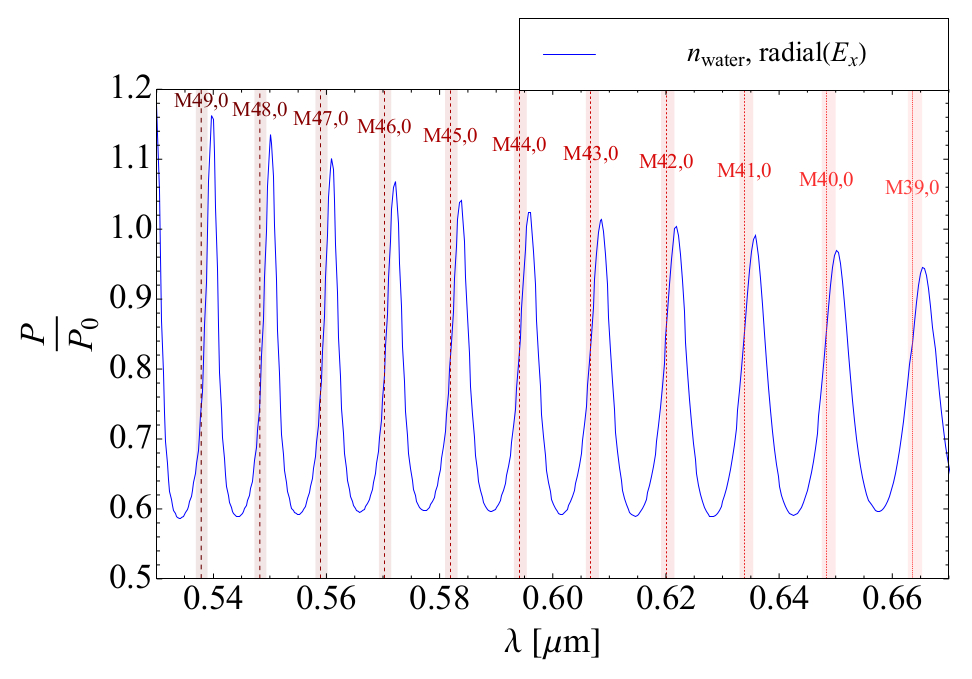}\\
\vspace{-5mm}
\hspace{14mm}\mbox{(b)}\\
\end{center}
\vspace{-6mm}
\caption[The effect of the dipole orientation is shown in a medium of water.]
{FDTD simulation of the normalised power spectrum of a 
polystyrene microsphere, $6$ $\mu$m in diameter, with a surrounding medium of 
water. (a) Whispering gallery modes are excited from a 
tangential or (b) radial electric dipole source with a central wavelength of 
$0.6$ $\mu$m. The width of the bands accommodates the 
systematic uncertainty due to the finite grid size of FDTD.}
\label{fig:WGM2}
\figrule
\end{figure}

Comparisons of the analytic 
model and the FDTD spectra for a microsphere are shown 
in Figs.~\ref{fig:WGM} and \ref{fig:WGM2} for tangentially and radially 
oriented dipole sources, in both air and water media, respectively. 
The WGM positions from the analytic model are shown 
as vertical bands, with TE$_{m,n}$ modes in green and TM$_{m,n}$ modes in red, 
for polar and radial mode numbers $m$ and $n$, respectively. 
The width of each band indicates the estimate of the systematic uncertainty 
due to the finite grid spacing of

\newpage
\noindent FDTD. Specifically, the Johnson analytic model, based on Mie scattering, 
is used to estimate the shift in the WGM positions due to uncertainty in the 
sphere diameter, $6$~$\mu$m~$\pm\Delta x /2$. 
In this case, the spatial 
resolution is held fixed at the value $\Delta x=22$ nm. 
For the temporal resolution, which is related to the wavelength range 
(in this case $500$ to $750$ nm), 
the spectral density of $800$ points yields an uncertainty of \\
$(750-500)$nm$/800 = 0.31$~nm. The two uncertainties are
 added in quadrature. 
Taking into account the finite grid resolution, the offset of each peak 
from the position expected from 
the analytic model may also be due to the fact that nearby modes, which  
can partially overlap with each other, 
are capable of shifting the peaks away from the central positions 
calculated from such an analytic model \cite{Hall:17a}. 

Note that the TE and TM modes cannot be completely 
decoupled due to the spherical symmetry in the case of a tangentially oriented 
dipole ($E_y$), as expected from a comparison of the forms of 
Eqs.~(\ref{eq:chewex}) and (\ref{eq:chewey}) occurring in the Chew model, for 
example. Thus contributions from both TE and TM modes, including higher order 
modes, are expected in the spectrum. 
Consider the case of Fig.~\ref{fig:WGM}(a), for a tangentially oriented 
dipole in a surrounding medium of air.  One might expect that the dominant 
WGMs excited are the lowest-order radial TE modes, since the electric field of 
each of the TM modes contains a radial component, which does not couple 
strongly to the tangential source. It is found, in fact, that the dominant 
peaks have an FSR consistent with a radial mode number of $n=1$. The 
fundamental radial modes ($n=0$) cannot be resolved for this index contrast 
at this finite grid size, since they are known to exhibit  
$Q$-factors of order $10^{3-4}$ or larger, 
both experimentally \cite{Francois:2011}, and in analytic models 
\cite{Little:99,Talebi,Min:2009a,Vahala:2004}. 
In FDTD, the discretisation of space leads 
to an effective surface roughness, which broadens the peaks, leading naturally 
to lower estimates of the cavity $Q$-factor, $Q_c$, as shown in 
Table~\ref{tab:qf}. 
In the case of the radial electric dipole, Fig.~\ref{fig:WGM}(b), 
the dominant peaks exhibit FSRs that match the TM modes for both $n=1$ and 
$n=2$, with no contribution from the TE modes. 

For a lower index contrast scenario, such as polystyrene microspheres in 
a surrounding medium of water, shown in Figs.~\ref{fig:WGM2}(a) and 
\ref{fig:WGM2}(b), the WGM peaks are broadened, and there is a reduced 
density of modes at wavelengths in the range 
$500$ to $750$ nm. As a result, the peaks observed in 
the FDTD simulation correspond to the fundamental radial TE or TM modes, for 
a tangential or radial dipole source, respectively. 

\begin{table*}[t]
  \caption[The $Q$-factors and positions are shown for prominent spectral 
peaks.]{A summary of the $Q$-factors and wavelength positions, 
$\lambda$ ($\mu$m), 
 for the four most prominent WGM peaks, 
for each plot displayed in Figs.~\ref{fig:WGM} and \ref{fig:WGM2}. 
The scenarios considered are: a surrounding medium of air or water   
with a tangential ($E_y$) or radial ($E_x$) dipole source. By comparing 
the $Q$-factors for collection times of $0.6$ ps and $2.0$ ps, 
a systematic uncertainty of up to $17\%$ in the $Q$-factors is estimated
 from the collection time. 
}
\vspace{-6pt}
  \newcommand\T{\rule{0pt}{2.8ex}}
  \newcommand\B{\rule[-1.4ex]{0pt}{0pt}}
  \begin{center}
    \begin{tabular}{lcccc}
      \hline
      \hline
      \T\B            
       Scenario \quad&\quad  $\lambda$ ($\mu$m),\,\, $Q_c$  \quad&\quad 
$\lambda$ ($\mu$m),\,\, $Q_c$  \quad&\quad 
$\lambda$ ($\mu$m),\,\, $Q_c$  \quad&\quad $\lambda$ ($\mu$m),\,\, $Q_c$  \\
      \hline     
      air,\,\, $E_y$ \quad&\quad$0.576$,\,\, $490$ \quad&\quad $0.588$,\,\, 
$510$ \quad&\quad $0.601$,\,\, $528$ \quad&\quad $0.615$,\,\, 
$559$\\
air,\,\, $E_x$ \quad&\quad$0.536$,\,\, $549$  \quad&\quad $0.546$,\,\, 
$572$ \quad&\quad $0.557$,\,\, $584$ \quad&\quad $0.569$,\,\, $584$\\
water,\,\, $E_y$ \quad&\quad$0.533$,\,\, $305$ \quad&\quad $0.543$,\,\, 
$286$ \quad&\quad $0.553$,\,\, $264$ \quad&\quad $0.564$,\,\, 
$250$ \\
      water,\,\, $E_x$ \quad&\quad$0.539$,\,\, $232$ \quad&\quad $0.550$,\,\, 
$218$ \quad&\quad $0.560$,\,\, $204$ \quad&\quad 
$0.572$,\,\, $191$  \\ 
      \hline
    \end{tabular}    
  \end{center}
  \label{tab:qf}
\end{table*} 

Note that in Fig.~\ref{fig:WGM2}(a), for a tangentially oriented source, 
the main contribution is from the TE modes, while there is a relatively 
small component 
originating from the TM modes -- the `shoulders' that appear to the left of the 
dominant TE peaks, which are more apparent 
at lower wavelengths where the peak positions are sufficiently separated. 

The separation of the spectrum into components that correspond directly to 
the TE and TM polarisations, derived from spherical symmetry, will be 
investigated more comprehensively in Chapter~\ref{chpt:mod}. However, 
for the purposes of this study of the FDTD method, it is sufficient 
to note that the total radiative contribution to the spectrum is included, 
and that FDTD allows for the introduction of a number of different 
resonator design, excitation and radiation collection scenarios 
to be simulated. 
In fact, in this investigation, the FDTD tool has already provided guidance 
for future designs of resonators, which can be adapted for biosensing
 applications. Figures~\ref{fig:WGM} and \ref{fig:WGM2} confirm that the 
dominance of TE or TM modes is dependent on the alignment of the dipole source 
on the surface of the sphere, as described in Ref.~\cite{Teraoka:06a}. 
Microsphere sensors are thus able to detect the 
orientation of an external biomolecule by recording the 
relative coupling strengths of TE to TM modes.

\subsection{\cdb Connection to Mie scattering}
\label{sec:mie}

\begin{figure}[H]
\begin{center}
\includegraphics[width=0.8\hsize]{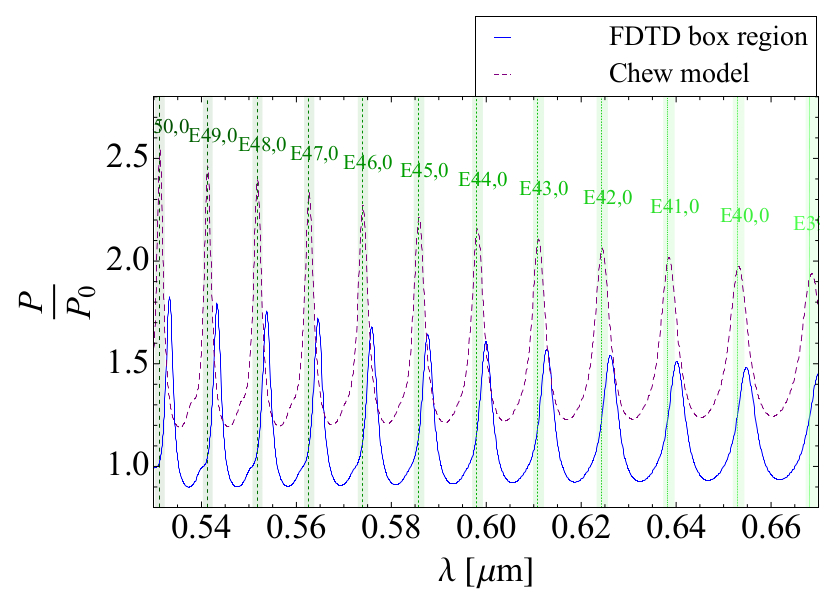}\\
\vspace{-5mm}
\hspace{17mm}\mbox{(a)}\\
\includegraphics[width=0.8\hsize]{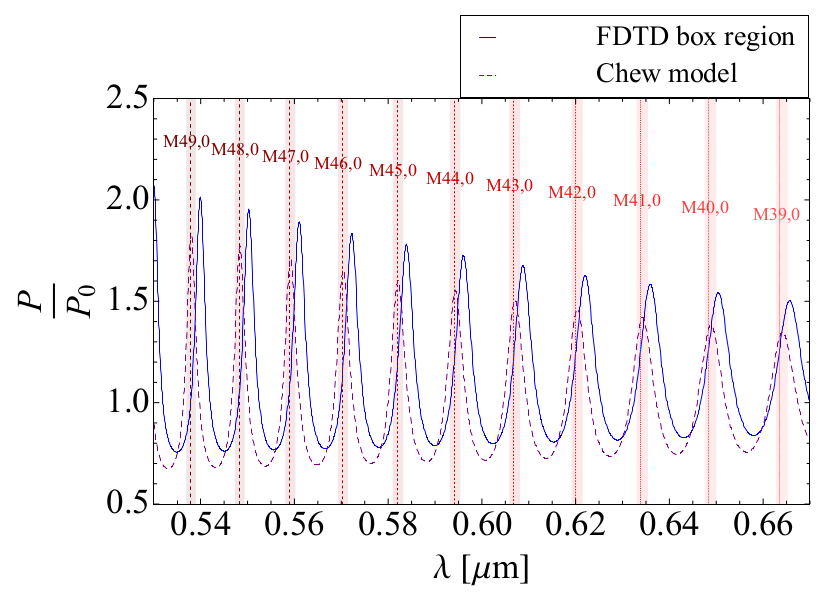}\\
\vspace{-5mm}
\hspace{16mm}\mbox{(b)}\\
\end{center}
\vspace{-6mm}
\caption[FDTD simulations are verified by the Chew model.]{
A comparison of the Chew model of Eqs.~(\ref{eq:chewex}) and (\ref{eq:chewey}) 
(dashed \color{purple} \textit{purple} \cbl line) with the Mie scattering 
analytic model (vertical dashed lines) and the FDTD simulation, with total 
power collected from a box that surrounds the microsphere (solid \color{blue} 
\textit{blue} \cbl line). 
A surrounding medium of water is used for a $6$ $\mu$m diameter polystyrene 
sphere. WGMs are excited from (a) a tangential or (b) a radial dipole source. 
The vertical \color{darkgreen}\textit{green} \color{black} lines are the 
fundamental radial TE modes, and the \color{red}\textit{red} \color{black} 
lines are the corresponding TM modes. The width of the bands accommodates the 
systematic uncertainty due to the finite grid size of FDTD.}
\label{fig:Chew}
\figrule
\end{figure}

Comparisons of the Chew model, the Johnson model and 
the FDTD simulations are shown in Fig.~\ref{fig:Chew}, for a surrounding 
medium of water. In Fig.~\ref{fig:Chew}(a), 
a tangential dipole source is used, and the fundamental radial TE modes 
from the Johnson model, corresponding to Mie scattering (green vertical lines), 
exactly match the peaks of the Chew model (dashed blue line), 
as expected, by construction \cite{Chew:88}. 
The peaks of the FDTD simulation also correspond to these TE modes, 
and the FSR matches that of the Chew model. 
Fig.~\ref{fig:Chew}(b) shows the result for a radial 
dipole. The peaks of both the Chew model and the FDTD simulation 
correspond to the fundamental radial TM modes.
Note that in both Figs.~\ref{fig:Chew}(a) and (b) there is 
 an offset in both the peak positions and the total power  
due to the finite grid resolution, as explained in Section~\ref{sec:spec}. 

Although this analytic model is better suited to simulating scenarios  
that involve dipole sources at a variety of positions and alignments, 
the FDTD method is able to collect flux from any desired collection 
region, in the near or far field, for any length of time. This is an important 
point, as this collection scenario will affect the measured power spectrum 
profile. It is useful to be able to estimate the size of the effect due to 
collection region, and this will be investigated in the specific case of the 
angular distribution of the flux, in the next section.

\subsection{\cdb Angular distribution of the modes}
\label{sec:ang}

The simulation can also provide information about the 
angular distribution of the flux, both directly and indirectly. 
First, the configurations of the electric and magnetic fields in the simulation 
are stored for each point in space, at every time slice, providing direct 
access to the behaviour of the fields. Second, a less direct but  
physically insightful method for assessing the angular distribution of the 
fields is to 

\newpage
\noindent examine several modes in the spectrum, 
in this case four modes which lie within the simulated wavelength window, 
TE$_{37,1}$, TE$_{38,1}$, TE$_{39,1}$ and TE$_{40,1}$,  for a variety of flux 
collection scenarios. This is of interest, particularly when comparing a variety of 
different methods of collecting the emitted power,
which have been shown to influence the structure of the spectrum 
\cite{Riesen:15a}.
Since the FDTD simulation records the electromagnetic field values, the 
flux density, $\mathbf{S}(\mathbf{r})$, may be projected onto the circular 
region of the $z$-$y$ plane (with normal vector $\hat{\mathbf{n}}$) for any 
wavelength value. An analysis of this type is helpful for visualizing the 
distribution, distinguishing the polar and radial modes at different time 
slices, and  identifying transient resonant features. By integrating the flux 
over a collection time of $1.0$ ps, the distribution over the collection plane 
indicates the angular dependence of the mode at a particular wavelength. 

\begin{figure}[H]
\begin{center}
\includegraphics[width=0.47\hsize]{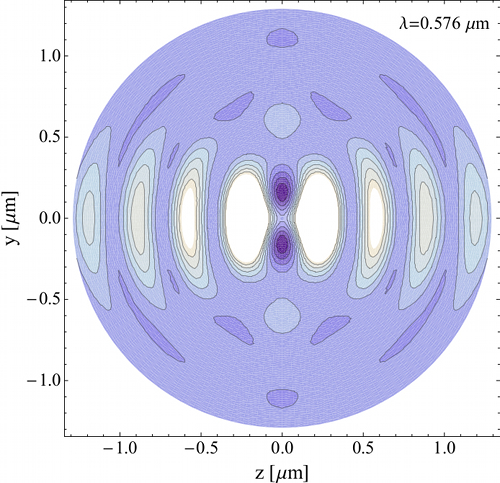}
\hspace{-1.0mm}
\includegraphics[width=0.47\hsize]{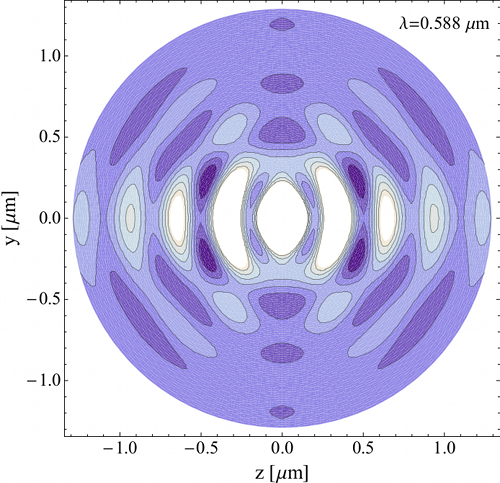} 
\\
\vspace{-3mm}
\mbox{\hspace{0.8cm}(a)\hspace{6.3cm}(b)}\\
\vspace{3mm}
\hspace{-1.0mm}
\includegraphics[width=0.47\hsize]{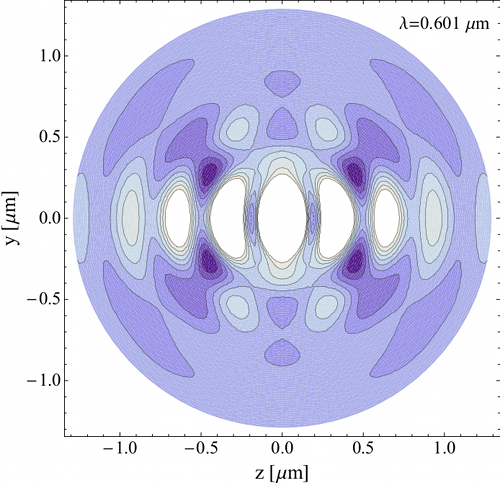}
\hspace{-1.0mm}
\includegraphics[width=0.47\hsize]{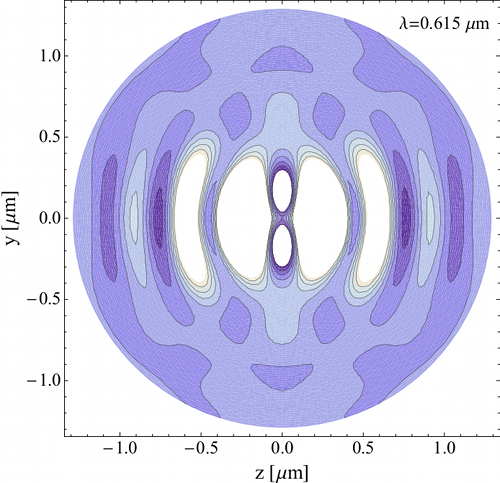}
\\
\vspace{-3mm}
\mbox{\hspace{0.8cm}(c)\hspace{6.3cm}(d)}
\vspace{-3mm}
\end{center}
\caption[Spatial distributions of the modes are contrasted.]{Spatial 
distributions of the flux density over the collection region for several modes, 
$\mathbf{S}(\mathbf{r})$, integrated 
over a flux collection period of $1.0$ ps, and projected onto the circular 
flux collection region. The modes considered are (a)  TE$_{40,1}$: 
$0.576$ $\mu$m, (b) TE$_{39,1}$: $0.588$ $\mu$m, (c) 
TE$_{38,1}$: $0.601$ $\mu$m  and (d) TE$_{37,1}$: $0.615$ $\mu$m.  
The axes are defined in the same way as in Fig.~\ref{fig:Sphere}, for 
$x$-coordinate: $3.24$ $\mu$m. Lighter colour corresponds to larger 
magnitudes of $\mathbf{S}(\mathbf{r})\cdot\hat{\mathbf{n}}$.}
\label{fig:WGMrad}
\figrule
\end{figure}

As an example, Fig.~\ref{fig:WGMrad} shows the flux distribution over the 
collection region, for a tangential dipole source, and a surrounding medium of 
air. The four panels display the four most prominent WGM peaks occurring in 
that wavelength region, corresponding to values of $0.576$, $0.588$, $0.601$ 
and $0.615$ $\mu$m, respectively. It is apparent that the modes at wavelengths 
of $0.601$ $\mu$m (TE$_{38,1}$) and $0.615$ $\mu$m (TE$_{37,1}$) exhibit a flux 
distribution with peaks spread out in the flux measuring region over a 
comparatively wide angle. The total power output for these modes is likely 
to vary significantly if the diameter of the region is reduced. 
In contrast, the mode occurring at $0.588$ $\mu$m (TE$_{39,1}$) has a more 
focused concentration of flux in the centre of the measuring region, 
corresponding to a narrower angular distribution of flux. The power output for 
this mode is consistent over a wider range of flux region diameters. 
In summary, it is clear that as the measuring region changes, the simulations 
indicate that the radiation coupling to each mode is changed by different 
amounts, altering the structure of the spectrum. 

Note that each circular plot of the flux density 
in Fig.~\ref{fig:WGMrad} contains alternating bright bands 
aligned horizontally along the $z$-axis, 
rather than a concentric circular pattern. This is a 
consequence of the use of a single dipole placed on the surface of the 
resonator, which breaks the symmetry of the distributions of the fields. 

\begin{figure}[H]
\begin{center}
\includegraphics[width=0.8\hsize]{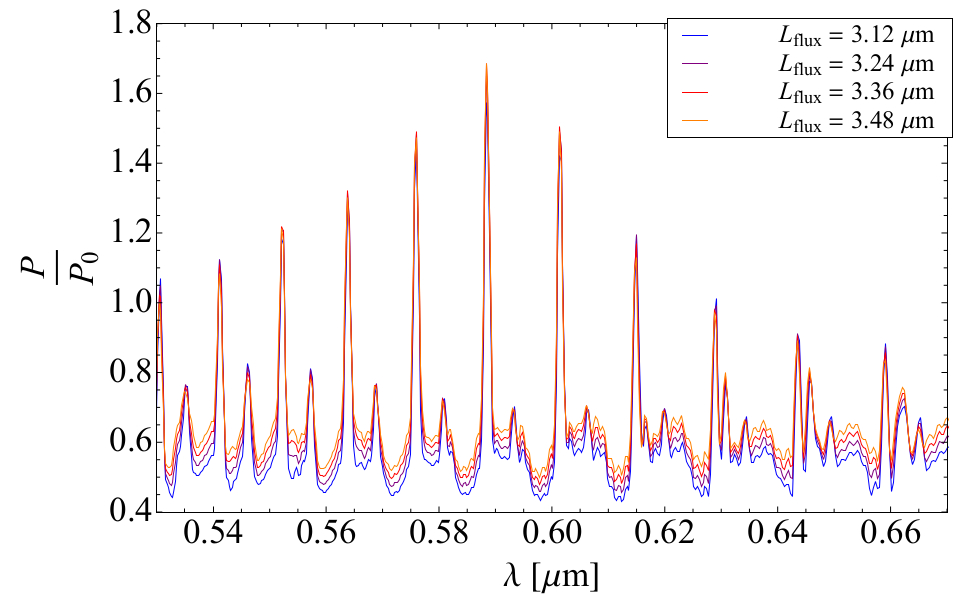}\\
\vspace{-5mm}
\hspace{14mm}\mbox{(a)}\\
\includegraphics[width=0.8\hsize]{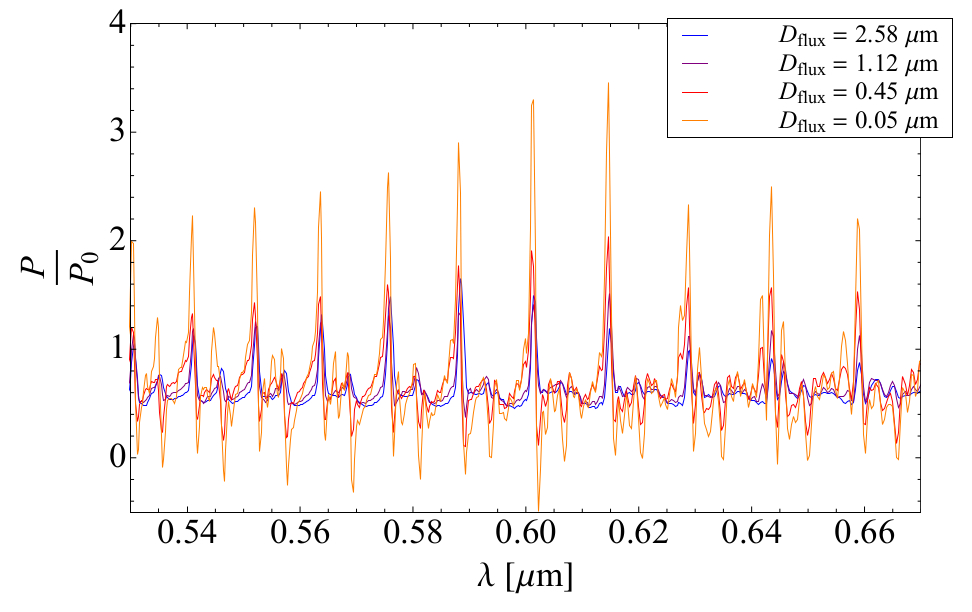}\\
\vspace{-5mm}
\hspace{10mm}\mbox{(b)}\\
\end{center}
\vspace{-6mm}
\caption[Spectra for different flux collection regions are compared.]
{A comparison of the power spectra of $6$ $\mu$m diameter microspheres with a 
tangential source ($\lambda = 0.6$ $\mu$m), (a) for flux collection regions at 
different distances $L_{\mathrm{flux}}$ from the centre of the sphere, and 
(b) for different flux region diameters $D_{\mathrm{flux}}$. Note that a 
resolution of $\Delta x=22$ nm has been used.}
\label{fig:WGMang}
\figrule
\end{figure}

\newpage
As a consequence of the nontrivial angular distributions of the modes, 
changes in the power spectrum are also detected.  
Figure~\ref{fig:WGMang} shows the spectrum for the $6$ $\mu$m sphere 
for flux planes at different distances from the sphere, as shown in 
Fig.~\ref{fig:WGMang}(a), and different diameters, shown in 
Fig.~\ref{fig:WGMang}(b). In each case, a tangential electric dipole is used, 
and the collection time is held fixed at $1.0$ ps. 
By comparing the spectra from several flux collection plane sizes and 
positions, changes in the peak heights are observed, indicating the different 
angular behaviour of the modes. Changes in the distance of the flux region, 
$L_{\mathrm{flux}}$, result in a similar mode structure, but the overall 
power output, especially for the dominant WGM peaks, 
changes as the flux is sampled differently 
in each region. Alterations in the diameter of the flux region 
result in more noticeable 
changes to the mode heights, demonstrating the highly angular-dependent nature 
of the relative contribution of each mode to the total flux.

\section{\cdb Introducing layered spheres and microbubbles}

It can be seen from this chapter that crucial insights into resonator 
behaviour can be gleaned from a careful examination of microspheres, the 
models that describe their resonances, and the behaviour of WGMs under a 
variety of conditions. Specifically, alterations in the size, refractive 
index, surrounding environment, excitation and collection methods all impact 
on the resonance positions or their $Q$-factors. 

Although microspheres serve as an important test case for understanding the 
behaviour of modes in resonators, they have some crucial limitations. Though 
many biological cells in nature have a symmetrical, near-spherical or circular 
cross section shape, as will be seen in Chapter~\ref{chpt:cel}, they also 
contain complex internal and external features, inhomogeneities and structures 
that need to be accounted for. 

As a way forward for the specific application of this thesis, 
the next development will be to take the general principles of resonator 
design explored in this chapter and apply this knowledge to a novel resonator 
design -- \emph{microbubbles}. These resonators differ from microspheres, in 
that they contain a layer of material, constraining the modes to lie within 
the narrow wall of the bubble. 

\newpage 
\noindent Microbubbles serve as an improved analogue for biological cells, 
which typically contain a layer or membrane. In the context of the vision of 
the project, enriching and then testing the modelling capabilities in an 
experimental setting brings us one step closer to addressing the question of 
whether a biological cell can ever be a viable candidate for a resonator. 

Understanding how the introduction of these architectural features affects the 
behaviour of the modes will occupy a significant portion of the next chapter. 
Apart from their applications in a variety of technological and sensing 
contexts, the use of layered resonators or microbubbles allows one to 
investigate the effect of an \emph{internal cavity of different refractive 
index} to that of the resonator wall or shell. How this affects 
the mode positions, free spectral range and $Q$-factors using a variety of 
excitation strategies will be of interest in assessing the feasibility of a 
biological cell resonator, and will lead to a new technique for the 
determination of the geometric parameters of these resonators purely from an 
analysis of their spectra. 

\chapter{\cdb Fluorescent Microbubbles as Cell Analogues}
\label{chpt:bub}

As a way forward in investigating the possibility of a biological 
cell resonator, the principles examined for microsphere 
sensing technology are extended to include resonators with a single dielectric 
layer of finite thickness. One particular example of this layered resonator 
is the microbubble, which consists of a shell of dielectric material and a 
hollow interior. One of the main goals in these next two chapters is to build 
a picture of the behaviour of multilayer resonators, which are analogous 
to certain types of cells \cite{Chew:76}, and a single-layer scenario
 serves as an important test case for illustrating the complexities 
and features that arise from multiple boundaries within a 
symmetrical object. 

There is extensive literature on the topic of microbubbles, 
which are a specific example of a single-layer resonator 
(with a hollow interior). It is therefore worth deliberating on this 
particular type of resonator, not only to cross-check the soundness of 
the modelling capabilities that will be introduced in 
the next two chapters, but also as a way of introducing new features, 
 such as layers of dipole sources, which can mimic 
fluorescent coatings used in experiments 
\cite{Kaufman2012,Tao21062016,C1JM10531K,6062374,1063-7818-44-3-189,
Francois:15a,Francois:15b}.

\section{\cdb Microbubbles as an emerging sensing platform}
  
Like their microsphere counterparts, microbubble resonators are also 
able to support WGMs \cite{Henze:11,Yang:15}, however, the advantages of 
microbubbles over solid microsphere resonators are numerous. 
Microbubbles exhibit improved refractive index sensitivity 
\cite{Yang:14}. That is, the shift in the mode positions due to the presence 
of a nearby particle, which breaks the symmetry of the electric and magnetic 
fields, is enhanced in resonators that have thin 
walls, such as microbubbles. A hollow interior (of lower refractive index) 
impedes the fields entering the cavity, and the field distribution is 
therefore more strongly biased towards the outer region of the resonator, 
leading to a more extensive evanescent field. 
One consequence of this is that microbubbles also typically have smaller mode 
volumes than microspheres \cite{Riesen:16}. This indicates that the modes are 
more tightly constrained to the outer region of the resonator, 
serving to reduce the lasing threshold \cite{Spillane2002}, 
as well as providing greater scope for applications of nonlinear optics 
\cite{Schliesser2010207} and QED \cite{PhysRevA.83.063847}. 
The ability of microbubbles to stretch and deform mechanically has 
also been used to tune their FSRs across a broad range
\cite{Sumetsky:10}, allowing all characteristic modes within a spectral 
interval to be accessed. 

To make use of these advantages, microbubbles are in development for a variety 
of applications, which include refractive index sensing \cite{Yang:14}, lasing 
\cite{Lee:06} and frequency comb generation \cite{Li:13,Riesen:15}. 
In fact, the hollow cavity inherent in a microbubble can be used to develop 
new, compact apparatus, such as integrated microfluidics, which has 
demonstrated applicability in chemical and temperature sensing 
\cite{Sumetsky:10a,Berneschi:11}. 
Realising these benefits usually requires one to be able to manufacture 
microbubbles that have comparatively thin shell walls.   
However, in decreasing the shell-wall thickness, 
one encounters a trade-off between the refractive index sensitivity and the 
maximum values of the $Q$-factor \cite{Yang:14}. 
Achieving an \emph{optimal} trade-off in the 
sensitivity and the 
$Q$-factors has therefore been the subject of intensive study through a 
variety of methods in recent times \cite{Preston:15,7120938}, 
and still remains an open area of investigation. 

In this chapter, the FDTD  approach described in Chapter~\ref{chpt:sph} is 
extended to involve a single-layer architecture, including a layer of dipole 
sources whose density can be freely varied. This work is presented in the 
publication, Ref.~\cite{Hall:17}, listed as \textbf{Item 2} in 
Section~\ref{sec:jou}. 
Accurate determination of the FSR plays an important 
role in identifying the mode numbers from scattering theory \cite{Preston:15}  
and the assessment of dispersion \cite{Ristic:14,Riesen:16a}, and these 
attributes cannot be obtained from the wavelength shift in the modes alone, 
which depends on a number of factors, as described in Section~\ref{sec:nut}.
Therefore, the behaviour of the FSR and the $Q$-factors is examined as the 
diameter ($D$) and shell thickness ($d$) of a microbubble are changed, and the
 effect of the power collection region is also considered. 
Finally, the new insights into the behaviour of the microbubble WGM spectrum 
are incorporated into a new, \emph{non-destructive} method for determining the 
geometric parameters of a given microbubble. The method is then tested 
for a real-life scenario of a silica microbubble.

\section{\cdb Simulating fluorescent microbubble resonators}
\label{sec:sim}

Early models for simulating WGMs in microbubble resonators tend to 
focus on the wavelength positions of the resonances excited by a single 
embedded dipole \cite{PhysRevA.13.396}. 
By adopting a simple generalisation of Mie theory, it was found that the 
spectra could be simulated by making use of the transfer-matrix approach 
\cite{Liang:04}. Other approaches exist, such as the multilayer version of the 
well known Chew model \cite{Chew:76}, or the extraction of the mode positions 
from the characteristic equation of a microsphere with a single-layer coating 
\cite{Teraoka:06,Teraoka:07}. However, it is important to note that a unified 
description of resonators with arbitrary numbers of layers for a variety of 
excitation scenarios, including active layers, has not been developed until 
now, and its description and explanation will occupy the majority of 
Chapter~\ref{chpt:mod}.  

Using the FDTD approach, the simulation of microsphere resonators can be 
extended to include an outer shell of dielectric material  with greater 
flexibility in the ellipticity and homogeneity of the layer, allowing for 
greater scope for the inclusion of deformations than for competing models. 
However, because the simulations can suffer from artefacts derived from the 
finite grid size, which typically take the form of a systematic shift in the 
wavelength positions and $Q$-factors of the simulated modes (as discussed in 
Section~\ref{sec:FDTD}), in this chapter, the grid resolution, $\Delta x$, is 
varied depending on the physical scale of the simulation, as set by the 
resonator diameter. The diameter, from this point on, refers to the outermost 
diameter of the microbubble, so that the outer boundary remains constant as 
the shell thickness is changed. For each diameter simulated, a resolution is 
chosen such that the positions of the modes reach convergence within a 
tolerance of $1\%$.\footnote{For the smallest and largest diameters considered 
($8$ $\mu$m and $16$ $\mu$m), the grid size varies from $\Delta x = 26.9$ nm 
to $\Delta x = 35.0$ nm, respectively. This range of diameters is chosen to 
overlap with the estimated diameter range of the silica microbubbles used in 
the experiment described in Section~\ref{sec:nondest}.}

As mentioned in Section~\ref{sec:FDTD},  the simulation of large resonator 
diameters becomes computationally challenging, particularly in terms of RAM 
usage,\footnote{For the most demanding scenario considered in this study in 
terms of RAM usage, the largest diameter simulated, $D=16$ $\mu$m, with a grid 
size of $\Delta x = 35.0$ nm, typically requires 235 GB for an unnormalised 
spectrum collection for a microbubble resonator. The normalisation run 
(no microbubble present) requires 219 GB. For $D=12$ $\mu$m ($\Delta x = 30.4$ 
nm) RAM usages are 156 GB (microbubble) and 145 GB (normalisation), and for 
$D=8$ $\mu$m ($\Delta x = 26.9$ nm), the requirements are 84 GB (microbubble) 
and 79 GB (normalisation), respectively. These values are based on resources 
from the Raijin machine at the National Computational Infrastructure (NCI) 
Facility, The Australian National University, Canberra, Australian Capital 
Territory 0200, Australia 
(\href{http://nci.org.au/systems-services/national-facility/peak-system/raijin}
{\cb http://nci.org.au/systems-services/national-facility/peak-system/raijin
\cbl}).} due to the difference in magnitude between the wavelength and the 
diameter of the resonator. Although FDTD is in principle extendable to larger 
diameters, as commonly used in the literature \cite{doi:10.1117/12.2037992,
Henze:11,Foreman:14}, the limitations of computing power, which scale 
proportionally with $D/\lambda$ to the fourth power \cite{Hall:15}, need to be 
taken into consideration.

\subsection{\cdb Uniform dipole coatings as an analogue for fluorescence}
\label{subsect:flu}

\begin{figure}
\begin{center}
\includegraphics[height=156pt]{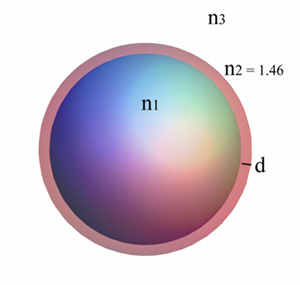}
\vspace{-3mm}
\includegraphics[height=146pt]{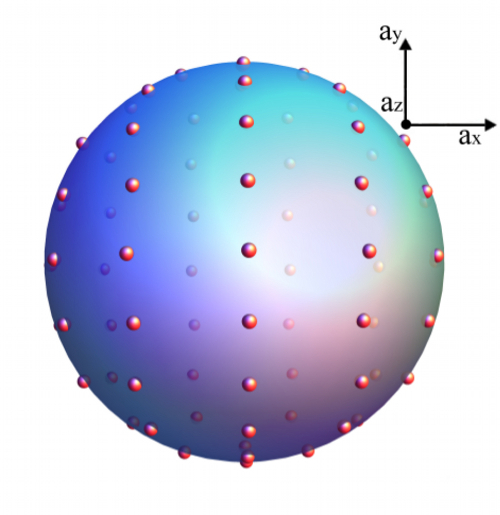}\\
\mbox{\hspace{0.35cm}(a)\hspace{5.15cm}(b)}
\end{center}
\vspace{-4mm}
\caption[The structure of the shell resonator is illustrated.]{{(a) An 
illustration of the shell structure of concentric spheres, as 
used in the FDTD simulation, with shell thickness $d$. 
(b) The distribution of electric dipole sources placed on the outer surface of 
the microbubble is shown accurately for $75$ sources as used in the simulation. 
The alignment of each dipole source contains components in the $x$, $y$ and 
$z$ directions, with amplitudes given by $a_x$, $a_y$ and $a_z$, respectively. 
Each dipole source has a random alignment, but the same total amplitude.}}
\label{fig:Shell}
\figrule
\end{figure}

Let us consider how a fluorescent coating may be modelled in the context of 
FDTD. First, a microbubble resonator is simulated by placing two spheres 
concentrically, with refractive indices as shown in Fig.~\ref{fig:Shell}(a). 
A distribution of electric dipole sources is then simulated on the surface of 
the microbubble resonator. The density of dipoles on the surface is kept 
constant for each simulated microbubble, as illustrated in 
Fig.~\ref{fig:Shell}(b), although varying the density is straightforward, and 
a variety of different densities will be considered in Section~\ref{sec:bubbehav}. 

The algorithm for the placement of the dipoles is as follows. In spherical 
coordinates, the polar angle ($\theta$) is split into $N$ increments, with 
$\theta_i = \pi\,i/N$, for $i=0,\ldots,N$. The azimuthal angle ($\phi$) is 
chosen to take the values  $\phi_j=2\pi\,j/(1 + [N\sin\theta])$, for 
$j=0,\ldots,[N\sin\theta]$, where the square brackets indicate 
rounding to the closest integer. That is, to keep the density of the sources 
uniform over the surface of the resonator, the number of sources must increase 
as the value of $\theta$ moves away from the pole at $\theta=0$. 
The total number of sources placed on the surface, $\mathcal{N}$, is then 
simply a function of $N$
\begin{equation}
\mathcal{N} = \sum_{i=0}^{N} [N \sin(\pi\,i/N)+1]. 
\end{equation}

It can be seen that the surface density ($\sigma$) is uniform by showing that 
the number of sources per infinitesimal solid angle 
($\mathrm{d}\mathcal{N}(\theta,\phi)/\mathrm{d}\Omega$) is constant as $N$ 
becomes large
\begin{align}
\mathcal{N}(\theta,\phi) &= \frac{1}{\pi}\mathrm{d}\theta\times
\frac{\sin\theta}{2\pi}\mathrm{d}\phi,\\
\frac{\mathrm{d}\mathcal{N}(\theta,\phi)}{\mathrm{d}\Omega} &\equiv \sigma = 
\frac{1}{2\pi^2}\frac{\mathrm{d}}{\mathrm{d}\Omega}
(\sin\theta\,\,\,\mathrm{d}\theta\mathrm{d}\phi) = 
\mbox{constant}.
\end{align}

Each source is placed on the surface of the resonator, with diameter $D$, at 
the locations $(\frac{D}{2}\sin\theta\,\cos\phi,\frac{D}{2}\sin\theta\,
\sin\phi,\frac{D}{2}\cos\theta)$. The amplitude of each source 
is represented by a single complex number premultiplying the source 
current of Eq.~(\ref{eqn:src}). The total amplitude $A$ 
is normalised 
to the total number of sources placed on the surface of the resonator, 
and is kept constant throughout the simulation. 

In order to model a \emph{fluorescent} coating, which has no preferred 
orientation, the alignment of each source must be randomly chosen. For an 
amplitude of $A$, the $x$, $y$ and $z$ components of each source take the form
\begin{align}
a_x &= A\cos\hat{\theta}\,\sin\hat{\phi},\\
a_y &= A\sin\hat{\theta}\,\sin\hat{\phi}, \\
a_z &= A\cos\hat{\theta},
\end{align}
where $\hat{\theta}$ and $\hat{\phi}$ take values from a uniform, 
random distribution
of polar and azimuthal angles in the alignment space, 
over the ranges $[0, \pi)$ and $[0, 2\pi)$, respectively. 

This method of assigning the locations of dipole sources provides 
a useful way of automating the simulation of an even coating of a fluorescent 
material on a resonator, or the autofluorescence of a cell membrane, 
and exploring the effect of different densities of the sources on the emitted 
spectrum of a resonator. One advantage of the FDTD method over the analytic 
models described in Chapter~\ref{chpt:sph} is the fact that interactions 
between the sources are automatically handled. This allows one to probe 
more general properties of the electromagnetic spectra derived from 
resonators with unconventional or anisotropic source distributions 
that are otherwise difficult to model.

\subsection{\cdb Customising the fluorescent emitter density in FDTD}

As an initial example, consider the polystyrene microsphere described in 
Section~\ref{sec:spec} such that the implementation of the new mode excitation 
method of Section~\ref{subsect:flu} can be directly compared with the 
results in Chapter~\ref{chpt:sph}. 
The microsphere has a diameter of $6$ $\mu$m, is surrounded by water, 
and the radiation is collected from a box surrounding the entire resonator. 
The dipole-coating algorithm is now applied to this microsphere for a variety 
of different densities of sources placed on the surface of the sphere. 
In this scenario, each dipole emits a Gaussian pulse 
with a central wavelength of $700$ nm, and an emission width of $80$ nm, which 
is chosen to match the emission range of (4-(2-Carboxyphenyl)-7- 
diethylamino-2-(7-diethylamino-2-oxochroman-3-yl)-chromylium perchlorate) 
\cite{Klantsataya:16} as closely as possible, which is the organic dye used in 
the experiment described in Section~\ref{sec:case}. 

The power is then aggregated in the wavelength domain, in the same manner as 
Chapter~\ref{chpt:sph}. 
Recalling that the total radiation collection time necessary to achieve a 
converged WGM spectrum depends on the resolution, it was found 
in Section~\ref{sec:spec} that a 
collection time of approximately $1.0$ ps was sufficient to achieve 
a converged $Q$-factor for a $6$ $\mu$m diameter microsphere, 
using a shorter wavelength of $600$ nm. 
In order to achieve the same level of convergence 
as the diameter and wavelength are varied from these values, 
the collection time 
must be scaled in proportion to $D/\lambda_0$, where $\lambda_0$ is the 
central wavelength of the dipole sources. For a resonator with a 
diameter of $8$ $\mu$m excited at a wavelength of 
$700$ nm, the collection time is $571$ wavelengths (equivalent to $1.33$ ps). 
For a diameter of $D=16$ $\mu$m, a collection time of $1143$ wavelengths 
($2.67$ ps) is required. Since the larger diameter resonators exhibit larger 
$Q$-factors and thus longer ring-down times, a longer collection time is 
required in order to reach a converged spectrum. 

\begin{figure}[H]
\begin{center}
\includegraphics[width=0.65\hsize]{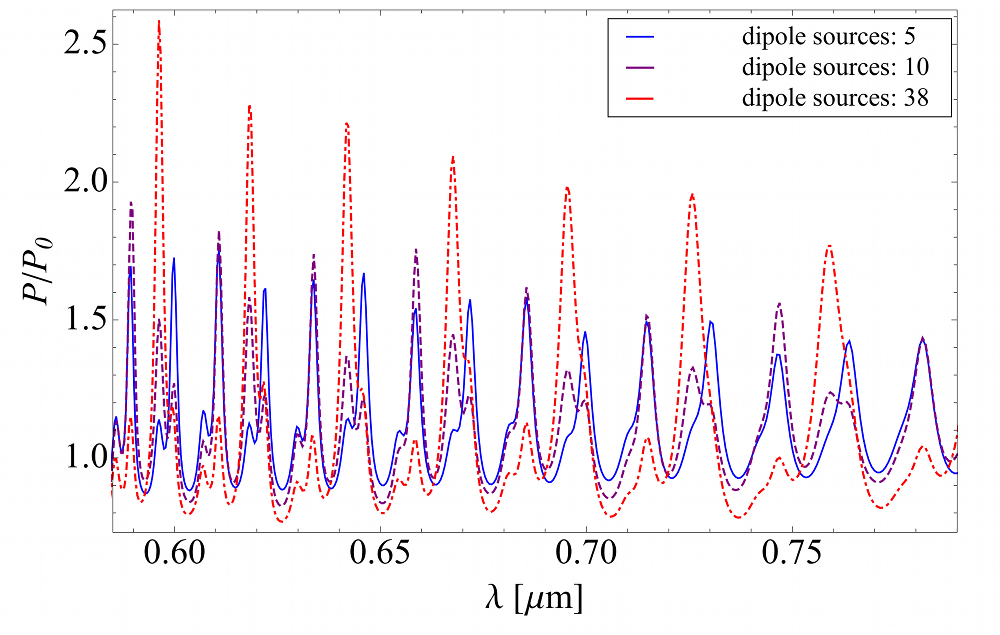}\\
\vspace{-5mm}
\hspace{6mm}\mbox{(a)}\\
\includegraphics[width=0.65\hsize]{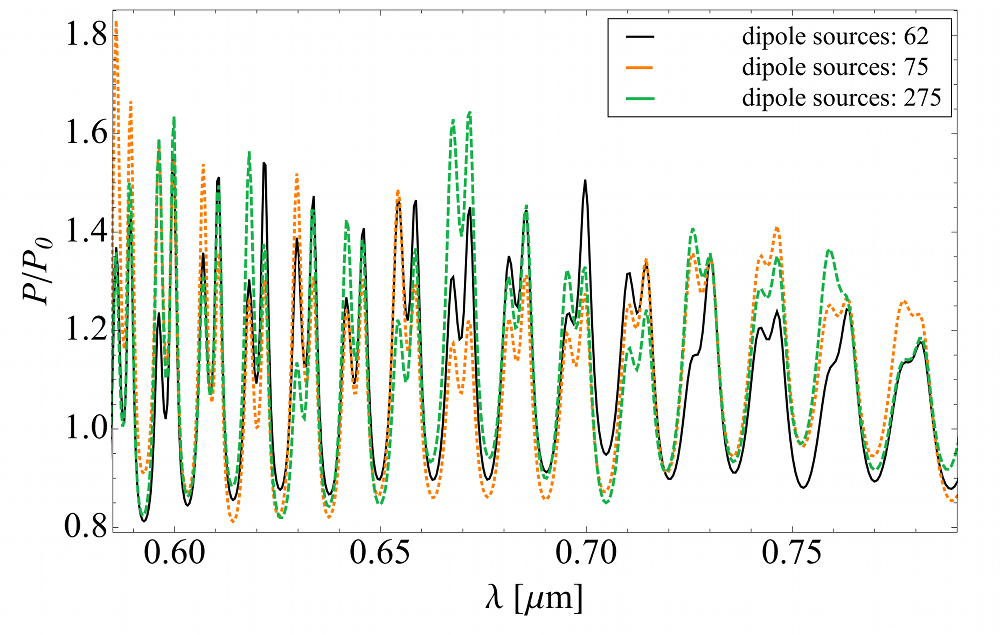}\\
\vspace{-5mm}
\hspace{6mm}\mbox{(b)}\\
\vspace{2mm}
\includegraphics[width=0.6\hsize]{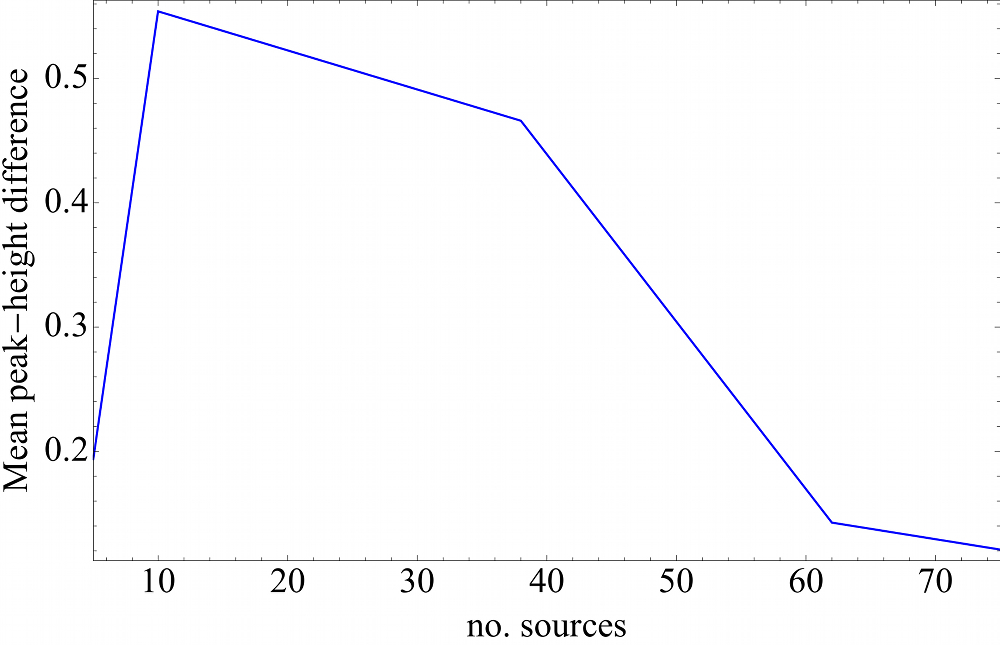}\\
\vspace{-3mm}
\hspace{6mm}\mbox{(c)}\\
\vspace{-3mm}
\end{center}
\caption[Spectra are compared for different numbers of dipole sources.]
{{A polystyrene microsphere, $6$ $\mu$m in diameter, in a 
surrounding medium of water, is coated with a distribution of electric 
dipole sources. (a) The normalised spectra are shown for numbers of sources 
from $5$ to $38$, and (b) from $62$ to $275$. (c) The difference in the height 
of a given peak as the number of sources is increased is repeated for $20$ 
prominent peaks and the mean value taken.}}
\label{fig:sourceconv}
\figrule
\end{figure}

The behaviour of the WGM spectrum as a function of the number of sources, and 
hence the source density, is important for developing a reliable model of a 
uniform fluorescent coating. Figures~\ref{fig:sourceconv}(a) and (b) show the 
spectra within a wavelength window of $500$ to $750$ nm, and illustrate how 
the relative peak heights are altered as the number of sources is changed. 
This effect is due to the fact that randomly 

\newpage 
\noindent oriented sources will exhibit 
a preference for WGMs depending upon their specific alignment. For example, 
as was shown in Section~\ref{sec:spec}, electric dipole sources with 
a large tangential component to the surface of the sphere will couple 
more strongly to the TE modes, whereas electric dipole sources with 
 a large radial component will couple only to the TM modes, as shown in 
Figs.~\ref{fig:WGM} and \ref{fig:WGM2}. 
As the number 
of sources is increased, this effect is averaged out, with 
additional sources of any orientation making little difference to the spectrum 
if the number of existing sources is sufficiently high. 

Figure~\ref{fig:sourceconv}(c) shows the trend in the difference in the height 
of a given peak as the number of sources is changed to a higher value. This 
procedure is carried out for $20$ prominent peaks within the wavelength window, 
and the mean value is then taken. It is found that the difference drops below 
$10$\% for $N>9$, corresponding to more than $62$ sources. Therefore, from now 
on, a value of $N=10$ is chosen, corresponding to $75$ sources, which 
represents a sufficiently even coating for the purposes of studying the 
behaviour of coated resonators. Note that, for a small number of sources, 
there is a sharp increase in the mean value of the peak-height difference, 
which is an artefact of the coincidental alignment of small numbers of sources, 
resulting in similar peak heights for the prominent modes. 

\begin{figure}[H]
\begin{center}
\includegraphics[width=0.8\hsize]{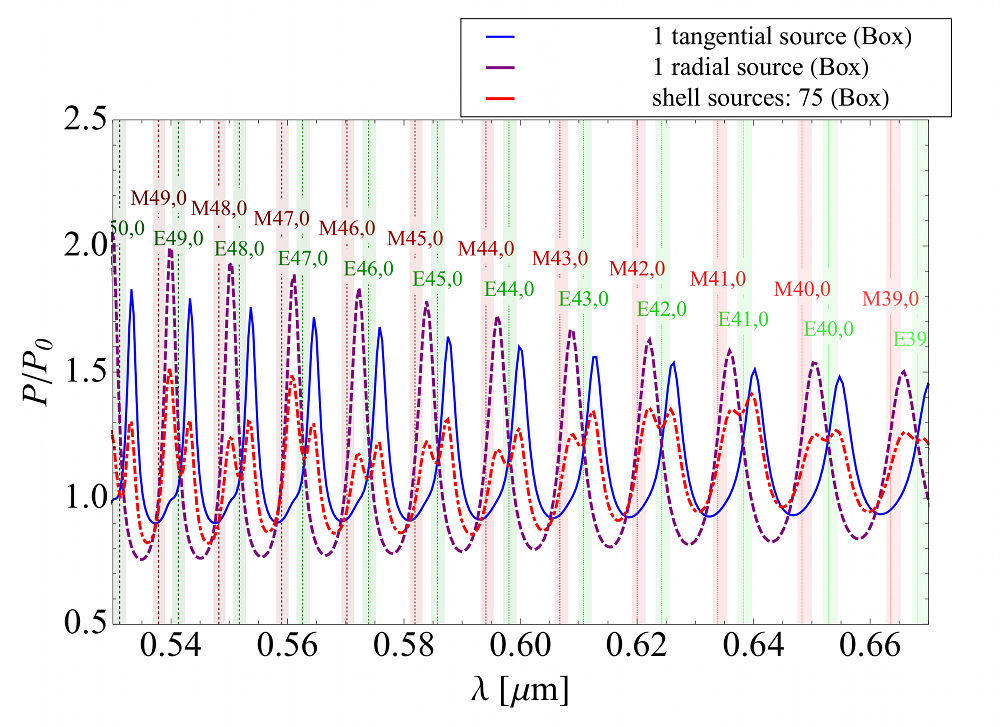}\\
\vspace{-5mm}
\hspace{4mm}\mbox{(a)}\\
\includegraphics[width=0.8\hsize]{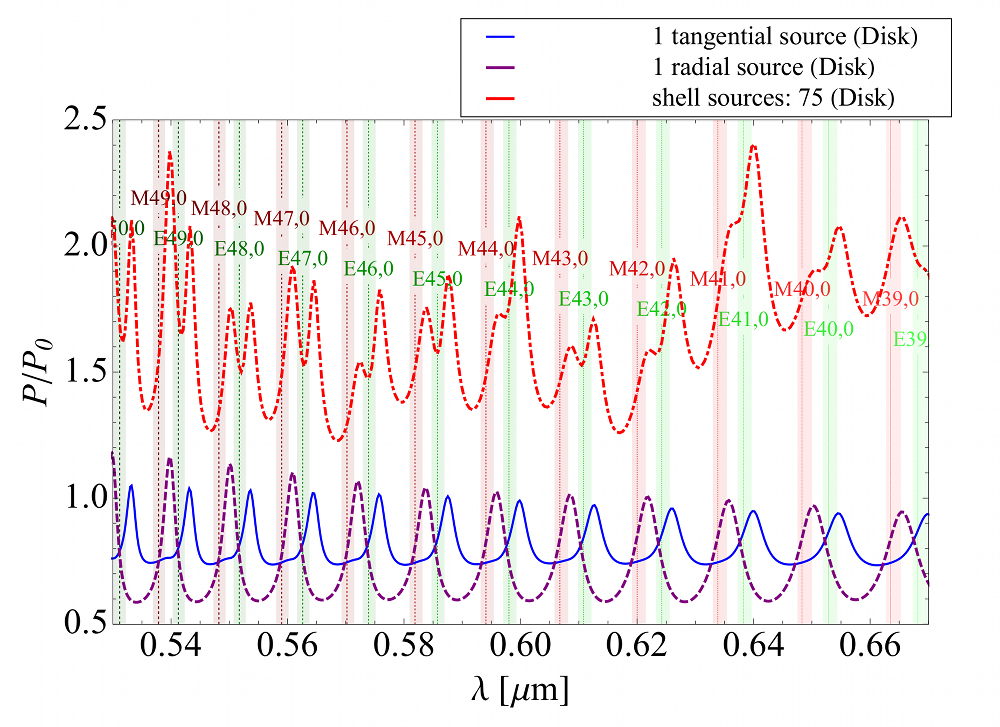}\\
\vspace{-5mm}
\hspace{4mm}\mbox{(b)}\\
\end{center}
\caption[The effect of the flux collection region on the spectrum is 
examined.]{{The spectra and mode positions of a polystyrene 
microsphere, $6$ $\mu$m in diameter, in a surrounding medium of water, for 
a single tangential or radial electric dipole source, or a distribution of 
$75$ sources placed using the algorithm of Section~\ref{subsect:flu}. 
(a) The total power is collected 
from a box that surrounds the microsphere. (b) The power is collected from a 
circular flux region with a diameter of $2.58$ $\mu$m, placed a distance of 
$240$ nm from the surface of the sphere in the $x$-direction, with its normal 
aligned along the same axis, as in Chapter~\ref{chpt:sph}.}}
\label{fig:sourcecomp}
\figrule
\end{figure}

A comparison of the spectra resulting from a distribution of $75$ electric 
dipole sources, and a single dipole in a tangential or radial orientation, is 
shown in Fig.~\ref{fig:sourcecomp}. The power is collected from a circular or 
box region, as defined in Section~\ref{sec:spec}. It can be seen that the 
distribution of sources, each with random orientation, couples to both the TE 
and TM modes, while the single dipole source preferentially couples to one 
particular polarisation exactly as described above.  
%The radial source only couples to the TM modes, 
%whereas the tangential source, which couples to both TE and TM modes, provides 
%a much greater contribution to the TE mode. 

Although it has been shown that the flux collection region affects the 
relative coupling strengths of the modes, for the purposes of analysing the 
behaviour of the spectrum with respect to the geometric variables 
or the refractive index contrast, the focus from here on will be on the total 
radiation collected from a box region. 

\subsection{\cdb Understanding the mode behaviour of microbubbles}
\label{sec:bubbehav} 

\begin{figure}[H]
\begin{center}
\includegraphics[width=0.8\hsize]{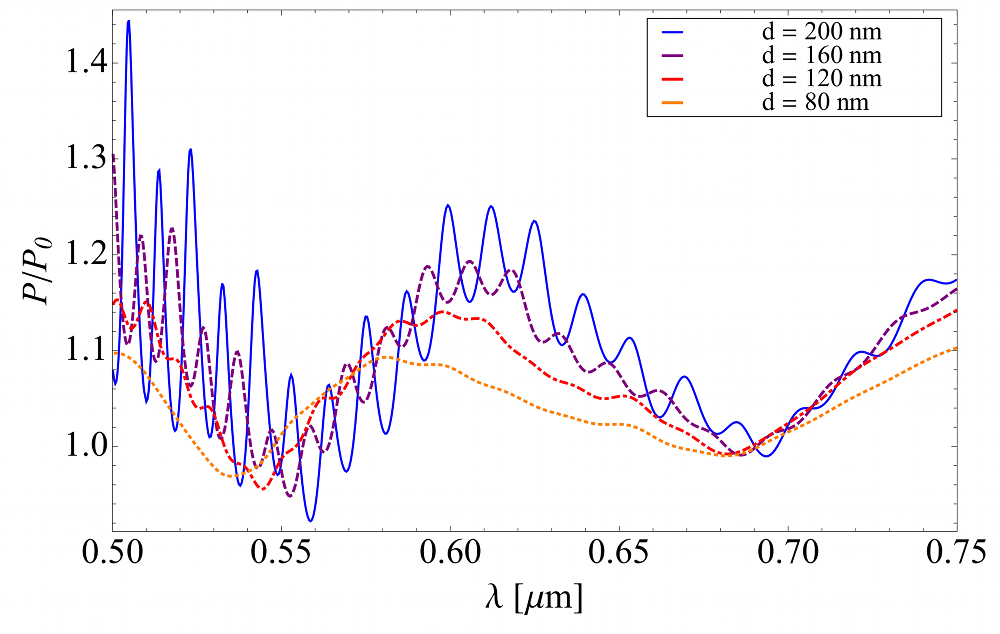}\\
\vspace{-5mm}
\hspace{6mm}\mbox{(a)}\\
\vspace{2mm}
\includegraphics[width=0.8\hsize]{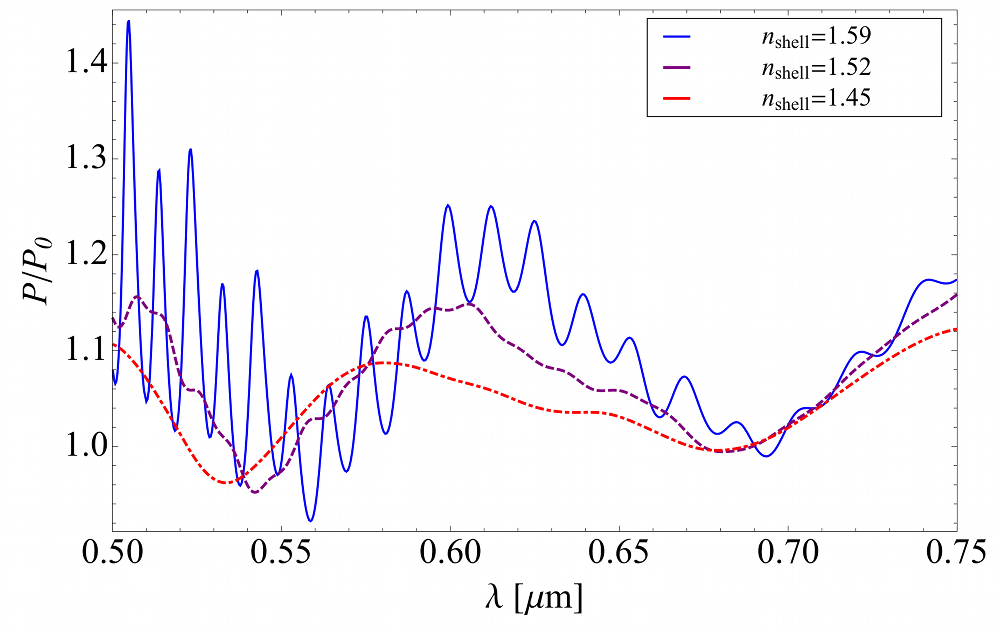}\\
\vspace{-5mm}
\hspace{6mm}\mbox{(b)}\\
\end{center}
\caption[The geometric parameters for microbubble spectra are modified.]
{{The behaviour of the WGM spectra of a $6$ $\mu$m diameter polystyrene 
microbubble, surrounded by water, and coated with a uniform layer of dipole 
sources. (a) The spectra are shown for a range of shell thicknesses.  
It is evident that mode loss occurs for $d \lesssim 120$ nm. 
(b) Spectra are shown for three values of refractive index 
of the shell layer. As the refractive index 
of the material is changed, mode loss is apparent for $\mathrm{n}_2 = 
\mathrm{n}_{\mathrm{shell}} \lesssim 1.52$.}}
\label{fig:d6comp}
\figrule
\end{figure}

The behaviour of WGMs as a function of microbubble shell thickness 
is now analysed for a number of examples. Consider the case of the  
 polystyrene microbubble with an outer diameter of $6$ $\mu$m, with water both 
inside and outside the resonator, and coated with a fluorescent layer. 
As the thickness of the shell wall is varied across a range of $80$ to 
$200$ nm, as shown in Fig.~\ref{fig:d6comp}(a), several effects are apparent. 
The mode positions do not remain the same as the thickness changes, 
and this behaviour is not linear, but crucially dependent on the structure 
of the characteristic equation, as seen in the Teraoka-Arnold model 
introduced in Section~\ref{sec:an}, and explored in more detail 
in the case of the multilayer model in Section~\ref{sec:struc}.
In addition, the peak height dramatically changes as the thickness decreases 
-- an effect that will be investigated further in Section~\ref{sec:demo}. 
Note that, for a sufficiently thin shell wall, there is a total loss of modes, 
as the dielectric layer is not able to contain the radiation and WGMs cannot 
therefore be sustained. 

In Fig.~\ref{fig:d6comp}(b), the shell thickness is kept fixed at $200$ nm, 
but it is the refractive index of the dielectric layer that is varied. As the 
refractive index decreases, there is also a reduction of the 
peak heights, resulting in mode loss if the index is too close to that of the 
surrounding medium. This provides a preliminary indication of the effect of 
the refractive index contrast from different materials in sustaining WGMs.

These two simple illustrations serve to highlight an important feature of 
layered resonators, namely, that the thickness of the shell and the refractive 
index contrast function as dual criteria in determining the ability of a 
structure to sustain WGMs. These effects will now be investigated more 
quantitatively, which will then become an important consideration in 
Chapter~\ref{chpt:cel} when the results are applied to real-life cells.

\section{\cdb Fluorescent silica microbubble case study}
\label{sec:case}

Although polystyrene microresonators have been the focus thus far, serving as 
a template to which additions and augmentations to the FDTD method have been 
added, the focus is now shifted to microbubbles, choosing a specific 
material that is easily 

\newpage
\noindent obtained for experimental purposes. Hollow silica 
glass beads\footnote{Polysciences, Inc., $500$-$548$ Valley Rd, Warminster, 
PA $18974$, USA, \textit{Hollow Glass Beads}, Catalogue Number: $19823$.} will 
be used in this section in order to verify the modelling capabilities explored 
up to this point, and to aid the investigation into the spectral properties of 
microbubble resonators. 

\begin{figure}[t]
\begin{center}
\includegraphics[width=0.95\hsize]{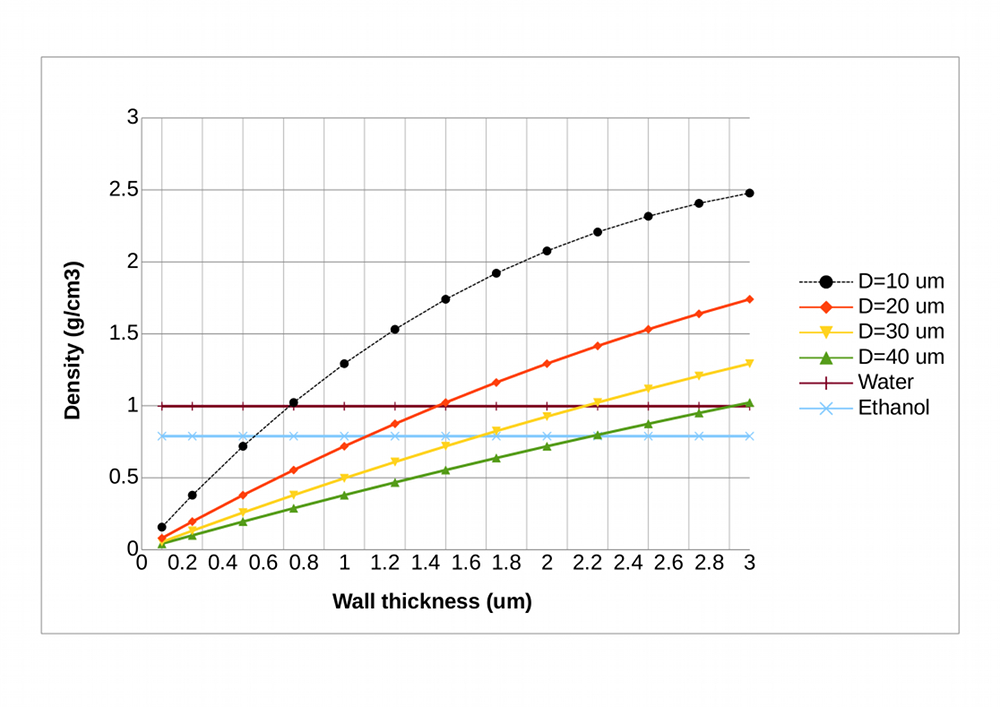}\\
\vspace{-3mm}
\end{center}
\caption[The density of silica microbubbles is measured experimentally.]
{{The density of the silica microbubble resonators as a function of the 
shell-wall thickness. The results are shown for a number of outer diameter 
values, $D$. The density of water and ethanol are both shown, indicating the 
bounds for the shell thicknesses of the microbubbles collected at the start of 
the experiment.}}
\label{fig:dens}
\figrule
\end{figure}

Initially, batches of silica microbubbles, with refractive index 
$\mathrm{n}_2=1.46$ at $\lambda\approx 700$ nm, are 
inspected in order to obtain approximate ranges for the size and shell 
thickness. They are then placed alternately in liquids of different densities 
in order to provide bounds on the shell thicknesses. The microbubbles are then 
suspended in ethanol, which has a density of $0.789$ g/cm$^3$ near room 
temperature. Those that sink to the bottom of the container are re-suspended 
in water, with a density of $0.997$ g/cm$^3$, and those that float to the top 
are thus considered to have a density between that of ethanol and water. 
Using this procedure, the 
behaviour of the microbubble density as a function of the shell-wall thickness 
may be plotted, as shown in Fig.~\ref{fig:dens} 
for a range of diameter values. 

The results show that the shell thicknesses of the microbubbles 
lie in the range $d=0.10\,D$ to $0.15\,D$. 
In addition to this experimental measurement, 
a visual inspection of a selection of microbubbles using confocal 
microscopy yielded a similar result, 
with a slightly higher upper bound of $0.18\,D$. As a result, a conservative 
estimate of the shell thickness is $0.10\,D$ to $0.18\,D$. This particular 
range of shell thicknesses is considered in order to restrict the parameter 
space to the values expected from the density measurements.

The microbubbles are then coated with the carboxyl-functionalised organic dye 
(4-(2-Carboxyphenyl)-7- diethylamino-2-(7-diethylamino-2-oxochroman-3-yl)-\\
chromylium perchlorate \cite{Klantsataya:16}, 
and a single droplet containing the microbubbles in solution is 
placed onto a microscope coverslip, into an oven at $80^\circ$C, and left to dry.
 
\begin{figure}[t]
\begin{center}
\includegraphics[width=0.8\hsize]{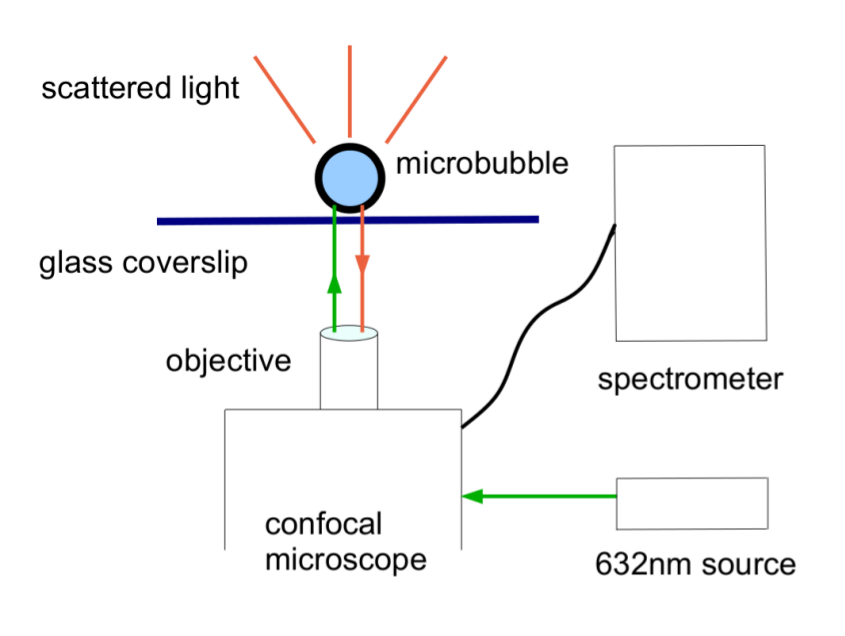}\\
\vspace{-3mm}
\end{center}
\caption[The experimental setup for measuring spectra is illustrated.]
{{The experimental setup used for the collection of the WGM spectra 
of a silica microbubble coated with a carboxyl-functionalised organic dye.}}
\label{fig:exp}
\figrule
\end{figure}

The microbubble WGMs are excited in free space by a $5$ mW, $632$ nm 
wavelength, continuous-wave laser, using an inverted confocal microscope 
setup as shown in Fig.~\ref{fig:exp}. 
The emitted power is 
collected in free space with the same microscope, and passed to a spectrometer. 

The fluorescence excites modes in multiple planes of symmetry in the resonator. 
Note that since only a small number of modes interacts with the coverslip 
beneath, their contribution to the total spectrum collected is minimised, 
and thus there is a negligible effect on the spectrum as measured using 
this technique \cite{s90906836}. 

\begin{figure}[t]
\begin{center}
\includegraphics[width=0.8\hsize]{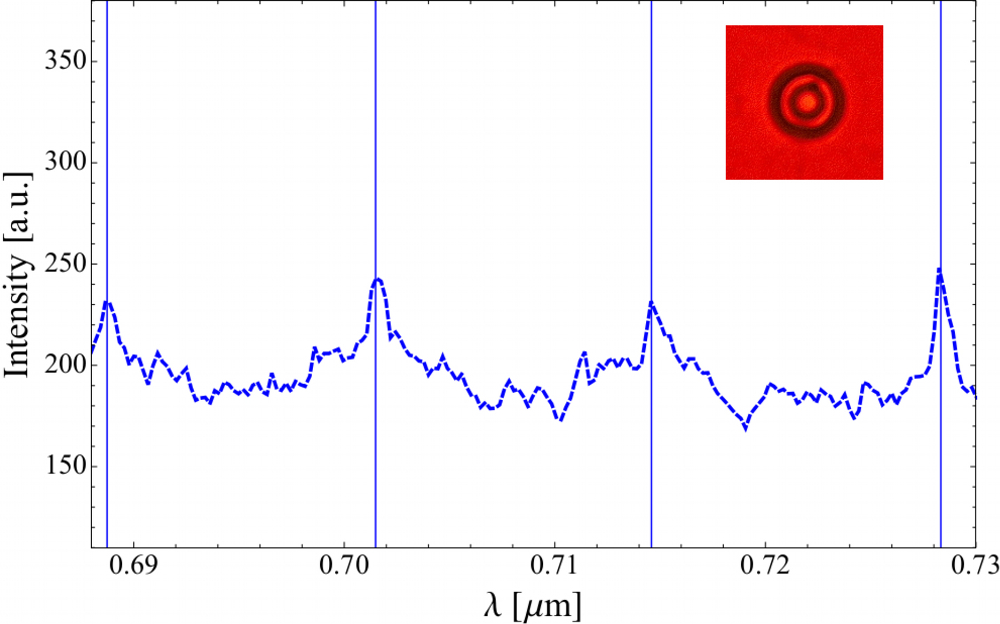}\\
\vspace{-3mm}
\end{center}
\caption[The measured spectrum from a dye-coated microbubble is shown.]
{{The WGM spectrum measured from a dye-coated silica microbubble. 
The vertical lines indicate the positions of the prominent modes from which 
the FSR can be estimated to be $0.0130 \pm 0.0006$ $\mu$m. 
\emph{Inset:} Silica glass microbubble image, manufactured with a diameter 
between $5$ and $20$ $\mu$m. The shell thickness is estimated 
to be between $0.10\,D$ and $0.18\,D$. Note that the results have been 
smoothed in order to remove a portion of the noise.}}
\label{fig:FSR}
\figrule
\end{figure}

The measured WGM spectrum is shown in Fig.~\ref{fig:FSR}, 
with vertical lines 
placed over the prominent modes to assist in the measurement of the FSR, which 
is found to be $0.0130 \pm 0.0006$ $\mu$m. 

\begin{figure}[t]
\begin{center}
\includegraphics[width=0.335\hsize]{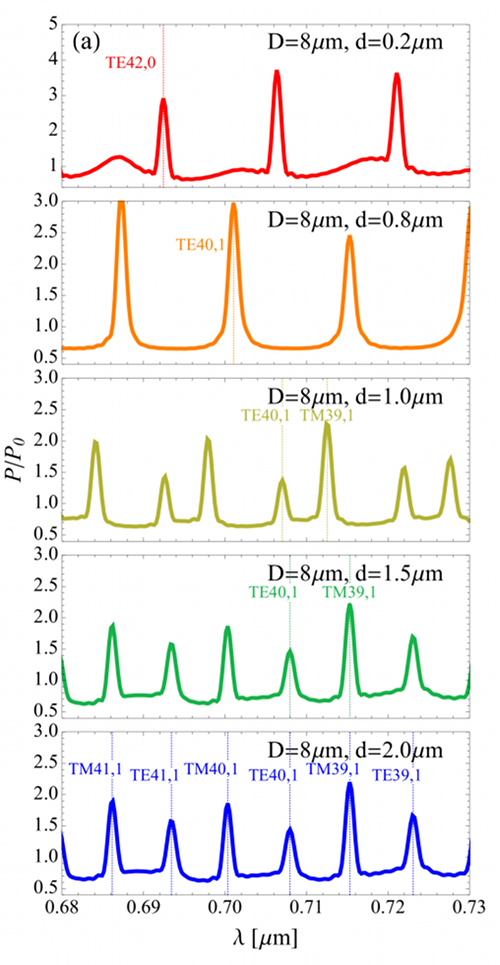}
\hspace{-2.5mm}
\includegraphics[width=0.335\hsize]{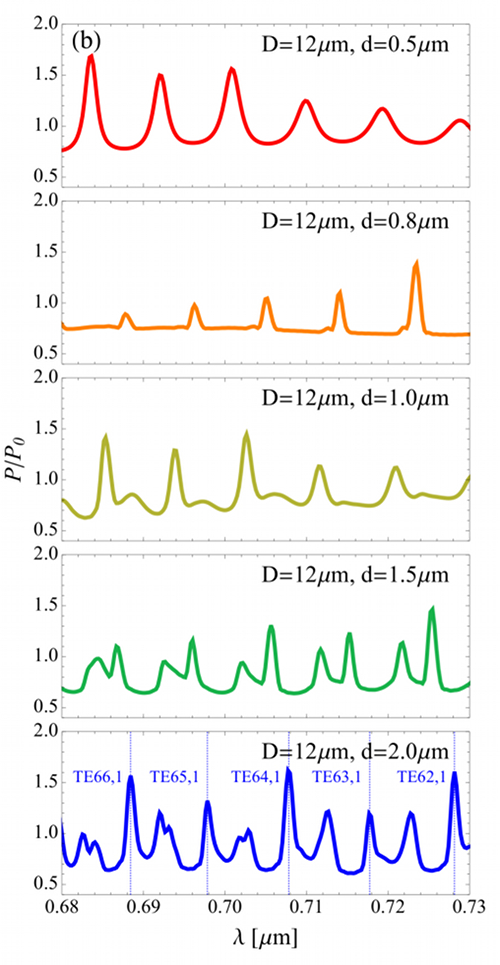}
\hspace{-2.5mm}
\includegraphics[width=0.335\hsize]{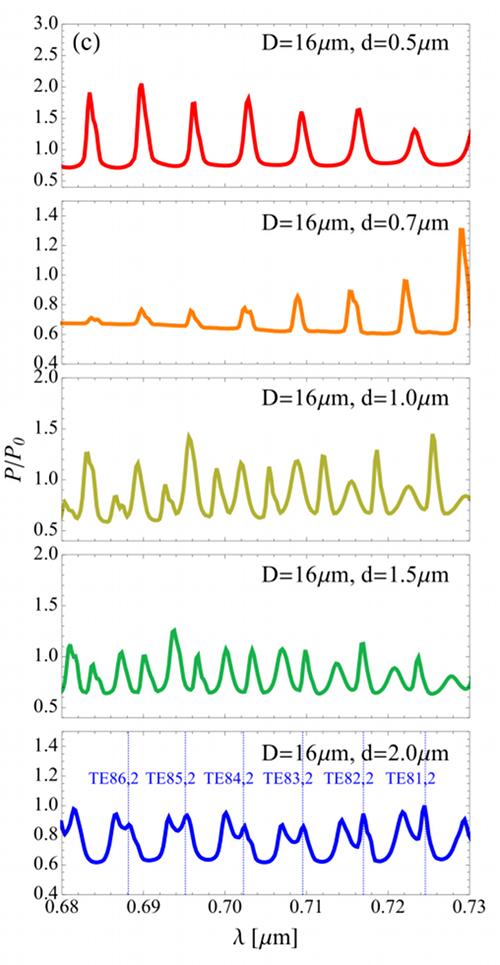}\\
\vspace{-3mm}
\end{center}
\caption[The behaviours of simulated microbubble spectra are explored.]
{{The behaviour of a sample of normalised WGM spectra for fluorescent-coated 
silica microbubbles in a surrounding medium of air. The resonances are 
plotted over a restricted wavelength window of $680$ to $730$ nm, 
for clarity. Multiple spectra are plotted in order to track the changes as a 
function of shell thickness, $d$, for a selection of diameters: 
(a) $D=8$ $\mu$m, (b) $D=12$ $\mu$m, and 
(c) $D=16$ $\mu$m. For the largest thickness, $d=2.0$ $\mu$m, vertical lines 
indicate the positions of the prominent TE modes. The TM modes are the 
dominant peaks in between the TE modes, and are suppressed as $d$ decreases.}}
\label{fig:dcomp}
\figrule
\end{figure}

The FDTD method is now used to simulate silica microbubbles of comparable size 
to the aforementioned experiment, and investigate the properties of the WGMs. 
A number of representative diameters and shell thicknesses is selected, and 
the change in the power spectrum  with respect to the shell thicknesses is 
shown in Fig.~\ref{fig:dcomp}. The positions of the modes are indicated in the 
bottom row ($d=2.0$ $\mu$m), from which the FSR can be calculated. The 
assignment of the mode numbers is achieved by selecting the closest mode to
 each peak, as obtained from a single-layer coated microsphere analytic model 
used to represent the microbubble \cite{Teraoka:06,Teraoka:07}. 

For the smallest diameter value $D=8$ $\mu$m in Fig.~\ref{fig:dcomp}(a), the 
modes in the spectra are well spaced, and there are no higher radial harmonics 
present, unlike the spectra for diameters $12$ and $16$ $\mu$m in 
Figs.~\ref{fig:dcomp}(b) and (c), respectively, which exhibit a number of 
higher order modes, especially for larger shell thicknesses. This larger 
quantity of nearly-overlapping modes makes the extraction of the mode 
positions, and especially the $Q$-factor, problematic. When examining purely 
the scaling behaviour of the $Q$-factor in Section~\ref{sec:fQ}, it is convenient 
to focus only on the geometric component of the $Q$-factor prior to the 
introduction of the radiative contributions, which affect the $Q$-factor 
through the choice of excitation method and the relative coupling strength of 
the emitted power to the modes. However, an extraction of the $Q$-factor 
directly from the spectrum, if it can be achieved, represents a more realistic 
assessment of the $Q$-factor than the geometric component alone, and these 
effects are incorporated into the $Q$-factor assessments performed in 
Section~\ref{sec:nondest}. 

For the larger shell thicknesses, the spectra are expected to converge to the 
microsphere results, since the outer diameter remains fixed as the thickness 
is increased. In addition, little variation in the WGM spectrum is 
apparent for thicknesses greater than $1.5$ $\mu$m; however, for $d<1$ $\mu$m, 
there is a sudden onset where the mode positions become highly sensitive to 
the shell thickness. 

Note that, as in Section~\ref{sec:spec}, the fundamental radial modes ($n=0$), 
which typically exhibit high $Q$-factors, are not able to be resolved at this 
grid resolution \cite{Hall:15}, unless the $Q$-factor drops within a measurable 
range. This can be seen in the top plot of Fig.~\ref{fig:dcomp}(a), where the 
diameter and shell thickness are both sufficiently low that the fundamental 
radial mode can be resolved. For thin shell thicknesses (top row), it is the 
fundamental \emph{TE modes} that become the most prominent mode for small 
shell thicknesses, and the TM modes become more difficult to resolve. 
In fact, the suppression of the TM modes for small shell thicknesses has been 
noted in a previous study \cite{Yang:14}, where the rapid reduction in the 
$Q$-factors in thin shell microbubble resonators was identified as an 
intrinsic property of resonator geometry. As will be explored in more 
depth with the benefit of the multilayer model developed in 
Chapter~\ref{chpt:mod}, particularly in Section~\ref{sec:demo}, 
it will become apparent that the dominance of the TE modes originates 
from the structure of the matrix element components that comprise the 
separate polarisation components of the total emitted power spectrum. 
Effectively, the component of the power in the radial direction becomes 
suppressed as the thin shell thickness can no longer support modes that 
oscillate with this orientation, which are TM modes. 
Similar mode behaviour has also been 
predicted for the radiation of dipoles in the vicinity of optical fibres 
\cite{AfsharV:14}.

\subsection{\cdb Free spectral ranges and $Q$-factors}
\label{sec:fQ}

The free spectral ranges (FSRs) of microbubbles with the selection of 
diameters and shell thicknesses 
considered are shown in Table~\ref{tab:FSR}. Note that the FSR should not be 
considered a \emph{constant} quantity over a large wavelength range, and 
care must be taken to ensure that this $\delta$FSR is consistent between 
simulation and experiment. This is addressed in the following section.

For large shell thicknesses relative to the diameter of the microbubble, the 
FSR does not vary significantly as the shell thickness changes, since the WGMs 
travel almost entirely within the shell, and are insensitive to the presence 
of the hollow interior region. As the shell thickness decreases, however, the 
evanescent field extends partially into the interior of the microbubble, and 
the FSR then begins to deviate from the microsphere value. 

\begin{figure}[t]
\begin{center}
\includegraphics[width=0.5\hsize]{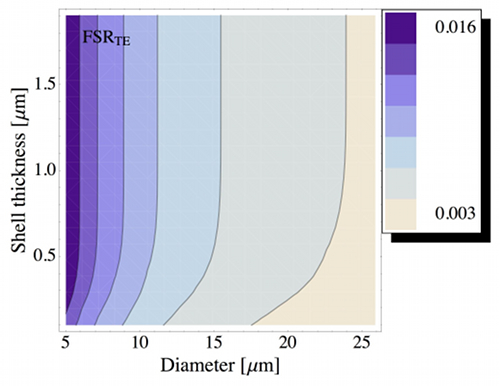}
\hspace{-3mm}
\includegraphics[width=0.5\hsize]{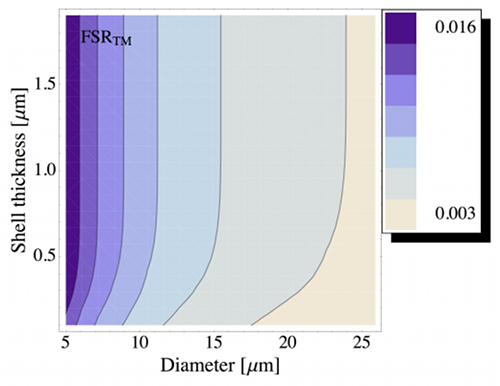}\\
\vspace{-2.5mm}
\hspace{-10mm}\mbox{(a)\hspace{6.5cm}(b)}\\
\includegraphics[width=0.5\hsize]{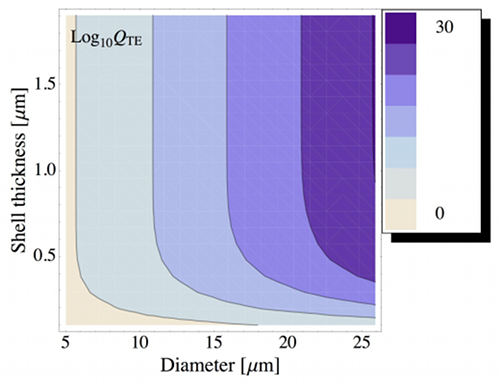}
\hspace{-3mm}
\includegraphics[width=0.5\hsize]{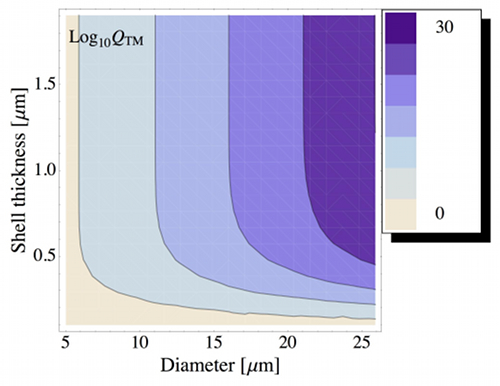}\\
\vspace{-2.5mm}
\hspace{-10mm}\mbox{(c)\hspace{6.5cm}(d)}\\
\vspace{-2mm}
\end{center}
\caption[Contour plots are shown for the free spectral ranges and $Q$-factors.]
{{Contour plots generated for a range of diameters and shell thicknesses of 
silica microbubbles, using a single-layer analytic model \cite{Teraoka:06,
Teraoka:07}, for the fundamental equatorial modes centred around $700$ nm. (a) 
The FSR for the TE modes is calculated by comparing the positions of nearby 
modes, and (b) similarly for the TM modes. The FSRs are in units of $\mu$m. 
(c) The geometric $Q$-factors are simulated for the TE modes and (d) for the 
TM modes. 
}}
\label{fig:contours}
\figrule
\end{figure}

To assist in characterising the behaviour of microbubbles within this 
parameter space, contour plots for both the FSRs and the $Q$-factors are 
presented in Fig.~\ref{fig:contours}. For each quantity, the TE and TM modes 
are considered separately. 

In order to facilitate the efficient calculation of these contours, the 
single-layer analytic model \cite{Teraoka:06,Teraoka:07} is used to validate 
the FDTD results, as a complementary approach. The fundamental equatorial 
modes only are selected, as described in Section~\ref{subsect:gq}, 
centred around a wavelength of $700$ nm. 
The reason for this is the fact that the FSRs are 
dependent only on the mode positions, not the full spectrum, unless the 
deviating effect from nearby overlapping modes is taken into account. 
In addition, the most prominent modes within a spectrum 
originate from different radial mode numbers as the thickness is changed, 
as noted in Fig.~\ref{fig:dcomp}.
Such effects, while useful for characterising the spectrum, 
are disregarded for the time being in order to build a picture of the 
analytic behaviour of the FSR and $Q$-factors of fundamental modes by 
themselves. 

In Fig.~\ref{fig:contours}(a) and (b), it is clear that the behaviours of the 
FSRs in the TE and TM cases are almost identical. It is also apparent that, for 
large shell thicknesses, the FSR is not strongly dependent on the shell 
thickness, which is consistent with the analysis of the spectra. As the 
diameter increases, the minimum shell thickness before encountering deviations 
also increases, and the curvature shows a reduction in the FSR from the value 
of a microsphere of the same diameter. 

Figures~\ref{fig:contours}(c) and (d) show the purely geometric portion of the 
$Q$-factor, from Eq.~(\ref{eqn:qg}), which can be calculated independently 
from the spectrum by finding the roots of the characteristic equations 
\cite{Teraoka:06,Teraoka:07}. In this case, the 
magnitude of the $Q$-factor scales 
very quickly, as described in Section~\ref{subsect:gq}, and so the logarithm of 
the $Q$-factor is plotted. The contours show a fairly stable magnitude of 
$Q_g$ with respect to the shell thickness, but that $Q_g$ 
rapidly decreases if the shell thickness becomes too small. 

\begin{table*}
  \caption[The free spectral ranges of simulated microbubbles are compared.]
{The mean simulated free spectral ranges (in $\mu$m) for 
several microbubble diameters. The subscripts indicate the shell thickness 
used in the FDTD microbubble simulations. For each diameter, 
a resolution is chosen so that the mode positions converge within $1\%$ 
tolerance.} 
\vspace{-6pt}
  \newcommand\T{\rule{0pt}{2.8ex}}
  \newcommand\B{\rule[-1.4ex]{0pt}{0pt}}
  \begin{center}
    \begin{tabular}{ccccccc}
      \hline
      \hline
      \T\B            
       D [$\mu$m] \,&\,  $\mathrm{FSR}_{2.0\,\mu\mathrm{m}}$ \,&\, 
$\mathrm{FSR}_{1.5\,\mu\mathrm{m}}$ \,&\, $\mathrm{FSR}_{1.0\,\mu\mathrm{m}}$ \,&\, 
$\mathrm{FSR}_{0.8\,\mu\mathrm{m}}$ \,&\, $\mathrm{FSR}_{0.5\,\mu\mathrm{m}}$ \\
      \hline     
      $8$ \,&\,  $0.0145$ \,&\, $0.0145$ \,&\, $0.0141$ \,&\, $0.0140$ \,&\, 
$0.0132$  \\
      $10$ \,&\, $0.0118$ \,&\, $0.0117$ \,&\, $0.0109$ \,&\, $0.0108$ \,&\, 
$0.0108$  \\
      $12$ \,&\, $0.0094$ \,&\, $0.0093$ \,&\, $0.0088$ \,&\, $0.0089$ \,&\, 
$0.0088$  \\ 
      \hline
    \end{tabular}    
  \end{center}
  \label{tab:FSR}
\end{table*}

\subsection{\cdb Critical values of geometric parameters for sustaining WGMs}
\label{subsec:loss}

While it is useful to characterise the mode behaviour for silica microbubbles 
as a test-case, when the knowledge is applied to the possibility of biological 
cells, which are unlikely to function as ideal resonators, it is useful to 
investigate under what conditions the structure of a cell 
is unable to support WGMs. While it is clear that 
sufficiently thin shell walls for silica microbubbles cannot support modes, 
which is also the case for small outer diameters, as indicated by the rapid 
reduction of the $Q$-factors, the critical values of these geometric parameters 
for sustaining modes for a given refractive index contrast 
will now be investigated. 

Here, a silica microbubble resonator is considered as a cell analogue, 
due to its 
refractive index being similar to most proteins associated with the 
outer membranes of some cells \cite{Voros2004553}. Since cells are usually 
suspended in a liquid medium, the scenario in the previous section is altered 
so that the microbubble is placed in water. This also allows greater scope for 
the FDTD method to probe the range of shell thicknesses required 
for sustaining WGMs, 
including the fundamental modes \cite{doi:10.1117/12.2078526}. 
Furthermore, a larger diameter is considered, $D=15$ $\mu$m, in order to 
compensate for the ensuing decrease in the index contrast, while remaining an 
example of a resonator that is close to the mode loss condition. 

\begin{figure}[H]
\begin{center}
\includegraphics[width=0.8\hsize]{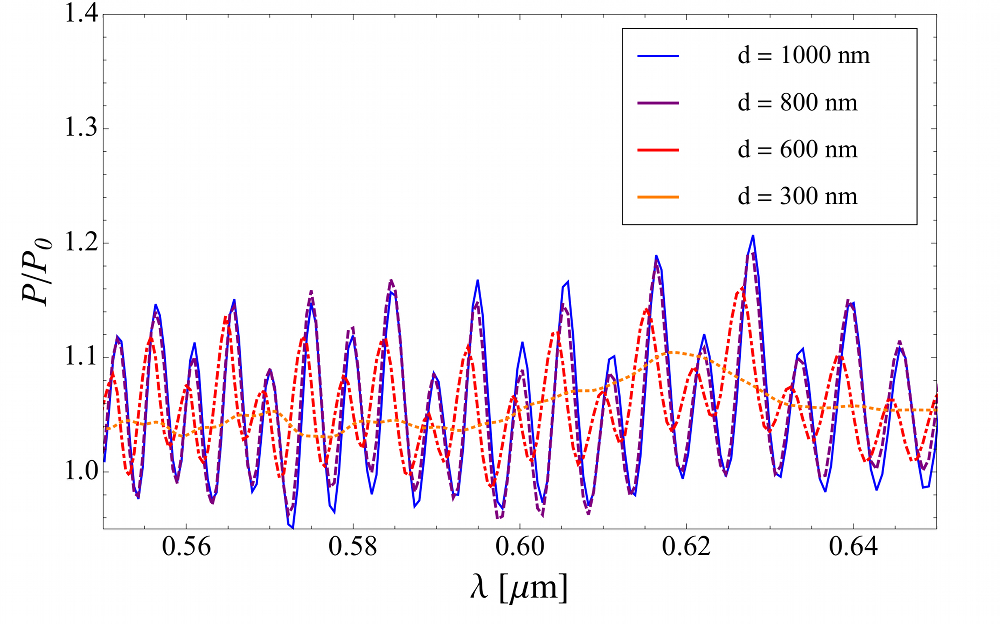}\\
\vspace{-5mm}
\hspace{4mm}\mbox{(a)}\\
\includegraphics[width=0.8\hsize]{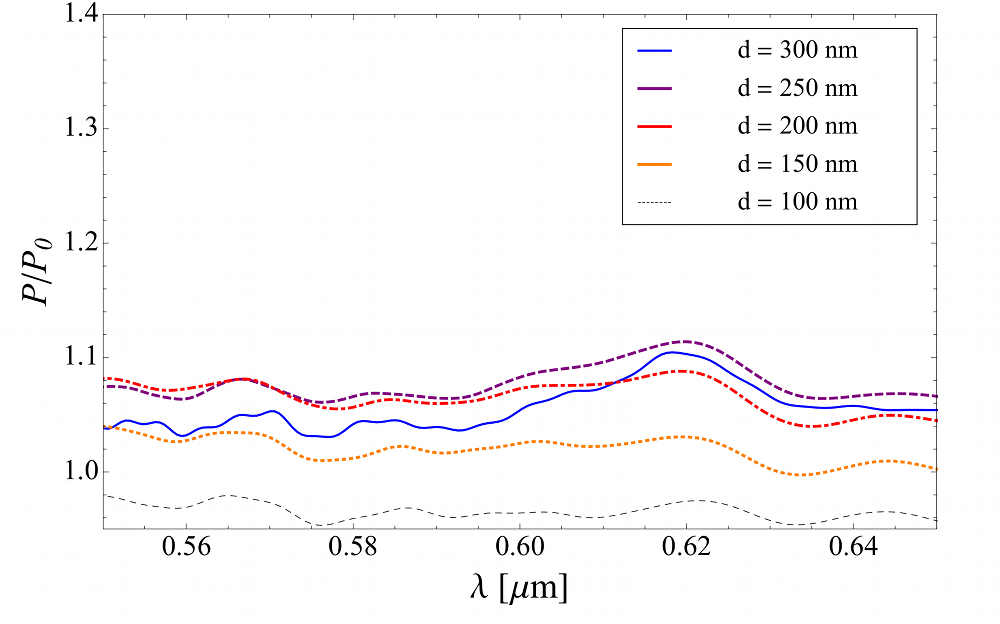}\\
\vspace{-5mm}
\hspace{4mm}\mbox{(b)}\\
\end{center}
\caption[The critical shell thickness for microbubble spectra is determined.]
{{Normalised WGM spectra for fluorescent-coated silica glass microbubbles, 
$15$ $\mu$m in diameter, with a variety of shell thicknesses, over a 
wavelength range of $550$ to $650$ nm, in a surrounding medium of water. (a) 
There is little change in the spectra for large shell thicknesses, and a 
critical value, $d_c$, for which there is sudden decrease in the ability 
of the resonator to sustain WGMs. (b) As $d<d_c$, 
the spectra contain little WGM information -- as the $Q$-factors reduce, 
adjacent modes overlap and are no longer distinguishable.}}
\label{fig:d15d25comp}
\figrule
\end{figure}

WGM spectra for a variety of shell thicknesses are shown in 
Fig.~\ref{fig:d15d25comp}. For larger values of shell thickness, there is 
little change to the spectra, as expected. However, as the shell thickness is 
decreased, there is a critical value, $d_c$, below which no modes can be 
sustained. This transition appears to occur rapidly, as a function of the 
shell thickness. Representative values of the shell thickness are shown in 
Fig.~\ref{fig:d15d25comp}(a), illustrating the transition. For $D=15$ $\mu$m, 
and a refractive index contrast of $\mathrm{n}_2/\mathrm{n}_3=1.10$ within this wavelength range, 
the critical value is close to $300$ nm. For shell thicknesses below $d_c$, 
the scattered spectra are quite similar in shape, with little WGM structure 
apparent, as shown in Fig.~\ref{fig:d15d25comp}(b). 
The reduction of the $Q$-factors as the shell thickness decreases allows 
adjacent modes to overlap, which are no longer distinguishable.

The implications of this regime of the geometric parameters 
where modes cannot be sustained 
are significant with regard to 
the search for a biological resonator. While it has been established 
in Section~\ref{sec:arch} that certain classes of biological cells can exhibit 
degrees of symmetry, in order to sustain resonances, such cells must also 
fall within the required geometric parameter regime, for their 
refractive index, diameter and layer thicknesses. 
If this is not the case, then cells can only be used as resonators with 
the addition of artificial components or equipment that can facilitate 
the trapping of modes, such as mirrors \cite{Gather2011}, 
or the use of pre-existing resonators incorporated into the cell
\cite{Himmelhaus2009418,Humar2015}. An estimation of the bounds 
of the geometric parameters required to sustain resonances in the context 
of cells is included in Section~\ref{sec:selec}. 

\section{\cdb Non-destructive determination of the geometry}
\label{sec:nondest}

The ability to extract the geometric parameters of a given microbubble 
resonator in a non-destructive manner is an important step toward 
\emph{reusable} sensing apparatus. In particular, the spectra obtained from 
WGM devices used for sensing are more easily interpreted if 
the outer diameter and the thickness of the shell wall are well known. 
Specific examples of apparatus that benefit from detailed knowledge of the 
resonator geometry include microbubble-based aerostatic pressure sensing 
\cite{Henze:11,Yang:16}, and frequency comb generation technologies 
\cite{Li:13,Riesen:15}. Such techniques are sensitive to the precise value of 
the shell thickness of the resonator, and make the development of an accurate 
method of extracting this information without destroying the apparatus itself 
an attractive prospect. To date, modeling-based methods of extracting the 
shell thickness rely heavily on assuming that the surface area of the 
resonator is preserved during the melting and expansion stages of fabrication, 
and their accuracy is limited to about $\pm 50\%$ for a $400$ $\mu$m diameter 
microbubble \cite{Cosci:15,Henze:11}. 
 
While alternative non-destructive methods exist for the determination of these 
parameters, such as confocal reflectance microscopy \cite{Cosci:15}, they are 
limited by the axial resolution of the microscope. Here, a new 
spectrum-matching method is developed, which utilises the sensitivity of the 
mode positions and the behaviour of the spectrum to the geometric parameters. 
The value of such a method is the ability to perform on-site identification of 
the parameters, as well as the identification of both the mode number and 
polarisation of each mode within a specific wavelength range. 

\begin{figure}[H]
\begin{center}
\includegraphics[width=0.8\hsize]{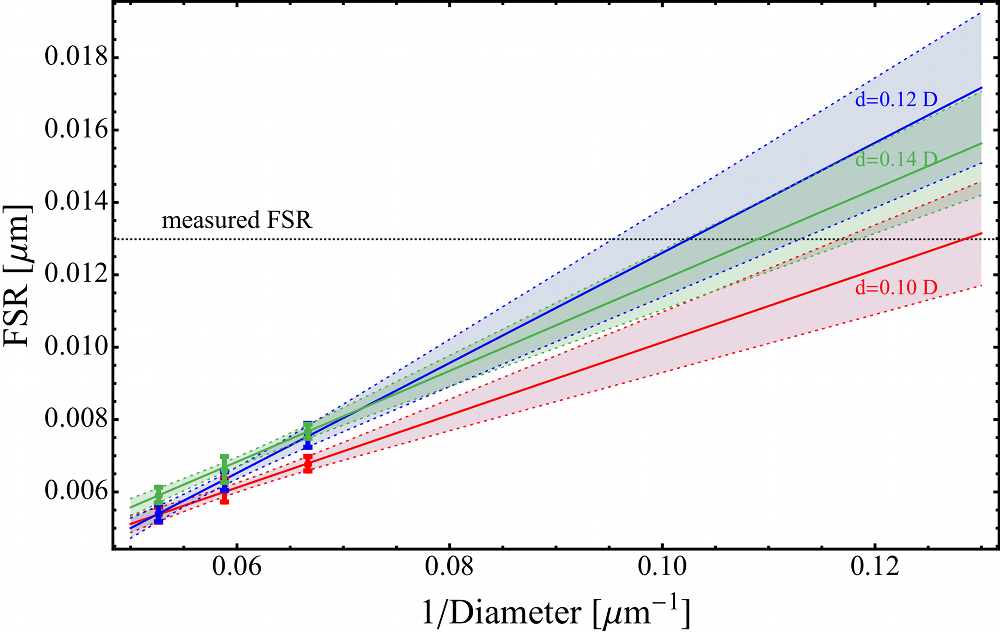}\\
\vspace{-3mm}
\hspace{6mm}\mbox{(a)}\\
\includegraphics[width=0.8\hsize]{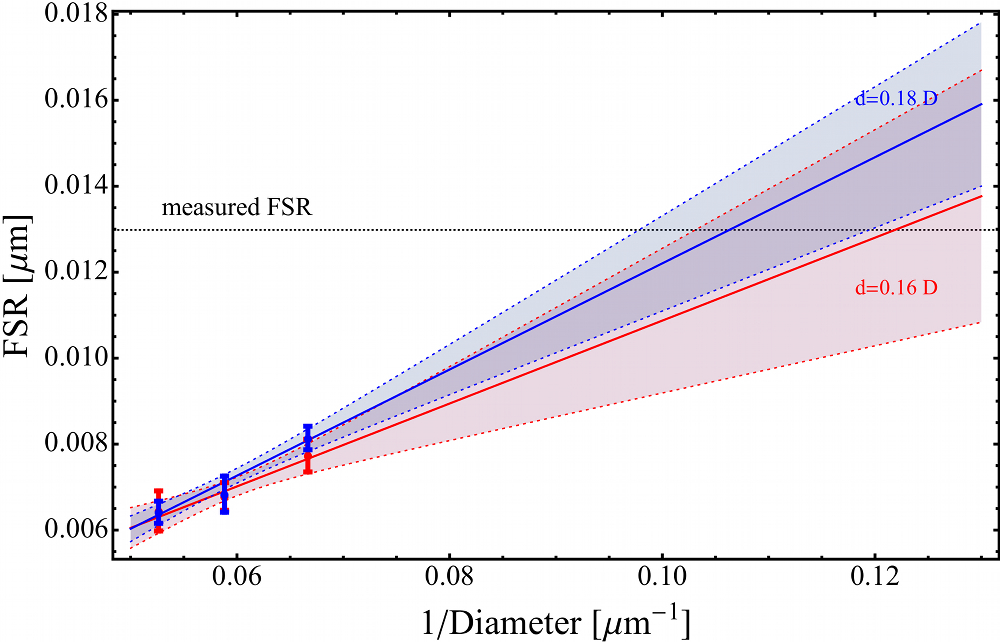}\\
\vspace{-3mm}
\hspace{6mm}\mbox{(b)}\\
\end{center}
\caption[The free spectral range is fitted to obtain the geometric parameters.]
{{The linear extrapolation of the FSR as a function of 
diameter $D$, for a range of shell thicknesses (a) $d=0.10\,D$ (red), 
$d=0.12\,D$ (\cb \textit{blue}\cbl), $d=0.14\,D$ 
(\color{darkgreen}\textit{green}\cbl), and (b) $d=0.16\,D$ 
(\color{red}\textit{red}\cbl), $d=0.18\,D$ (\cb \textit{blue}\cbl). 
See Table~\ref{tab:fit} for best fit values of diameter $D$ and shell 
thickness $d$.}}
\label{fig:extrap}
\figrule
\end{figure}

Using a linear form, $\mathrm{FSR} = a_1 + a_2\frac{1}{D}$ \cite{Astratov2010}, 
fits are produced for a range of simulated shell thicknesses, as shown in 
Fig.~\ref{fig:extrap}. By performing a sufficient number of simulations, the 
mode positions of the simulated spectra can be fitted to those of the measured 
spectrum. The shaded band indicates the statistical uncertainty at each value 
of $D$, obtained from the covariance matrix of the linear fit.

\begin{table*}[t]
  \caption[The best fit results for microbubble parameters are summarised.]
{Summary of best fit results for a range of simulated microbubble shell 
thicknesses and diameters.}
\vspace{-6pt}
  \newcommand\T{\rule{0pt}{2.8ex}}
  \newcommand\B{\rule[-1.4ex]{0pt}{0pt}}
  \begin{center}
    \begin{tabular}{ccc}
      \hline
      \hline
      \T\B            
       thickness/diameter ($d/D$) \qquad&\qquad thickness [$\mu$m] 
\qquad&\qquad best fit diameter 
[$\mu$m] \\
      \hline     
      $0.10$ \qquad&\qquad\,\, $0.8$ \qquad&\qquad\,\, $7.8$  \\
      $0.12$ \qquad&\qquad\,\, $1.2$ \qquad&\qquad\,\, $9.8$   \\
      $0.14$ \qquad&\qquad\,\, $1.3$ \qquad&\qquad\,\, $9.2$   \\
      $0.16$ \qquad&\qquad\,\, $1.3$ \qquad&\qquad\,\, $8.2$   \\ 
      $0.18$ \qquad&\qquad\,\, $1.7$ \qquad&\qquad\,\, $9.4$   \\
      \hline
    \end{tabular}    
  \end{center}
  \label{tab:fit}
\end{table*}

Combinations of $d$ and $D$ for which the extrapolations in 
Fig.~\ref{fig:extrap} attain the measured value of FSR are summarised in 
Table~\ref{tab:fit}. The simulated spectrum, whose prominent resonance 
positions match those of the experimental results most closely, corresponds to 
a diameter of $9.2$ $\mu$m, and a shell thickness of $1.3$ $\mu$m. 
The criteria used in determining the best match is as follows. 
For each of the trends considered in Fig.~\ref{fig:extrap}, 
the values of $d$ and $D$ corresponding to the measured FSR are selected. 
The spectra for each of these combination of $d$ and $D$ are compared to the 
measured spectrum, and the simulated spectrum that corresponds to the 
minimum difference in the positions of the prominent modes is selected. 

\begin{figure}[t]
\begin{center}
\includegraphics[width=0.8\hsize]{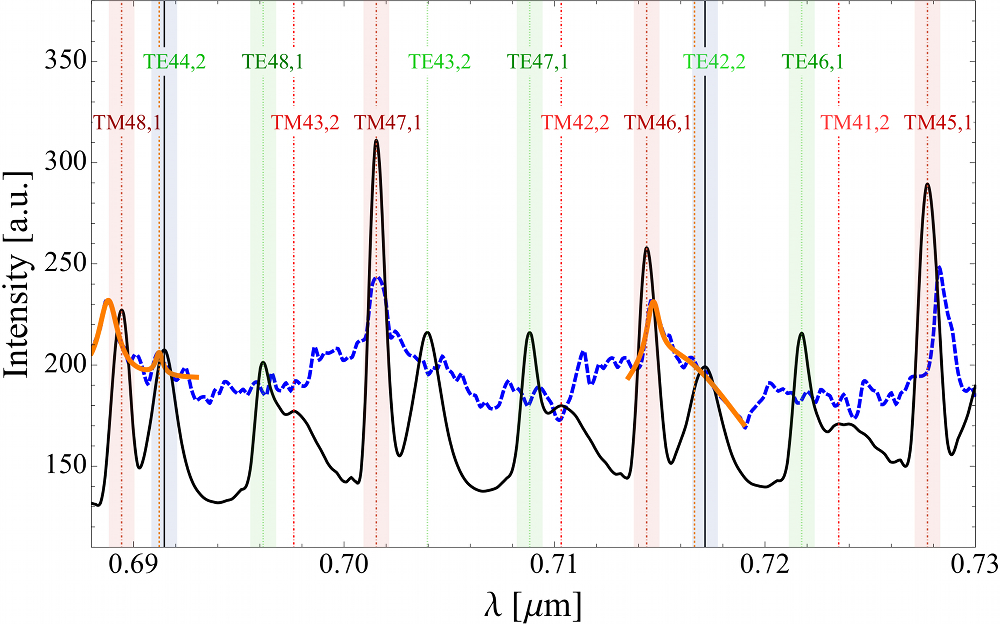}\\
\vspace{-3mm}
\end{center}
\caption[The fitted FDTD spectrum is compared to the measured spectrum.]
{{The whispering gallery mode spectrum measured from a silica 
microbubble (dotted line), and the fitted FDTD simulated spectrum (solid line), 
shown in arbitrary units (a.u.). The extrapolated values 
of diameter and shell thickness are $D=9.2$ $\mu$m, and $d=1.3$ $\mu$m, 
respectively. The TE and TM mode numbers and their positions are 
marked as dotted and dash-dotted vertical lines, respectively. The solid lines 
overlaying the measured spectrum indicate fits to two Lorentzians, resolving 
underlying higher order modes. The peak positions of these modes are marked as 
dashed and solid vertical lines for the experimental spectrum and simulated 
spectrum, respectively. The most prominent modes obtained from the dye-doped 
microbubble align closely with the first radial harmonic TM modes. The 
experimental results have been smoothed to match the wavelength resolution of 
the FDTD simulation and to remove a portion of the noise.}}
\label{fig:fit}
\figrule
\end{figure}

In Fig.~\ref{fig:fit}, the measured spectrum is represented by the dotted 
line, and the simulated spectrum is represented by the solid line. 
The resolution used in the simulation corresponds to $\Delta x = 0.028$ $\mu$m, 
with a wavelength resolution of $0.2286$ nm. Note that the experimental 
results have been smoothed to match the wavelength resolution of the FDTD 
simulation, to remove part of the noise. Unlike the experimental results, the 
simulation results are normalised, and so the $y$-axis is offset appropriately 
for the simulated spectrum. 

The mode positions are identified using the single-layer 
microsphere model \cite{Teraoka:06,Teraoka:07}. The positions of the prominent 
TE and TM modes are shown by dotted and dash-dotted vertical lines, 
respectively. The finite-width vertical bands indicate the uncertainty derived 
from the spatial and temporal resolutions added in quadrature. 
For two selected higher order modes, TE$44,2$ and TE$42,2$, a fit of the 
measured spectrum, using two Lorentzians, is able to resolve the underlying 
doublet mode structure as shown by a thick solid line placed over the 
experimental spectrum. 
The

\newpage
\noindent  peak positions from these fits lie within the systematic uncertainty 
bands of the simulated higher order modes, denoted by dashed and solid 
vertical lines, respectively. This indicates that the underlying mode structure 
of the experimental results is consistent with the FDTD simulations for these 
modes. 

The most prominent modes excited by a fluorescent dye are not easily predicted. 
In this experiment, it appears that the first radial TM modes 
received the largest coupling. The first radial TE modes and higher order 
modes that appear  in the simulation are not readily apparent in the 
experiment. Deviations from perfect sphericity in the experimental 
case can reduce the $Q$-factors to the point where these modes are not easily 
measurable. This occurs due to the generation of extra overlapping modes, i.e. 
mode-splitting, corresponding to deviations in the mean diameter 
in different axes of rotation of the resonator 
\cite{Reynolds:15,Foreman:15,Riesen:15a}. 

The $\delta$FSR over the wavelength range $690$ to $730$ nm is expected to be 
non-zero, and should be taken into account when performing the fit, since it 
affects the value of the measured and simulated FSR. The $\delta$FSR can be 
estimated for the two spectra displayed in Fig.~\ref{fig:extrap}. The results 
are shown in Table~\ref{tab:spread}. In both the experiment and the simulation, 
there is a systematic trend for the FSR to increase as the wavelength 
increases. The $\delta$FSR is larger in the case of the FDTD simulation, with 
a shift of $9.37\%$ in the FSR over this wavelength range. This difference in 
the $\delta$FSR between the measured and simulated spectra makes a precise 
matching of all the mode positions problematic at first glance. However, 
it should be noted that the systematic uncertainty in the mode positions of 
the FDTD simulation 
overlaps the peak positions 
from the experimental results, and thus agrees within the uncertainty. 

A comparison of the $Q$-factors of the prominent modes is given in 
Table~\ref{tab:Q}. The mean $Q$-factors from FDTD are statistically consistent 
with those of the experiment. Furthermore, the $Q$-factors are fairly 
comparable with those expected from a finite element model (FEM) 
\cite{Kaplan:13,Shirazi:13,Yang:14}, $Q \approx 10^3$, for a diameter of $10$ 
$\mu$m. 

The non-destructive extraction of the geometric parameters of a given 
microbubble purely from its spectrum is a valuable technique in resonator 
design. While it is possible to refine the estimation of the geometric 
parameters further by considering an expanded range of initial diameter values
 to increase the tightness of the extrapolation, requiring the 
use of more computing 
resources, the study presented in this chapter serves as an example 
of the practical use of the method. 
It is clear that FDTD simulation represents an important tool for 
predicting the performance and behaviour of microbubbles prior to 
fabrication, allowing for realistic imperfections to be incorporated.

\section{\cdb Microbubbles as a prelude to biological cells}

In this chapter, the properties of microbubble resonators have been explored, 
with a view to understanding the fundamental features of whispering gallery 
modes, and the limitations on sustaining them. 
A model was developed to include a distribution of dipoles placed on the 
surface of the resonator, to allow for the simulation of the 
mode-coupling expected from fluorescent dyes used in recent biosensing 
experiments. Furthermore, a method for extracting the geometric parameters of 
microbubble resonators was presented. 

\begin{table*}[t]
  \caption[The free spectral ranges from simulation and experiment are 
compared.]{The FSRs of the four prominent TM modes occurring in the 
wavelength range $690$ to $730$ nm, shown in Fig.~\ref{fig:fit}, in 
units of $\mu$m. The $\delta$FSR  is shown in terms of a percentage shift from 
the lower wavelength value.}
\vspace{-6pt}
  \newcommand\T{\rule{0pt}{2.8ex}}
  \newcommand\B{\rule[-1.4ex]{0pt}{0pt}}
  \begin{center}
    \begin{tabular}{lcccc}
      \hline
      \hline
      \T\B            
       Spectrum & TM $48,1-47,1$ & TM $47,1-46,1$ & TM $46,1-45,1$ & 
$\%$ shift \\
      \hline     
      Silica shell & 0.0128 & 0.0131 & 0.0137 & 7.21 \\
      FDTD  & 0.0121 & 0.0129 & 0.0133 & 9.37\\
      \hline
    \end{tabular}    
  \end{center}
  \label{tab:spread}
\end{table*}
\begin{table*}[t]
  \caption[The $Q$-factors from simulation and experiment are compared.]
{The $Q$-factors, (mean and standard deviation), derived from  
prominent modes of the spectra, shown in Fig.~\ref{fig:fit}, in the wavelength 
range $690$ to $730$ nm.}
\vspace{-6pt}
  \newcommand\T{\rule{0pt}{2.8ex}}
  \newcommand\B{\rule[-1.4ex]{0pt}{0pt}}
  \begin{center}
    \begin{tabular}{lcccccc}
      \hline
      \hline
      \T\B            
       Spectrum & TM $48,1$ &  TM $47,1$ &  TM $46,1$ &  TM $45,1$ & 
Mean & std. dev. \\
      \hline     
      Silica shell & 663 & 877 & 768 & 783 & 773 & 88 \\
      FDTD  & 733 & 911 & 760 & 707 & 778 & 92 \\
      \hline
    \end{tabular}    
  \end{center}
  \label{tab:Q}
\end{table*}

As an application for modelling biological cells, the FDTD method itself 
presents some opportunities, but also limitations. The general nature of FDTD 
allows arbitrary shapes, deformations, inhomogeneities and radiation 
collection regions to be incorporated into the resonator structure. 
Investigating how these different scenarios impact the measured spectra is a 
worthwhile exercise, especially for understanding the key criteria required of 
a resonator, in particular, a \emph{biological} resonator, which will be 
explored in Chapter~\ref{chpt:cel}. However, it is clear that the range of 
diameters and shell thicknesses that can be feasibly simulated presents a 
crucial limitation. It may be that biological cell resonators required a 
larger diameter than is easily simulated in FDTD, especially, as it will become 
apparent, as the refractive index contrast of the protein structures 
surrounding many cells compared to their native media is not large. 
Therefore, in the next chapter, a new model is developed, which incorporates 
the most useful features of each of the modelling approaches explored thus 
far. The 
new model is able to handle an arbitrary number of dielectric layers and 
excitation methods, including dipole sources, and any number of uniform layers 
of sources, as well as the underlying fields that produce the spectra. 
The model unifies the formulae of previous approaches in the literature, and 
encompasses a number of scenarios such as source placement, and dipole 
polarisation, expressed as a single equation. Furthermore, the code produced 
for Chapter~\ref{chpt:bub} has been structured so that the algorithm runs 
efficiently over a wide range of wavelength values, as described in 
Appendix~\ref{sec:scal}. 

It is the development of this general, multilayer model that will form the 
cornerstone for the modelling side of the investigation 
for the remainder of this thesis. Since the model is able to encapsulate a 
wide variety of scenarios and parameters, it is the ideal tool for 
assessing the feasibility of a biological cell resonator, and for assisting in 
the task of narrowing the search for a cell to a few most likely 
candidates. 

\chapter{\cdb A Unified Model for Active Multilayer Resonators}
\label{chpt:mod}

While the principal focus of this thesis is to understand the behaviour 
of whispering gallery modes in the context of biological resonators, 
and under what conditions a cell could support these modes, 
it is worthwhile to explain in detail the construction of a new, general 
model \cite{Hall:17a} that combines many useful aspects of the existing 
techniques described in the previous chapters, while remaining fast and 
efficient to use.  

This model allows for an arbitrary number of concentric, spherical 
dielectric layers, and any number of embedded dipole
sources or uniform distributions of dipole sources, to be included in the 
resonator. These methods for exciting the modes are analogous to embedded 
nanoparticles, or fluorescent dye coatings, respectively. The latter of these 
options may also potentially be used to model the natural autofluorescence of 
a cell. 
While the focus of the model is limited to spherically-shaped 
multilayer resonators, it represents a suitable choice for the 
most promising candidate cells to be investigated, introduced in 
Chapter~\ref{chpt:cel}. 

Included in this model is the ability to simulate the spectrum, which has been 
shown to be a powerful and instructive technique for analysis of WGMs  
in Chapters~\ref{chpt:sph} and \ref{chpt:bub}. 
The efficient generation of power spectra, together with accurate 
calculation of the mode coupling efficiencies that are comparable with 
experiment in a multilayer scenario, represents an important step forward for 
tools that aid microresonator design.
In each mode coupling scenario, the emitted power is expressed
conveniently as a function of wavelength, 
making use of the transfer matrix approach, and 
incorporating improvements to its stability, resulting in a
reliable, general set of formulae for calculating WGMs. 

In the specific cases of the dielectric microsphere and the 
single-layer coated microsphere, 
the derivations of the formulae 
are shown explicitly in Appendix~\ref{chpt:app2}. 
In these cases, the model has been verified by known formulae in the 
literature, as described in Appendix~\ref{chpt:app3}. 
This work is presented in the publication, Ref.~\cite{Hall:17a}, listed as 
\textbf{Item 3} in Section~\ref{sec:jou}.

\section{\cdb Motivation for a unified description}

The derivation of a model that contains multiple layers of concentric spheres
has, until now, not been treated comprehensively. The 
WGM spectrum model developed by Chew \textit{et al.} \cite{Chew:76,Chew:87a} 
considered spherical
dielectric particles with embedded dipole sources, with the motivation being the
 modelling of Raman and fluorescence scattering. The Chew model was
then extended to include a uniform distribution of dipole sources placed on
the surface of a sphere \cite{Chew:88}. A multilayer variant of the Chew
model exists, but contains no method for calculating the power spectrum 
\cite{Chew:76} and 
uses a different notation to other methods in the field \cite{Moroz05}. 
Meanwhile, a generalisation of Mie scattering theory, which is constrained to 
model the excitation of modes by an external ray, has been developed for 
spherical concentric `onion' resonators \cite{Xu:03,Xu:04,Liang:04}. 
While the long-range scattering behaviour of an external ray is an important 
test case, this latter approach does not provide an emitted power 
spectrum with which to compare with \textit{measured} WGM resonances. 

Analytic models typically make use of the
transfer matrix approach \cite{Moroz05}, which is faster and more convenient
to construct mathematically than the multilayer Chew model. However, this
approach suffers from numerical instabilities for certain parameter values, in 
particular, resonators with very thin shell layers \cite{Yang:03}, and thus 
needs to be treated carefully by computing the Wronskian for 
transfer matrices with determinants near zero. More recent studies on
single-layer coated microspheres clearly separate the spectrum into TE and TM
polarisations, allowing for greater flexibility in analysing the mode behaviour 
\cite{Teraoka:06,Teraoka:07}. Although this Teraoka-Arnold model 
can accurately calculate the resonance positions for a single-layer coated 
microsphere, the spectrum is not derived for this case, and therefore 
quantities such as the $Q$-factor can only be estimated by considering their 
geometric components, $Q_g$. 

To summarise, the main advancement of the present work is the derivation of a 
multilayer microsphere model, which may include dipole sources in any layer, 
or an active layer of sources, and which is able to generate the wavelength 
positions of the TE and TM modes. The model can also be used to calculate the 
emitted power spectrum from a formalism that is easily extended to include a 
number of other excitation strategies.

\section{\cdb Defining the multilayer model}
\label{sec:mod}

In this section, the methodology for calculating the quantities needed to find 
the wavelength positions of the resonances and the radiation power spectrum of 
a multilayer microsphere is stated. In general, one must calculate 
coefficients that determine the relative contributions of the components of 
the electric and magnetic fields in a specific region of space -- either in the 
central part of the resonator, one of the internal layers, or in the outermost 
medium. The coefficients in each of these regions will be shown to be related 
to one another through the use of boundary conditions 
at the interfaces between layers. Therefore, by knowing a set of field 
coefficients in one specific layer, and knowing the boundary conditions 
required for a multilayer scenario, the coefficients, and their corresponding 
fields, can be calculated for any other layer. 

The coefficients encode the behaviour of the electric and magnetic fields 
within the structure, from which the power spectrum can also be calculated by 
integrating the Poynting vector over the total solid angle. That is, 
the coefficients can be used to access the geometric attributes of the 
resonances. This is useful in itself -- while the absolute values of the 
coefficients may not be known in the absence of radiation, it will be 
discovered that resonance positions of the WGMs are determined by 
\emph{ratios} of the coefficients, as described in Section~\ref{sec:rec}. 
The general formalism of this model also allows radiation to be included, 
which is incorporated into the coefficients as well. This feature 
imbues the model with a powerful and general functionality.

\subsection{\cdb Geometry}

\begin{figure}
\hspace{1cm}\includegraphics[height=220pt]{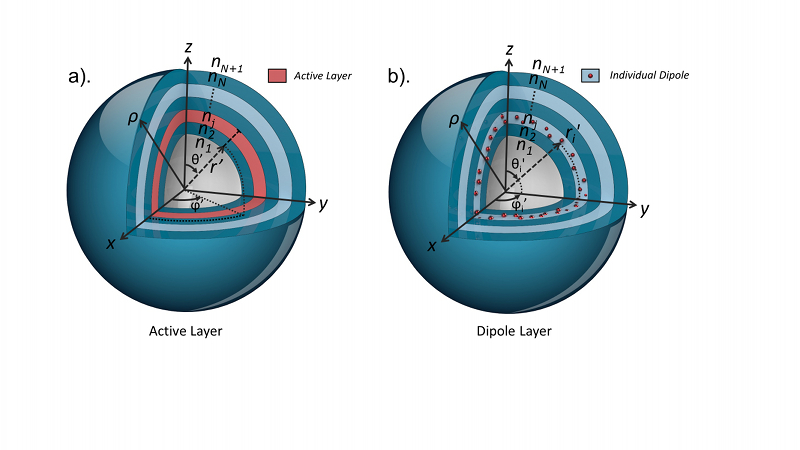}
\vspace{-2.5cm}
\caption[The active multilayer resonator geometry is illustrated.]
{The geometry of a spherical resonator with $N$ layers, with the 
outer medium being labelled as $N+1$ in the convention described in 
Section~\ref{sec:mod}. (a) A single layer contains a uniform 
distribution of dipoles to represent an active layer. (b) One or more 
individual dipoles can be placed in a given layer, to represent 
one or more embedded nanoparticles.}
\label{fig:multi}
\figrule
\end{figure}

In describing the model in detail, first, the geometry of the multilayer 
scenario will be explained. 
Consider a microsphere with an arbitrary number of concentric layers, $N$. 
The refractive index distribution and thicknesses are illustrated in 
Fig.~\ref{fig:multi}. Let us assume that each layer includes a dipole emitter 
located at position $\mathbf{r}_{j}^{\prime}$\,, where $j$ is the layer number 
and the prime symbol is specifically used for the position of the 
\emph{sources}. The radial coordinate $\mathbf{r}$ will be used to represent 
the fields at an arbitrary point in space, and the quantity $r_{j}$ will be 
used to specify unambiguously the radius of the boundary between the 
$j^{\mathrm{th}}$ and $j+1^{\mathrm{th}}$ region. The outermost region, $N+1$, is 
extended to infinity, and the innermost region is denoted as $1$. 
Hence, as an example, the radius of the inner region is called $r_{1}$, and 
the outer boundary of region $N$ is $r_{N}$. 

It is required that the model describe solutions to 
Maxwell's Equations, which take the following form in Gaussian units 
\begin{equation}
\mathbf{E}    =\frac{\mathrm{i} c}{\omega\varepsilon}(\bm{\nabla}
\times\mathbf{H}),
\quad\mathbf{H}    =-\frac{\mathrm{i} c}{\omega\mu}(\bm{\nabla}\times
\mathbf{E}),
\label{R001}
\end{equation}
and also take into account the appropriate boundary conditions of the 
scenario. The solutions for the fields can generally be decomposed into radial 
and angular components using vector spherical harmonics (VSH). While there is 
a number of different notations for the VSH occurring throughout the 
literature \cite{edmonds1996angular,Jackson89,Liang:04,Moroz05}, 
the notation used here is $\mathbf{Y}_{lm}(\theta,\phi)$, $\mathbf{\bm{\Psi}
}_{lm}(\theta,\phi),$ and $\bm{\Phi}_{lm}(\theta,\phi)$, the properties of which, 
and their relations with the other notations, are described in 
Appendix~\ref{chpt:app1}. These three functions form  an orthonormal, complete 
set. That is, any vector field in spherical coordinates can be expanded based on
these functions. For example, the electric field can be expanded as 
\begin{equation}
\mathbf{E}(r,\theta,\phi)=\sum_{l=0}^{\infty}\sum_{m=-l}^{m=l}[E_{lm}
^{r}(r)\mathbf{Y}_{lm}(\theta,\phi)+E_{lm}^{(1)}(r)\bm{\Psi}
_{lm}(\theta,\phi)+E_{lm}^{(2)}(r)\bm{\Phi}_{lm}(\theta,\phi)],
\label{C002}%
\end{equation}
where the coefficients $E_{lm}^{r},E_{lm}^{(1)}$ and $E_{lm}^{(2)}$ can be found
by using orthogonality relations (see Appendix~\ref{chpt:app1}). Note that 
the vector $\mathbf{Y}_{lm}$ is oriented in the radial direction and 
$\bm{\Psi}_{lm}$ and $\bm{\Phi}_{lm}$ are oriented in the transverse plane, 
perpendicular to the unit vector $\mathbf{\hat{r}}$, and hence $E_{lm}^{(1)}$ 
and $E_{lm}^{(2)}$ represent the coefficients of the transverse field. 

The total electric and magnetic field in each region $j$ can be written as the 
sum of the fields -- first, those originating from a dipole within the layer 
$j$ itself, denoted $\mathbf{E}_{jd}$ and $\mathbf{H}_{jd}$, and 
second, those associated with the reflection and transmission of the fields 
from other layers, denoted $\mathbf{E}_{j}$ and $\mathbf{H}_{j}$. Hence, the 
total fields may be found simply by taking the superposition of these two 
contributions for each field, $\mathbf{E}_{j}^{\text{total}
}=\mathbf{E}_{j}+\mathbf{E}_{jd}$ and $\mathbf{H}_{j}^{\text{total}
}=\mathbf{H}_{j}+\mathbf{H}_{jd}.$ 

The general form of the fields $\mathbf{E}_{j}$ and $\mathbf{H}_{j}$ 
in an arbitrary layer $j$ 
can be written in terms of a radial component and a transverse component, each 
with a related incoming and outgoing wave, described by $j_l$ and $h_l^{(1)}$, 
respectively \cite{PhysRevA.13.396}
\begin{align}
\mathbf{E}_{j}  &  =\sum_{l,m}\left(  \frac{\mathrm{i} c}{\mathrm{n}_{j}^{2}
\omega}\right)  A_{j}\bm{\nabla}\times\lbrack j_{l}(k_{j}
r)\bm{\Phi}_{lm}(\theta,\phi)]+\left(  \frac{\mathrm{i} c}{\mathrm{n}_{j}^{2}
\omega}\right)  B_{j}\bm{\nabla}\times\lbrack h_{l}^{(1)}(k_{j}
r)\bm{\Phi}_{lm}(\theta,\phi)]\nonumber\\
&  +C_{j}j_{l}(k_{j}r)\bm{\Phi}_{lm}(\theta,\phi)+D_{j}h_{l}
^{(1)}(k_{j}r)\bm{\Phi}_{lm}(\theta,\phi),\label{R002}\\
\mathbf{H}_{j}  &  =\sum_{l,m}-\left(  \frac{\mathrm{i} c}{\mu_{j}\omega
}\right)  C_{j}\bm{\nabla}\times\lbrack j_{l}(k_{j}r)\bm{\Phi
}_{lm}(\theta,\phi)]-\left(  \frac{\mathrm{i} c}{\mu_{j}\omega}\right)
D_{j}\bm{\nabla}\times\lbrack h_{l}^{(1)}(k_{j}r)\bm{\Phi}
_{lm}(\theta,\phi)]\nonumber\\
&  +(\frac{1}{\mu_{j}})A_{j}j_{l}(k_{j}r)\bm{\Phi}_{lm}(\theta
,\phi)+(\frac{1}{\mu_{j}})B_{j}h_{l}^{(1)}(k_{j}r)\bm{\Phi}
_{lm}(\theta,\phi).\label{R003}
\end{align}
One may then show explicitly how the fields can be written in a form
consistent with Eq. (\ref{C002}) in terms of the orthonormal functions 
$\mathbf{Y}_{lm}$, $\bm{\Psi}_{lm}$ and $\bm{\Phi
}_{lm}$, by appealing to the properties from Appendix~\ref{chpt:app1} for the 
VSH 
\begin{align}
\mathbf{E}_{j}  &  =\sum_{l,m}\Bigg[-\left(  \frac{\mathrm{i} c}{\mathrm{n}_{j}
^{2}\omega}\right)  \frac{\sqrt{l(l+1)}}{\mathrm{i}}[A_{j}\frac{1}{r}j_{l}
(k_{j}r)+B_{j}\frac{1}{r}h_{l}^{(1)}(k_{j}r)]\mathbf{Y}_{lm}(\theta
,\phi)\nonumber\\
&  -\left(  \frac{\mathrm{i} c}{\mathrm{n}_{j}^{2}\omega}\right)  
\left\{  A_{j}\frac{1}
{r}\frac{\mathrm{d}}{\mathrm{d}r}[rj_{l}(k_{j}r)]+B_{j}\frac{1}{r}
\frac{\mathrm{d}}{\mathrm{d}r}[rh_{l}^{(1)}(k_{j}r)]\right\}  \bm{\Psi
}_{lm}(\theta,\phi)\nonumber\\
&  +\left\{  C_{j}j_{l}(k_{j}r)+D_{j}h_{l}^{(1)}(k_{j}r)\right\}
\bm{\Phi}_{lm}(\theta,\phi)\Bigg],\label{R006}\\
\mathbf{H}_{j}  &  =\sum_{l,m}\Bigg[\left(  \frac{\mathrm{i} c}{\mu_{j}\omega
}\right)  \frac{\sqrt{l(l+1)}}{\mathrm{i}}[C_{j}\frac{1}{r}j_{l}(k_{j}
r)+D_{j}\frac{1}{r}h_{l}^{(1)}(k_{j}r)]\mathbf{Y}_{lm}(\theta,\phi
)\nonumber\\
&  +\left(  \frac{\mathrm{i} c}{\mu_{j}\omega}\right)  \left\{  C_{j}\frac{1}
{r}\frac{\mathrm{d}}{\mathrm{d}r}[rj_{l}(k_{j}r)]+D_{j}\frac{1}{r}
\frac{\mathrm{d}}{\mathrm{d}r}[rh_{l}^{(1)}(k_{j}r)]\right\}  \bm{\Psi
}_{lm}(\theta,\phi)\nonumber\\
&  +(\frac{1}{\mu_{j}})\left\{  A_{j}j_{l}(k_{j}r)+B_{j}h_{l}^{(1)}
(k_{j}r)\right\}  \bm{\Phi}_{lm}(\theta,\phi)\Bigg].\label{R007}
\end{align}
Here, $A_{j}$, $B_{j}$, $C_{j}$ and $D_{j}$ are coefficients 
that are determined through the boundary conditions of the multilayer 
structure. 
Note that $A_{j}$ and $B_{j}$ describe the transverse component of 
$\mathbf{H}_{j}$ and thus the TM modes, whereas $C_{j}$ and $D_{j}$ describe the 
TE modes, as defined by Jackson \cite{Jackson89}. In this notation, the TM 
modes are defined as the modes for which $\mathbf{H}_{j}$ has no component in 
the radial direction, whereas the TE modes are those for which $\mathbf{E}_{j}$ 
has no such radial component. 

At this point, it is important to emphasise that the forms of 
Eqs.~(\ref{R002}) through (\ref{R007}) could be written using alternative sets 
of linearly independent Bessel and Hankel functions to describe the incoming 
and outgoing waves. Here, the functions $j_{l}(k_{j}r)$ and $h_{l}^{(1)}(k_{j}r)$ 
have been chosen throughout so that, as far as possible, 
the form of the fields and the contributions to them from radiation sources 
are in the same format as each other. This ensures that appropriate functions 
of $r$ can be constructed for any layer, including the innermost and outermost 
regions, in a unified way. This will become apparent when the contributions 
from dipole sources are included in the field coefficients, in 
Eq.(\ref{R020}). But first, the transfer matrix method will be explained for 
an arbitrary layer $j$ in the absence of dipole sources.

\subsection{\cdb Transfer matrix method}
\label{subsec:trans}

\subsubsection{\cdb Fields in an arbitrary layer}

The transfer matrix method is a way of computing the boundary conditions for 
an arbitrary number of interfaces, and particularly useful for adaptation in 
the multilayer microsphere scenario. The boundary conditions at the interfaces 
of the layers imply that the transverse components of the fields are 
continuous, while there is a discontinuity in the normal components of the 
fields. In order to enforce continuity, the focus must then be on transverse 
components of the fields in layer $j$, indicated by the superscripts on 
$E_{lm}^{(1)}$ and $E_{lm}^{(2)}$ in Eq.~(\ref{C002}). These components may be 
written in matrix form
\begin{equation}
\mathbf{EH}_{j}=M_{j}(r)\mathbf{A}_{j}, \label{R009}%
\end{equation}
where $\mathbf{EH}_{j}$ is a four-component vector of the transverse 
components of the electric and magnetic fields, and $\mathbf{A}_{j}$ includes 
the four field coefficients as follows:  
\begin{align}
\mathbf{EH}_{j}(r)  &  =\left(
\begin{array}
[c]{c}%
E_{j}^{(1)}(r)\\
H_{j}^{(2)}(r)\\
E_{j}^{(2)}(r)\\
H_{j}^{(1)}(r)
\end{array}
\right)  ,\quad\mathbf{A}_{j}=\left(
\begin{array}
[c]{c}%
A_{j}\\
B_{j}\\
C_{j}\\
D_{j}%
\end{array}
\right). \label{R010}%
\end{align}
The matrix $M_{j}(r)$ is a block diagonal matrix, and 
can be derived from Eqs.~(\ref{R006}) and (\ref{R007}) 
\begin{align}
M_{j}(r)  &  =\frac{1}{k_{j}r}\left(
\begin{array}
[c]{cccc}%
-\left(  \frac{\mathrm{i}}{\mathrm{n}_{j}}\right)  \psi_{l}^{\prime}(k_{j}r) & 
-\left(
\frac{\mathrm{i}}{\mathrm{n}_{j}}\right)  \chi_{l}^{\prime}(k_{j}r) & 0 & 0\\
(\frac{1}{\mu_{j}})\psi_{l}(k_{j}r) & (\frac{1}{\mu_{j}})\chi_{l}(k_{j}r) &
0 & 0\\
0 & 0 & \psi_{l}(k_{j}r) & \chi_{l}(k_{j}r)\\
0 & 0 & \left(  \frac{\mathrm{i} \mathrm{n}_{j}}{\mu_{j}}\right)  \psi_{l}^{\prime}%
(k_{j}r) & \left(  \frac{\mathrm{i} \mathrm{n}_{j}}{\mu_{j}}\right)  
\chi_{l}^{\prime
}(k_{j}r)
\end{array}
\right).\label{R0101}
\end{align}
In arriving at Eq.~(\ref{R0101}), the relation 
$c/(\mathrm{n}_{j}\omega)=1/k_{j}$ has been used, as well as the 
Riccati-Bessel and Riccati-Hankel functions, 
$\psi_{l}(k_{j}r)=k_{j}rj_{l}(k_{j}r)$ and 
$\chi_{l}(k_{j}r)=k_{j}rh_{l}^{(1)}(k_{j}r),$ and their
derivatives with respect to their arguments, $\psi_{l}^{^{\prime}}(k_{j}r)$ and 
$\chi_{l}^{^{\prime}}(k_{j}r)$. 

Consider the top left block of the matrix $M_j(r)$. It is clear from 
Eq.~(\ref{R009}) that these terms have coefficients of 
$A_j$ or $B_j$, and thus represent the 
field contributions to the TM modes, whereas the bottom right block 
corresponds to the TE modes, 
$M_j(r)=\left(\begin{array}[c]{cc}M_{j}^{TM} & 0\\0 & 
M_{j}^{TE}\end{array}\right)$. 
The terms in the top row, from left to right, correspond to the incoming and 
outgoing contributions to a transverse component of the electric field 
$E_j^{(1)}$, respectively. The second row represents the equivalent 
contributions to the magnetic field component of the TM modes, $H_j^{(2)}$. 

In the transfer matrix method, it is necessary to invert the matrices 
$M_j(r)$, and therefore it is important to ensure that the determinant of the 
matrices is well characterised, so that parameter regimes that are 
problematic can be identified ahead of time. It is already well known that a 
 calculation of the 
transfer matrix method that is not sensitive to this effect 
can lead to instabilities for very thin layers, and that these can be avoided 
numerically by pre-calculating the known, analytic Wronskian form of the 
determinants \cite{Yang:03}. 
For the Wronskian defined as 
\begin{equation}
W_{r}[f(kr),g(kr)]\equiv f(kr)g^{\prime}(kr)-f^{\prime}(kr)g(kr),
\end{equation}
with the derivative with respect to $r$, 
and using the following equation derived from the properties of the 
Riccati-Bessel and Riccati-Hankel functions \cite{Moroz05} 
\begin{equation} 
W_{k_{j}r}[\psi_{l}(k_{j}r),\chi_{l}(k_{j}r)]=\mathrm{i},
\end{equation}
the determinants of $2\times2$ blocks, $\ M_{j}^{TM}$ and $M_{j}^{TE}$, can then 
be simplified in the following manner
\begin{align}
\det(M_{j}^{TM}(r))  &  =\frac{\mathrm{i}}{\mu_{j}\mathrm{n}_{j}k_{j}^{2}r^{2}}
W_{k_{j}%
r}[\psi_{l}(k_{j}r),\chi_{l}(k_{j}r)]=-\frac{1}{\mu_{j}\mathrm{n}_{j}k_{j}^{2}r^{2}%
},\label{R011}\\
\det(M_{j}^{TE}(r))  &  =\frac{1}{k_{j}^{2}r^{2}}\left(  \frac{\mathrm{i}\, 
\mathrm{n}_{j}%
}{\mu_{j}}\right)  W_{k_{j}r_{j}}[\psi_{l}(k_{j}r),\chi_{l}(k_{j}%
r)]=-\frac{\mathrm{n}_{j}}{\mu_{j}k_{j}^{2}r^{2}}. \label{R0111}%
\end{align}
This concludes the discussion of the form of the matrix $M_j(r)$, 
which operates in the absence of radiation from dipole or other sources, 
in this notation. In order to incorporate the radiation effects into the 
field coefficients, 
the form of the contributions from the dipole sources must be explained. 
While this choice of source has become the focus of the investigation 
because of its usefulness in considering fluorescent layers or 
embedded nanoparticles, as mentioned in Section~\ref{sec:sim}, 
it does not preclude the use of alternative methods of excitation, the 
contributions from which can be obtained in a straightforward manner 
using the following 
technique for modifying the field equations.

\subsubsection{\cdb Fields generated by dipole sources}

The contributions to the electric and magnetic fields from a single 
dipole located in layer $j$ and at position $\mathbf{r}_j^{\prime}$ 
are denoted $\mathbf{E}_{jd}(r)$ and $\mathbf{H}_{jd}(r)$, in a similar way to 
Eq.~(\ref{R009}) 
\begin{equation}
\mathbf{EH}_{jd}(r)=\theta(r_{j}^{\prime}-r)M_{j}(r)\mathbf{a}_{jL}%
+\theta(r-r_{j}^{\prime})M_{j}(r)\mathbf{a}_{jH}. \label{R012}
\end{equation}
Here, $\theta(r)$ is the Heaviside step function ($\theta(r)=0$ for $r<0$ and
$\theta(r)=1$ for $r\ge 0$), which ensures that the correct Bessel or Hankel 
function, 
$j_{l}(k_{j}r)$ or $h_{l}^{(1)}(k_{j}r)$, is selected when evaluating the fields at
point $r$. 
In the case where $r<r_{j}^{\prime}$, a subscript $L$ is used, such as in
$\mathbf{a}_{jL}$, whereas, for $r>r_{j}^{\prime}$ a subscript $H$ is used. 
The vectors corresponding to the dipole source, $\mathbf{EH}_{jd}$,
$\mathbf{a}_{jL}$ and $\mathbf{a}_{jH}$ take the following forms
\begin{equation}
\mathbf{EH}_{jd}=\left(
\begin{array}
[c]{c}%
E_{jd}^{(1)}\\
H_{jd}^{(2)}\\
E_{jd}^{(2)}\\
H_{jd}^{(1)}%
\end{array}
\right)  ,\mathbf{a}_{jL}=\left(
\begin{array}
[c]{c}%
a_{jEL}\\
0\\
a_{jML}\\
0
\end{array}
\right)  ,\mathbf{a}_{jH}=\left(
\begin{array}
[c]{c}%
0\\
a_{jEH}\\
0\\
a_{jMH}%
\end{array}
\right). \label{R013}%
\end{equation}
The quantities that occur inside the dipole vectors, 
$a_{jEL}$, $a_{jML}$, $a_{jEH}$ and $a_{jML}$ are special coefficients that 
correspond to the dipole radiation Ansatz \cite{Chew:87a}:
\begin{align}
\label{R014}
a_{jEL}(r_{j}^{\prime})  &  =4\pi k_{j}^{2}\sqrt{\frac{\mu_{j}%
}{\epsilon_{j}}}\mathbf{P}\cdot\bm{\nabla}_{j}^{\prime}\!\!\times\!\lbrack 
h_{l}^{(1)}\!\!%
(k_{j}r_{j}^{\prime})\bm{\Phi}_{lm}^{\ast}(\theta_{j}^{\prime}%
,\phi_{j}^{\prime})]\mathbf{,}\\
a_{jML}(r_{j}^{\prime})&=4\pi\mathrm{i}
k_{j}^{3}\frac{1}{\epsilon_{j}}h_{l}^{(1)}\!\!(k_{j}r_{j}^{\prime})\mathbf{P}%
\cdot\bm{\Phi}_{lm}^{\ast}(\theta_{j}^{\prime},\phi_{j}^{\prime
}),\\
a_{jEH}(r_{j}^{\prime})  &  =4\pi k_{j}^{2}\sqrt{\frac
{\mu_{j}}{\epsilon_{j}}}\mathbf{P}\cdot\bm{\nabla}_{j}^{\prime}\mathbf{\times
\lbrack}j_{l}(k_{j}r_{j}^{\prime})\bm{\Phi}_{lm}^{\ast}\mathbf{(}%
\theta_{j}^{\prime},\phi_{j}^{\prime}\mathbf{)],}\\
a_{jMH}(r_{j}%
^{\prime})&=4\pi\mathrm{i} k_{j}^{3}\frac{1}{\epsilon_{j}}j_{l}(k_{j}r_{j}^{\prime
})\mathbf{P}\cdot\bm{\Phi}_{lm}^{\ast}(\theta_{j}^{\prime},\phi
_{j}^{\prime}). \label{R015}%
\end{align}
Since the coefficients $a_{jEL}$ and $a_{jEH}$ correspond to an electric 
field in the radial direction, they contribute only to the TM modes, whereas 
$a_{jML}$ and $a_{jMH}$ contribute only to the TE modes \cite{Jackson89,
BohrenHuffman}, a fact that has sometimes caused the labels to be 
interchanged in the literature \cite{Moroz05}. In this notation, the 
subscripts $E$ and $M$ of these dipole coefficients indicate their origin in 
the electric and magnetic multipole expansions, respectively. 
The $\bm{\nabla}_{j}^{\prime}$ symbol indicates derivatives with 
respect to the position $\mathbf{r}_{j}^{\prime}$, and the symbol $\mathbf{P}$ 
refers to the dipole moment vector, which may be freely chosen. 

Now that the field components in the layer $j$ due to a dipole have been 
established in Eq.~(\ref{R012}), and noting that it has essentially the same 
format as Eq.~(\ref{R009}) where a vector of coefficients is pre-multiplied 
by a matrix $M_j(r)$, the total field can be found by simply adding the two 
equations together
\begin{equation}
\mathbf{EH}_{j}^{\text{total}}(r)=M_{j}(r)\mathbf{A}_{j}+\theta(r_{j}^{\prime
}-r)M_{j}(r)\mathbf{a}_{jL}+\theta(r-r_{j}^{\prime})M_{j}(r)\mathbf{a}_{jH}. 
\label{R016}
\end{equation}
With the total fields now specified above, the continuity 
condition across each boundary can be enforced, and performing 
this process will lead to the definition of the transfer matrix.

\subsubsection{\cdb The continuity equation}

The continuity of the transverse components of the electric and magnetic 
fields at the interface of the adjacent regions $j$ and $j+1$ leads to the 
following relationships among their respective field coefficients 
\begin{align}
\mathbf{EH}_{j}^{\text{trans}}(r_{j},\theta,\phi)  &  =\mathbf{EH}_{j+1}
^{\text{trans}}(r_{j},\theta,\phi),\\
M_{j}(r_{j})(\mathbf{A}_{j}+\mathbf{a}_{jH})  &  =M_{j+1}(r_{j})(\mathbf{A}
_{j+1}+\mathbf{a}_{j+1L}),\label{R017}\\
\mathbf{A}_{j+1}  &  =M_{j+1}^{-1}(r_{j})M_{j}(r_{j})\mathbf{A}_{j}
+M_{j+1}^{-1}(r_{j})M_{j}(r_{j})\mathbf{a}_{jH}-\mathbf{a}_{j+1L}.
\label{Ajrec}
\end{align}
Equation (\ref{R017}) can be used recursively to connect the coefficients in
the outermost region, $N+1$, to the innermost region, $1$.  
This encodes the continuity condition. 
By applying this recursive process, the coefficients in the inner and 
outermost regions can be related to each other, using the following 
compact notation, which includes a new composite matrix $T$, and a 
constant vector $\mathbfcal{D}$ that includes the contributions from the dipole 
radiation 
\begin{equation}
\mathbf{A}_{N+1}=T(N+1,1)\mathbf{A}_{1}+\mathbfcal{D}. \label{R018}%
\end{equation}
The matrix $T(N+1,j)$, which 
relates the coefficients in the outermost region $N+1$ to a layer $j$ 
(for which $j=1$ is a special case), is known as the scattering matrix, 
and is defined by the repeated matrix multiplication occurring in the 
first term of the right hand side of Eq.~(\ref{Ajrec}) 
\begin{align}
T(N+1,j)  &  =M_{N+1}^{-1}(r_{N})M_{N}(r_{N})M_{N}^{-1}(r_{N-1})M_{N-1}
(r_{N-1})M_{N-1}^{-1}(r_{N-2})M_{N-2}(r_{N-2})\ldots\nonumber\label{R0201}\\
&  M_{j+2}^{-1}(r_{j+1})M_{j+1}(r_{j+1})M_{j+1}^{-1}(r_{j})M_{j}(r_{j}),\text{
for }1\leq j<N+1,\\
T(N+1,j)  &  =I_{4\times4},\text{ for }j=N+1.
\end{align}
%NEW: 
Note that the scattering matrix $T(N+1,j)$ is always composed of repetitive 
blocks of matrices in the form of $M_{j+1}^{-1}(r_{j})M_{j}(r_{j})$. These blocks 
are the transfer matrices, as they encode the continuity condition 
across the boundary at a radius $r_j$ for two adjacent media with wave-numbers 
$k_j$ and $k_{j+1}$. Explicitly writing the form of these matrices, it can be 
found that
\begin{align}
M_{j+1}^{-1}(r_{j})M_{j}(r_{j})
&=\frac{\mathrm{i}}{\mathrm{n}_{j+1}\mu_{j}}\frac{k_{j+1}
}{k_{j}}G(j+1,j)\\
&=\frac{\mathrm{i}}{\mathrm{n}_{j+1}\mu_{j}}\frac{k_{j+1}}{k_{j}}
\begin{pmatrix}
G_{11} & G_{12} & 0 & 0\\
G_{21} & G_{22} & 0 & 0\\
0 & 0 & G_{33} & G_{34}\\
0 & 0 & G_{43} & G_{44}
\end{pmatrix}
_{(j+1,j)}\!\!\!\!\!. \label{R0191}
\end{align}
As expected, the transfer matrices are block diagonal matrices, since 
they are composed entirely from the block diagonal matrices.  
The blocks $G^{TM}=\begin{pmatrix}
G_{11} & G_{12}\\
G_{21} & G_{22} 
\end{pmatrix}$ and $G^{TE}=\begin{pmatrix}
G_{33} & G_{34}\\
G_{43} & G_{44} 
\end{pmatrix}$ can be computed directly from Eq.~(\ref{R0101}) and written 
in the same way for both TE and TM polarisations
\begin{equation}
G^{TE/TM}=
\begin{pmatrix}
\!\!G_L\psi_{l}^{\prime}
(z)\chi_{l}(z_+)-G_R\psi_{l}(z)\chi
_{l}^{\prime}(z_+) & G_L\chi_{l}^{\prime}
(z)\chi_{l}(z_+)-G_R\chi_{l}(z)\chi
_{l}^{\prime}(z_+)\\
G_L\psi_{l}^{\prime
}(z_+)\psi_{l}(z)-G_R\psi_{l}(z_+
)\psi_{l}^{\prime}(z) & G_L\psi_{l}^{\prime
}(z_+)\chi_{l}(z)-G_R\psi_{l}(z_+
)\chi_{l}^{\prime}(z)\label{R0192}\!
\end{pmatrix}
\end{equation}
where the shorthands for the size parameters $z\equiv k_j r_j$ and 
$z_+\equiv k_{j+1} r_j$ are introduced, as well as 
$G_L=\mu_{j}\mathrm{n}_{j+1}^2/\mathrm{n}_j$ and 
$G_R=\mu_{j+1}\mathrm{n}_{j+1}$ in the case of $G^{TM}$, whereas  
$G_L=\mu_{j+1}\mathrm{n}_{j}$ and $G_R=\mu_{j}\mathrm{n}_{j+1}$ in the case of 
$G^{TE}$. 

The vector $\mathbfcal{D}$, which includes the dipole contributions,  
can be written in a way that is general for any choice of $j$ from $1$ to 
$N+1$, given by
\begin{equation}
\mathbfcal{D}=\left(
\begin{array}
[c]{c}%
\mathcal{D}_1\\
\mathcal{D}_2\\
\mathcal{D}_3\\
\mathcal{D}_4
\end{array}
\right)=
\sum_{j=1}^{N+1}T(N+1,j)(1-\delta_{j,N+1})\mathbf{a}_{jH}
-T(N+1,j)(1-\delta_{j,1})\mathbf{a}_{jL}. \label{R020}
\end{equation}
The Kronecker delta functions ensure that the limiting cases of $j=1$ and 
$j=N+1$ have the correct form. Taking the example of $j=1$, the dipole would be 
in the innermost medium, and it is thus strictly the case that $r_1>r_1'$, and 
therefore only the $\mathbf{a}_{jH}$ term contributes to $\mathbfcal{D}$. 
In contrast, for $j=N+1$, the dipole is in the outermost medium, 
and since there is no radius of the structure greater than the radial location 
of the dipole, $r_{N+1}<r_{N+1}'$, only the $\mathbf{a}_{jL}$ term 
contributes. In all the other intermediate values of $j$, there is a 
contribution to the field coefficients in layer $j$, $\mathbf{A}_j$, 
from dipole sources placed at radii above $r_j$ and sources placed at radii 
below $r_j$, if such sources exist within the structure. 
The sum in Eq.~(\ref{R020}) is effectively over the regions that contain 
dipoles, since terms associated with regions with no dipoles are zero. 

Equation~(\ref{R018}) can be inverted to obtain $\mathbf{A}_{1}$ in terms of 
$\mathbf{A}_{N+1}$ as follows, 
\begin{equation}
\mathbf{A}_{1}=T^{-1}(N+1,1)\mathbf{A}_{N+1}+T^{-1}(N+1,1)\mathbfcal{D}.
\label{R021}%
\end{equation}
Note that the vector $\mathbfcal{D}$ contains both information about the
structure through the scattering matrix $T$, and information about the 
dipole sources through
$\mathbf{a}_{jH}$ and $\mathbf{a}_{jL}$. However, the matrices $T(N+1,1)$ and
$T^{-1}(N+1,1)$ are independent of any source, depending \emph{only} on the
parameters of the structure. 
To avoid confusion between the elements of matrices $T(N+1,1)$
and $T(N+1,j)$, from now on, the notations 
$T\equiv T(N+1,1)$ and $T^{j}\equiv T(N+1,j)$
for $j\neq1$ are used. 

The matrices of the form $T^{j}$ are block diagonal matrices, 
since they also have been built based on block diagonal matrices $M$. 
One can define an inverse scattering matrix 
$S\equiv\left(\begin{array}[c]{cc}S^{TM} & 0\\0 & S^{TE}\end{array}\right)  = 
T^{-1}(N+1,1)=T^{-1}$, 
which allows the equations for the 
resonances positions and the power spectra to be put in the most convenient 
form
\begin{equation}
S=\left(
\begin{array}
[c]{cccc}%
T_{11} & T_{12} & 0 & 0\\
T_{21} & T_{22} & 0 & 0\\
0 & 0 & T_{33} & T_{34}\\
0 & 0 & T_{43} & T_{44}%
\end{array}
\right)  ^{-1}=\left(
\begin{array}
[c]{cccc}%
S_{11} & S_{12} & 0 & 0\\
S_{21} & S_{22} & 0 & 0\\
0 & 0 & S_{33} & S_{34}\\
0 & 0 & S_{43} & S_{44}%
\end{array}
\right). \label{R022}%
\end{equation}
In the innermost region, the coefficients of the outgoing waves in 
Eq.~(\ref{R002}) and (\ref{R003}) are zero, and hence the vector 
$\mathbf{A}_{1}$ has the form $\mathbf{A}_{1}=(A_{1},0,C_{1},0)^{\mathrm{T}}$. Using 
this observation, 
the recursive formula in Eq.~(\ref{R021}) can be simplified and then 
solved to find $\mathbf{A}_{N+1}$ in terms of $A_{1}$ and $C_{1}$:
\begin{align}
A_{N+1}  &  =\mathcal{D}_{1}+\frac{S_{22}}{(-S_{21}S_{12}+S_{11}S_{22})}A_{1},
\label{R024}\\
B_{N+1}  &  =\mathcal{D}_{2}-\frac{S_{21}}{(-S_{21}S_{12}+S_{11}S_{22})}A_{1},
\label{R023}\\
C_{N+1}  &  =\mathcal{D}_{3}-\frac{S_{44}}{(S_{43}S_{34}-S_{33}S_{44})}C_{1},
\label{R026}\\
D_{N+1}  &  =\mathcal{D}_{4}+\frac{S_{43}}{(S_{43}S_{34}-S_{33}S_{44})}C_{1}.
\label{R0241}
\end{align}
Since the coefficients $\mathbf{A}_{N+1}=(A_{N+1},B_{N+1},C_{N+1},D_{N+1})^{\mathrm{T}}$
determine the fields in the outermost region, they 
are the scattering coefficients of the \emph{whole system of the microsphere and
its sources}. By way of comparison, these coefficient are equivalent to 
the $a_n$ and $b_n$ coefficients of Mie scattering ($B_{N+1}$ corresponds to 
$a_n$, and $D_{N+1}$ corresponds to $b_n$), and represent the different 
polarisations of the dipole moments. 

Once the values of the field coefficients in the outermost layer are known by 
calculating Eqs.~(\ref{R024}) through (\ref{R0241}), the coefficients in any 
other layer, $\mathbf{A}_{j}$, can be obtained from the recursive formula of 
Eq.~(\ref{Ajrec}). Then, the coefficients $\mathbf{A}_{j}$ may be substituted 
back into Eqs.~(\ref{R006}) and (\ref{R007}), in order to obtain the values 
of the total fields in any layer $j$, including dipole radiation, 
$\mathbf{E}_{j}^{\text{total}}$ and $\mathbf{H}_{j}^{\text{total}}$.  

In this section, the transfer matrix approach has been explained. The 
form of the transfer matrix has been shown explicitly, and its use in 
obtaining the field coefficients in any layer of a multilayer microsphere 
resonator has been described. For reference, a MATLAB code for the numerical 
calculation of these quantities, and the formulae in the next section, has 
been placed online for public availability-\footnote{
\href{http://www.photonicsimulation.net}
{\cb http://www.photonicsimulation.net\cbl}} see Appendix~\ref{sec:code}. 
While access to the field coefficients allows a wide range of quantities of 
potential interest to be calculated, it is important to enumerate some 
examples that are of particular relevance to this thesis. In the next section,  
recipes will be provided on how these quantities may be extracted from the 
general equations laid out in this section, and how to interpret them.

\section{\cdb Simulation recipes}
\label{sec:rec}

In simulating physical processes that are of phenomenological interest, such 
as WGMs, one must be clear in determining what aspect of the process 
to draw out from the formulae. Two particular examples of this, 
encountered in Chapter~\ref{chpt:sph}, are the positions of the WGMs 
and the emitted power spectrum. In the former case, the method of excitation 
of the modes is irrelevant -- the positions of the modes are dependent purely 
on the structure of the resonator in a given medium. 
However, in the case of the spectrum, the relative coupling of energy to 
different modes is highly dependent on the technique used to excite them. 
In fact, the $Q$-factor that is extracted from a measured spectrum 
takes into account this radiation component, as well as the intrinsic 
geometric $Q$-factor, and other sources of loss described in 
Section~\ref{subsect:gq}. 

In the first recipe, the method for extracting the positions of the structural 
resonances from the transfer matrix approach is stipulated. Then, the formulae 
for the scattered power as a function of wavelength are shown for several 
excitation scenarios (focusing on dipole sources in particular), which can be 
used to simulate spectra. While the results are summarised here for reference, 
full derivations of these formulae can be found in 
Appendix~\ref{chpt:app2}. 
But first, the use of Eqs.~(\ref{R024}) through (\ref{R0241}), which are 
general, is explained for a range of scenarios:\\

\vspace{-2mm}
\noindent\textbf{\cdb Scenario 1. No sources are present:}
If there are no dipoles in the structure, and no
incident wave on the resonator, then one can assign $\mathbfcal{D}=0$. 
In this case, one may calculate the ratios of the far-field scattering 
coefficients $B_{N+1}/A_{N+1}$ and $D_{N+1}/C_{N+1}$ from Eqs.~(\ref{R024}) through 
(\ref{R0241}), which removes the dependence of these quantities on the unknown 
coefficients $A_1$ and $C_1$ for the innermost layer. From these ratios, the 
denominators $A_{N+1}$ and $C_{N+1}$ specify the structural resonances since 
they become zero at the resonance wavelength.  By searching for roots in these 
two denominators, one can find the TM and TE mode positions, respectively, 
as described in Section~\ref{sec:struc}.\\

\vspace{-1mm}
\noindent\textbf{\cdb Scenario 2. Dipole sources are present:}
If there are dipole sources in the structure, then, 
without loss of generality, one may choose
$A_{N+1}=C_{N+1}=0$, solve for $A_{1}$ and $C_{1}$
from Eqs. (\ref{R024}) and (\ref{R026}), and then find $B_{N+1}$ and $D_{N+1}$
from Eqs. (\ref{R023}) and (\ref{R0241}), respectively. 
The reason why it is permissible to do this is due to 
the fact that the total emitted power must be the same in both the near and 
far fields. Therefore, by freely selecting the values of $A_{1}$ and $C_{1}$, 
the total power is not altered; however, it is more convenient to select a 
notation whereby the radiation, in the absence of structure, is simply a plane 
wave with a Bessel function. If a structure is present, then the extra 
scattering component is entirely incorporated into the coefficients $B_{N+1}$ 
and $D_{N+1}$, rather than being distributed arbitrarily across all four 
coefficients. This flexibility is essentially a consequence of the extra 
degrees of freedom afforded by the model in order to keep the notation of 
Eqs.~(\ref{R002}) through (\ref{R007}) in the same, general format for any 
layer, including the innermost and outermost regions.\\

\vspace{-1mm}
\noindent\textbf{\cdb Scenario 3. There is an incident field only:}
If there is only an incident plane wave interacting with the structure, then 
once again one may set $\mathbfcal{D}=0$. In addition, the coefficients $A_{N+1}$
and $C_{N+1}$ are known from the incident wave expansion. This provides enough 
information to solve the system of Eqs. (\ref{R024}) and (\ref{R026}), 
including $A_{1}$, $C_{1}$, $B_{N+1}$ and $D_{N+1}$.

\subsection{\cdb Structural resonances}
\label{sec:struc}

Consider the case where there are no sources -- neither plane waves nor 
dipoles, as described in \textbf{Scenario 1}. By setting $\mathbfcal{D}=0$ in 
Eqs.~(\ref{R023}) through (\ref{R026}), one finds
\begin{align}
B_{N+1}  &  =-\frac{S_{21}}{S_{22}}A_{N+1},\\
D_{N+1}  &  =-\frac{S_{43}}{S_{44}}C_{N+1}. 
\end{align}
Recall that, in the expansion of the electric field in terms of the VSH in 
Eq.~(\ref{R002}), $B_{N+1}$ is the coefficient of the $h_{l}^{(1)}(kr)$ term and 
$A_{N+1}$ is the coefficient of the $j_{l}(kr)$ term in the outermost region 
$N+1$. Hence, it is expected that the ratio $B_{N+1}/A_{N+1}$ should approach 
infinity near a resonance with a transverse magnetic component only, 
corresponding to the TM modes. Similarly, in the magnetic field of 
Eq.~(\ref{R003}), $D_{N+1}$ is the coefficient of the $h_{l}^{(1)}(kr)$ term and 
$C_{N+1}$ is the coefficient of the $j_{l}(kr)$ term, in the outer layer $N+1$. 
In this case, the ratio of $D_{N+1}/C_{N+1}$ should approach infinity near a 
resonance with a transverse electric component only, corresponding to the TE 
modes. As a result, both $S_{21}/S_{22}$ and $S_{43}/S_{44}$ can be used to 
calculate the TM and TE modes, respectively, meaning
\begin{align}
T_{11}  &  =0\text{ for TM resonances,}\label{R0261}\\
T_{33}  &  =0\text{ for TE resonances.} \label{R0262}
\end{align}
Note that these equations are, in general, multivalued. 
For a given \emph{azimuthal} quantum number $l$, the solutions  
to Eqs.~(\ref{R0261}) and (\ref{R0262}) form the fundamental \emph{radial} 
modes and their harmonics. Since the higher order harmonics have larger values 
of the Bessel function argument ($z$) at their root values than the fundamental 
modes, \emph{all} modes that appear within a given wavelength range, 
including higher order modes, can be obtained.
The numerical code associated with this thesis (Appendix~\ref{sec:code})
uses the optical and geometrical properties of any structure, to find the 
scattering matrix $T$, and numerically solves Eqs.~(\ref{R0261}) and 
(\ref{R0262}) for any range of wavelengths. Note that in the single-layer 
case, where $N=1$, these equations become exactly equivalent to the 
corresponding characteristic equations of the Johnson model \cite{Johnson:93}  
introduced in Section~\ref{sec:an}, as shown in Appendix~\ref{sec:john}.

\subsection{\cdb Scattered power}

In order to be able to simulate a spectrum, a method for explicitly 
calculating the total radiated power of the system including any dipole 
sources, as described in \textbf{Scenario 2}, must be outlined. 
The first step is to note the behaviour of the electric and magnetic fields in 
the far-field region, that is, for $r\rightarrow\infty$. In the outermost 
region, for $r\gg r_{N+1}^{\prime}$, the forms of the scattered 
fields in Eqs.~(\ref{R002}) and (\ref{R003}) may be simplified in the same 
manner as in Ref.~\cite{PhysRevA.13.396}, using the property that the Bessel 
function of the first kind is suppressed at long range. 

By performing the angular integration over the Poynting vector, constructed 
from the scattered $\mathbf{E}$ and $\mathbf{H}$ fields, the total radiated 
power takes the following form
\begin{align}
P_{\text{total}}&=\frac{c}{8\pi}\sqrt{\frac{\epsilon_{N+1}}{\mu_{N+1}}}\frac
{1}{k_{N+1}^{2}}\times\nonumber\\
&\sum_{l,m}\Bigg[\Big(\frac{1}{\mathrm{n}_{N+1}^{2}}\Big)\left\vert B_{N+1}
+a_{N+1EH}(r_{N+1}^{\prime})\right\vert ^{2}+\left\vert
D_{N+1}+a_{N+1MH}(r_{N+1}^{\prime})\right\vert ^{2}\Bigg]. \label{Pgen}
\end{align}
This equation is \emph{general}, 
and is valid for any number of dipole sources 
present within the multilayer structure. 
When there is no dipole in the 
outermost region, the formula can be further simplified, using 
$a_{N+1EH}(r_{N+1}^{\prime})=a_{N+1MH}(r_{N+1}^{\prime})=0$. 
However, to unpack this equation, a range of special cases will 
now be considered in order to demonstrate how the formula 
can be simplified to a given form.

\subsubsection{\cdb An embedded nanoparticle}

For the specific case where only a single dipole is placed in one of the 
regions $j$, which may be the innermost or outermost region, the scattering 
coefficients of Eq.~(\ref{Pgen}) can be written in a convenient form
\begin{align}
\label{eqn:BN}
B_{N+1}+a_{N+1EH}(r_{N+1}^{\prime}) &= \alpha_{l}a_{jEH}(r_{j}^{\prime})-\beta_{l}
a_{jEL}(r_{j}^{\prime}),\\
\label{eqn:DN}
D_{N+1}+a_{N+1MH}(r_{N+1}^{\prime}) &= \gamma_{l}a_{jMH}(r_{j}^{\prime})-\zeta_{l}
a_{jML}(r_{j}^{\prime}),\\
\text{where }\alpha_{l}\equiv(T_{22}^{j}+\frac{S_{21}}{S_{22}}T_{12}^{j})
&(1-\delta_{j,N+1})+\delta_{j,N+1},\,\,
\beta_{l}\equiv(T_{21}^{j}+\frac{S_{21}}{S_{22}}T_{11}^{j})
(1-\delta_{j,1}), \\
\text{and } \gamma_{l}\equiv(T_{44}^{j}+\frac{S_{43}}{S_{44}}T_{34}^{j})
&(1-\delta_{j,N+1})+\delta_{j,N+1},\,\, 
\zeta_{l}\equiv(T_{43}^{j}
+\frac{S_{43}}{S_{44}}T_{33}^{j})(1-\delta_{j,1}).
\end{align}
Using this notation, it is clear that $\alpha_{l}$ and $\beta_{l}$ correspond to 
the contributions from the \emph{TM modes}, whereas $\gamma_{l}$ and $\zeta_{l}$ 
correspond to the contributions from the \emph{TE modes}. 

By substituting in the forms of the dipole coefficients from 
Eqs.(\ref{R014}) through (\ref{R015}), radial (normal) and tangential 
(transverse) components of the total power can now be specified separately:
\begin{align}
&\frac{P_{\bot}}{P_{\bot}^{0}} = \frac{1}{2}\sqrt{\frac{\epsilon_{N+1}}
{\mu_{N+1}}}\frac{\mathrm{n}_{j}^{2}}{\mathrm{n}_{N+1}^{2}}\frac{3\mathrm{n}_{j}}
{\epsilon_{j}}\sum
_{l}(\frac{\mathrm{n}_{j}^{2}}{\mathrm{n}_{N+1}^{2}})(2l+1)l(l+1)\frac{\left\vert
\mathbf{\mathbf{[}}\alpha_{l}j_{l}(k_{j}r_{j}^{\prime})-\beta_{l}h_{l}
^{(1)}(k_{j}r_{j}^{\prime})\mathbf{]}\right\vert ^{2}}{k_{j}^{2}r_{j}
^{\prime2}},\label{Pbot}\\
&\frac{P_{\Vert}}{P_{\Vert}^{0}} = \frac{1}{4}\sqrt{\frac{\epsilon_{N+1}
}{\mu_{N+1}}}\frac{\mathrm{n}_{j}^{2}}{\mathrm{n}_{N+1}^{2}}\frac{3\mathrm{n}_{j}}
{\epsilon_{j}} \sum_{l}(2l+1)
\Bigg\{\Bigg[\Big(\frac{\mathrm{n}_{j}^{2}}{\mathrm{n}_{N+1}^{2}}\Big)
\frac{\left\vert
\{\alpha_{l}\frac{\mathrm{d}}{\mathrm{d}r_{j}^{\prime}}[r_{j}^{\prime}
j_{l}(k_{j}r_{j}^{\prime})]-\beta_{l}\frac{\mathrm{d}}{\mathrm{d}r_{j}
^{\prime}}[r_{j}^{\prime}h_{l}(k_{j}r_{j}^{\prime})]\}\right\vert ^{2}}
{k_{j}^{2}r_{j}^{\prime2}}\nonumber\\
  &+\left\vert [\gamma_{l}j_{l}(k_{j}r_{j}^{\prime})-\zeta_{l}h_{l}
^{(1)}(k_{j}r_{j}^{\prime})]\right\vert ^{2}\Bigg]\Bigg\}.\label{Pvert}
\end{align}
The normalisation of the emitted powers have been carried out with respect to 
the medium that contains the dipole, such that 
\begin{align}
P_{\bot}^{0}&=ck_{j}^{4}\left\vert \mathbf{P}_{r}\right\vert ^{2}/(3\epsilon_{j}
\mathrm{n}_{j}) \\
P_{\Vert}^{0}&=ck_{j}^{4}(\left\vert \mathbf{P}_{\theta}\right\vert
^{2}+\left\vert \mathbf{P}_{\phi}\right\vert ^{2})/(3\epsilon_{j}\mathrm{n}_{j}),
\end{align}
where $\mathbf{P}_{r},$ $\mathbf{P}_{\theta},$ and $\mathbf{P}_{\phi}$ are the
polar components of the polarisation vector $\mathbf{P}$. 
Equations~(\ref{Pbot}) and (\ref{Pvert}) are essentially 
more generalised versions of Eqs.~(\ref{eq:chewex}) and (\ref{eq:chewey}) from 
Chapter~\ref{chpt:sph}, respectively. It is shown in Appendix~\ref{sec:chew} 
that in the special case of a single layer $N=1$, corresponding to a 
microsphere, these sets of equations are equivalent. 

Note that the formulae for the emitted power shown thus far, 
Eqs.~(\ref{Pgen}), (\ref{Pbot}) and (\ref{Pvert}), are functions of the 
wave-number $k_j\equiv \mathrm{n}_j k$, and thus functions of the wavelength 
$\lambda=2\pi/k$. The emitted power may be computed as the wavelength 
$\lambda$ is input for a range of values. The wavelength values that coincide 
with the resonances will encounter strong contributions to the emitted power, 
corresponding to the WGMs.

\subsubsection{\cdb An active layer}

While the inclusion of any number of separate dipole sources represents an 
important, general functionality of the transfer matrix model, it will be 
useful to be able to replicate the fluorescent-layer approach, developed in 
Chapter~\ref{chpt:bub} for FDTD, in this general multilayer setting. 

Consider a multilayer structure where one of the layers consists 
of an active material. By applying the same technique employed in the 
Chew model \cite{Chew:88}, a uniform distribution of randomly-oriented 
dipoles, represented by a density function $\rho(r_j^{\prime})=1$, is introduced 
into a layer $j$. Since it is assumed that the orientations of the constituent 
dipoles are averaged over the distribution, the emitted power can be written 
in the form
\begin{equation}
\left\langle \frac{P_{\text{total}}}{P^{0}}\right\rangle =\frac{1}{3}\left\langle
\frac{P_{\bot}}{P_{\bot}^{0}}\right\rangle +\frac{2}{3}\left\langle
\frac{P_{\Vert}}{P_{\Vert}^{0}}\right\rangle, 
\end{equation}
the greater weighting of the transverse component originating from the 
two degrees of freedom associated with tangential oscillations \cite{Chew:88}. 
The normal and transverse components can be calculated separately by 
integrating the power with respect to the parameter $r_j^{\prime}$ over the 
shell region $j$
\begin{align}
\label{Pbotav}
\left\langle \frac{P_{\bot}}{P_{\bot}^{0}}\right\rangle 
&=\frac{1}{2}\sqrt{\frac{\epsilon_{N+1}}{\mu_{N+1}}}\frac{\mathrm{n}_{j}^{2}}
{\mathrm{n}_{N+1}^{2}}\frac{3\mathrm{n}_{j}}{k_{j}^{2}\epsilon_{j}V_{j\text{shell}}}
4\pi\sum_{l}(\frac{\mathrm{n}_{j}^{2}}{\mathrm{n}_{N+1}^{2}})l(l+1)I_{l}^{(1)},\\
\left\langle \frac{P_{\Vert}}{P_{\Vert}^{0}}\right\rangle 
&=\frac{1}{4}\sqrt{\frac{\epsilon_{N+1}}{\mu_{N+1}}}\frac{\mathrm{n}_{j}^{2}}%
{\mathrm{n}_{N+1}^{2}}\frac{3\mathrm{n}_{j}}{k_{j}^{2}\epsilon_{j}V_{j\text{shell}}}4\pi
\sum_{l}(\frac{\mathrm{n}_{j}^{2}}{\mathrm{n}_{N+1}^{2}})I_{l}^{(2)}+I_{l}^{(3)}. 
\label{Pvertav}
\end{align}
The shell region has a volume 
$V_{j\text{shell}}=4\pi(r_{j}^{\prime2} - r_{j-1}^{\prime2})$, noting that 
 if $j=1$, that is, the innermost layer is chosen to be populated with 
dipoles, then by convention, $r_0$ is taken to be zero as the volume is simply 
a sphere bounded by the innermost radius. 
The integral components of the above equations take the forms
\begin{align}
I_{l}^{(1)}  &\equiv (2l+1)\int_{j\text{shell}}\left\vert \mathbf{\mathbf{[}}
\alpha_{l}
j_{l}(k_{j}r_{j}^{\prime})-\beta_{l}h_{l}^{(1)}(k_{j}r_{j}^{\prime}
)\mathbf{]}\right\vert ^{2}\mathrm{d}r_{j}^{\prime},\\
I_{l}^{(2)}  &\equiv (2l+1)\int_{j\text{shell}}\left\vert \{\alpha_{l}
\frac{\mathrm{d}}{\mathrm{d}r_{j}^{\prime}}[r_{j}^{\prime}j_{l}(k_{j}
r_{j}^{\prime})]-\beta_{l}\frac{\mathrm{d}}{\mathrm{d}r_{j}^{\prime}}
[r_{j}^{\prime}h_{l}(k_{j}r_{j}^{\prime})]\}\right\vert ^{2}\mathrm{d}r_{j}^{\prime},\\
I_{l}^{(3)}  &\equiv (2l+1)\int_{j\text{shell}}k_{j}^{2}r_{j}^{\prime2}\left\vert
[\gamma_{l}j_{l}(k_{j}r_{j}^{\prime})-\zeta_{l}h_{l}^{(1)}(k_{j}r_{j}^{\prime})]
\right\vert ^{2}\mathrm{d}r_{j}^{\prime}. 
\end{align}
Note that the TE modes are entirely contained within $I_{l}^{(3)}$, 
with the remaining components containing the TM modes.  
These integrals can be computed by appealing to the properties of the 
Bessel and Hankel functions, using the functional form in Eq.~(\ref{R044}) 
of Appendix~\ref{sec:act}. The expressions that follow 
constitute a tractable form, which has been included in the 
code online (see Appendix~\ref{sec:code}). Estimates of the
computing time for the normalised emitted power at a single wavelength 
are discussed in Appendix~\ref{sec:scal}. 
With the expressions $I_{l}^{(1)}$, $I_{l}^{(2)}$ and $I_{l}^{(3)}$ now known, 
the total averaged power can be calculated by combining Eqs.~(\ref{Pbotav}) 
and (\ref{Pvertav})
\begin{equation}
\label{Ptot}
\left\langle \frac{P_{\text{total}}}{P^{0}}\right\rangle 
= \frac{1}{2}\sqrt{\frac{\epsilon_{N+1}}{\mu_{N+1}}}\frac{\mathrm{n}_{j}^{2}}%
{\mathrm{n}_{N+1}^{2}}\frac{\mathrm{n}_{j}}{k_{j}^{2}\epsilon_{j}V_{j\text{shell}}}
4\pi\sum_{l}\left[  (\frac{\mathrm{n}_{j}^{2}}
{\mathrm{n}_{N+1}^{2}})l(l+1)I_{l}^{(1)}+(\frac{\mathrm{n}_{j}^{2}%
}{\mathrm{n}_{N+1}^{2}})I_{l}^{(2)}+I_{l}^{(3)}\right].
\end{equation}
In the special case of a microsphere, a numerical calculation testing that 
this formula correctly reproduces a spectrum from the Chew model \cite{Chew:88} 
is shown in Appendix~\ref{sec:chew}.

\section{\cdb Demonstration of a coated microsphere including dispersion}
\label{sec:demo}

In order to demonstrate the functionality of the multilayer model, 
a scenario of contemporary relevance is considered, which  can be uniquely 
treated by the model. 
Consider a microsphere with a solid core of silica, which is then coated with 
a thin polymer (PMMA) layer \cite{LPOR:LPOR201200074}, 
leading to an outer diameter of 
$25$ $\mu$m. This scenario corresponds to the number of layers, $N=2$. 
PMMA is simulated in this example because it can act as an active layer 
\cite{Francois:13} and is a straightforward way of testing the functionality 
of the model. Both materials include dispersion through their respective 
Sellmeier equations \cite{Malitson:65,Kasarova20071481}, and across the 
range of wavelengths simulated, $590$ to $610$ nm, the refractive index of 
silica varies from $1.4584$ to $1.4577$, whereas the index of PMMA varies from 
$1.4913$ to $1.4902$. The resonator is also surrounded by water 
($\mathrm{n}_{3}=1.3300$, without dispersion). 

\begin{figure}[H]
\begin{center}
\includegraphics[width=0.8\hsize,angle=0]
{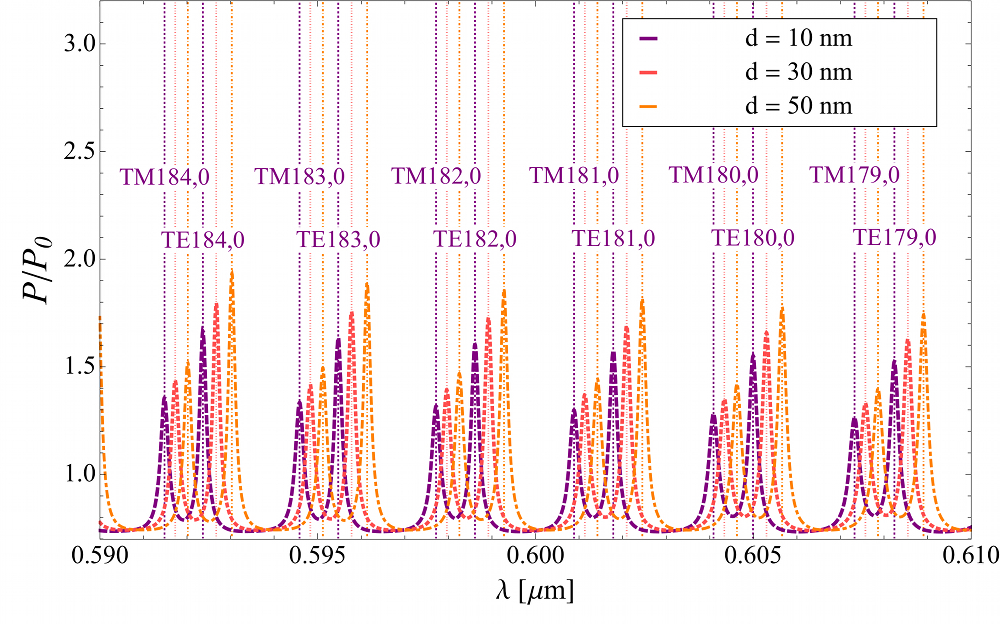}\\
\vspace{-3mm}
\hspace{6mm}\mbox{(a)}\\
\vspace{3mm}
\includegraphics[width=0.8\hsize,angle=0]{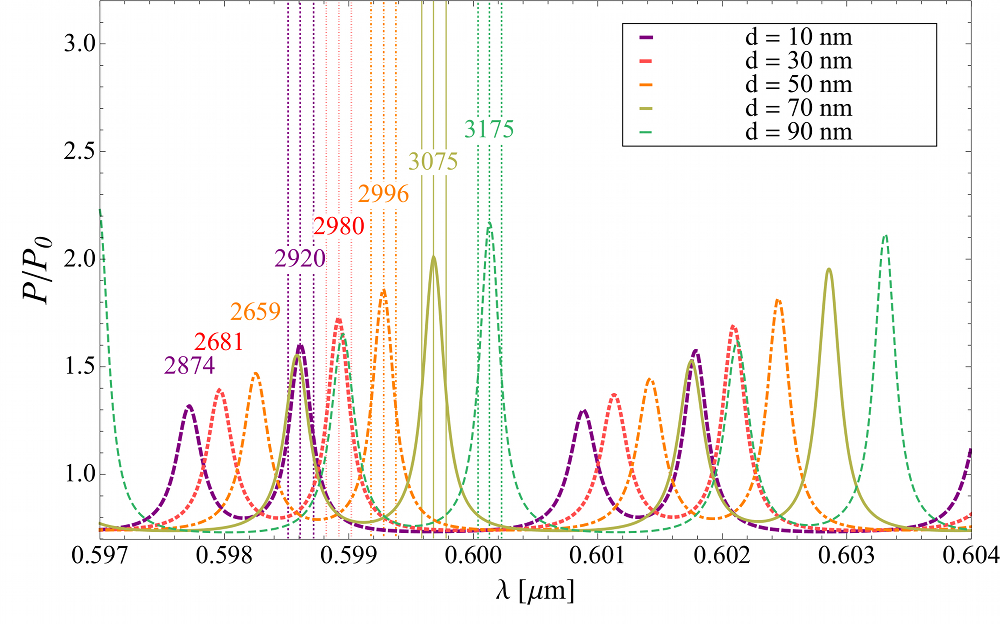}\\
\vspace{-3mm}
\hspace{6mm}\mbox{(b)}\\
\end{center}
\vspace{-9mm}\caption[PMMA-coated microspheres are simulated using multilayer 
model.]{{\protect{Spectra for a silica microsphere coated with a polymer layer 
(PMMA), both of which include dispersion. The polymer layer functions as an 
active layer. (a) Both TE and TM modes are excited. (b) A zoomed plot showing 
the FWHM of several peaks and their $Q$-factors.}}}
\label{fig:PMMA}
\figrule
\end{figure}

The behaviour of the WGM spectra as the thickness of the coating $d$ is varied 
is shown in Fig.~\ref{fig:PMMA}. 
At the diameter considered, higher order modes are not strongly coupled to, 
and so both the TE and TM modes remain distinct, as can be seen in 
Fig.~\ref{fig:PMMA}(a). The refractive index contrast between silica and PMMA 
is relatively small, and so the dependence of the mode positions on the 
thickness of the coating is more mild. A close-up view of the plot is shown in 
Fig.~\ref{fig:PMMA}(b). The vertical lines indicate the full-width at 
half-maximum (FWHM) positions for a variety of peaks, from which the 
$Q$-factors can be extracted, using Eq.~(\ref{eqn:qc}). $Q$-factors 
corresponding to a selection of TM and TE modes are also shown above each peak. 
Note that for the TM modes (the left three peaks) there is a decrease in the 
$Q$-factor as $d$ increases, but not for the TE modes. 
This is a consequence of the structure of the matrix elements that 
comprise the scattering coefficients of  
Eqs.~(\ref{eqn:BN}) and (\ref{eqn:DN}). 
This observation is 
consistent with the results in Section~\ref{sec:case}, and in the literature, 
which note the reduction of the $Q$-factors  of the TM modes for resonators 
with thin dielectric layers \cite{Yang:14,Hall:17}. 

\begin{figure}[H]
\begin{center}
\includegraphics[width=0.8\hsize,angle=0]{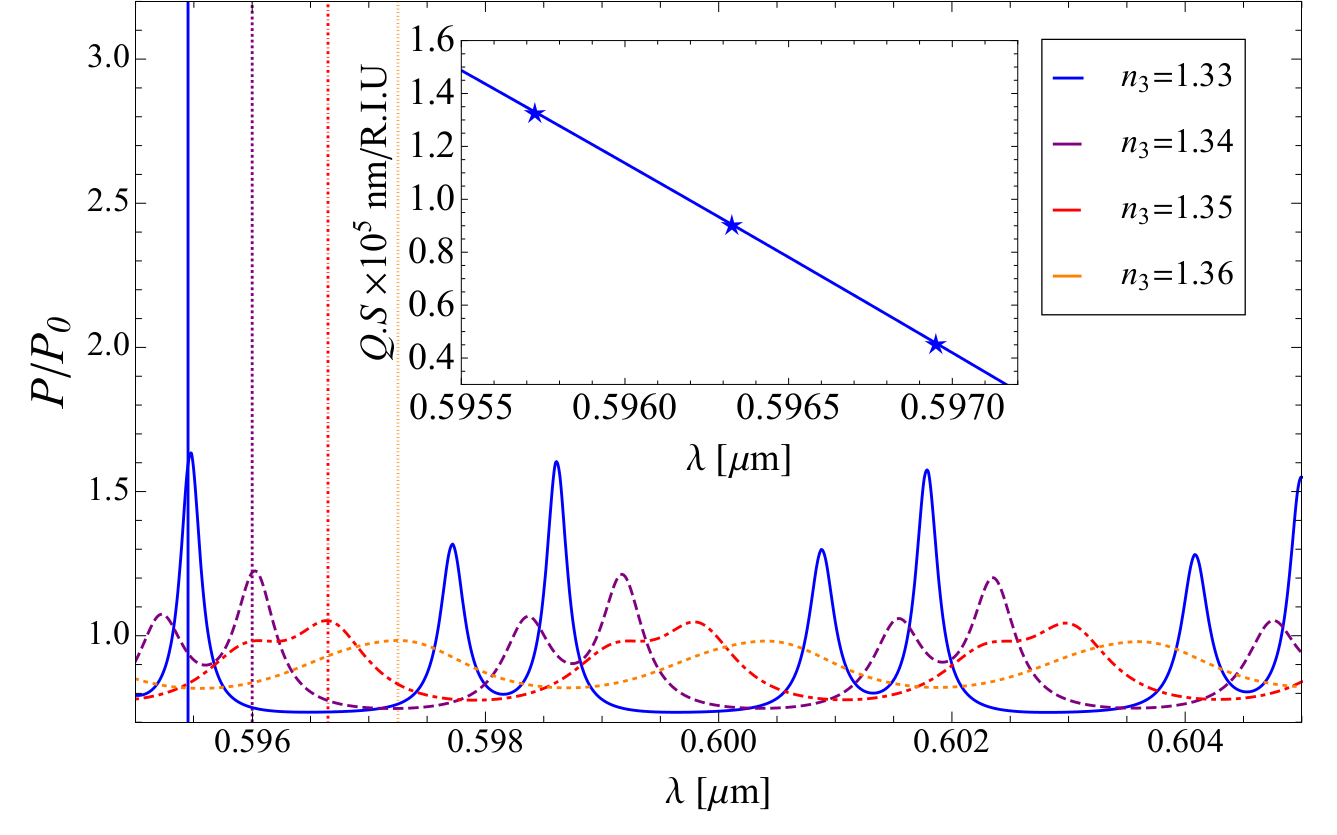}\\
\vspace{-3mm}
\hspace{6mm}\mbox{(a)}\\
\vspace{3mm}
\includegraphics[width=0.8\hsize,angle=0]{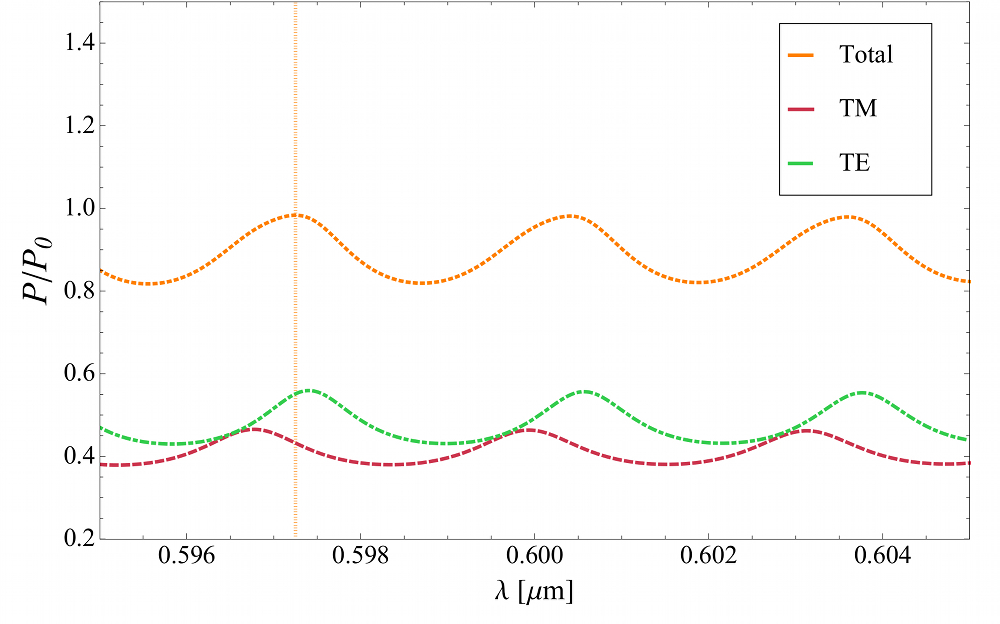}\\
\vspace{-3mm}
\hspace{5mm}\mbox{(b)}\\
\end{center}
\vspace{-9mm}\caption[The sensitivity of coated microspheres is simulated for 
$d=10$ nm.]{{\protect{The sensitivity of silica microspheres coated with a 
layer of PMMA of thickness of $d=10$ nm, (a) as a function of the surrounding 
refractive index. \textit{Inset:} a figure of merit, $Q.\mathcal{S}$, in units 
of $10^5$ nm/RIU, as a function of $\lambda$. (b) The total spectrum for 
$\mathrm{n}_3=1.3600$ is decomposed into its separate TE and TM components.}}}
\label{fig:demo}
\figrule
\end{figure}

\newpage
The sensitivity of the WGM peaks can be examined by varying the refractive 
index of the surrounding medium, $\mathrm{n}_3$. In Fig.~\ref{fig:demo}(a), 
the thickness of the layer is $d=10$ nm, and the mean TE peak shift leads to a 
sensitivity of 
$\mathcal{S}\equiv \mathrm{d}\lambda/\mathrm{d}\mathrm{n}_3 = 60.0$ nm per 
refractive index unit (RIU). 
A figure of merit (FOM), here defined as $Q.\mathcal{S}$ \cite{Reynolds:15}, 
is also provided to assess the sensing performance of the microspheres. 
For a given sensitivity $\mathcal{S}$, the value of $Q$ chosen is the mean 
value of the peak as it shifts to its new position as a function of 
$\mathrm{n}_3$. The inset of Fig.~\ref{fig:demo}(a) shows a decrease in the 
FOM as a function of $\lambda$.  
The total spectrum for  $\mathrm{n}_3=1.3600$ 
is shown in Fig.\ref{fig:demo}(b), 
along with the separate TE and TM contributions. 
By looking at the position of a particular mode in the total power spectrum, 
it is clear that it is affected by the close 
proximity and broad width of the TE and TM modes. This shows that in the 
simulation it is sometimes instructive to decompose the modes and analyse the 
structure in terms of the separate polarisation components, as the resonances 
that appear in a spectrum may indeed be superpositions of multiple modes. 

\begin{figure}[H]
\begin{center}
\includegraphics[width=0.8\hsize,angle=0]{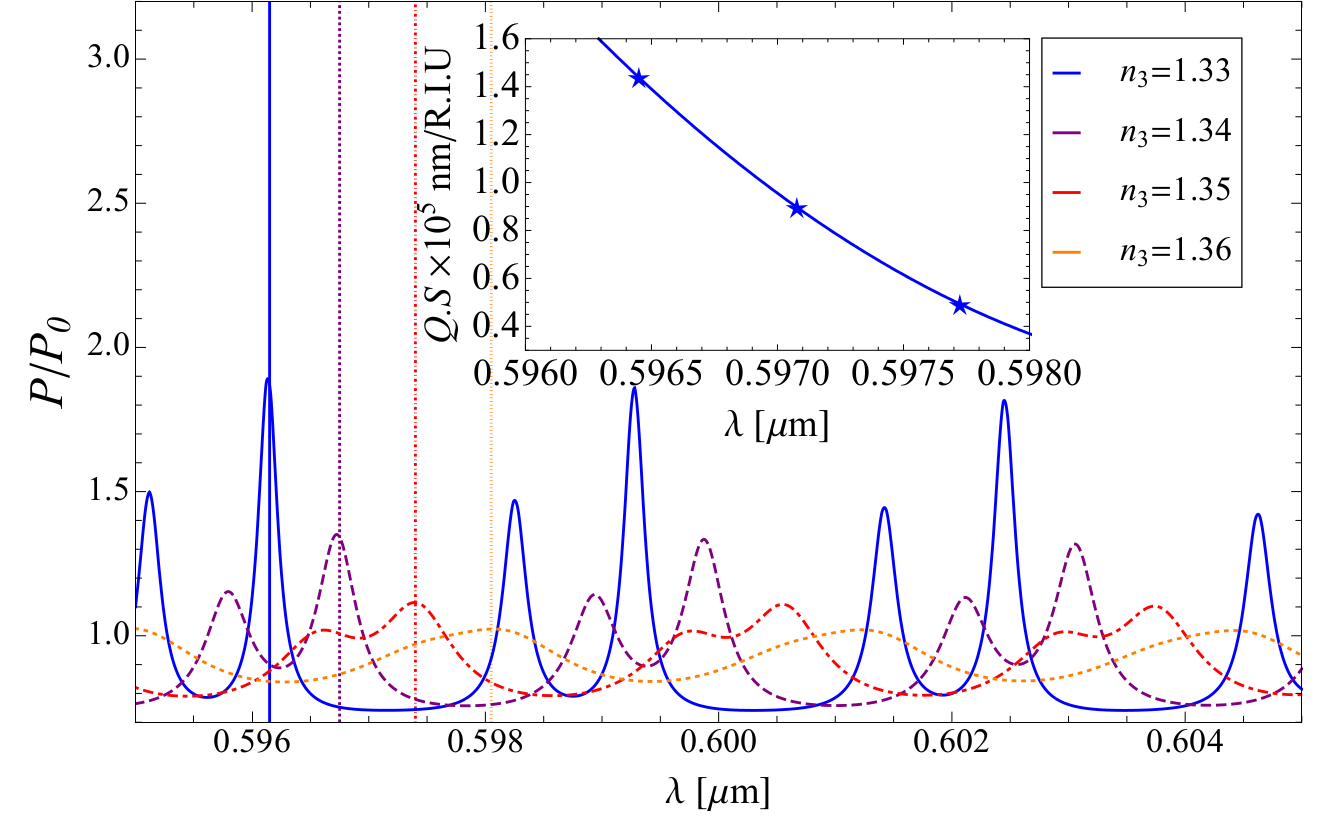}\\
\vspace{-3mm}
\hspace{6mm}\mbox{(a)}\\
\vspace{3mm}
\includegraphics[width=0.8\hsize,angle=0]{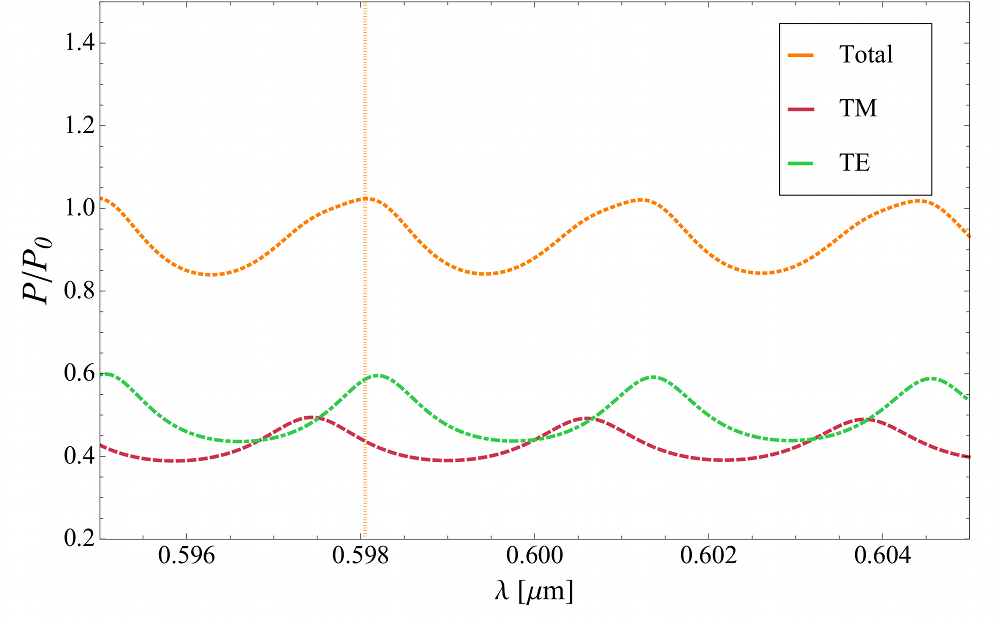}\\
\vspace{-3mm}
\hspace{5mm}\mbox{(b)}\\
\end{center}
\vspace{-9mm}\caption[The sensitivity of coated microspheres is simulated for
$d=50$ nm.]{{\protect{The sensitivity of silica microspheres coated with a 
layer of PMMA of thickness of $d=50$ nm, (a) as a function of the surrounding 
refractive index. \textit{Inset:} a figure of merit, $Q.\mathcal{S}$, in units 
of $10^5$ nm/RIU, as a function of $\lambda$. (b) The total spectrum for 
$\mathrm{n}_3=1.3600$ is decomposed into its separate TE and TM components.}}}
\label{fig:demo2}
\figrule
\end{figure}

In Figs.~\ref{fig:demo2}(a) and (b), a thicker layer is simulated, $d=50$ nm, 
and both the sensitivity and the FOM are larger, with 
$\mathcal{S} = 63.3$ nm/RIU. 
For both choices of layer thickness, the TE and TM modes are broad and 
overlapping in the case of a surrounding index of $1.3600$, 
and an extraction of the $Q$-factor using the FWHM approach will necessarily 
include contributions from both mode polarisations. This is a consequence of 
the fact that the $Q$-factors obtained from the spectrum include both 
intrinsic, non-radiative contributions from the geometry of the structure, and 
contributions from the relative coupling of the active layer to the modes. 
This makes it more difficult to separate nearby modes and define independent 
$Q$-factors, as illustrated in this analysis.

\section{\cdb Broader capacity of the model}

The development of a method for the modelling of whispering gallery modes in 
multilayer resonators represents an important tool for assessing the 
feasibility of a cell resonator. The model is able to handle an arbitrary 
number of concentric,

\newpage
\noindent spherical dielectric layers, and extract the resonance 
positions from the characteristic equation, and the emitted power, to obtain 
the spectrum. Because of the attributes of the specific cells considered in 
the next chapter, these features provide the best chance of giving a reliable 
assessment of a cell's ability to sustain resonances. While the focus of this 
thesis is to explore this possibility, first by developing the tools necessary 
for a systematic inquiry into the topic, including the functionalities 
required for this task, and then, by applying them to the search for a cell, 
the tools are able to contribute more widely to the field.

Various excitation scenarios that closely mirror the experimental
techniques are included in the model, and the model is cast in a framework 
that is easily extendable to consider other excitation strategies, which is a 
vital step toward facilitating the design of novel resonator architectures. 
Improvements to stability issues inherent in the transfer-matrix approach 
have also been addressed. Furthermore, alongside the development of the model 
and its formulae for the power spectrum, an efficient numerical algorithm was 
also produced, which may be of interest to the scientific community by adapting 
it for use in other scenarios that require multilayered dielectric structures. 

\addtocontents{toc}{\protect\newpage}
\chapter{\cdb The Search for a Cell}
\label{chpt:cel}

\textit{``But as soon as we touch the complex processes that go on in a 
living thing, be it plant or animal, we are at once forced to use the 
methods of this science. No longer will the microscope, the kymograph, 
the scalpel avail for the complete solution of the problem. 
For the further analysis of these phenomena which are in the flux and flow, 
the investigator must associate himself with those who have 
labored in the fields where molecules and atoms, rather than multicellular 
tissues or even unicellular organisms, are the units of study.''} 
\vspace{5mm}\\
\indent
J. J. Abel, ``Experimental and Chemical Studies of the Blood with an 
Appeal for More Extended Chemical Training for the Biological and Medical 
Investigator. {II}'', pp.~176-177 (1915) \cite{10.2307/1639865}.\\

The modelling methods developed in Chapters~\ref{chpt:sph} through 
\ref{chpt:mod} are integrated with the experimental investigations 
and methodologies reported in Chapters~\ref{chpt:cel}
 and \ref{chpt:wgm} in order to obtain a 
complete multi-layered 
solution of the study of resonances within biological cells. 
Upon arriving at a robust understanding of resonator behaviour through the 
development of sophisticated models, the goal of determining the existence of 
a biological candidate that can function as a resonator comes within reach. 
This is an attractive prospect, since the discovery of biological resonators 
would allow the useful properties of whispering gallery modes, especially 
their sensing applications, to be applied in a new setting. By producing 
light directly at the site of inquiry, a new level of access to the 
properties of the cells themselves, and their immediate environments, can 
thus be obtained. 

In the search for a cell resonator, one must first develop an understanding 
of precisely which physical qualities are required for resonances to be 
realised. Thus, in this chapter I seek to address a particular 
aspect of the vision of the project, as outlined in the Prologue, 
encapsulated by the question: `How \emph{imperfect} can a resonator be?'. 
To make this research question more precise, 
the goal of this chapter is to develop an original set of selection criteria, 
which may be used to narrow the search for a prospective biological resonator, 
and provide guidance as to the maximum amount of deviation from an ideal 
resonator such a candidate can exhibit before modes can no longer be sustained 
at a measurable level.

\section{\cdb Outline of the challenge}

As a preliminary step towards classifying biological cells in terms of their 
viability as resonators, the challenges of this task are now outlined. 
First, the 
methods and tools available for the measurement of the physical qualities 
of a prospective biological resonator are documented, and their roles 
explained. A wide variety of physical qualities is considered with 
regard to their impact on the ability of a given cell to sustain WGMs. 
Second, various physical 
properties that could in principle affect the resonances are enumerated. 
The next step is to narrow the 
list of conceivable physical influences on the resonance behaviour to a few 
key measurable features, which distil what is required of a resonator into 
a checklist of selection criteria. This list can be used to assess the 
feasibility of a candidate biological resonator for both this project and 
future research activities embarking on a similar path. 

Upon development of the selection criteria, a variety of candidate cells 
is analysed, and their likelihood of being able to support WGMs is 
subsequently assessed. This is an important exercise in the context of 
this thesis, because it not only demonstrates how the selection criteria 
should work in practice with real-life examples, but it also shows that a 
range of options has been considered in the event that the most viable 
candidate cells encounter challenges in sustaining WGMs that require more 
concerted research focus in future endeavours.

\section{\cdb Resonator assessment methods and tools}
\label{sec:pred}

The physical parameters of a cell can be quantified by the following methods 
and tools. 
Typically, the biological cells discussed herein exhibit some transparency in
the visible spectrum. 
As a result, confocal microscopy and scanning electron 
microscope (SEM) imagery can resolve the geometric properties of the cell 
\cite{doi:10.1095/biolreprod62.2.463}, such as the diameter and the 
relative membrane thicknesses. 
Accurate knowledge of the structure of the cell, and its approximate 
physical parameters, allows 
 modelling-based methods described in Chapters~\ref{chpt:sph} through 
\ref{chpt:mod} to make predictions for the refractive indices and other 
properties relating to the scattering signal. 
For example, calculating the \emph{dependence} of the mode positions
and $Q$-factors on the thickness of the outermost layer of a cell is particular 
suited to modelling-based approaches, while the absolute values of 
the $Q$-factors are known to be more difficult to estimate 
\cite{Little:99,Talebi,Min:2009a,Vahala:2004}. 

The symmetry of cross sections of a candidate cell can be similarly measured 
using conventional microscopy. Although there are a variety of shapes of cell 
with at least one axis of symmetry, such as the biconcave (lozenge-shaped) 
erythrocyte \cite{ery}, or animal red blood cell, the focus will primarily be 
on those of approximately spherical shape. The reason for this focus is that 
the majority of cells of the appropriate size for sustaining modes (see 
Section~\ref{sec:selec}) exhibit a level of symmetry predominantly due to 
forces such as surface tension, which are applied isometrically to the outer 
surface of the cell, and as such, demonstrate no preferred orientation in the 
development of the cell. 

In estimating the surface roughness of a resonator, the precision of 
conventional microscopy is generally insufficient, as it is limited to the 
scale of a micron \cite{Cosci:15}, whereas it has been shown that
 the positions and $Q$-factors of the WGM modes in, for example, a microsphere, 
are sensitive to deformations in diameter down to the nanometre scale 
\cite{Rahachou:03,Reynolds:15}. Thus, atomic-force microscopy (AFM), 
in which the surface roughness is measured by a mechanical probe attached to a 
piezoelectric filament, represents a more suitable method for studying the 
surface properties of microscopic resonators \cite{Ruan:14}. 

The `stiffness', or elasticity of the surface, presents an interesting 
challenge to resolving whether or not WGMs can be sustained. 
Conventional artificial resonators are typically 
constructed from substances that are highly unmalleable at room temperature, 
such as crystals \cite{doi:10.1021/nn5061207,PhysRevA.91.023812,
PhysRevLett.92.043903,Grudinin200633,Grudinin:07,PhysRevA.83.063847,
PhysRevLett.114.093902}, 
glass \cite{Vollmer:02a,Baaske2014,Ruan:14,Wildgen:15},  metal layers 
\cite{Ruppin:82,Talebi,Joshi:08,PhysRevLett.111.203903,Lou:14} 
and even certain types of polymers 
\cite{:/content/aip/journal/apl/83/8/10.1063/1.1605261,Francois:15b,C5LC00670H},
as mentioned in Section~\ref{sec:arch}. 
One benefit gained as a result of these choices is the subsequent minimisation 
of the distorting effects of a resonant wave on the geometry of the resonator, 
which could disrupt the path of the electromagnetic waves near the surface, 
leading to mode-splitting, and thus reducing the $Q$-factors 
\cite{Reynolds:15,Riesen:15a,LPOR:LPOR201600265}. 
However, the specific proteins from which cells are constructed differ widely 
in their mechanical properties \cite{Papi2010},  and depend on their chemical 
configuration.  To what extent 
the surface elasticity affects the positions and behaviour 
of the modes has been studied in the field of 
cavity optomechanics \cite{Kippenberg:07,RevModPhys.86.1391}. 
While it has been noted that the introduction of optomechanical 
behaviour in shell resonators yields acoustic modes with 
strong velocity dispersion characteristics causing deformations at the 
inner surface of the shell \cite{1367-2630-14-11-115026}, 
the critical value beyond which WGMs can no longer be sustained has not been 
considered exhaustively in the literature. 
A study of the surface elasticity of a prospective cell will 
help to place bounds on the $Q$-factor anticipated for potential resonances 
present during passive or active excitation. 
It is reported that one can minimise the elasticity 
by selecting a cell at a specific stage in its development 
\cite{Papi2010,Yanez2016}, discussed in more detail in in Section~\ref{sec:oo} 
for the cells that form the primary focus of this thesis as most viable 
resonator candidate.  
Alternatively, the elasticity can be altered by placing the cell in a medium 
that exhibits chemical properties that alter the stiffness of the 
surface layer. Examples of such handling media and the reasons for this 
effect on the elasticity are studied in detail in 
Section~\ref{subsec:surr}, and their recipes included in 
Appendix~\ref{chpt:app4}. The effect of the surrounding medium on the 
ability for a resonator to sustain WGMs 
forms the basis of the experiments described in Section~\ref{subsec:sil}.  

The refractive index experienced by WGMs will also be determined by the 
specific protein, or mixture of proteins, that constitutes the outermost layers 
of the cell. 
Measuring the refractive index of a protein can be achieved by collecting a 
quantity of the protein 
and precisely measuring the angle of the transmitted light at a specific 
wavelength \cite{Hand,doi:10.1021/ba-1964-0044.ch004,Voros2004553}. 
However, there is a practical limitation in collecting a sample of 
protein from a specific region of a cell in order to be able to carry out the 
measurement -- the largest cell considered in this thesis is approximately 
$150$ microns in 
diameter, and other kinds of cells of near-spherical shape are up to two 
orders of magnitude smaller. More elaborate methods of measuring the 
refractive index have also been attempted, such as the use of two adjacent 
fibres to emit and collect light from a low-power white-light source, in 
conjunction with a micromanipulator holding-pipette 
to guide the cell into position \cite{biodevices10,Wacogne:13}. 

Refractive indices can potentially be measured using WGMs themselves. 
WGMs provide precise information about the structure of a resonator 
while simultaneously presenting an opportunity to observe the optical 
behaviour of a cell, for which extant information in the literature is 
limited. As an example, a high refractive index coupling prism can be used to 
tune the coupling condition required for generating WGMs, and the angle of 
incidence depends on the refractive index experienced by the WGMs, 
explained in Section~\ref{subsec:pris}. It 
is important to keep in mind that the proteins that comprise the layers of 
real-life cells may not be uniform, and that the value of the refractive index 
ultimately measured through the use of WGMs will likely be an \emph{averaged} 
value, as the detection of modes requires multiple trips of the 
electromagnetic waves returning in phase within the cell, 
and any inhomogeneities in the refractive index will serve to broaden these 
modes and reduce the measured $Q$-factors. 

In addition to the refractive index of the biological material itself, the 
confinement of radiation within the cell is dependent on its 
surrounding environment. The behaviour of the WGMs as a function of the 
surrounding refractive index is a measure of the \textit{sensitivity} of the 
resonator, which in turn depends on the refractive index of the cell itself 
and its geometric attributes, as described in Chapter~\ref{chpt:intro}. 
Equally, a measurement of the sensitivity of a cell of known diameter and 
layer thickness can be used to estimate the refractive index of the cell. 
This technique is explored in Chapter~\ref{chpt:fdi}, which 
describes the flow of a 
glycerol-water solution over the cell as WGMs are being measured. 
This increases the refractive index of the surrounding environment by a known 
amount. 

The presence of contaminants in 
the surrounding medium can confer practical limitations to the experimental 
measurement of radiation modes within a cell. Detritus and dust that 
accumulates within a droplet of medium can interfere with or damage delicate 
apparatus such as fibre tapers. 
These contaminants must be removed, 
minimised, or their effects taken into account without compromising the other 
physical properties of the cell, in order to optimise 
achieving WGMs. 

The generation of WGMs and values of their $Q$-factors are, of course, 
dependent on the excitation strategy used. 
For a given method, it is possible to tune the coupling efficiency 
in order to improve the $Q$-factors and the clarity of the modes in the 
spectrum. 
Prism coupling \cite{Gorodetsky:99} 
introduces a reusable method for testing cell candidates, as the prism and 
laser setup does not require realignment after each test. This method is mode 
selective, and the coupling condition is extremely sensitive to both the exact 
placement of the laser spot on the surface of the prism and the distance of 
the cell above the prism. Coupling into a specific mode can 
therefore become a challenging procedure, especially for a resonator with 
geometric parameters that are not well known prior to mode coupling. 
Furthermore, the spot size of the laser must be constrained so 
that the propagation constant is phase-matched at the boundary of the prism, 
as  described in Eq.(~\ref{eq:kwgm}) of Section~\ref{sec:an}, 
in order to achieve tunnelling into the resonator. 
Since laser spot diameters 
below $10$ $\mu$m become difficult to collimate, relaxing the requirement of 
beam collimation becomes mandatory; however, it poses an additional difficulty 
in the placement of bulky detectors in a confined region around the resonator. 
As an alternative, one may use free-space optics to stimulate autofluorescence 
of the cell \cite{ELPS:ELPS200406070}, fluorescent emission of incorporated 
nanoparticles \cite{C1JM10531K,6062374,1063-7818-44-3-189},  or an active 
material coating \cite{Francois:15a}. Such active resonators typically exhibit 
lower $Q$-factors than those expected using passive interrogation, as 
discussed in Sec~\ref{subsec:act}, but in return, they provide some distinct 
practical advantages. In the specific case of free-floating resonators, such 
as microspheres or cells, these advantages include the fact that the coating 
of fluorescence emitters can be prepared prior to the experimental measurement, 
and requires no special alignment in order to generate WGMs. Each of these 
methods is applied in Chapter~\ref{chpt:wgm} in order to build a picture of 
the viability of the chosen cell candidate as a biological resonator. 

The coupling efficiency of the prism may be tuned by altering 
both the spot size of the beam on the upper surface of the prism, and the 
angle of incidence of the incoming beam, so that the propagation constant near 
the surface of the prism matches those of the WGMs near the interface of the 
resonator and prism, fulfilling the condition in Eq.~(\ref{eq:kwgm}) of 
Section~\ref{sec:an}. This process achieves the phase-matching 
condition necessary for the experiments reported in Section~\ref{subsec:pris}. 

Optical waveguides \cite{Astratov:04,0022-3727-39-24-006} or
fibre tapers \cite{Armani2003,1674-1056-17-3-047,6525394,Riesen:15a} provide 
coupling to a range of modes for sufficiently small waveguide or taper waists 
that ensure phase matching at the resonator boundary. 
While alterations to the taper during measurement are impractical, this is 
effectively achieved by tuning the distance of the taper to the resonator 
until they are phase-matched, as demonstrated in Section~\ref{subsec:fib}. 
Fibre tapers have been 
shown to lead to particularly high $Q$-factors relative to other mode coupling 
methods \cite{Knight:97,Dong:08,Riesen:15a,Jin:15}, since 
placement of the taper tangentially in the vicinity of the resonator 
allows radiation from WGMs along a single axis of revolution to be isolated, 
as illustrated in Fig.~\ref{fig:tapbubex} of Chapter~\ref{chpt:intro}. 
However, the manufacturing 
process of the tapers is time-consuming, and being fragile, their longevity is 
somewhat limited, and thus present practical restrictions on testing a large 
number of candidate cell resonators. 

In the case of an active layer coating, 
there is little control over the phase-matching condition. A layer of 
fluorophores will encompass a range of different orientations so that the 
excitation occurs at all points on the resonator uniformly. While  a simple 
free-space power collection 
will not be able to isolate modes from a single axis of symmetry, it is 
nevertheless possible to use excitation and collection methods in 
combination, such as a resonator including an active coating that is excited, 
with power collected via a fibre taper. 

The absorption and scattering of light by a resonator represent crucial 
limiting factors that must be taken into account in the determination of a 
candidate cell resonator. 
For an amount of energy $E$ either transmitted or incident upon a surface, 
the radiant flux is defined as $\Phi_e = \partial E_e/\partial t$. 
The absorbance, a dimensionless quantity, denoted $\mathcal{A}$, 
can be defined in terms of the ratio of the radiant flux transmitted 
through a surface ($\Phi_e^t$) to radiant flux received by a surface 
($\Phi_e^i$) \cite{mcnaught1997compendium}
\begin{equation}
\mathcal{A} = \mathrm{log}_{10} \left(\frac{\Phi_e^i}{\Phi_e^t}\right). 
\end{equation}
$\mathcal{A}$ can be measured using 
an ultraviolet--visible--near-infrared (UV--Vis--NIR) spectrophotometer, 
making use of the Beer-Lambert Law 
\begin{equation}
\mathcal{A} = \sum_{i=1}^{\mathscr{N}} \epsilon_i \int_0^{\mathcal{L}} 
\mathrm{c}_i(z)
\mathrm{d}z,
\end{equation}
for a mixture of $\mathscr{N}$ protein types with molar attenuation 
coefficients of $\epsilon_i$, and concentrations of $\mathrm{c}_i$ along a 
path length $\mathcal{L}$. 
The attenuation coefficients, $\epsilon_i$, represent an intrinsic property 
of a given chemical species, with SI units of cross sectional area per mole. 
The concentrations are defined in terms of moles of solute per unit volume. 
While contemporary developments in the measurement of 
the properties of biological tissue have yielded interesting behaviour in the 
UV region \cite{0031-9155-58-11-R37,doi:10.1117/12.2080408,Levenson:16}, DNA, 
and its constituent proteins, are strongly absorbing in the UV 
\cite{Clydesdale2001,Jackson2009}. In the context of this investigation, 
one consequence of this absorbance profile is the difficulty in sustaining WGMs 
in this wavelength range. 
While two-photon fluorescence microscopy has been used to determine 
architectural features of cells, particularly for photosensitive varieties 
such as mammalian embryos \cite{Squirrell1999}, the existence 
of underlying modes within such cells and their spectroscopic properties 
remains uncharted territory. 

Scattering, on the other hand, can be related back to the surface roughness of 
the resonator. Both the scattered and transmitted light can be measured using 
the prism coupler technique, described in more detail in 
Section~\ref{subsec:pris}, and this simultaneous measurement provides 
critical real-time information on how the cell is behaving as a resonator. 
If the transmitted signal is very weak, it indicates that the cell is absorbing 
or scattering the light to a high degree. Careful monitoring of the collection 
of the scattered light thus serves to identify the source of the low 
transmission, and possible causes for poor resonator performance such as 
unoptimised coupling, the presence of surface defects, or contamination of the 
surrounding medium. 

Mie scattering models have long been used to 
estimate scattering effects \cite{cellscatt}, and more recently, FDTD has been 
applied to assess the general properties of free-floating cells 
\cite{Drezek:00,1492942}. 
Other methods that have been used in the 
literature to consider scattering effects in cells include Fourier Transform 
Light Scattering (FTLS) \cite{5443519} and Fourier Transform Infrared (FTIR) 
Spectroscopy \cite{Ami20111220}. While power spectra associated with 
scattering from mammalian oocytes have been reported in the literature 
\cite{An:15}, they are typically adopted for medical imaging of tissue.  
Studies that explore coupling methods for generating  
resonances, such as WGMs, within cells such as oocytes are practically 
nonexistent. With the aid of the sophisticated modelling tools 
developed in Chapters~\ref{chpt:sph} and \ref{chpt:mod}, and the 
non-destructive 
method for the determination of the geometric parameters of a resonator 
developed in Chapter~\ref{chpt:bub}, the work in this thesis is ideally poised 
to address the problem of matching the theoretical predictions of scattered 
spectra with experimental results. 

The physical parameters described above represent the main features 
to be investigated in the search for a cell resonator. Now that these 
parameters and their methods of measurement have been enumerated, the specific 
selection criteria for a candidate cell resonator, which has been narrowed 
down to seven key requirements, can be summarised.

\section{\cdb Selection criteria}
\label{sec:selec}

In this section, the key selection criteria are stated and explained based on 
the properties required of a resonator, as well as the range of complexities 
introduced by biological material that can serve to frustrate, enhance or 
otherwise modify these properties. \\

\begin{figure}[t]
\begin{center}
\includegraphics[width=0.8\hsize]{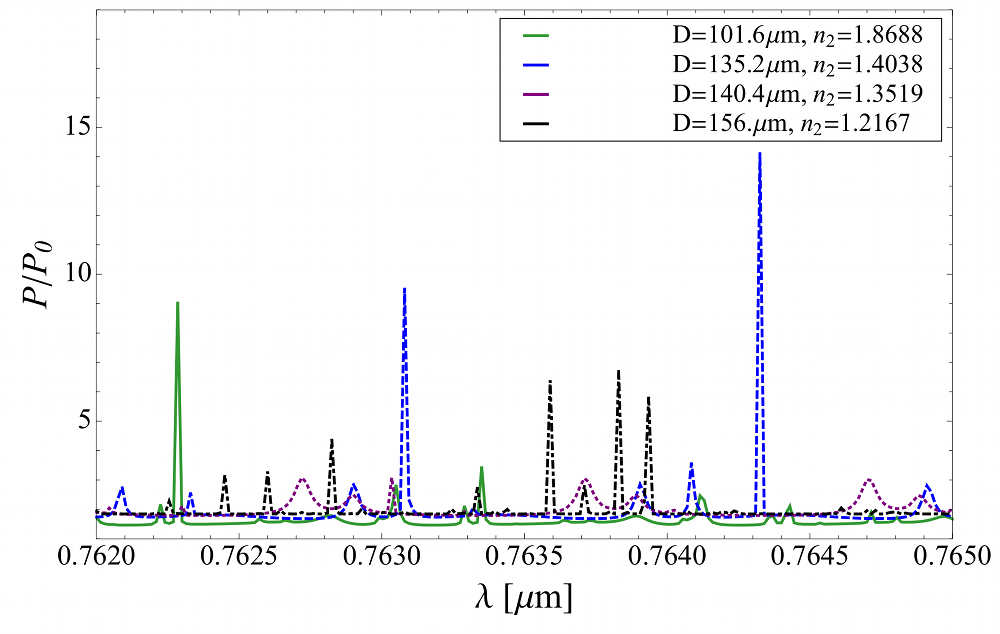}
\\
\vspace{-3mm}
\end{center}
\caption[Spectra are shown for a range of 
diameters and index contrasts.]{{The emitted power from a 
cell-analogue microsphere resonator in a surrounding medium of water, 
as obtained from the Chew model \cite{Chew:87a}. Both the TE and TM 
modes are excited by a dipole source with both radial and tangential 
components. Results are shown for a range of diameter values, $D$, 
and resonator refractive index values,  $\mathrm{n}_2$, 
such that the combination $\mathrm{n}_2\,D$ is held fixed. This example 
illustrates the competing effects of diameter and refractive 
index contrast on the mode positions within a spectrum,
as well as the ability of a resonator to sustain WGMs.}}
\label{fig:oo}
\figrule
\end{figure}

\noindent\textbf{\cdb Criterion 1. Size and index contrast:} The ability of a 
resonator to sustain modes depends on the relative values of the size and 
refractive index contrast, as introduced in 
Section~\ref{sec:bubbehav}. The relationship between size and index contrast 
has been well studied \cite{Reynolds:15}. A large refractive index contrast 
will typically allow higher order modes to be sustained \cite{Hall:15}, and the 
same is true for large diameter resonators \cite{Hall:17}. 
The two are not directly equivalent, however, and an \emph{increase} in size 
with a corresponding proportional \emph{decrease} in index contrast 
does not result in an 
identical mode pattern. This is illustrated in Fig.~\ref{fig:oo}, 
where the emitted power from a microsphere in water is plotted for a range of 
diameters $D$ and refractive indices $\mathrm{n}_2$, 
such that the quantity $\mathrm{n}_2\,D$ remains constant. The diameter and 
index values chosen in this example correspond closely to those of the most 
likely cell candidate described in Section~\ref{sec:oo}. This effect is a 
consequence of the mathematical form of the emitted power formulae 
of Eqs.~(\ref{Pbot}) and (\ref{Pvert}), derived from the matrix 
elements of Eq.~(\ref{R0192}), which are functions of size parameters 
$z \equiv k_2\,r = \pi\,\mathrm{n}_2\,D/\lambda_0$, and 
$z_+ \equiv k_3\,r_j = \pi\,\mathrm{n}_3\,D/\lambda_0$. 
While the spectrum remains unaltered in the case where the length scale 
$D/\lambda$ is held constant, changes in $D$ and $\mathrm{n}_2$ such that 
$z$ remains constant will still result in a different value of $z_+$, unless 
the index of the surrounding medium, $\mathrm{n}_3 = \mathrm{n}_{\text{water}}$, 
is also altered by the same proportional factor. 

Of particular interest is the limiting case of the \emph{minimum} viable 
size for a typical range of refractive indices, which for cells is generally 
dependent on the 
optical properties of the proteins that constitute them.  A conservative 
estimate for these proteins suggests the range to be $\mathrm{n}_2=1.36$ to 
$1.51$, with some proteins approaching $1.55$ \cite{Voros2004553}. These 
estimates are important for the discussion of glycoprotein layers in
 Section~\ref{subsec:gri}. 
For example, using a nominal criterion that the $Q$-factor must be greater than 
$10^3$ to be 
adequately detected, the minimum viable diameter for these lower and upper 
values of the index are calculated from the multilayer model of 
Chapter~\ref{chpt:mod} to be $D=86$ $\mu$m and $16.8$ $\mu$m, respectively. 
Note that analytic models typically overestimate the measurable $Q$-factors 
\cite{Little:99,Talebi}, and that 
without the inclusion of surface roughness \cite{Ilchenko:13} or 
asphericity \cite{Riesen:15a} it is difficult to match the $Q$-factors 
derived from models with those obtained experimentally 
\cite{Min:2009a,Vahala:2004}. As a result, it is important to note that 
the minimum viable diameters obtained from the multilayer model are 
conservative, and realistically, it is expected that diameters should be 
substantially larger than the minimum values calculated below 
for the adequate detection of WGMs experimentally. 

\begin{figure}[H]
\begin{center}
\includegraphics[width=0.8\hsize]{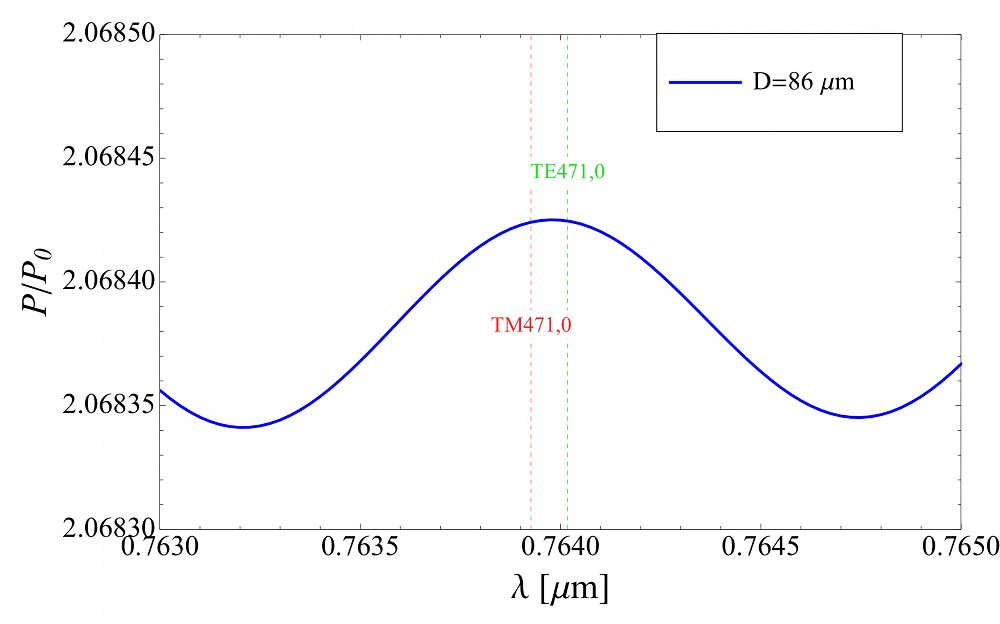}\\
\vspace{-5mm}
\hspace{12mm}\mbox{(a)}\\
\hspace{8mm}\includegraphics[width=0.8\hsize]{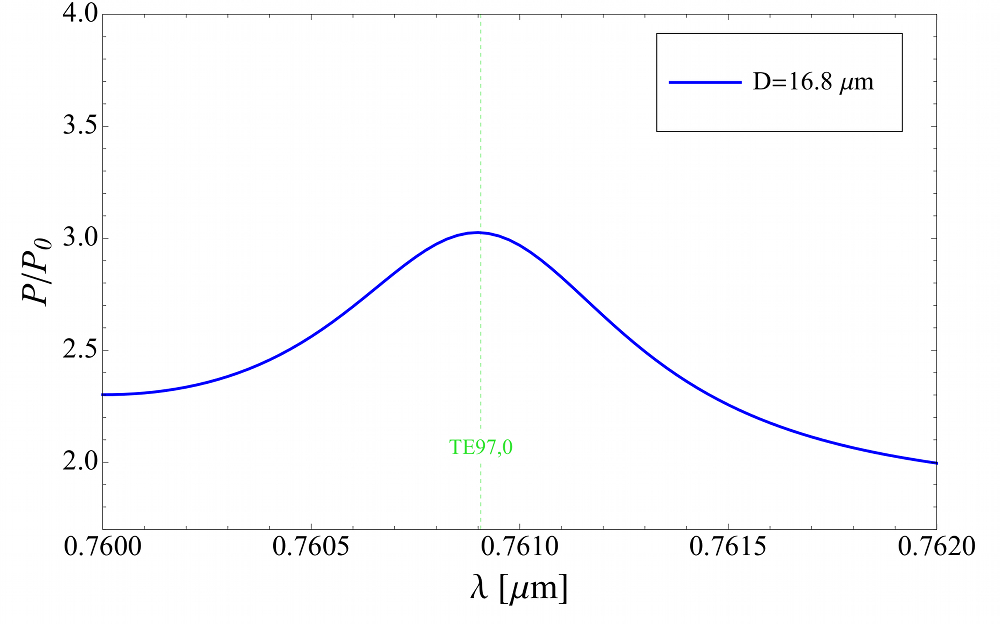}\\
\vspace{-5mm}
\hspace{11mm}\mbox{(b)}\\
\end{center}
\vspace{-5mm}\caption[The limiting diameters for mode detection are explored 
for proteins.]{Spectra corresponding to the limiting diameter values for mode 
detection are shown, estimated from the multilayer model of 
Chapter~\ref{chpt:mod} for a microsphere in water, for wavelengths within 
the range of $760$ to $765$ nm. 
The cell refractive indices considered are (a) the lower limit 
$\mathrm{n}_2=1.36$ corresponding to a minimum diameter 
of $D=86$ $\mu$m, and (b) the upper limit $\mathrm{n}_2=1.51$ 
corresponding to a minimum diameter of $D=16.8$ $\mu$m. 
In the former case, the fluctuation in the emitted power 
is relatively small, indicating the impact of a low index contrast on the 
spectrum and the ability to detect modes. 
}
\label{fig:lim}
\figrule
\end{figure}

The spectra derived from the two limiting cases from the multilayer model, 
$D=86$ $\mu$m and $16.8$ $\mu$m, are shown in Fig.~\ref{fig:lim}, 
as estimated from the central modes of TE/TM$_{471,0}$ and TE$_{97,0}$, 
respectively. Note that, in the former case, the resonance envelope contains 
contributions from nearby overlapping TE and TM modes of the same azimuthal 
order. 
In both of these plots, the wavelength window is chosen to coincide with $760$ 
nm,  for consistency with the distributed feedback (DFB) laser used 
for the experiments described in Chapter~\ref{chpt:wgm}. 
Note that in the case of Fig.~\ref{fig:lim}(a), the lower refractive index 
contrast leads to a mode with a relatively small fluctuation in the emitted 
power. This is a consequence of the fact that a low index 
contrast impacts strongly on the ability to detect modes. While it is possible 
to resolve such small power fluctuations, the purpose of this exercise is 
simply to establish the extreme limits from the $Q$-factor derived from the 
multilayer model. 
It is clear that the minimum 
viable diameter varies widely for this range of possible index values. To 
ensure the best possible chance of selecting a viable cell resonator 
candidate, without knowing the specific value of the refractive index in 
advance, the diameter must be larger than the value of $86$ $\mu$m.
\\
 
\noindent\textbf{\cdb Criterion 2. Sphericity:} Deformations of the resonator 
that impact upon the sphericity have been shown to have an effect on the 
$Q$-factors of the WGMs \cite{Fujii:2005a,Fujii:2005b,Riesen:15a}. 
According to one study, FDTD simulations of a prolate polystyrene 
resonator excited by an incident beam exhibit a decrease in $Q$-factor 
that is roughly proportional to the ellipticity of the resonator (measured in 
terms of its aspect ratio) \cite{Fujii:2005b}. 
In the case of microspheres with only small asphericities, it is useful to 
incorporate this limiting factor into the $Q$-factor by adding an elliptical 
contribution, $Q_e$, in parallel, as in Eq.~(\ref{eqn:Q}). 
While this poses a potential difficulty in sourcing a resonator that is 

\newpage
\noindent  
sufficiently spherical to sustain modes, there is a number of
techniques that can serve to mitigate this effect. In particular, excitation 
of WGMs preferentially along one equatorial plane 
can minimise the contributions from the 
different path lengths experienced by the internal fields that 
contribute to mode-splitting and mode-broadening, as described in 
Section~\ref{subsect:gq}. This may be achieved using fibre taper couplers, 
which are able to isolate a single axis of revolution of a resonator in order 
to achieve higher $Q$-factors and coupling efficiencies 
\cite{Riesen:15a,Knight:97,Dong:08}, as mentioned in Section~\ref{sec:pred}. 
This represents an attractive feature of 
fibre tapers, particularly when considering resonators that are 
unlikely to exhibit a perfectly spherical shape, such as biological many cells. 
\\

\noindent\textbf{\cdb Criterion 3. Surface roughness:} It has been shown that 
the $Q$-factors of WGMs are extremely sensitive to surface roughness 
\cite{Rahachou:03,Reynolds:15}, with deviations down to the scale of a single 
nanometre significantly affecting the wavelength positions of the modes 
\cite{Ruan:14}. A study of the behaviour of the WGMs within a microdisk cavity 
has demonstrated a reduction of the normally high $Q$-factors by several 
orders of magnitude for an edge roughness of less than $\lambda/30$ 
\cite{Rahachou:03}. This makes the 
isolation of any one single mode potentially difficult task, placing additional 
burden upon this criterion as an important selection characteristic  
to be addressed for the range of cells considered in Section~\ref{sec:cand}. 
\\

\noindent\textbf{\cdb Criterion 4. Elasticity:} The mechanical properties of 
resonators, particularly of the surface or outermost layer, can affect its 
ability to support WGMs for the following reasons. 
Bound states of radiation located near the surface of a resonator will 
tend to dissipate heat and cause 
mechanical stress. The fluctuations in the power corresponding to the modes 
thus introduce perturbations that are capable of disrupting the path travelled 
by the bound waves, especially for malleable materials. This can lead to 
shorter ring-down times, shifts in the wavelength positions of the modes 
\cite{Himmelhaus2009418} or even total mode loss, which may occur sooner than 
anticipated compared to a rigid resonator, as explained in 
Section~\ref{sec:pred}. 
It is therefore important to consider the mechanical properties of biological 
cells, and in particular, the protein structures that form the outer layers of 
the viable cell candidates. This will be discussed in Section~\ref{sec:oo}. 
\\

\noindent\textbf{\cdb Criterion 5. Osmolarity:} The total concentration of 
soluble particles, or \emph{osmolarity} of the surrounding medium, affects the 
ability to measure WGMs in a resonator. 
Lasers can serve to ionise the salts in the solution, 
which can accumulate on the surface and compound the contamination of the 
medium. This effect is particularly detrimental to the functioning of fibre 
tapers. While this issue 
can be alleviated by reducing the laser power, 
the binding of non-soluble particles to a fibre taper can contribute 
a non-specific background signal as well as lowering the overall transmission 
through the taper, making the detection of WGM signals more challenging 
\cite{s150408968}. 
This is 
particularly important when considering the cells described in 
Section~\ref{sec:oo}, 
since the media used to store the cells 
typically contain certain quantities of salts (see Appendix~\ref{chpt:app4}), 
and even a relatively small osmolarity ($280$ mOsm) can serve to 
hamper measurement. Furthermore, in small volumes of liquid medium, the heat 
caused by the energy from the laser will eventually evaporate the liquid, 
occurring within minutes for large power outputs such as those 
above $20$ mW. 
As a result, the 
increased concentration and subsequent crystallisation of residual salt 
within the medium 
can cause physical damage to the cell and to a fragile coupler such as a taper. 
Before performing a measurement, it is vital to 
ensure that the medium is as dilute as possible without indirectly altering 
the morphology of the cells. This is handled on a case-by-case basis, as the 
chemical composition of media can differ widely, as shown in 
Appendix~\ref{chpt:app4}. In the specific case of the most viable cell 
candidate, the methods outlined for handling this effect are described in 
Sections~\ref{subsec:surr} and \ref{sec:samp}. 

As an additional point, it is interesting to note that the chemical 
composition of the surrounding medium chosen for a specific biological cell 
can affect the elastic properties of its outermost layer, such as in the case 
of the bovine oocyte \cite{fetuin}. 
\\

\noindent\textbf{\cdb Criterion 6. Excitation method:} The successful 
measurement of WGMs is intimately connected to the method of introducing 
radiation into the resonator. As has been explained in the case of the fibre 
taper in \textbf{Criterion~2}, one may isolate specific 
resonances, or classes of resonances (such as those stemming from a single 
axis of symmetry), by carefully choosing an appropriate coupling method. 
As an example, the approximate $Q$-factors for the fundamental modes in silica 
microspheres, in a surrounding medium of air, have been quantified as a 
function of diameter \cite{Gorodetsky:99}. In the case of a diameter within
a broad range of $100$ to $150$ micron, noting that the cells considered as 
optimal candidates in Section~\ref{sec:oo} fall within this range, 
it is expected that waveguide 
couplers lead to $Q$-factors of less than $10^3$. Prism couplers, on the other 
hand, are expected to provide $Q$-factors in excess of $10^4$. Finally, tapers 
are potentially able to access the highest $Q$-factors, in excess of $10^7$ 
for resonators of this size. 
Using the phase-matching methods described in Section~\ref{sec:pred}, 
the prism coupler, taper coupler, and fluorescent coating methods are 
explored in Chapter~\ref{chpt:wgm} in order to assess the ability of a cell 
to sustain WGMs. 
\\

\noindent\textbf{\cdb Criterion 7. Loss (absorption and scattering):} It is 
known that some biological cells exhibit changes in their electrical 
capacitance from absorption of light in the IR spectrum \cite{Shapiro2012} 
due to the fact that water, which dominates the spectral behaviour of the 
surrounding medium, is highly absorbing in this region, generating heat that 
can affect the cells. It is also known that the DNA in cells is strongly 
absorbing in the UV spectrum \cite{Clydesdale2001,Jackson2009}. 
As a result, the visible and NIR regions represent the optimal windows for 
achieving resonance. 

Both scattering and absorption directly affect the $Q$-factor, as is evident 
in the first two terms on the right hand side of Eq.~(\ref{eqn:Q}). Clearly, 
absorption of energy due to material properties will prevent electromagnetic 
radiation from circulating continually, thus preventing clear modes from being 
observed. In this case, the challenge is to achieve a sufficient 
signal-to-noise 
ratio in the spectra for determining the presence of modes. 
Similarly, if the surface properties of a given cell allow scattering 
effects to dominate, any underlying WGM spectrum will be difficult to measure. 

\begin{table}[t]
\begin{mdframed}[backgroundcolor=boxcol,hidealllines=true]
{
\begin{center}
\begin{mdframed}[backgroundcolor=boxcold,hidealllines=true]
\textbf{\color{white} \large Selection criteria summary \cbl}\vspace{1mm}
\end{mdframed}
\end{center}
\cdb
\vspace{3mm}
\noindent\textbf{1.  \quad Size and index contrast} -- 
for a protein index in the range $1.36<\mathrm{n}<1.51$, 
diameters must be in the range $16.8$ $\mu$m $<D<86$ $\mu$m 
for $Q$-factors $>10^3$.\vspace{4mm}\\
\noindent\textbf{2.  \quad Sphericity} -- decrease in the $Q$-factor 
is roughly \textit{proportional} to the ellipticity 
unless modes along a preferred orientation can be selected. 
The ellipticity of a range of candidate cells is 
investigated in Section~\ref{sec:cand}. 
\vspace{4mm}\\
\noindent\textbf{3.  \quad Surface roughness} -- 
decrease in the $Q$-factor of the fundamental modes by several orders of 
magnitude for roughness $< \lambda/30$. Candidate cell surface roughness 
investigated in Section~\ref{sec:cand}.
\vspace{4mm}\\
\noindent\textbf{4.  \quad Elasticity} -- optomechanical behaviour 
causes deformations on the inner surface of the resonator, 
adding a secondary source of surface defects.\vspace{4mm}\\
\noindent\textbf{5.  \quad Osmolarity} -- non-soluble particles 
can cause damage to fibre tapers and instigate loss, 
whereas salt content must be 
extremely dilute to avoid ionisation.\vspace{4mm}\\
\noindent\textbf{6.  \quad Excitation method} -- passive interrogation 
such as prism/taper can lead to higher $Q$-factors by selecting a preferred 
orientation, whereas active interrogation requires no alignment 
in order to generate WGMs.\vspace{4mm}\\
\noindent\textbf{7.  \quad Loss (absorption and scattering)} --
high absorption prevents radiation from circulating, 
lowering the signal-to-noise ratio, whereas high scattering 
can indicate inadequate surface properties.
\cbl
\vspace{0mm}
}
\end{mdframed}
\end{table}

\section{\cdb Candidate cells}
\label{sec:cand}

In this section, a range of candidate cells is explored, and their 
suitabilities as resonators are assessed using the newly-developed 
selection criteria. Upon selecting a candidate cell, the refractive 
index and size are estimated, since it is this criterion to which it 
is most difficult to make significant alterations. If such kinds of cells 
lie in the viable range of geometric parameters established in 
Fig.~\ref{fig:lim}, 
then the surface properties are subsequently considered. As will be seen, 
careful consideration of surface effects is critical. However, it is not 
always straightforward to be able to assess in advance whether surface 
deformations will prevent the observation of WGMs. 
The type of medium used to store or handle the cells can also play 
a pivotal role in facilitating or preventing the generation of modes, 
through contaminants, high salt content and effects on the surface 
properties of the cells themselves. 
While it is important to establish that a number of 
candidates has been considered carefully, the purpose of this section 
is to identify the most viable candidate, which is then used for a more 
in-depth study in Chapter~\ref{chpt:wgm} using the techniques described in 
Section~\ref{sec:pred}. 
\\

\begin{figure}[t]
\centering
\includegraphics[height=100pt]{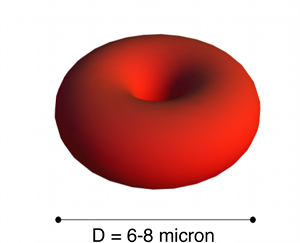}
\includegraphics[height=100pt]{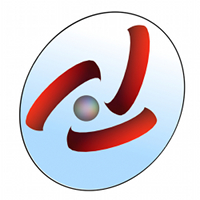}
\includegraphics[height=110pt]{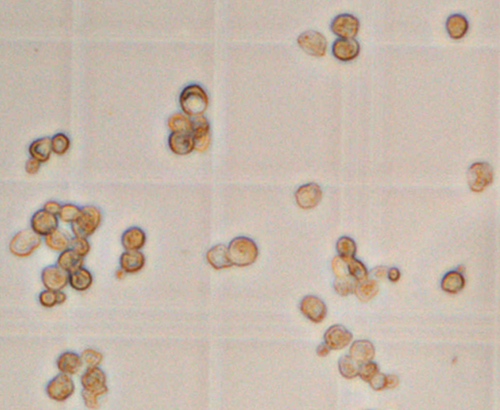}\\
\hspace{1mm}(a)\hspace{3.5cm}(b)\hspace{3.7cm}(c)\\
\vspace{-3mm}
\caption[A range of cells is examined using the selection criteria.]{
A range of biological cells is examined using the selection criteria. (a) 
The erythrocyte (animal blood cell), (b) the \textit{Cryptosporidium} oocyst 
with \textit{granule} (\color{grey30}\textit{grey}\cbl) and 
\textit{sporozoites} (\color{red}\textit{red}\cbl) as marked, and (c) 
genetically modified \textit{eukaryota} of the species 
\textit{Saccharomyces cerevisiae}, denoted $cdc28$, a variety of yeast. 
\textit{Image (c):} 
produced with the assistance of Mr. Steven Amos, 
The University of Adelaide.
}
\label{fig:cells1}
\figrule
\end{figure}

\noindent\textbf{\cdb Erythrocytes:} The first cell considered in this thesis 
is the animal red blood cell, which resembles a biconcave disk, depicted in 
Fig.~\ref{fig:cells1}(a). 
One particularly attractive feature of this cell is the disk shape, which 
could potentially assist in constraining the modes to a more limited geometric 
orientation, leading to increased $Q$-factors \cite{Boyd:01,Armani2003}. 
Considering \textbf{Criterion~1}, erythrocytes have diameters of the order of 
$6$ to $8$ micron -- approximately the size of the smallest resonators 
considered in Chapters~\ref{chpt:sph} and \ref{chpt:bub} \cite{ery}. 
Cells within this range of diameters are unlikely to be able to support 
modes, since in order to do so, the refractive index must be sufficiently 
large, as determined by the material properties of the cells. 
It is found that erythrocytes are composed of a 
combination of proteins and lipids, with a refractive index in the range 
$1.39$ to $1.42$ \cite{Semyanov:00,Yurkin:05,Park16092008}. 
The limiting values estimated in Fig.~\ref{fig:lim} indicate that an 
erythrocyte would require a diameter approximately one order of magnitude 
larger in order to sustain modes, and thus cannot be considered a viable 
biological resonator. It should be noted, however, 
that while this choice of cell will not become the basis of study in this 
thesis, it should not be ruled out for future technological developments that 
may be able to overcome the limitations of the diameter. 
\\

\noindent\textbf{\cdb \textit{Cryptosporidium} oocysts:} The next cell to be 
examined is that of \textit{Cryptosporidium}. This eukaryotic cell, while 
hazardous to humans \cite{surrey729747}, forms a distinctly spherical oocyst, 
as illustrated in Fig.~\ref{fig:cells1}(b). The size of the oocyst varies 
depending on the species of \textit{Cryptosporidium}. For example, 
\textit{C. parvum}, a smaller variety, is reported with planar dimensions of 
$5.0 \times 4.5$ $\mu$m. The ratio of the major and minor axes 
corresponds to a \emph{shape index} 
(or shape factor) of $1.11$. On the other hand, 
\textit{C. muris}, a slightly larger but less symmetrical variety is reported 
to have dimensions of $7.6 \times 5.6$ $\mu$m, corresponding to a shape index 
of $1.3$ \cite{MERCADO1995}. As with the erythrocite, an application of 
\textbf{Criterion~1} indicates that these cells are too small to be likely to 
sustain WGMs given the range of refractive indices exhibited by proteins 
\cite{Voros2004553}. While the range of sizes provides some indication that 
genetic variations in certain cells can provide increased diameter values, 
in this case, the asymmetry increases with size, limiting the opportunity for 
investigation into resonance behaviour. However, in other cases, it is indeed 
possible to genetically modify cells so that they are less limited in size 
\cite{Dungrawala2012}. 
One particular example of such a cell, which is both commonly used and 
relatively straightforward to source, is yeast. \\

\noindent\textbf{\cdb Genetically modified yeast:} The eukaryotic fungus known 
as yeast, of the species \textit{Saccharomyces cerevisiae}, is both widely 
available, and reasonably well understood genetically compared to other cells 
\cite{Dungrawala2012}. In cell biology, cells that exhibit one complete set of 
chromosomes are called \emph{haploids}, whereas those that incorporate two 
complete sets of chromosomes are known as \emph{diploids}. 
Diploid yeast cells can exhibit twice the cell volume of haploids 
\cite{yeastsize}. 
Furthermore, it is possible to cultivate particularly large yeast cells by 
altering their genetic makeup, typically through gene deletions 
\cite{Dungrawala2012}. These modified cells are usually labelled, by 
convention, with the gene that has been deleted. 
For example, the strains that have had the genes deleted that express $Ecm9$ 
(denoted $ecm9\Delta$)  or $Ctrr9$ (denoted $ctr9\Delta$) are described in 
Ref.~\cite{Dungrawala2012}. As an example, the $S288C$ strain has a median 
cell volume of $42\pm2$ $\mu$m$^3$. Of these strains, the one that is able to 
produce the largest and most spherical cell type, with diameters 
of the order of $10$ $\mu$m is known as $cdc28$. 

\begin{figure}[H]
\begin{center}
\includegraphics[width=0.8\hsize,angle=0]{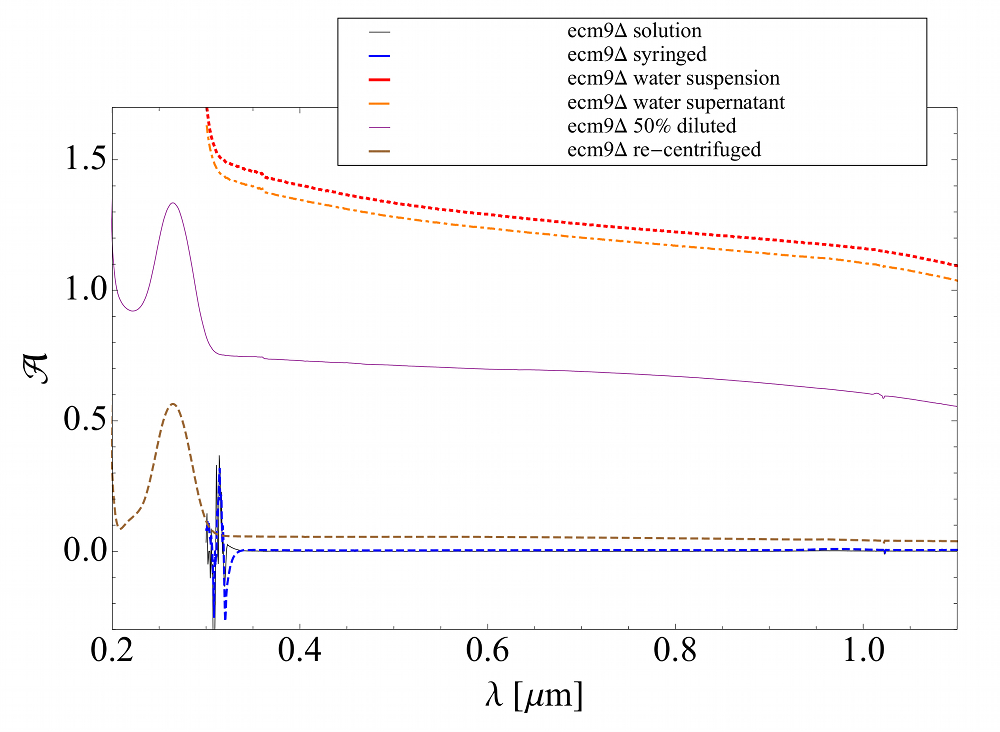}\\
\vspace{-5mm} 
\hspace{8mm}\mbox{(a)}\\
\includegraphics[width=0.8\hsize,angle=0]{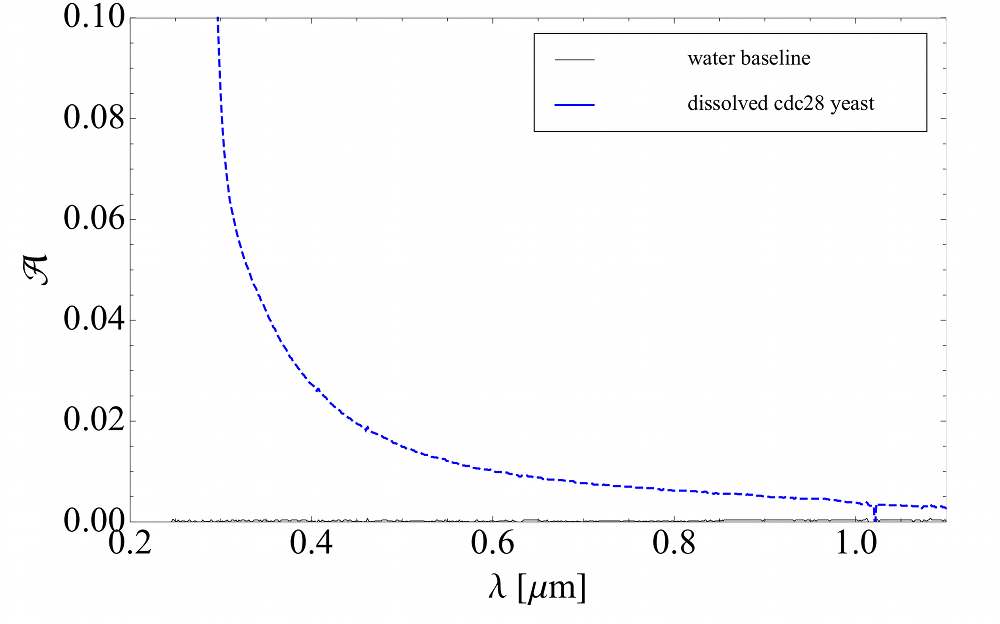}\\
\vspace{-5mm}
\hspace{8mm}\mbox{(b)}\\
\end{center}
\vspace{-6mm}
\caption[The absorbance spectra of yeast show dominance from DNA.]
{The spectra obtained for the absorbance, $\mathcal{A}$ (dimensionless), 
of two strains of 
genetically modified yeast. (a) The strain $ecm9\Delta$ is crushed using a 
hypodermic syringe, the debris is removed via centrifuge, and the resultant 
supernatant liquid is then diluted, and re-centrifuged. The spectra of each 
stage show a strong absorbance peak corresponding to presence of DNA. (b) The 
strain $cdc28$ is dissolved using the same method.}
\label{fig:yeast}
\figrule
\end{figure}

Figure~\ref{fig:cells1}(c) shows an image of the $cdc28$ strain, with each 
square marked on the scale as $50$ $\mu$m in length. While cells of the order 
of $10$ to $15$ $\mu$m in diameter are still relatively small for the purposes 
of sustaining modes according to \textbf{Criterion~1}, the upper diameter 
values are close to the minimum limiting diameter considered in 
Fig.~\ref{fig:lim} so long as the refractive index is sufficiently large 
\cite{Voros2004553}. While this cell is not optimal, it is worth investigating 
its other properties with respect to the selection criteria. While several of 
the $cdc28$ cells are near spherical, satisfying \textbf{Criterion~2}, it is 
difficult in general to gauge the surface attributes, 
\textbf{Criteria~3} and \textbf{4}; these will often represent the last criteria to be 
considered in each case of cell. The media, excitation and loss properties 
(\textbf{Criteria~5} through \textbf{7}), however, are more easily examined. 
First, the absorption properties of these cells are considered. \\
\noindent\textbf{\cdb Method:\cbl} The yeast are streaked onto agar plates, 
grown for a period of $2$ to $3$ days and stored at $4^\circ$C with the plates 
inverted. Each of the three strains, $ecm9\Delta$, $ctr9\Delta$ and $cdc28$ 
is cultivated in standard, non-selective yeast media (YEPD), described 
in Appendix~\ref{chpt:app4}, and placed into aliquots. While this medium 
contains quantities of proteins, glucose/dextrose and yeast extract, little 
quantities of salt are present (which would impede the growth of yeast), 
nominally satisfying \textbf{Criterion~5}. 
An initial concentration of cells in media corresponding to a haemocytometer 
reading of $9\times 10^6$ cells/mL is used. Next, a hypodermic 
syringe and fine needle is used to agitate and distribute the cell 
constituents throughout a medium of water. 
The debris is removed via a centrifuge, and the remaining supernatant liquid 
is then diluted to a concentration of $50\%$ and re-centrifuged. 
The spectra corresponding to each stage of the process are shown in 
Fig.~\ref{fig:yeast}(a), as measured from solutions placed in $1$ mL cuvettes,  
by a UV--Vis--NIR spectrophotometer. The absorbance spectra of water is 
subtracted from these results. 
Acetone is not used in this case to assist the 
solvency, 

\newpage
\noindent as it can cause a proportion of the starches and sugars to 
precipitate, leaving only DNA and protein. The absorbance of the resultant 
solution is measured using a spectrophotometer, shown in Fig.~\ref{fig:yeast}, 
and corresponding to \textbf{Criterion~7}. 
Figures~\ref{fig:yeast}(a) and \ref{fig:yeast}(b) show 
that the DNA is nevertheless the dominant source of 
absorbance, with the principal peak occurring at $260$ nm, while there is 
little discernible effect in the visible region. 
The ribosome content, given 
by the ratio of the absorbance at $\lambda = 260$ nm to that at $280$ nm, 
$\mathcal{A}_{260/280}$ \cite{PMID:7702855},  leads to a value of 
$0.518/0.258 \approx 2$. Based on the known results for DNA 
for the quantity $\mathcal{A}_{260/280}$, as described in 
Ref.~\cite{PMID:7702855}, this result indicates that the solution  
consists primarily of DNA. 
Thus, any absorbance from sugars and starches 
left in the solution have a negligible effect on the absorbance compared to 
that of 
the DNA. As a result, the absorbance spectrum indicates that the choice of 
wavelength in this case is not strongly constrained, offering scope for the 
generation of WGMs. 

Finally, the excitation and scattering behaviour of 
\textbf{Criteria~6} and \textbf{7}, 
are considered by applying a fibre taper placed above a glass coverslip on an 
inverted confocal microscope, similar to the setup illustrated in 
Fig.~\ref{fig:exp}. A droplet of water is placed on top of the fibre taper to 
prevent breakage due to surface tension effects. The droplet is then 
inoculated with yeast cells of the $cdc28$ strain. It is found that the yeast 
cells do not interact with the fibre taper, nor produce any measurable 
autofluorescence. The reason for this is that the coupling of light into a 
resonator from a fibre taper is extremely sensitive to the phase-matching 
condition of Eq.~(\ref{eq:kwgm}), 
determined by the diameter of the taper waist, and the distance of 
the taper to the resonator. For cells of this diameter, approximately 
$10$ $\mu$m, a taper waist of 
$1$ $\mu$m is insufficient to achieve such a coupling. While the evanescent 
field of a thin taper extends further into the surrounding medium, it renders 
the taper extremely sensitive to contamination from dust, detritus or 
biological matter introduced into the droplet. As a result, it is unlikely 
that these cells represent a pathway toward a biological cell resonator that 
can be easily achieved at this time. \\

\begin{figure}[t]
\centering
\includegraphics[height=110pt]{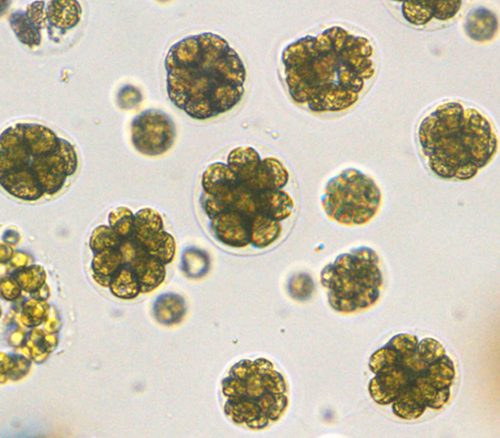}
\includegraphics[height=110pt]{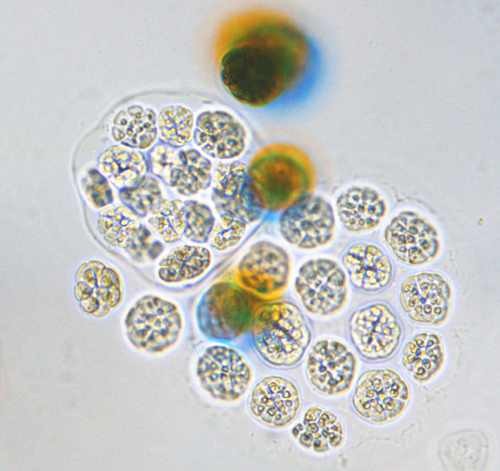}
\includegraphics[height=110pt]{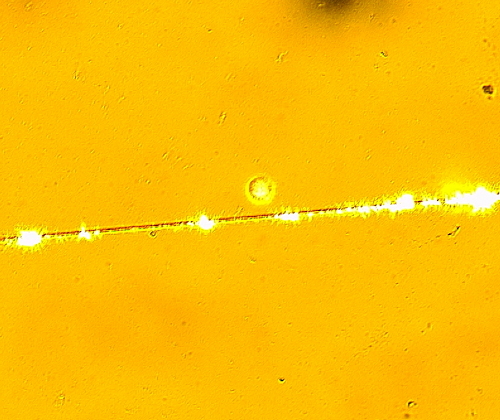}\\
\hspace{2mm}(a)\hspace{4cm}(b)\hspace{4cm}(c)\\
\vspace{-3mm}
\caption[Examples of the geometries formed by 
\textit{Eudorina}-\textit{Pandorina} algae.]{Examples of the geometries formed 
by algae of the \textit{Eudorina}-\textit{Pandorina} varieties. (a) Groups of 
cells form microspheres that conglomerate into macroscopic structures, 
which themselves are surrounded by a translucent layer. (b) Examples of more 
elaborate microstructures formed by these algae. (c) An algal 
microsphere is placed in the evanescent field of a nearby SMF$28$ optical 
fibre taper, and exhibits autofluorescence. 
\textit{Images (a) and (b):} 
produced with the assistance of Mr. Steven Amos, 
The University of Adelaide.}
\label{fig:algae}
\figrule
\end{figure}

\newpage
\noindent\textbf{\cdb \textit{Eudorina}-\textit{Pandorina} algae:} The next 
cell to be considered is a much larger variety of \textit{eukaryota}, from the 
family \textit{Volvocaceae}. These varieties of algae are known for their 
symmetrical geometric shapes, through conglomerations of cells into 
microspheres significantly larger in diameter than the aforementioned 
candidates, as well as macroscopic structures formed by the conglomerates 
themselves. Note that each structure can also be surrounded by a faint, 
translucent layer, roughly spherical in shape. Examples of these formations 
are displayed in Fig.~\ref{fig:algae} for the 
\textit{Eudorina}-\textit{Pandorina} genus. 
Figures~\ref{fig:algae}(a) and (b) depict examples of the structures and 
meta-structures formed by the algae. An image of an algal microsphere placed 
within the evanescent field of a fibre taper, with a waist diameter 
of $3$ $\mu$m, and using a wavelength of approximately $760$ nm, is shown in 
Fig.~\ref{fig:algae}(c) producing significant autofluorescence. 

Considering \textbf{Criterion~1}, 
an analysis of the geometric properties from a sample of algae (full results 
shown in Appendix~\ref{chpt:app4}) indicates a mean value of 
the equivalent circle diameter of $43.88$ $\mu$m, and a mean aspect ratio of 
$1.35$. Note that the maximum value of the equivalent circle 
diameter is over $100$ $\mu$m, which is a useful feature 
given that the refractive 
index of algal cells exceeds that of water by a factor of no 
more than $1.167$ \cite{doi:10.1093/plankt/18.12.2223}. In the case of 
\textbf{Criterion~2}, the mean sphericity is measured to be 
$0.61$; however, for some algal particles, the sphericity is as high as $0.96$. 
Finding an optimal combination of size and sphericity is 
achievable given the large quantity of algae available. It is these large 
spherical cell conglomerates that are selected preferentially for 
experiment. 

The \textit{Eudorina}-\textit{Pandorina} variant are successfully cultivated 
in a medium known as MLA, 
the recipe for which can be found in Ref.~\cite{Bolch1996}, 
reproduced in Appendix~\ref{chpt:app4}. While this media stock requires a 
range of salts and micronutrients, they are in relatively small quantities, 
and the resultant medium is very dilute from the point of view of the WGM 
coupling apparatus, such as the prism coupler or the fibre taper, thus 
providing little concern regarding \textbf{Criterion~5}. 

\begin{figure}[H]
\begin{center}
\includegraphics[width=0.8\hsize,angle=0]{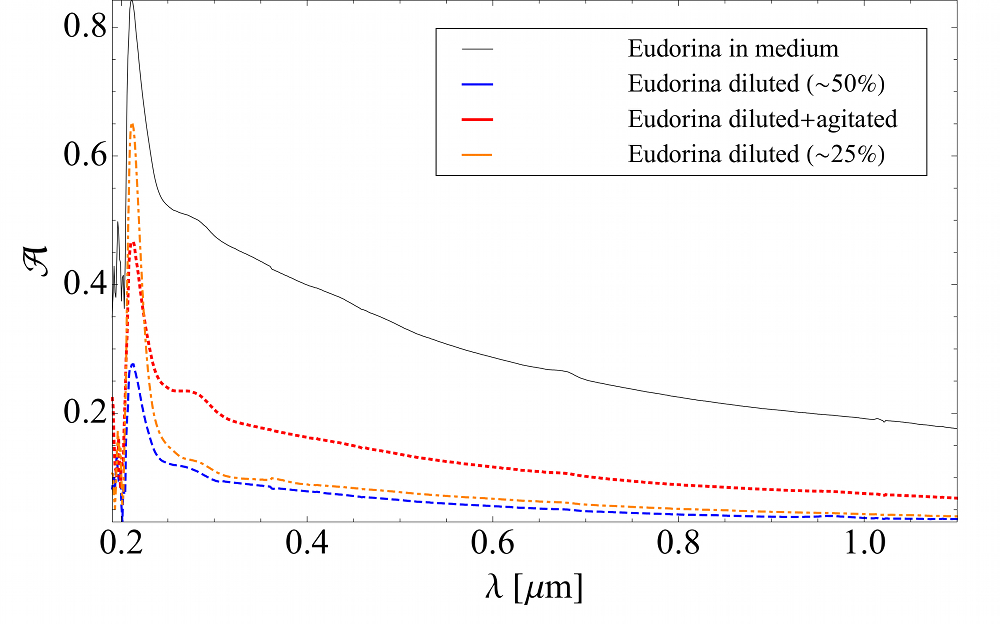}\\
\vspace{-5mm} 
\hspace{7mm}\mbox{(a)}\\
\includegraphics[width=0.8\hsize,angle=0]{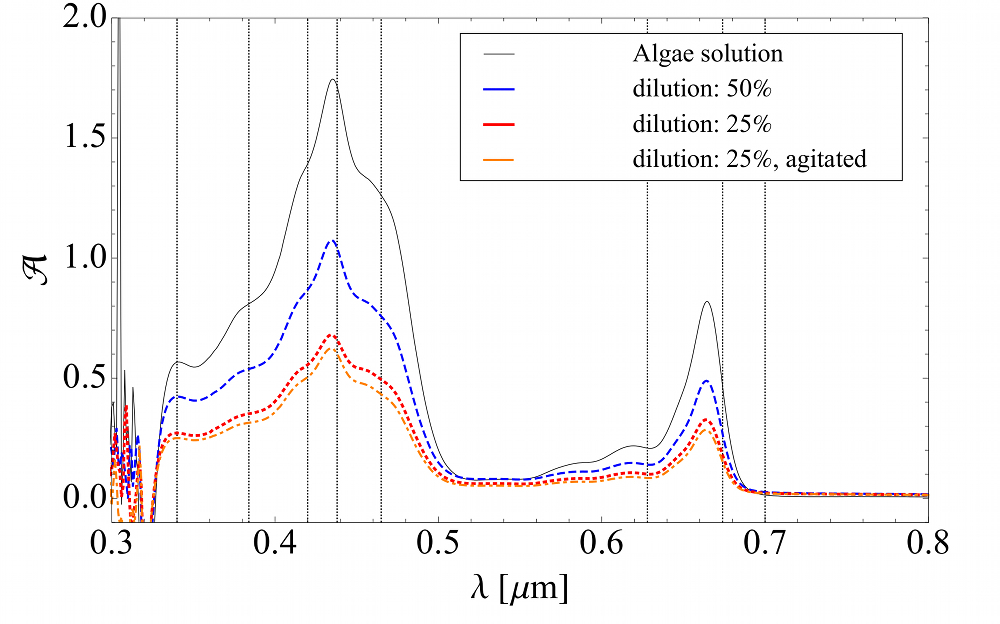}\\
\vspace{-5mm}
\hspace{5mm}\mbox{(b)}\\
\end{center}
\vspace{-6mm}
\caption[Absorbance spectra for \textit{Eudorina}-\textit{Pandorina} algae.]
        {Absorbance ($\mathcal{A}$) spectra are shown for
          \textit{Eudorina}-\textit{Pandorina} algae. 
(a) The suspended algal cells exhibit strong scattering behaviour, broadening 
the absorption peaks, and as a result the spectrum is dominated by that of the 
MLA media. (b) The constituent matter of the algae is distributed throughout 
a solution which displays strong 
peaks associated with chlorophyll.}
\label{fig:euabs}
\figrule
\end{figure}

Considering the absorption properties, \textbf{Criterion~7}, for these algae,  
the conglomerates are relatively bulky compared to the other cells 
examined thus far, and a sample of algal cells suspended in MLA (which has a 
refractive index close to that of water) is unlikely to be homogeneous.  
A measurement of the absorbance of this mixture yields a relatively smooth 
spectrum, increasing towards the UV region. This is due to the scattering of 
the light from the cells, which serves to broaden the peaks that would 
correspond to the algae, 
as shown in Fig.~\ref{fig:euabs}(a) for a number of dilutions. 
Note that the absorbance spectra of the MLA medium alone is subtracted 
from these results. 
Since the goal is not to activate a solution of many cells to explore 
the possibility of generating WGMs, but rather, to activate a single cell with 
a laser and mode coupler, it is the properties of the contents of the algae 
cells themselves that must be examined as follows. 

\noindent\textbf{\cdb Method:\cbl} First, samples of the algae solution with a 
volume of $1$ mL are centrifuged into pellets. Then the pellets are dissolved 
and combined in acetone in order to break down the biological material into a 
translucent, homogeneous fluid. Two pellets are added per $1$ mL of acetone, 
thus doubling the concentration. This solution is then agitated using a 
weighted test tube designed to distribute the constituents within the 
cells throughout the solution and remove the resulting debris. 
The translucent 

\vspace{1cm}

\newpage
\noindent solution is measured against a background of buffered acetone. 
As a result, the constituent chemicals of the algae are more homogeneously 
distributed through the media. 

The absorbance spectrum now contains much clearer peaks, as shown in  
Fig.~\ref{fig:euabs}(b) for a number of dilutions, with the dominant spectral 
peaks originating from chlorophyll shown as vertical dotted lines. In 
particular, the absorption peak at $663$ nm is known to originate from 
chlorophyll \cite{Govindjee74}. While not all algae display identical spectra, 
the shape is similar to Fig.~8 in Ref.~\cite{Wu2013}. Note that the absorbance 
becomes difficult to measure in the UV region due to the DNA present in the 
sample. It is clear that within the visible region, particularly for 
wavelengths between $500$ and $630$ nm, the absorbance is relatively 
low, and thus this region is the most optimal for attempting to generate WGMs. 
While the \textit{Eudorina}-\textit{Pandorina} algae represent an initially 
promising direction in the search for a cell resonator, the difficulty with 
sustaining modes stems from \textbf{Criterion~3}. While the groups of algal 
cells are predominantly round, the surface roughness and defects are 
significant -- approximately half the diameter of a single daughter colony 
within the conglomerate -- $10$ $\mu$m. However, some colonies are much 
more compact with reduced surface roughness, as shown in Fig.~\ref{fig:eugeo} 
of Appendix~\ref{chpt:app4}. 
Recalling that modes of microspheres are sensitive to surface 
roughness at the nanometre scale \cite{Rahachou:03,Reynolds:15}, it is thus 
difficult to obtain resonant behaviour of light returning in phase for any more 
than an extremely limited number of round-trips. 
\\

\noindent\textbf{\cdb Volvox algae:\cbl} This algae is a relative of the 
\textit{Eudorina}-\textit{Pandorina} variety, in the same family 
\textit{Volvocaceae}. What distinguishes this algae from its relatives is its 
propensity to form approximately-spherical macroscopic colonies up to a 
millimetre in diameter, constructed from large numbers of algal cell 
conglomerates. While this potentially provides scope to explore the resonance 
properties of these structures, the multicellular roughness is of a similar  
magnitude to that of \textit{Eudorina}-\textit{Pandorina}, limiting the 
likelihood that modes could be sustained, according to \textbf{Criterion~3}.
\\  

\newpage
\noindent\textbf{\cdb Other studies on WGMs in biological cells:\cbl}
Competing studies in the literature on cell based WGMs include the placement 
of cells expressing green fluorescent protein (GFP)
within a cavity, with the modes excited by the emission from the GFP within 
the cell. Lipid droplet based lasers have also been reported in 
Ref.~\cite{Humar2015}, in which spherical droplets of oil or natural lipids 
within porcine adipocyte cells are able to sustain WGMs. 
Alternatively, free-floating beads of polystyrene 
\cite{Himmelhaus2009418,Humar2015} or GFP coated BaTiO$_3$ \cite{Humar:17} 
have been inserted into living cells or allowed to be engulfed by phagocytes 
for the purposes of tagging and tracking. 
Of these studies, the closest counterpart to the research presented 
in this thesis is that of the Harvard-MIT group's porcine adipocyte based 
lasers, which demonstrate the fact that WGMs can be 
sustained within naturally occurring 
subcutaneous fatty cells. Adipocyte cells contain a single lipid droplet 
($\mathrm{n}=1.47$) that 
exhibits a high degree of spherical symmetry, and it is this droplet that 
is reported to act as an optical cavity with a sufficient refractive 
index contrast between the droplet and the surrounding tissue. 
The adipocytes were treated with a lipophilic dye (Nile red) and collagenase 
in order to release them from the surrounding subcutaneous tissue 
and to induce lasing \textit{in situ}. This experiment is of relevance 
to the present study, as it represents the first example of WGMs reported 
as sustainable within a completely natural cavity within a biological cell. 
While the focus on cells containing lipid 
droplets limits the ability to extend the study to a range of other cells, 
it nevertheless demonstrates an important validation for the concept of a 
naturally occurring optical cavity in nature. 

Recalling the formulation of the project in Section~\ref{sec:form}, the 
ability for a biological cell to act as a resonator without the introduction 
of an artificial cavity is of particular interest conceptually. 
The development of a sensing modality for providing information of the internal 
structure of an untampered cell is not only complementary to existing 
label-free 
technologies \cite{doi:10.1167/iovs.03-0628,Freudiger1857,ELPS:ELPS200406070,
4014191,Kita:11,doi:10.1021/ac0513459,BIOT:BIOT200800316} 
but has potentially 
broad implications on health sciences. A sensing modality based on 
cells that can sustain WGMs on their own may also avoid the immune responses 
of an organism from the introduction of foreign materials. 
\\

\noindent\textbf{\cdb Mammalian oocytes:\cbl} The final cell to be considered 
in this thesis is the mammalian oocyte. A variety of types of oocyte will be 
studied, and how each of the selection criteria are addressed will be examined 
in detail and expanded upon in the next section. 
The reasons for this particular kind of cell 
being an attractive choice are manifold. 
First, they exhibit significant diameter values 
of the order of $100$ $\mu$m (depending on species). 
Second, the life cycle of an oocyte, 
particularly after fertilisation, has been well studied in the health 
literature. As a result, there is some indication as to the surface roughness 
and elastic properties of the outer layers, which depend on the stage of the 
fertilised oocyte, or embryo. Finally, measuring the resonant properties of
embryos presents a major opportunity to make an impact 
on health sciences in the future. As WGM sensing technologies are 
developed, the ability to characterise the status of an 
embryo through WGM interrogation comes within reach.

\section{\cdb Oocyte and embryo structure}
\label{sec:oo}

\begin{figure}[H]
\begin{center}
\vspace{1.5cm}
\includegraphics[width=1.0\hsize]{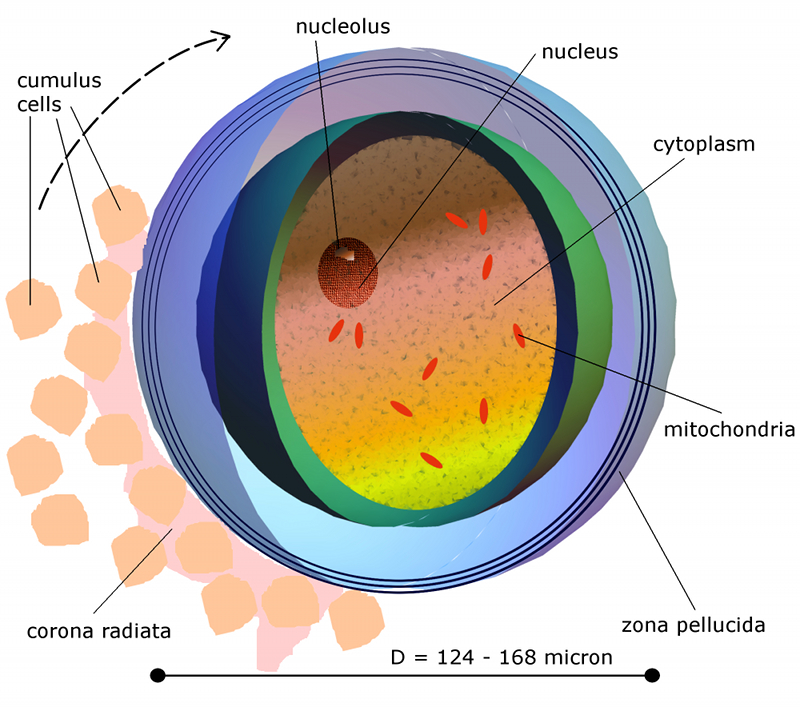}
\end{center}
\vspace{-5mm}\caption[A diagram of the mammalian oocyte.]{A diagram of the 
components of the mammalian oocyte, including the nucleus, the cytoplasm, the 
mitochondria, and the glycoprotein layers that comprise the 
\textit{zona pellucida}. An unfertilised oocyte may also be surrounded by 
cumulus cells and is known as a \emph{COC}. With these cells removed, the 
oocyte is then called \emph{denuded}. The outer diameter $D$ of the 
oocyte can vary significantly, depending on species. Here, the range 
quoted corresponds to the measurements taken for the bovine oocytes 
used in this thesis.}
\label{fig:cel}%
\figrule
\end{figure}

The structure of an oocyte can be divided into the following regions, as 
illustrated in Fig.~\ref{fig:cel}. The primary regions are known as the 
\textit{nucleus}, where the majority of the oocyte DNA is stored, the 
\textit{cytoplasm}, the \textit{mitochondria}, and the \textit{zona pellucida}, 
which is of particular importance in this work. Unfertilised oocytes may also 
be surrounded by cumulus cells, known as the \textit{corona radiata} when 
adjacent to the \textit{zona pellucida}, which are critically involved in the 
fertilisation process, and dissipate shortly afterwards. An oocyte surrounded 
by cumulus cells is known as a \emph{cumulus-oocyte complex (COC)}, and if the 
cumulus cells are removed, it is called \emph{denuded}. 

While the biological functions of these regions is a rich and complex subject, 
this thesis seeks to explore the \emph{physical} properties of the oocyte, and 
in particular, its ability to sustain electromagnetic resonances. Optical 
attributes of oocytes, such as the refractive index, have not been 
comprehensively studied in the literature, and as a result, this work 
represents a pioneering attempt to characterise some of the properties 
necessary for a biological cell to exhibit WGMs. 
Thus, to begin, the potential influence of each of the regions of an oocyte on 
the generation of WGMs

\newpage
\noindent will be studied. 

Recall that the behaviour of the modes of a layered resonator is 
highly dependent on the properties of the outermost layer, as demonstrated in 
in Chapters~\ref{chpt:bub} and~\ref{chpt:mod}. 
This is because of the fact that the fundamental radial modes 
 are confined to the surface of the resonator. If the outer layer becomes 
sufficiently thick, for a given refractive index and outer diameter, the mode 
behaviour becomes insensitive to further increases in the layer thickness, as 
described in the case study of fluorescent microbubbles in 
Section~\ref{sec:case}. 
As a result, the structure of the cytoplasm, which lies within the 
\textit{zona} region, has less influence on the structure of any modes 
confined within the \textit{zona}. For this reason, more attention is paid to 
the physical properties of the \textit{zona} for the purposes of optical 
resonance. The structure of the \textit{zona pellucida} will now be considered.
 
The \textit{zona pellucida}, illustrated in Fig.~\ref{fig:cel}, 
is a semi-transparent 
coating that surrounds mammalian 
oocytes and embryos, which allows for communication among 
oocytes, follicle cells and sperm, as well as providing mechanical 
protection during embryo development \cite{BLEIL1980185}. 
In embryology, the \textit{zona pellucida} is of particular interest 
because of its role in regulating the interactions of the oocyte 
with its surrounding environment \cite{JEZ:JEZ8}, especially 
at the moment of fertilisation from free-floating sperm, 
which represents a rich area of ongoing research \cite{Gupta15}. 

\begin{figure}
\begin{center}
\includegraphics[width=0.35\hsize]{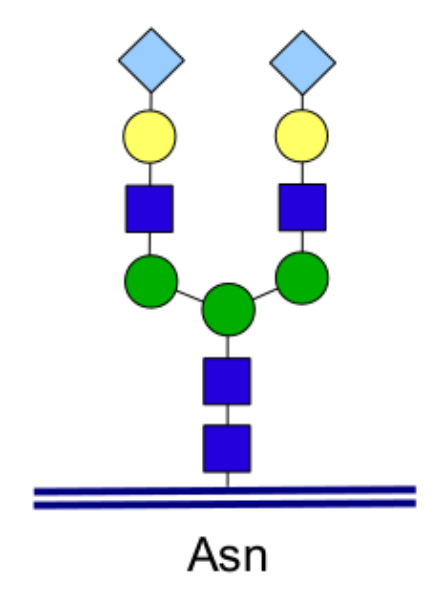}
\end{center}
\vspace{-5mm}\caption[A diagram of the molecular structure of a glycoprotein.]
{The molecular structure of a glycoprotein chain, showing the attached 
\textit{oligosaccharide} sugar group \textit{glycan} to a nitrogen atom in an 
amino acid sequence \textit{asparagine (\textbf{Asn})}. In this simplified 
example, the \cb \textit{blue} \cbl squares denote an \textit{aminohexose}-type 
polysaccharide such as \textit{N-acetylglucosamine (GlcNAc)}, the 
coloured circles denote a range of possible of saccharides, 
and the \textit{\color{pb}pale blue\cbl} diamonds indicate the presence of a 
\textit{sialic acid} molecule \cite{Imperiali1999643}.}
\label{fig:glyco}
\figrule
\end{figure}

The \textit{zona} comprises an extracellular matrix that consists of a 
mixture of glycoproteins 
\cite{JEZ:JEZ8}, and is held together in long, interconnected fibrils by 
non-covalent interactions. These proteins involve sugar groups (certain 
\textit{oligosaccharides} known as \textit{glycans}) attached to the 
constituent molecules, which affect their molecular weights 
\cite{Wassarman2008}. A simplified molecular structure of a nitrogen(N)-linked 
\textit{glycosylation}, or attachment of the glycan to a nitrogen atom, is 
shown in Fig.~\ref{fig:glyco}. The molecular weights vary across a significant 
range, from $55$ to $200$ kilodalton (kDa), depending on the protein type, and 
species to which it belongs \cite{Wassarman2008}. 
The glycoproteins are not laid down 
in uniform layers, but are intermixed with varying relative densities at 
different points in the \textit{zona}, as illustrated in Section~\ref{sec:oo}. 
The reduction of the more bulky protein structures yields a more compacted 
mechanical structure, which not only increases the refractive index 
\cite{Voros2004553} (\textbf{Criterion~1}), but also leads to reduced surface 
roughness (\textbf{Criterion~3}) and elasticity \cite{Papi2010}. 
One interesting phenomenon in the development of mammalian oocytes is that the 
elastic and surface properties change \emph{after} fertilisation. This is 
an important feature of the current work. In the 
case of bovine oocytes, AFM reveals the greatest density of networked 
glycoproteins, the most significant \textit{zona} hardening, and the minimum 
surface roughness all occurring immediately after fertilisation 
\cite{Papi2010}. Similar mechanical changes have been observed recently in 
murine and human oocytes \cite{Yanez2016}. For these reasons, the majority 
of the cell experiments presented in Chapter~\ref{chpt:wgm} will involve 
\emph{presumptive zygotes}, that is, oocytes that have undergone fertilisation, 
and are subsequently either placed in a fixing solution or otherwise 
immediately tested at this phase of their development. 

In the case of murine oocytes, the 
proteins are labelled in descending order of molecular weight: (m)ZP$1$ 
($200$ kDa), (m)ZP$2$ ($120$ kDa) and (m)ZP$3$ ($83$ kDa) \cite{Wassarman2008}. 
In the case of other mammalian oocytes, such as those of bovine, porcine and 
human, a fourth protein of intermediate weight is present, ZP$4$ 
\cite{yonezawa2012}. The molecular weights of these proteins differ from 
species to species, and each protein functions differently with regard to 
fertilisation \cite{yonezawa2012}. For the purposes of the experiments 
described in Chapter~\ref{chpt:wgm}, it is important to note that, while all 
available glycoproteins are present throughout the \textit{zona}, the 
distribution of the proteins within the \textit{zona} region is inhomogeneous 
\cite{Green01091997}. For example, a greater proportion of the protein ZP$3$ 
is found near the outer boundary of the \textit{zona}, since it is the protein 
that is predominantly involved in binding to sperm \cite{VanDuin19921064}. 
Furthermore, it has been well documented that the distribution of proteins is 
highly dependent on the stage of fertilisation 
\cite{Gupta2012,doi:10.1095/biolreprod66.4.866}. 
Therefore, it will become important to choose an oocyte or embryo from a stage 
in the fertilisation process that presents the best opportunity for sustaining 
WGMs. It is possible that this distribution of proteins results in a 
refractive index gradient that would serve to broaden the higher order radial 
modes, which extend further into the \textit{zona} region. While the optical 
properties of these proteins, especially in intact oocytes, is not 
well understood, the intention of this investigation is to elucidate how the 
physical parameters can affect the generation of WGMs.

\subsection{\cdb Size and topology}
\label{subsec:st}

\begin{figure}
\begin{center}
\includegraphics[height=0.3\hsize]{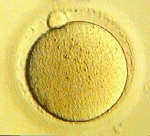}
\includegraphics[height=0.3\hsize]{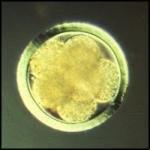}
\includegraphics[height=0.3\hsize]{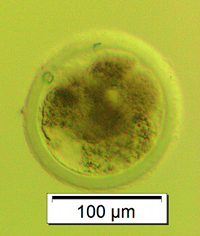}\\
\hspace{0.8cm}(a)\hspace{4cm}(b)\hspace{3.5cm}(c)\\
\end{center}
\vspace{-5mm}\caption[Microscope images of bovine embryos shown to scale.]
{Microscope images of bovine embryos. \textit{Images:} (a) and (b) courtesy of 
Assoc. Prof. J. G. Thompson, The University of Adelaide. (c) A scale image of a 
\emph{presumptive zygote} used in the experiments of Chapter~\ref{chpt:wgm}. 
The diameter is approximately $150$ $\mu$m including the 
\textit{zona pellucida}. This value can vary by up to $15\%$ among different 
samples.}
\label{fig:bov}
\figrule
\end{figure}

\begin{figure}
\begin{center}
\includegraphics[height=0.22\hsize]{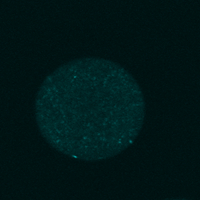}
\includegraphics[height=0.22\hsize]{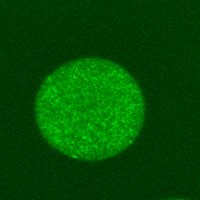}
\includegraphics[height=0.22\hsize]{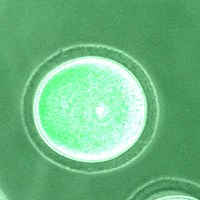}
\includegraphics[height=0.22\hsize]{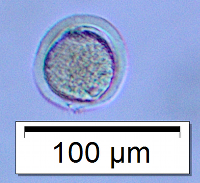}\\
\hspace{-3mm}(a)\hspace{2.6cm}(b)\hspace{2.7cm}(c)\hspace{3.3cm}(d)\\
\end{center}
\vspace{-5mm}\caption[Microscope images of murine embryos shown to scale.]
{Confocal microscope images of murine embryos. (a) Blue autofluorescence light 
is collected ($488$ nm). (b) Green autofluorescence light is collected 
($543$ nm). (c) Phase-contrast, blue and green fluorescence images are merged 
to produce a composite image. \textit{Images:} (a) through (c) courtesy of Dr. 
L. J. Ritter, The University of Adelaide. (d) A scale image of an unfertilised 
murine oocyte is shown. The diameter, including the \textit{zona} region, is 
approximately $50$ $\mu$m.}
\label{fig:mur}
\figrule
\end{figure}

At the most basic level, the range of diameters exhibited by mammalian oocytes 
satisfies \textbf{Criterion~1}. In the case of bovine oocytes, 
it is reported in the literature 
that the outer diameters (including the \textit{zona} regions) take values 
$150$ to $190$ $\mu$m \cite{Bo:13}, 
with a variation of $23.5$\%. The thickness of the 
\textit{zona} itself takes values $12$ to $15$ $\mu$m \cite{Bo:13}. 
In this study, measurements taken for the bovine oocytes used in this 
project yield smaller mean outer diameter values 
of $146.3 \pm 21.9$ $\mu$m, that is,  %$143.96$ and $148.66$ $\mu$m, 
with a variation of $15$\%. 
Excluding the 
\textit{zona} region, the diameter is consistently in the range $110$ to $116$ 
$\mu$m, in agreement with the literature \cite{MRD:MRD1080420410}. 
Figure~\ref{fig:bov} shows examples of bovine embryos in 
Figs.~\ref{fig:bov}(a) and (b), and a scale image of a presumptive zygote, a 
fertilised embryo that remains undeveloped, which has been chemically 
\textit{fixed}, in Fig.\ref{fig:mur}(c). 

A similar study is conducted on murine embryos, as shown in 
Fig.~\ref{fig:mur}. 
The images shown in Figs.~\ref{fig:mur}(a) through (c) correspond to the 
autofluorescence captured from a single murine embryo using a confocal 
microscope, for two channels ($488$ nm and $543$ nm), as well as a merged 
image of both channels and a phase-contrast image. The scale image in 
Fig.~\ref{fig:mur}(d) indicates an outer diameter of 
$50$ $\mu$m. 
The sphericity appears optimal for biological cells, particularly 
for the bovine samples, nominally satisfying \textbf{Criterion~2}. 
The effect of autofluorescence in oocytes and embryos will be discussed in
 more detail in Section~\ref{subsec:auto}. 

While the refractive index values of the glyocoproteins that comprise the 
\textit{zona} layers have not been clearly characterised in the literature, 
as discussed in the next section, a comparison of these diameter values 
with the limiting results of Fig.~\ref{fig:lim} indicate that the bovine 
oocytes and embryos are in the suitable region for sustaining WGMs 
according to \textbf{Criterion~1}.

\subsection{\cdb Glycoprotein refractive index}
\label{subsec:gri}

While the refractive index does not represent a property that 
is typically of immediate interest to practitioners in the field of embryology 
or cell biology, several measurements have been reported in the context of 
determining the optimal fertilisation time using optical microsystems 
\cite{biodevices10,Wacogne:13}. The refractive indices in this case are 
measured by collecting the transmission spectra from two adjacent fibres, one 
of which is connected to a low-power white-light source, in close proximity to 
an oocyte. The values obtained from this method vary widely depending on the 
stage of fertilisation, from $1.68$ to $2.47$, with statistical 
uncertainties ranging from $0.02$ to $0.06$. The index values above $2.0$ 
are, however, unrealistically high for organic 
matter, which does not exhibit any high-index crystalline structural 
properties that would be expected for refractive indices of this magnitude. 
This study represents one of the only attempts to 
obtain the refractive index directly from an intact oocyte, 
providing scope for an improved estimate using WGM-based techniques. 
Measurements of the 
refractive index of the glycoproteins that comprise the \textit{zona pellucida} 
lead to a more conservative range of $1.36$ to $1.51$ \cite{Voros2004553}. 
While this range of values is still comparatively wide given the sensitivity 
of the WGM spectra to changes in the index, it is possible to tune the coupling 
condition of the excitation method, particularly in the case of the prism 
coupler, to sweep this range of values.

\subsection{\cdb Surface properties}
\label{subsec:surf}

A consideration of the surface properties of oocytes, including the roughness 
(microscopic surface deviations) and the elasticity 
(\textbf{Criteria~3} and \textbf{4}) 
will become particularly important in assessing the feasibility of generating 
WGMs in these kinds of cells. 
Focusing on the most likely candidate, that of bovine oocytes, a topographic 
analysis of the \textit{zona} region has been conducted in 
Ref.~\cite{Papi2010}, leading to quantitative measures of several of the 
surface features. These include 
the density of the mesh of fibril bundles that comprise the \textit{zona}, 
the roughness as measured by the root mean square (RMS) values of cantilever 
deflections using AFM, and the plastic deformation force as estimated through 
Young's modulus. It is found that the surface of the \textit{zona} contains 
thicker and more compact fibril bundles, with diameters of $500$ to $1000$ nm, 
for oocytes that are either immature or only recently fertilised. The RMS 
deformations corresponding to the surface roughness reach a minimum for the 
recently fertilised oocytes, at $62$ nm \cite{Papi2010}. Comparing this result 
to the maximum surface roughness permitted for sustaining modes,  $200$ nm, 
leading to  $Q$-factors of $10^2$ \cite{Rahachou:03}, the minimum surface 
roughness exhibited by bovine oocytes indicates the possibility of 
measurable WGMs. Furthermore, the plastic deformation force corresponding to 
these oocytes is $3.9\pm 0.8$ nN, greater than that of matured oocytes at 
$2.1\pm 0.4$ nN, which indicates a reduction in elasticity immediately after 
fertilisation. Studies on human oocytes have yielded similar dependencies of 
the surface properties on the stage of development \cite{Yanez2016}. 
As a result, presumptive zygotes, oocytes that have been recently fertilised 
and then placed in a fixing solution, will be predominantly used in the 
experiments presented in Chapter~\ref{chpt:wgm} in order to ensure the optimal 
likelihood of measuring WGMs. Note that both the chemicals selected for the 
fixing procedure and the composition of the surrounding media will also 
influence the surface properties and potentially the size of the oocytes, and 
must be taken into account.

\subsection{\cdb Surrounding media}
\label{subsec:surr}

In the case of oocytes and embryos, the surrounding media is predominantly 
water based, and exhibits equivalent optical properties, as described under 
\textbf{Criterion~7}. However, a number of media variants exists in practice, 
depending on the storage method selected, and the concentration of salts must 
be monitored carefully, according to \textbf{Criterion~5}. 
The chemical composition of the two most common fixing procedures, and their 
corresponding media, can be found in Appendix~\ref{chpt:app4}; 
namely, the \textit{zona} salt solution -- phosphate-buffered saline (PBS), and 
the paraffin formaldehyde (PFA) method, which uses a special 
handling medium composed of 3-(N-morpholino) propanesulfonic acid (MOPS) 
buffered wash and bovine albumin serum (BSA). 
In the particular case of the fibre taper coupling method, any dissolved salts 
become strongly ionised by the evanescent radiation from the tapered section of 
the fibre, causing contamination to the taper, rendering it insensitive. 
Stray detritus from damaged cell fragments can also act as contaminants during 
the experiments. 
Subsequently diluting the surrounding medium imposes additional considerations, 
since a forced reduction of the osmolarity of the 
medium can impose osmotic pressure on the cell, altering its 
topology and size. 
It will become apparent in Chapter~\ref{chpt:wgm} that while the dilution 
of a droplet of media with Milli-Q$^{\text{\textregistered}}$ water for the purposes 
of measurement causes no major \emph{topological} defects of the 
\textit{zona pellucida}, it experiences uniform hypo-osmotic pressure on the 
inner wall, enlarging the cell volume by $40$\%, and 
the diameter by $12$\% or greater 
\cite{doi:10.1093/molehr/gam027,doi:10.1093/humupd/dmp045}. 
The discussion of Fig.~\ref{fig:lim} under \textbf{Criterion~1} 
indicates that the increased diameter may serve to facilitate the 
generation of modes. 

\subsection{\cdb Autofluorescence}
\label{subsec:auto}

Autofluorescence describes the process of light that is absorbed and re-emitted 
by certain biological materials \cite{ELPS:ELPS200406070} or crystals 
\cite{Kaufman2012,Ilchenko:13,Tao21062016}. 
It can be difficult to distinguish 
autofluorescence signals from scattering mechanisms, such as Mie scattering, 
and thus it is important to identify the extent to which the cell candidates 
autofluoresce. This directly affects \textbf{Criterion~7}. 
On the other hand, extensive autofluorescence also provides an opportunity to 
realise active biological resonators without the requirement of introducing an 
external fluorescent agent, such as a foreign organic dye, which can sometimes 
exhibit unwanted toxicity \cite{Gupta201515,C5OB00299K,
doi:10.1021/acs.joc.5b00077,Francois:15b,doi:10.1021/acsami.5b10710,
ADOM:ADOM201500776}. 
While the introduction of dyes or artificial emitters may represent a  
viable initial pathway to achieving a biological resonator, as argued in 
Chapters~\ref{chpt:bub} and \ref{chpt:mod}, 
if autofluorescence can be harnessed, it would present a conceptually 
elegant methodology for on-site cell interrogation, where measurement of resonances 
would necessarily report information pertaining to the internal and external 
environment of the cell in question. 

This method has already been trialled in the literature 
through two-photon fluorescence microscopy, 
which is able to determine architectural features of cells, particularly for 
 embryos, which are photosensitive \cite{Squirrell1999}. 
For an CW excitation laser with a wavelength of $532$ nm, and 
using the setup shown in Fig.~\ref{fig:exp}, 
the images taken in Figs.~\ref{fig:mur}(a) through (c) show that, in the case 
of murine embryos, there is indeed an underlying autofluorescence signal, 
particularly in the green portion of the visible spectrum ($543$ nm). 
In order to estimate the relative intensity and the spectrum of the 
fluorescence signal from the typical fixed cells that will be used in 
Chapter~\ref{chpt:wgm}, a single murine embryo in a droplet of PBS media is 
placed on a glass coverslip, and excited with a green ($532$ nm) CW laser, and 
collected with a spectrometer, in the same manner as in Fig.~\ref{fig:exp}. 
The result is shown in Fig.~\ref{fig:autofluorescence}. 
By examining Fig.~\ref{fig:autofluorescence}(a), it is clear that there is 
indeed an autofluorescence peak in the green visible region 
(near $600$ nm). The plateau region between $580$ and $660$ nm 
is derived from the nearby and overlapping peaks corresponding to 
the complex protein and lipid structure of the oocyte. 
Outside the region, however, the signal is negligible. 
In cases where it is required that the autofluorescence signal be removed for a 
clean measurement of the scattered radiation, such as in the case of passive 
interrogation, selecting a wavelength region such as the NIR, which does not 
encounter the absorption effects inherent in the UV or the far IR regions, will 
be mandatory. For comparison, a single \textit{Eudorina}-\textit{Pandorina} 
microsphere conglomerate of cells suspended in MLA medium is placed on the 
microscope under the same conditions. The algae exhibit a much larger 
autofluorescence signal, particularly at $606$ nm, as shown in 
Fig.~\ref{fig:autofluorescence}(b).

\subsection{\cdb Absorptive properties}

The final consideration for assessing the viability of mammalian oocytes 
as biological resonators is the extent to which they absorb light across 
the spectrum, as described in \textbf{Criterion~7}. 

The absorbance spectral profile of murine oocytes in the IR region is shown in 
Ref.~\cite{Diletta}, which includes spectral peaks corresponding to lipids and 
acyl chains, proteins with amide groups, nucleic acids, 
and carbohydrates. Similarly, FTIR spectra obtained from these oocytes 
\cite{Ami20111220} indicate that while the sources of absorbance 
are well characterised and suitable for detection of structural defects in 
oocytes, there is little control over the regions of high absorbance. Since 
amino acid chains involved in DNA are highly absorbing in the UV region 
\cite{Clydesdale2001,Jackson2009}, 
the focus of this investigation is constrained to the visible and NIR regions, 
for which wavelengths in the range $532$ to $769$ nm will be considered. 
The lower wavelength light is absorbed efficiently 
by a range of organic dyes, including those of the Rhodamine group 
\cite{pmid19950338}, which can be used as a fluorescent dopant, and will be 
explored briefly with respect to \textbf{Criterion 6} in 
Section~\ref{subsec:co}. 
As an alternative, the upper wavelength values correspond to a tunable DFB 
laser, which is used for the passive interrogation of a resonator. 

\section{\cdb Selecting a viable cell}
\label{sec:seleccand}

Upon consideration of a range of cells through Sections~\ref{sec:cand} 
and \ref{sec:oo}, mammalian oocytes, in particular, 
 bovine embryos in the \emph{presumptive zygote} 
phase of development, are selected for further study. 
While each of the cells considered has displayed a variety of compelling 
properties, in each case, except for the mammalian oocytes, 
one or more of the selection criteria are not able to be fulfilled. 

In the case of the erythrocite, the diameters are simply not 
large enough to sustain modes for the refractive index of their 
constituent material, despite their symmetrical 

\begin{figure}[H]
\begin{center}
\includegraphics[width=0.8\hsize,angle=0]{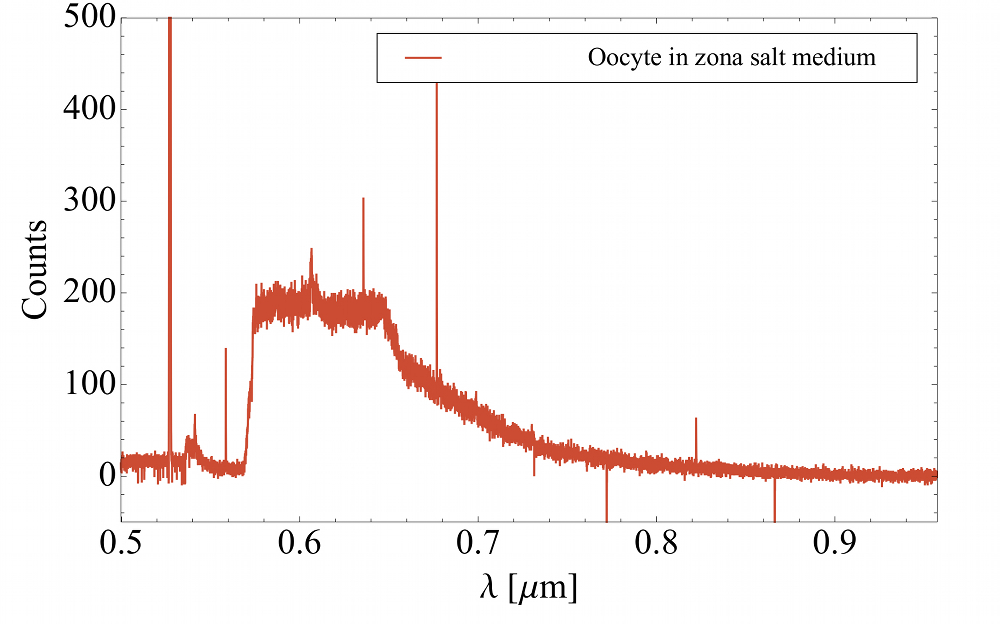}\\
\vspace{-5mm} 
\hspace{6mm}\mbox{(a)}\\
\includegraphics[width=0.8\hsize,angle=0]{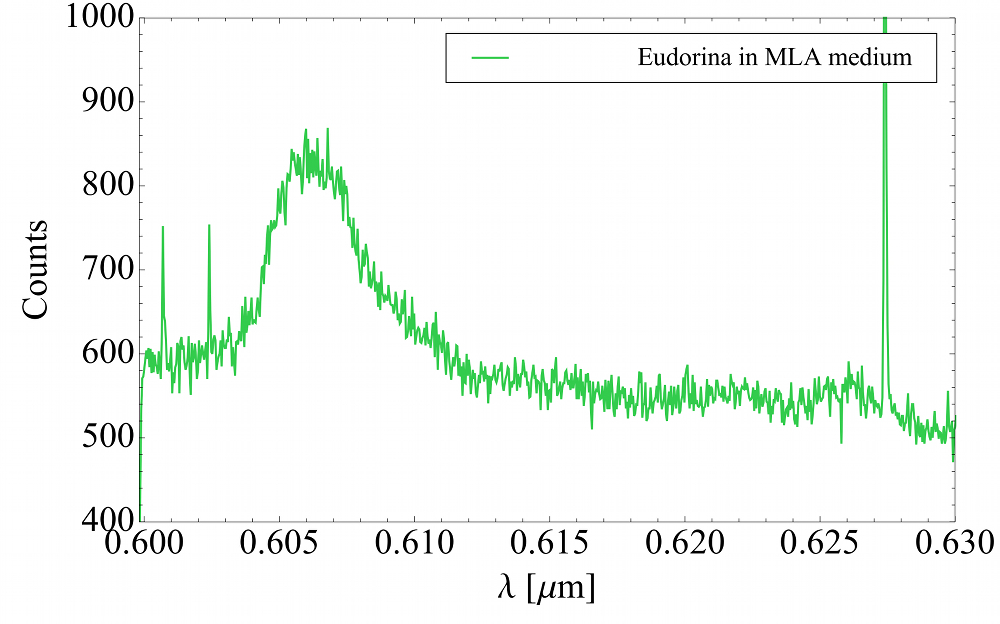}\\
\vspace{-5mm}
\hspace{9mm}\mbox{(b)}\\
\end{center}
\vspace{-6mm}
\caption[The relative autofluorescence spectra of murine oocytes and the 
\textit{Eudorina}-\textit{Pandorina} algae are compared.]
{The fluorescence spectra for a single biological cell are shown, 
excited by a $532$ nm CW laser, and collected 
by a spectrometer. (a) A Murine embryo in PBS medium indicates a relatively 
small autofluorescence signal. (b) A zoomed-in spectrum for a single 
\textit{Eudorina}-\textit{Pandorina} microsphere in MLA medium indicates more 
significant autofluorescence, particularly for green light corresponding to a 
wavelength of $606$ nm.}
\label{fig:autofluorescence}
\figrule
\end{figure}

\newpage
\noindent shape. 
Similarly, the \textit{Cryptosporidium} oocysts, while potentially larger than 
the erythrocytes, would not reasonably be able to support WGMs.
One possible avenue of investigation considered was the 
genetic modification of cells, taking the example of 
\textit{Saccharomyces cerevisiae}, or yeast, for which much 
literature on the topic exists. 
However, the largest variety of modified yeast did not respond to 
fibre taper excitation. 
The next most promising candidate, the varieties of algae, 
are able to produce compelling structural arrangements 
with a high degree of symmetry, but 
are unfortunately hampered by large surface defects 
that prevent resonances from being realised in their case. 
Lastly, the mammalian oocytes are found to be in the viable regime 
for their diameter and refractive index, display a high degree of 
sphericity and exhibit an acceptable degree of scattering and absorption loss. 
While the handling media used for oocytes can include a high 
degree of osmolarity and debris, with adverse effects on the ability 
to measure WGMs, these can be controlled by choosing an appropriate 
handling medium with a low osmolarity and a sufficient level of 
dilution. A cell candidate comparison matrix is provided on the next page. 

To what extent subtle microscopic deformations on the surface 
of the \textit{zona} or the inherent elasticity of this layer 
will play a role in the successful generation of modes is, at this point, 
largely unknown. The investigation of the surface properties, 
as well as the experimental confirmation of the viability of the oocyte 
with respect to the other selection criteria, 
will be the subject of intensive study in Chapter~\ref{chpt:wgm}.

\begin{table*}[t]
\begin{mdframed}[backgroundcolor=boxcol,hidealllines=true]
{
\begin{center}
\begin{mdframed}[backgroundcolor=boxcold,hidealllines=true]
\textbf{\color{white} \large \hspace{2mm} Cell candidate comparison 
matrix  \\}
\color{white}(\cmark satisfied,\, \xmark\, not satisfied,\, 
$\mathbf{\approx}$ borderline satisfied,\, \textbf{--} not determined) \cbl
\end{mdframed}
\end{center}
\cdb
}
\end{mdframed}
\vspace{-3mm}
\begin{mdframed}[backgroundcolor=boxcol,hidealllines=true,innerleftmargin=0pt,
innerrightmargin=0pt,skipabove=0pt]
{
\begin{center}
\begin{tabular}{cccccc}
      \cdb\textbf{Criterion} & \cdb\textbf{Erythrocite} & \cdb
\textbf{\textit{Cryptosporidium}} & \cdb\textbf{Yeast} 
& \cdb\textbf{Alga} & \cdb\textbf{Oocyte}\vspace{2mm}\\
\cdb\textbf{1.} & \xmark & \xmark & $\mathbf{\approx}$ & \cmark & \cmark 
\vspace{2mm}\\
\hline\vspace{-2mm}\\
\cdb\textbf{2.} & \cmark & \cmark & \cmark & \cmark & \cmark \vspace{2mm}\\
\hline\vspace{-2mm}\\
\cdb\textbf{3.} & \textbf{--} & \textbf{--} & \textbf{--} & \xmark & 
$\mathbf{\approx}$ \vspace{2mm}\\
\hline\vspace{-2mm}\\
\cdb\textbf{4.} & \textbf{--} & \textbf{--} & \textbf{--} & \xmark & 
$\mathbf{\approx}$ \vspace{2mm}\\
\hline\vspace{-2mm}\\
\cdb\textbf{5.} & \textbf{--} & \textbf{--} & \cmark & \cmark & \cmark 
\vspace{2mm}\\
\hline\vspace{-2mm}\\
\cdb\textbf{6.} & \textbf{--} & \textbf{--} & \xmark & \cmark & \cmark 
\vspace{2mm}\\
\hline\vspace{-2mm}\\
\cdb\textbf{7.} & \textbf{--} & \textbf{--} & \cmark & \cmark & \cmark 
    \end{tabular}
\end{center}         
}
\end{mdframed}
\end{table*}

\section{\cdb Sample preparation}
\label{sec:samp}

The extraction, preparation, storage, mounting and manipulation of the 
biological cell samples, particularly in the case mammalian oocytes, represent 
tasks that must be handled carefully according to the procedure outlined below. 

First, bovine or murine ovaries are sourced from the Robinson Research 
Institute -- The University of Adelaide. 
Upon extraction from an ovary, the oocytes must then be denuded in order to 
remove the residual cumulus cells, shown in Fig.~\ref{fig:cum} in the Prologue, 
increasing the smoothness of the outer surface of the \textit{zona pellucida}. 
They are then cleaned and fixed using either of the methods described in 
Appendix~\ref{sec:zona} and \ref{sec:pfa}. Note that the fixing method 
influences the osmolarity of the handling medium that is used in the 
experiment. Upon fixing, the oocytes can then be stored in a refrigerator at 
$4^\circ$C. 

Immediately prior to mounting the cells onto the experimental setup, either on 
the top surface of the prism coupler or on the glass coverslip (see 
Section~\ref{sec:meth}), the medium must be diluted in order to control the 
inhibiting effects of salt contamination, described under \textbf{Criterion~5}. 
It is found that a dilution by a factor of $3$ is sufficient 
for the MOPS+BSA medium, whereas a much greater dilution factor of $10$ is 
required for the PBS medium. 

\begin{figure}[t]
\begin{center}
\includegraphics[height=5cm]{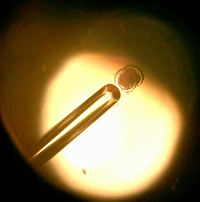}
\includegraphics[height=5cm]{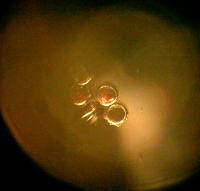}
\end{center}
\vspace{-5mm}\caption[Bovine presumptive zygotes in a droplet of medium on a 
prism coupler.]{As a prelude to the experimental methods described in 
Chapter~\ref{chpt:wgm}, bovine presumptive zygote oocytes in a droplet of 
specialised handling medium (see Appendix~\ref{sec:mops}) are placed on the 
surface of a high-index prism. The positions of the oocytes are 
manipulated with a suction holding-pipette. Scattered light is 
evident when the oocytes are placed directly above the location of the laser 
spot.}
\label{fig:oman}
\figrule
\end{figure}

In order to investigate the coupling of light into an oocyte, its position 
must be controlled precisely so that it is directly above the laser spot in 
the case of the prism, or within the evanescent field of the fibre taper. This 
is achieved through the use of a suction holding-pipette, as shown in 
Fig.~\ref{fig:oman}. 
The sample preparation procedure is summarised on the next page, while
the specific 
configurations of the experimental setups will be described in the 
Chapter~\ref{chpt:wgm}. 
\begin{table}
\begin{mdframed}[backgroundcolor=boxcol,hidealllines=true]
{
\begin{center}
\begin{mdframed}[backgroundcolor=boxcold,hidealllines=true]
\textbf{\color{white} \large Experimental preparation procedure \cbl}
\vspace{1mm}
\end{mdframed}
\end{center}
\cdb
\vspace{3mm}
\noindent\textbf{A.}  \quad Oocytes are cleaned, and fixed according to 
Appendix~\ref{chpt:app4}.\vspace{2mm}\\
\noindent\textbf{B.}  \quad Oocytes are stored in medium at $4^\circ$C.
\vspace{2mm}\\
\noindent\textbf{C.}  \quad Medium must be diluted with 
Milli-Q$^{\text{\textregistered}}$ water prior to mounting.\vspace{2mm}\\
\noindent\textbf{D.}  \quad Inoculated medium is added to the coupling prism 
or coverslip with pipette.\vspace{2mm}\\
\noindent\textbf{E.}  \quad The oocyte positions are manipulated using a 
suction holding-pipette. \cbl
\vspace{0mm}
}
\end{mdframed}
\end{table}
While it is by no means certain that clearly identifiable whispering gallery 
modes will be measured within a biological cell, in this chapter, a concerted 
effort has been made to identify the most likely challenges, and the physical 
parameters of the cell and its environment to which they pertain. 

Although a 
number of cell candidates has been examined, the primary focus of the 
experiments presented in the next chapter will be the bovine oocytes, 
presumptive zygotes and embryos, which are deemed to satisfy the selection 
criteria most closely. Thus difficulties in isolating WGM 
signals from the transmitted and scattered spectra can be clearly explained, 
and a pathway to overcoming them can be developed.

\chapter{\cdb Whispering Gallery Modes in an Embryo}
\label{chpt:wgm}

This chapter describes both the experimental setups and the results for the 
study of WGMs within embryos. 
After an introduction as to how embryos may feasibly function 
as resonators in Section~\ref{sec:eres}, 
the various experimental configurations and their respective methodologies 
and procedures are described in Section~\ref{sec:meth}. 
Modelling predictions of the FSR, $Q$-factors and sensitivity of a simulated 
cell are investigated in Section~\ref{sec:modpred}. 
Then, in Section~\ref{sec:exres}, the experimental results are explained and 
compared, and the new understandings of resonator behaviour 
that can be gleaned from each technique are summarised. 
Finally, the conclusions drawn from the results 
are stated in Section~\ref{sec:fin}. 
The experimental results and 
the model simulations are thus tied together, 
with a view towards the next key steps and future 
direction of research on this topic.

\section{\cdb Embryos as resonators}
\label{sec:eres}

In Chapter~\ref{chpt:cel}, the structure of mammalian oocytes 
and embryos was explored and, after an analysis of their 
potential resonance properties using the selection criteria 
developed in Section~\ref{sec:selec}, these cells 
were deemed the most viable candidates considered in 
Section~\ref{sec:seleccand}. 
These embryos are thus chosen for a more in-depth experimental analysis, 
presented herein. While biological cells are not \emph{ideal} resonators, 
in that they present challenges that can serve to frustrate 
the generation of WGMs, the mammalian oocyte exhibits some attractive features 
that make it particularly noteworthy as a resonator candidate. 

First, its size and refractive index combination are estimated to lie in a 
region of the parameter space that can ostensibly support modes. 
Bovine oocytes in particular have an outer diameter of 
approximately $150$ $\mu$m, and a refractive index of $\mathrm{n}=1.36$ to 
$1.51$ for the outermost layer, known as the \textit{zona pellucida} 
\cite{Voros2004553}.  
Next, the surface 
roughness, while presenting the potential for hampering the generation of WGMs, 
can nevertheless be potentially reduced  by selecting a specific period 
in the life-cycle of the cell, and then exposing it to a 
carefully selected fixing solution, as outlined in Section~\ref{subsec:surf}. 
Bovine embryos -- specifically \emph{presumptive zygotes}, 
are used predominantly in these experiments for precisely this reason. 
It is these cells, when denuded, as described in the Prologue, 
that exhibit the most favorable surface 
properties in surface roughness, elasticity, and also potentially higher 
refractive index values via their greater density of surface fibril bundles 
in the \textit{zona pellucida} region \cite{Papi2010}. 
These attributes, together, present the greatest likelihood for sustaining 
modes within a cell. 
As a note on nomenclature, since these kinds of oocytes are fertilised, 
the removal of the cumulus cells (depicted in Figs.~\ref{fig:cum} and 
\ref{fig:cel}) is expedited by the underlying biological processes that 
govern zygote development. 
For convenience, the term \emph{embryo} will be used from now on. 

The embryos are studied using a number of excitation methods, 
namely, the passive interrogation methods -- the prism coupler and the fibre 
taper technique, and the active interrogation methods -- fluorescent 
dye-doping, 
and quantum dot coating. By using these methods alongside one another, 
and in combination, a 
picture of the resonance properties of the embryos is unveiled, including the 
sources of both scattering and loss. As a result, the vision of a WGM spectrum 
detected in a self-supporting biological resonator in a realistic 
environment becomes a real possibility.

\section{\cdb Experimental methodology}
\label{sec:meth}

In this section, the experimental methodologies, preparation procedures and 
measurement apparatus and techniques are described in detail. 
Each technique requires a \emph{coupling} method, in which the radiation is 
introduced into the resonator, and energy is stored in the modes that 
are subsequently excited. 
 
Coupling represents a vital experimental consideration in the generation of 
WGMs, as it profoundly affects the ability to detect modes, as well as 
the structure of the spectrum. Thus, a significant portion of this chapter 
is dedicated to the presentation and comparison of a range of coupling methods. 

The coupling methods can be 
split into two categories: passive and active interrogation, as described 
in Section~\ref{sec:arch}. Recall that passive interrogation involves 
tunnelling radiation into a resonator by placing an evanescent field from a 
coupler, such as a prism or fibre, in the vicinity 
of the surface. The radiation from the coupler must be phase-matched with the 
WGMs at the surface. The first passive interrogation technique to be discussed 
is the prism coupler.

\subsection{\cdb Prism coupler method}
\label{subsec:pris}

The coupling of the light into a resonator through the use of a high-index 
coupling prism has long been used in the field 
\cite{BRAGINSKY1989393,Gorodetsky:99,Gotzinger2001,
:/content/aip/journal/apl/82/4/10.1063/1.1540242,Wildgen:15,Foreman:16}, 
and in this investigation it 
represents the first screening tool applied to the candidate cells passing 
the selection criteria of Chapter~\ref{chpt:cel}. The strength of this method 
is its reusability, in that samples of prospective biological resonators can be 
examined in quick succession. 
However, this method does require careful alignment to be 
maintained throughout the experiment, presenting practical challenges 
for the development of robust devices for WGM detection. 

\begin{figure}
\begin{center}
\includegraphics[width=0.9\hsize]{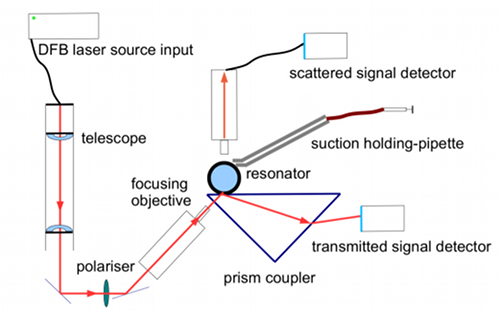}
\end{center}
\vspace{-5mm}\caption[The experimental setup for the prism coupler method is 
illustrated.]{The experimental setup for detecting the scattered and 
transmitted signal from a free-floating resonator, such as an embryo, using 
the prism coupler method.}
\label{fig:prissetup}
\figrule
\end{figure}

The experimental setup used for mounting, interrogating and measuring the 
scattered and transmitted signal from an embryo is shown in 
Fig.~\ref{fig:prissetup}. A tunable DFB laser is used to scan across a 
wavelength range of $762$ to $769$ nm, which lies in the NIR portion of the 
electromagnetic spectrum -- the optimal wavelength window for reducing 
absorbance in the embryo, as 
explained in Section~\ref{subsec:auto}. 

The beam is collimated with an objective ($\times 4$ magnification), and 
passed through a focusing telescope constructed from a convex followed by a 
concave lens, with an overall magnification of $1/5$. 
Then the beam is filtered through an adjustable polariser in order 
to provide some control on the excitation of WGMs. Since it is clear from the 
investigation into the decomposition of the power spectrum in 
Section~\ref{sec:demo} that the polarisation of the 
radiation couples selectively to certain modes only, the use of a polariser 
filter facilitates the acquisition of a cleaner spectrum and more convenient 
identification of any WGMs present in the signal. 

The material used for the prism coupler can be chosen according to purpose. 
In the case of the bovine embryos, 
both a borosilicate Schott BK$7$ glass prism\footnote{
Thorlabs, Inc., $56$ Sparta Ave, Newton, NJ $07860$, USA, 
\textit{N-BK7 Right-Angle Prism, 15 mm}, Catalogue Number: PS$915$.} 
and a titanium dioxide (TiO$_2$) or \emph{rutile} 
prism\footnote{Team Photon, Inc., $4653$ Carmel Mountain Rd, Suite $308$-$116$, 
San Diego, CA $92130$, USA, \textit{Rutile prism}, Catalogue Number: 
TiO$2$-$10$-$10$-$10$.} 
were tested. Since the ability of a prism to couple into WGMs requires that 
the refractive index of the prism exceed that of the prospective resonator, and 
since the precise value of the index of the glycoproteins that comprise the 
\textit{zona} region of the bovine embryo is not known, the more conservative 
option of a rutile prism is selected. This is done in 
order to accommodate the possibility of larger refractive indices exhibited by 
the \textit{zona}, suggested in Refs.~\cite{biodevices10,Wacogne:13}. 
While the transmittance of rutile is $68$\% compared to that of BK$7$ glass 
at $90$\% at a wavelength of $760$ nm, the reduction of input 
power is not an issue for the generation of WGMs using the DFB laser, 
which has a maximum output power of approximately $20$ mW. Power values of the 
order of milliwatts are sufficient to excite modes. 

\begin{figure}
\begin{center}
\includegraphics[width=0.5\hsize]{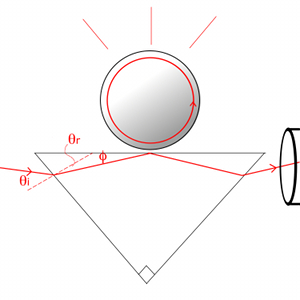}
\includegraphics[width=0.45\hsize]{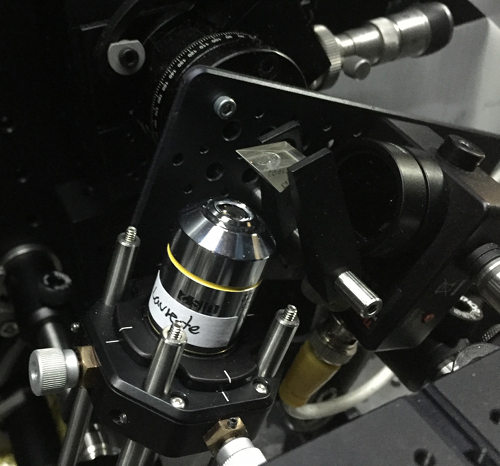}\\
\hspace{0.5cm} (a)\hspace{6.6cm} (b)
\end{center}
\vspace{-5mm}\caption[Images of the prism coupler and the phase-matching 
condition.]{The prism coupler method proceeds via frustrated total 
internal reflection, as shown in the 
following images. (a) A zoomed-in diagram of a prism and resonator, 
illustrating the path taken by an incident beam, and the angles used in 
defining the phase-matching condition. (b) The experimental setup is shown, 
with the focusing objective (\textit{centre left}), the 
rutile prism attached to a 
rotating actuator (\textit{centre top}) and the transmission detector 
(\textit{centre right}).}
\label{fig:pris}
\figrule
\end{figure}

In order to fulfill the phase-matching condition, the propagation constant of 
the portion of the incident beam that is parallel to the surface of the prism
 must match that of the WGM modes, $\beta_{\Vert} = \beta_{\mathrm{WGM}}$. 
Within the prism with refractive index $\mathrm{n}_{\mathrm{pris}}$,
 the incident beam has a propagation constant 
$\beta_{i} = k_0 \mathrm{n}_{\mathrm{pris}}$. 
According to Fig.~\ref{fig:pris}, the parallel portion of the beam 
corresponds to $\beta_{\Vert} = k_0 \mathrm{n}_{\mathrm{pris}} \mathrm{cos}(\phi)$. 
The corresponding propagation constant of the WGMs can be estimated 
for the fundamental modes of a microsphere with mode numbers $l=m$, 
from the derivation in Section~\ref{sec:an} leading to 
Eq.~(\ref{eq:kwgm}),  
$\beta_{\mathrm{WGM}}= m/R = k_0\mathrm{n}_{\mathrm{res}}$. Thus, the phase-matching 
condition 
can now be specified in terms of the angles of refraction and incidence 
from the entry point of the beam  into the prism, 
$\theta_r$ and $\theta_i$, respectively, using Snell's Law
\vspace{-5mm}
\begin{align}
\phi &= \mathrm{arccos}(\mathrm{n}_{\mathrm{res}}/\mathrm{n}_{\mathrm{pris}}),\\
\theta_r &= \phi - \pi/4,\\
\theta_i &= \mathrm{arcsin}\left(\frac{\mathrm{n}_{\mathrm{pris}}}
{\mathrm{n}_{\mathrm{air}}} \, \sin\theta_r\right).
\label{eq:prisang}
\end{align}
Note that, in these experiments, the prism is always surrounded by air, 
$\mathrm{n}_{\mathrm{air}}=1$. 

Using Eq.~(\ref{eq:prisang}), the angle of the incident laser 
beam upon the prism can be adjusted, and the coupling in the WGMs 
of the resonator thus optimised. 
Values of the angle $\theta_i$ are shown in Table~\ref{tab:RIprism} 
for a range of typical values of the resonator refractive index 
$\mathrm{n}_{\mathrm{res}}$. By performing 
incremental adjustments to $\theta_i$, the propagation constant 
$\beta_{\Vert}$ can be swept across a range of values, providing the optimal 
chance of observing WGMs in a cell. 
This technique will be used in particular in Sections~\ref{subsec:sil} and 
\ref{subsec:pass}.

\subsection{\cdb Fibre taper method}
\label{subsec:fib}

The fibre taper represents an 
alternative coupling method to the prism coupler for 
the passive interrogation of modes in a resonator. 
While the prism coupler is highly mode selective, thus making the detection of 
a single mode within an imperfect resonator 
a particular challenge, the fibre taper has been shown to couple effectively 
to a large number of modes across a wide wavelength range 
with comparatively high $Q$-factors, in the case of microspheres 
\cite{Riesen:15a,Knight:97,Dong:08}. 
However, the prolific use of tapers for large-scale 
batch measurements is limited, as mentioned in Section~\ref{sec:pred}, by the 
fact that the effective life-times of fibre tapers are typically less than a 
few hours after fabrication, 
due to their fragility, and by the ease of collecting contaminants that can 
greatly affect their signal-to-noise ratios.

\begin{table*}
  \caption[Rutile prism coupler angles for a range of resonator refractive 
indices.]{Values in degrees for the angle $\theta_i$, as shown in 
Fig.~\ref{fig:pris}(a), 
for which the matching condition of the propagation constants is fulfilled for 
a TiO$_2$ (\textit{rutile}) prism coupler, $\mathrm{n}_{\mathrm{pris}}=2.53$ at 
$\lambda= 760$ nm, and 
a range of resonator refractive indices. Results are provided for a 
surrounding medium of air.}
\vspace{-6pt}
  \newcommand\T{\rule{0pt}{2.8ex}}
  \newcommand\B{\rule[-1.4ex]{0pt}{0pt}}
  \begin{center}
    \begin{tabular}{cc}
      \hline
      \hline
      \T\B            
       $\mathrm{n}_{\mathrm{res}}$ & \qquad\qquad\qquad angle of incidence 
$\theta_i$ (degrees) \\ 
      \hline     
      $1.35$ &  \qquad\qquad\qquad$34.0$ \\
      $1.40$ &  \qquad\qquad\qquad$30.0$ \\
      $1.45$ &  \qquad\qquad\qquad$26.2$ \\
      $1.50$ &  \qquad\qquad\qquad$22.4$ \\
      $1.55$ &  \qquad\qquad\qquad$18.6$ \\
      \hline
    \end{tabular}    
  \end{center}
  \label{tab:RIprism}
\end{table*}

The manufacture of the fibre tapers may proceed in a number of ways. The 
method outlined in this chapter makes particular use of a 
Vytran,\footnote{Thorlabs, Inc., $56$ Sparta Ave, Newton, NJ $07860$, USA, 
\textit{Vytran$^{\text{TM}}$ Filament Fusion Splicer}, Catalogue Number: 
LFS$4100$.} 
 which uses a high-power filament to anneal a section of fibre a few 
centimetres in length down to a diameter of approximately $1$ $\mu$m. 

In producing a fibre taper of sufficient quality for the measurement 
of WGMs in resonators, a strict procedure must be followed. 
A filament that is capable of annealing an optical fibre 
evenly for an extended period (approximately $5$ minutes), 
such as graphite or iridium, must be loaded 
into the Vytran, and the corresponding software settings selected. 
A sufficient argon flow is mandatory to avoid damaging the filament. 
After activating the Vytran and its attached vacuum pump, 
a connectorised optical fibre, such as an SMF$28$, must be prepared 
by `window-stripping' the outer cladding, and cleaned  
using isopropyl alcohol and abrasion. The fibre is then mounted 
on the Vyrtran, and secured either side of the filament 
in $250$ $\mu$m $v$-grooves, ensuring that less than $18$ grams of tension 
is applied prior to activating the filament. 
While the program executes, the filament will evenly reduce the diameter 
of the stripped section of fibre to a prescribed value, while keeping 
the fibre taut. Upon completion of the fabrication process, 
it is important to adjust the tension of the taper to avoid breakage or 
loss due to deviations in the tapered section. 
Once this has been done, a fork-shaped brace is raised up underneath the 
taper, which is then clamped securely in place. After deactivating 
the vacuum pump, the brace and its attached taper are then 
transported to the experimental setup, and connected to the 
tunable DFB laser.  

\begin{figure}
\begin{center}
\includegraphics[width=0.8\hsize]{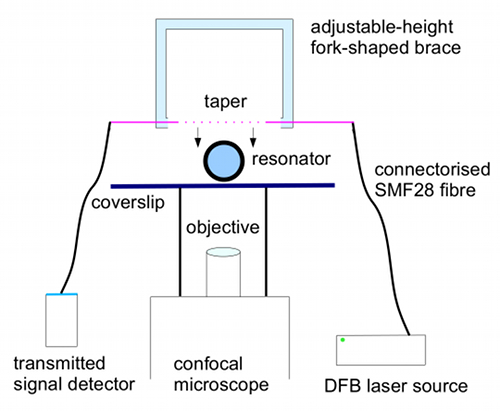}
\end{center}
\vspace{-5mm}\caption[The experimental setup for the taper method is 
illustrated.]{The experimental setup for exciting WGMs within 
a resonator using a fibre taper, and collecting the resultant transmitted 
signal. The taper is fixed to an adjustable-height brace so that it remains 
intact, and can be moved into close proximity with the resonator in order to 
fulfill the phase-matching condition.}
\label{fig:taper}
\figrule
\end{figure}

A diagram of this setup is shown in Fig.~\ref{fig:taper}. 
A confocal microscope setup is used, similar to Fig.~\ref{fig:exp}, 
with the addition of a fork-shaped brace mounted above the 
glass coverslip and resonator, as shown. The brace is attached 
to a stage of adjustable height so that the taper and its 
evanescent field can be lowered accurately in order to couple 
light into the resonator. The height of the taper above the resonator 
controls the phase-matching condition, and consequently, fine adjustments 
are required in order to efficiently couple radiation into WGMs. 
This will be explored in Section~\ref{subsec:pass}.

\subsection{\cdb Fluorescence methods}
\label{subsec:fl}

The excitation of WGMs may occur through use of an active layer 
or dopant, as introduced in Chapter~\ref{chpt:intro}. 
While the uses of fluorescent materials such as organic 
fluorophores \cite{LPOR:LPOR201200074,C5LC00670H,Linslal:16,Vanga2015209,
Chandrahalim2015,Saito:08,Himmelhaus2009418,OPPH:OPPH201600006}, 
crystals \cite{Kaufman2012,Ilchenko:13,Tao21062016} 
or quantum dots \cite{C1JM10531K,6062374,1063-7818-44-3-189} 
have been explored extensively in the literature, the majority 
of WGM resonators studied in the literature nevertheless rely 
predominantly on passive interrogation techniques. 
While the advantages of active resonators are manifold, as described in 
detail in Section~\ref{subsec:act}, of particular note is their convenience 
in allowing free-space excitation, alleviating the practical 
challenges involved in the alignment of complex apparatus. 
The key disadvantage of active interrogation also poses a quandary in the 
case of \emph{biological} resonators, namely, the reduction in $Q$-factor 
inherent in the inability to specify a preferred orientation along 
a single axis of symmetry. 
The $Q$-factors of biological resonators 
are expected to be orders of magnitude lower than those of 
artificial resonators due to surface imperfections and 
asphericity. Additional reduction of the $Q$-factor 
due to the method of excitation, \textbf{Criterion~6}, may 
prevent the detection of any underlying modes altogether. 
 
Assessment of the $Q$-factor is thus important in this investigation as it 
acts as an indicator as to whether WGMs have been measured. 
The feasibility studies presented 
in Chapter~\ref{chpt:cel} place bounds on the values of the $Q$-factors 
and FSRs expected from a resonator with given physical characteristics. 
Thus, by analysing the $Q$-factors of features within a signal, as well as 
the FSR and the index sensitivity, evidence for or against the 
identification of WGMs can be established.

The presence of a uniform dye also has the effect of reducing the
 $Q$-factors, when compared to excitation via a dipole with 
 a single orientation. This is a consequence of closely-spaced or overlapping 
modes, as shown in the FDTD simulations of Fig.~\ref{fig:sourcecomp}, which 
are excited simultaneously, thus making the distinction of separate modes 
more difficult. 
The isolation of a fundamental radial mode, for example, 
using the selective prism method, also presents a challenge in the absence 
of well documented material properties, such as the glycoprotein refractive 
index. 
This, however, can be ameliorated with the introduction of an adjustable 
polariser, depicted in Fig.~\ref{fig:prissetup}, as discussed in 
Section~\ref{subsec:sil}. The measurement of modes within a 
biological resonator provides the added advantage of providing a new 
methodology for the characterisation of the refractive index of glycoproteins. 
 
In the case of biological cells, the exhibition of autofluorescence 
provides an additional method for the exploration of active mode coupling. 
While the genetic modification of cells to exhibit enhanced 
fluorescence properties is an intriguing prospect, 
discussed in Chapter~\ref{chpt:fdi}, the natural autofluorescence 
of mammalian oocytes and embryos is typically insufficient 
for the purposes of generating WGMs, as evident from
 Fig.~\ref{fig:autofluorescence} in Chapter~\ref{chpt:cel}. 

Since the principal intention of this thesis is to investigate
 the possibility of realising a biological resonator, 
both passive and active techniques are brought to bear, and the insights 
drawn from each experiment are used to establish a proof-of-concept. 
Thus, future work focused on developing this concept into practical 
applications may be informed, 
and scientific insight into biological systems may be revealed.

\subsection{\cdb ICSI dye injection and co-culturing method}
\label{subsec:co}

The introduction of an organic dye into a embryo can allow it to perform as 
an active resonator, and can proceed in a number of ways. 
Two particular methods that are used in this thesis are the intracytoplasmic 
sperm injection (ICSI) method, 
and the co-culturing method. In each case, an organic dye is selected that can 
be efficiently excited at 
$532$ nm wavelength, and also emits in the visible region -- in this case, 
[9-(2-carboxyphenyl)-6-diethylamino-3-xanthenylidene]-diethylammonium chloride, 
otherwise known as Rhodamine B. The peak excitation and emission wavelengths 
of this dye are $\lambda_{\mathrm{ex}} = 543$ nm and $\lambda_{\mathrm{em}}= 565$ 
nm, respectively \cite{KUBIN1982455}. 
Note that in the case of active resonators, which exhibit a broad range of 
emitted wavelengths, 
there is greater flexibility in selecting a suitable wavelength region, since 
a tunable laser is no longer a requirement in order 
to scan across a wavelength range. 

\begin{figure}
\begin{center}
\includegraphics[width=0.8\hsize]{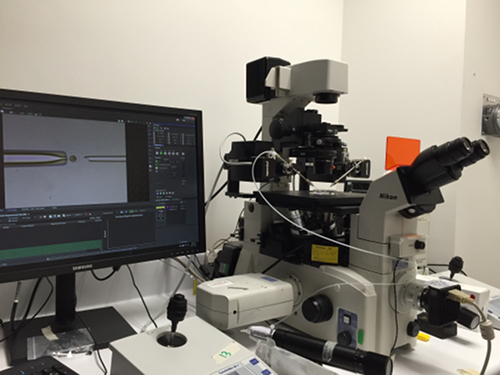}
\end{center}
\vspace{-5mm}\caption[The ICSI procedure in progress on an embryo.]{The 
intracytoplasmic sperm injection (ICSI) method in progress, for a quantity of 
high-concentration organic dye, Rhodamine B. The embryo is placed in 
conjunction with a holding-pipette, and a specialised ICSI needle is guided 
into position using a micromanipulator (\textit{centre right}). The needle 
pierces the \textit{zona pellucida} of the embryo and is used to inject the 
dye.}
\label{fig:dye}
\figrule
\end{figure}

\begin{figure}
\begin{center}
\includegraphics[height=0.29\hsize]{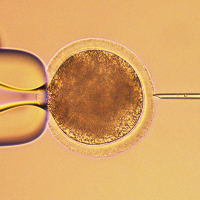}
\includegraphics[height=0.29\hsize]{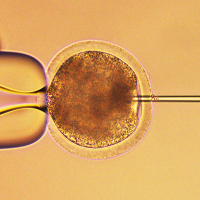}
\includegraphics[height=0.29\hsize]{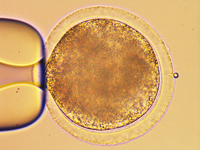}\\
\hspace{-0.4cm}(a)\hspace{3.7cm}(b)\hspace{4.7cm}(c)\\
\end{center}
\vspace{-5mm}\caption[A bovine embryo undergoing ICSI is shown.]
{A bovine embryo is 
placed adjacent to a holding-pipette (\textit{centre left}). (a) The 
micromanipulator is used to adjust the position of the needle 
(\textit{centre right}) prior to ICSI. (b) ICSI is performed by injecting up 
to $5$ picolitre of $100$ $\mu$mol concentrated organic fluorescent dye, 
in this case Rhodamine B. 
(c) The puncture in the \textit{zona pellucida} made by the 
ICSI needle can be observed (\textit{far right}). \textit{Images:} 
produced with the assistance of Mr. Avishkar Saini, 
The University of Adelaide.}
\label{fig:dyeinj}
\figrule
\end{figure}

ICSI must be performed on an embryo prior to fixing. This is due to the fact 
that the fixing procedure alters the mechanical properties of the 
\textit{zona pellucida} region 
in such a way that the ICSI needle can no longer penetrate its outer layer. 
This introduces a time constraint, in that fresh cells must be sourced 
and ICSI performed immediately 
upon arrival. ICSI proceeds by placing an embryo on a micromanipulator setup, 
as shown in Fig.~\ref{fig:dye}. The embryo is placed in conjunction with a 
suction holding-pipette, 
and a specialised ICSI injection needle is manoeuvred into position using the 
micromanipulator. This configuration is shown in Fig.~\ref{fig:dyeinj}(a). 
The ICSI needle is then used to puncture the \textit{zona} region and inject 
a quantity of dye into the cytoplasm (see Fig.~\ref{fig:cel}). 
The volume of dye solution that is able to be injected is relatively small, 
less than $5$ picolitre. Attempting to inject a greater volume typically 
results in rupturing 
of the \textit{zona}, causing the embryo to lose structural cohesion, and the 
leaking of both cytoplasm and dye into the surrounding medium. 
Therefore, a concentrated solution of dye, $100$ $\mu$mol, is used in order to 
offset the small volume. The injection stage is shown in 
Fig.~\ref{fig:dyeinj}(b). 
Upon completion of ICSI, the embryo is then fixed in PFA using the method 
described in Appendix~\ref{sec:pfa}. 

The behaviour of the dye within the cytoplasm represents an important 
consideration. Rhodamine B diffuses quickly, so that it is evenly 
distributed throughout the cytoplasm within a few minutes. 
One potential difficulty is that the dye is not adequately contained within 
the \textit{zona} structure and quickly diffuses into the surrounding medium. 
The puncture region of the embryo 
post-ICSI is 
shown in Fig.~\ref{fig:dyeinj}(c). 
While this diffusion is a critical factor, it is well documented that 
intracellular mitochondria absorb and retain quantities of dye of the 
Rhodamine group 
\cite{Johnson01021980,Baracca2003137}. The co-culturing method relies on this 
fact in order to achieve uptake of dye within the embryo. 

The co-culturing method must proceed prior to fixing. Living 
embryos are placed in a culture medium that contains the dye, and left to 
develop until the presumptive zygote stage. 
Then, the embryos undergo fixing in PFA, as in Appendix~\ref{sec:pfa}. The 
uptake of Rhodamine dye is investigated for a variety of concentrations. 
It is found that the small quantities of dye retained in the mitochondria are 
not sufficient to generate substantial fluorescence for the generation of modes.
However, sources of fluorescent materials other than organic dyes, 
such as quantum dots, present some unique advantages that will now be explored. 

\subsection{\cdb Quantum dot coating method}

One potential method for facilitating the retention 
of an active material, either within or 
attached to the outer surface of the cell, 
is to make use of fluorescent nanoparticles that have been 
functionalised for such a purpose. Quantum dots represent a convenient form 
of fluorescent particle, which have proved 
successful in generating WGMs
\cite{Beier2009,C1JM10531K,6062374,1063-7818-44-3-189}. 
They generally have a broad excitation range 
and can thus be readily excited using the $532$ nm CW laser. 
Furthermore, quantum dots 
can be produced with a range of 
chemical functional groups. 
Two such examples, amino (PEG) 
group\footnote{Life Technologies Australia Pty Ltd, PO Box $4296$, Mulgrave, 
VIC $3170$, Australia, \textit{Qdot$^{\text{\textregistered}}$ 705 ITK Amino (PEG) 
Quantum Dots}, Catalogue Number: Q$21561$MP.} and carboxyl 
functionalised\footnote{
\textit{ibid. Qdot$^{\text{\textregistered}}$ 705 ITK Carboxyl Quantum Dots}, 
Catalogue Number: Q$21361$MP.} quantum dots are used 
in Section~\ref{subsec:ac}. 
It will become apparent that the use of the functionalised 
quantum dots alone does not result in an emitter density 
on the surface of the \textit{zona} sufficient for the generation of modes. 
However, the functional groups of the quantum dots are able to be attached 
to a polyelectrolyte layer that can be added to the embryo. This 
new layer is able to support a much higher density of quantum dots, and 
a sufficient fluorescence signal can be generated. 
In this manner, the problem of dye retention within 
a cell can be averted, 
providing the most reliable test scenario for the observation of WGMs through 
active interrogation.

\subsection{\cdb Polyelectrolyte layers and crosslinking for quantum dots}
\label{subsec:poly}

Polyelectrolytes are electrically charged polymers that are able 
to form thin films that can be attached to a surface of opposite charge. 
Positively and negatively charged polyelectrolytes can be combined 
to form alternating layers to which floating particles in solution can be 
adsorbed \cite{Antipov2003175}. Quantum dots 
can be immobilised by transforming the amino/carboxyl functional groups 
in order to facilitate \textit{crosslinking}. This process involves the 
bonding of one polymer chain to another, or to a functional group 
\cite{Fischer2010}.
In this case, the negatively charged polyelectrolyte is polyallylamine 
hydrochloride (PAH), and the positively charged polyelectrolyte is 
poly acrylic acid (PAA) \cite{doi:10.1021/la202267p}. 

The process for adding a polyelectrolyte layer to each 
embryo is achieved by adding either a negatively or positively charged 
polyelectrolyte solution to the suspended embryos in media. 
The choice depends upon the electric charge of the surface of the embryo. 
Alternatively, if the electric charge of a surface is not known in advance, 
as is the case for embryos, one of each polyelectrolyte will be attempted, 
ensuring that the suspended embryos are thoroughly rinsed between each 
attempt. 
This procedure is summarised in the box on the next page. 

First, $50$ $\mu$L of the cationic PAH is added to the embryos in solution. 
The concentration of the PAH is $2$ mg/mL in $1$ mol/L of NaCl. 
The mixture is agitated, and allowed to settle for $15$-$20$ minutes. 
Second, the solution is rinsed, by removing the supernatant,  
refilling with water, and agitating. This rinsing process is repeated until 
a dilution of $200$ times is achieved, which was found to be sufficient 
to limit unintended interaction between the polyelectrolytes PAH and PAA. 
This process is then repeated for a $50$ $\mu$L volume of PAA 
at the same concentration, rinsing thoroughly. 
At this point, the embryos are coated with a polyelectrolyte layer, 
which can be crosslinked with the functional groups of the quantum 
dots to achieve the coating. 

Volumes of $20$ $\mu$L of amino group functionalised quantum dots, 
as well as carboxyl group activating reagents, 
$1$-ethyl-$3$-($3$-dimethylaminopropyl) carbodiimide (EDC),  
and N-hydoxysuccinimide, are then added to the solution. 
The compounds EDC and NHS act as crosslinking reagents that 
are able to adsorb the floating quantum dots to the polyelectrolyte layer. 
After the solution is left for $1$ hour to interact, the embryos are 
rinsed, and the fluorescence properties analysed. 
The study of the 
generation of WGMs using this form of active interrogation 
will be integrated with predictions derived from 
the multilayer model presented in Chapter~\ref{chpt:mod}, 
and will be implemented experimentally in Section~\ref{subsec:ac}. 

\begin{table}
\begin{mdframed}[backgroundcolor=boxcol,hidealllines=true]
{
\begin{center}
\begin{mdframed}[backgroundcolor=boxcold,hidealllines=true]
\textbf{\color{white} \large Quantum dot / polyelectrolyte coating 
procedure \cbl}
\vspace{1mm}
\end{mdframed}
\end{center}
\cdb
\vspace{3mm}
\noindent\textbf{A.}  \quad $50$ $\mu$L of (cationic) polyallylamine 
hydrochloride (PAH) is added to embryos suspended in media, 
agitated via pipette, and left for $15$-$20$ minutes. \vspace{2mm}\\
\noindent\textbf{B.}  \quad The embryos are then rinsed 
 by extracting the supernatant, 
and resuspending the embryos in 
Milli-Q$^{\text{\textregistered}}$ water. A dilution factor of $200$ times 
was found to be sufficient. 
\vspace{2mm}\\
\noindent\textbf{C.}  \quad \textbf{Steps A} and \textbf{B} 
are repeated, substituting the PAH for $50$ $\mu$L of 
(anionic) poly acrylic acid (PAA). This ensures that a polyelectrolyte coating 
has taken place regardless of the electric charge of the surface of the 
\textit{zona pellucida}.
\vspace{2mm}\\
\noindent\textbf{D.}  \quad $20$ $\mu$L of the amino group functionalised 
quantum dots are added to the suspended embryos, 
as well as $20$ $\mu$L of EDC and $20$ $\mu$L of NHS, in order to 
facilitate carboxyl-to-amine crosslinking. The mixture is agitated, 
and left to stand for $1$ hour. \vspace{2mm}\\
\noindent\textbf{E.}  \quad The rinsing procedure, \textbf{Step B}, 
is then carried out to ensure all unattached quantum dots are removed 
from the solution. \cbl
\vspace{0mm}
}
\end{mdframed}
\end{table}

\section{\cdb Modelling predictions}
\label{sec:modpred}

Having established the bovine embryo as the most viable candidate cell for the 
generation of WGMs, it is important 
to perform an assessment of the most likely spectral features using the 
modelling techniques developed in Chapters~\ref{chpt:sph} through 
\ref{chpt:mod}. 
In particular, the precise values of the FSR, the $Q$-factors and the 
refractive index sensitivity $\mathcal{S}$ for a given diameter, 
\textit{zona} thickness and refractive index profile of the components of the 
embryo, shown in Fig.~\ref{fig:cel}, remain an open question. 

An analysis of the scale image in Fig.~\ref{fig:bov}(c) reveals an 
approximate outer diameter of $150$ $\mu$m, with a variation of $15\%$ across 
a range 
of embryos. Recalling Section~\ref{subsec:st}, 
the thickness of the \textit{zona} region is typically 
$13$ $\mu$m, and does not deviate below $10$ $\mu$m unless specialised chemical 
or enzyme-based annealing procedures are used, described in 
Section~\ref{sec:anneal}. 

\begin{figure}[H]
\begin{center}
\includegraphics[width=0.8\hsize]{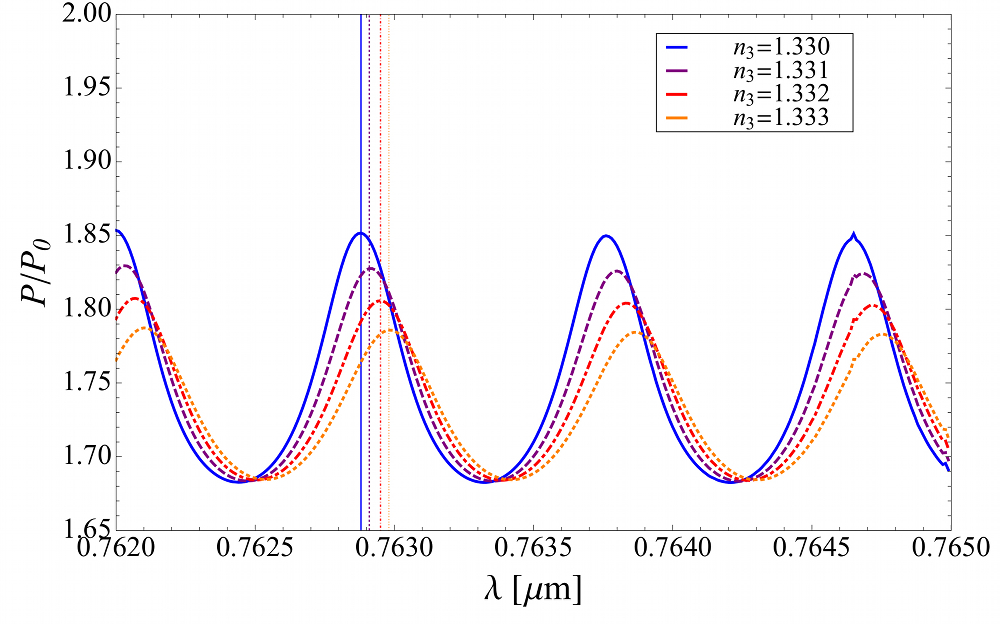}\\
\vspace{-5mm}
\hspace{6mm}\mbox{(a)}\\
\vspace{3mm}
\includegraphics[width=0.8\hsize]{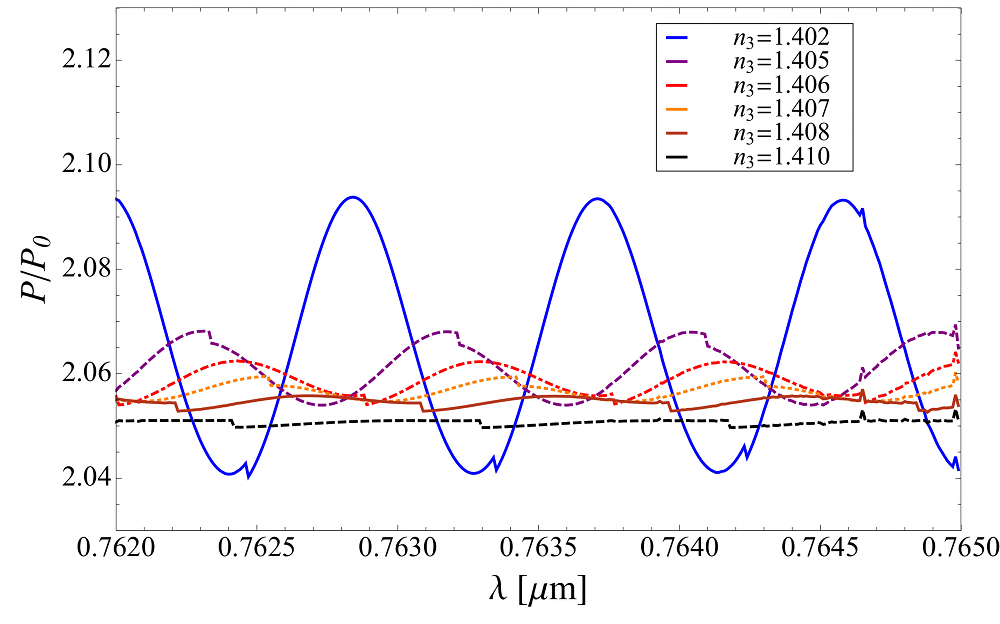}\\
\vspace{-5mm}
\hspace{6mm}\mbox{(b)}\\
\end{center}
\vspace{-5mm}\caption[Simulated examples of WGM spectra for bovine 
embryos.]
{Examples of a simulated WGM spectrum for a bovine embryo with $D=150$ 
$\mu$m, $\mathrm{n}_1 = 1.360$ (cytoplasm), $\mathrm{n}_2 = 1.435$ 
(\textit{zona pellucida}) and 
$\mathrm{n}_3=1.330$ (MOPS+BSA handling media), normalised 
to the surrounding medium. Modes are excited by a dipole source with both 
radial and tangential components placed on the surface of the resonator. 
Variation in the thickness of 
the outer layer has a negligible effect on the behaviour of the spectrum 
for values corresponding to those of typical bovine embryos. (a) This example 
corresponds to an FSR of $0.885$ nm, a $Q$-factor of $2.5\times10^3$ 
(excluding surface effects), 
and a sensitivity of $33.3$ nm/RIU. (b) 
The index $\mathrm{n}_3$ is changed over a broad range 
approaching that of glycerol, $1.467$, until modes can no longer be 
sustained. This upper value of $\mathrm{n}_3$ is $1.408$.}
\label{fig:tapersens}
\figrule
\end{figure}

Simulated spectra are shown in Fig.~\ref{fig:tapersens}, using 
a nominal diameter of $D=150$ $\mu$m, and cytoplasm and \textit{zona} 
refractive index values of $\mathrm{n}_1 = 1.36$ and $\mathrm{n}_2 = 1.435$, 
respectively, as an example. 
Since the cytoplasm is less compacted and more granulated than 
the \textit{zona} region \cite{Papi2010}, measurements reported in 
the literature suggest it has a lower 
refractive index than the \textit{zona} \cite{Wacogne:13}, and thus the lower 
value for the glycoprotein refractive index range from Chapter~\ref{chpt:cel} 
is chosen. The \textit{zona} region, on the other hand, is estimated as 
the mean value of the refractive index range, conservatively, 
$\mathrm{n}_2=(1.36+1.51)/2=1.435$. 
The surrounding medium is set to that of water 
$\mathrm{n}_3=1.330$, since the use of a 
specialised handling medium does not make a significant alteration to this 
value, and modes are excited by a dipole source with both radial and tangential 
components. The wavelength range is chosen to fit within the sweep range of the 
DFB laser, $762$ to $765$ nm. 
Note that each resonance in Fig.~\ref{fig:tapersens} comprises both TE 
and TM modes, which in this case are broad and overlapping, in the same manner 
as reported in Figs~\ref{fig:demo}(b) and \ref{fig:demo2}(b). 

It is found that the behaviour of the spectrum is 
extremely insensitive to the thickness of the \textit{zona pellucida} 
for $d \gtrsim 8$ $\mu$m. This effect has been noted earlier, in the 
discussion of Fig.~\ref{fig:d15d25comp} in Section~\ref{subsec:loss}. 
In this case, while the refractive index of the outer layer is less than that 
explored in Section~\ref{subsec:loss}, 
the diameter of the resonator is significantly larger, 
leading to a more tightly constrained evanescent field, as described in 
Section~\ref{sec:bio}. 
Radiation corresponding to WGMs does not penetrate the \textit{zona} region 
sufficiently to feel the effects of the cytoplasm. 
Thus, the outer diameter and the refractive index can now be obtained exactly 
from measurements of the FSR and the sensitivity. Note that this 
discovery crucially depends upon the 
multilayer model developed in Chapter~\ref{chpt:mod}, which is ideally 
placed for automating the non-destructive determination 
of the geometric parameters of a cell, developed in Section~\ref{sec:nondest}. 

The example in Fig.~\ref{fig:tapersens}(a) also provides guidance 
on how to estimate the 
approximate values of the FSR, the $Q$-factors and the sensitivity as follows. 
The FSR as measured across this wavelength range has a mean value of $0.885$ 
nm, consistent with the value estimated from Eq.~(\ref{eqn:FSR}): 
$0.860$ nm near the central peak. 

The $Q$-factors may also be extracted from a spectrum. 
Recall that the matching of simulated $Q$-factors with those of experiment 
is typically a difficult prospect, especially without the 
inclusion of surface roughness and asphericity 
\cite{Little:99,Talebi,Min:2009a,Vahala:2004}. 
While the multilayer model includes the radiation component to the $Q$-factor 
described in Section~\ref{sec:rec}, values of the $Q$-factor should be 
understood only as a guide to the order of magnitude expected. 
The sample spectrum of 
Fig.~\ref{fig:tapersens} yields a $Q$-factor of $2.5\times10^3$, 
which lies at the lower limit of detectability. 
This is due to the fact that 
$Q$-factors below $10^3$ are similar in magnitude to the FSR, and begin 
to overlap, thus making adequate detection of WGMs a challenge -- a point which 
is explored in more detail in Section~\ref{subsec:pass}. 

Experimentally, the addition of surface effects, such as those 
described under \textbf{Criteria~3} and \textbf{4} in Section~\ref{sec:selec}, 
are expected 
to reduce the $Q$-factors of the fundamental modes by more than an order of 
magnitude, reaching values as low as $10^2$ for RMS deformations in the 
diameter of $200$ nm, according to one study \cite{Rahachou:03}.  
Higher values of the \textit{zona} refractive index may be accessible, 
however, by 
carefully selecting embryos that have been fixed at a certain developmental 
stage. In the next section, bovine embryos in the presumptive zygote phase have 
been measured with an nominal roughness of $62$ nm \cite{Papi2010}. The 
introduction of osmotic pressure from the surrounding medium can also serve to 
increase the diameter of the embryos by a small amount, as explained with 
respect to \textbf{Criterion~5} in Section~\ref{subsec:surr}. 
As a result, a conservative estimate places the $Q$-factors measure in the 
laboratory in the order of $10^3$. 

By systematically altering the refractive index simulated for the surrounding 
medium, $\mathrm{n}_3$, in regular intervals, Fig.~\ref{fig:tapersens}(a) 
yields a sensitivity value of $\mathcal{S}=33.3$ nm/RIU. 
Having access to a 
prediction of the sensitivity can assist in determining whether the 
experimental shift is indicative of the presence of WGMs, and in 
assessing the likelihood of spectra features to be 
modes. 
In Section~\ref{subsec:ac}, the refractive index sensitivity is 
used specifically 
to identify whether modes are present. The surrounding index is increased 
with the addition of a droplet of glycerol, thus quenching any modes present
in the cell. 
Furthermore, in Section~\ref{sec:RI}, an experimental procedure is summarised 
whereby the refractive index of the local environment around the embryo is 
increased systematically using a more precise glycerol injection method. 
Thus, it is useful to determine the behaviour of the spectrum 
as the surrounding index approaches that of glycerol, $\mathrm{n}_3=1.467$, 
and at what value of $\mathrm{n}_3$ modes can no longer be sustained. 
Figure~\ref{fig:tapersens}(b) shows that a surrounding index of $1.408$ is 
sufficient to quench the modes present in the signal completely for 
resonators at $D=150$ $\mu$m. Note that in Figure~\ref{fig:tapersens} 
both plots exhibit small highly localised peaks within the main envelopes 
of the spectra, which are part of the underlying mode structure as simulated 
by the multilayer model. 

\begin{figure}[H]
\begin{center}
\includegraphics[width=0.8\hsize]{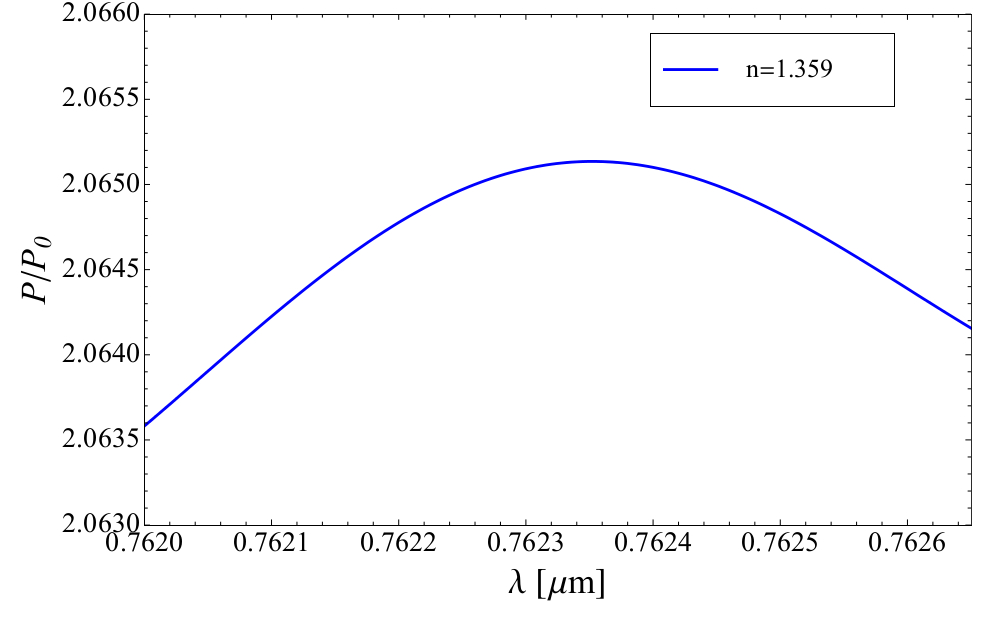}\\
\vspace{-5mm}
\hspace{12mm}\mbox{(a)}\\
\hspace{0mm}\includegraphics[width=0.8\hsize]{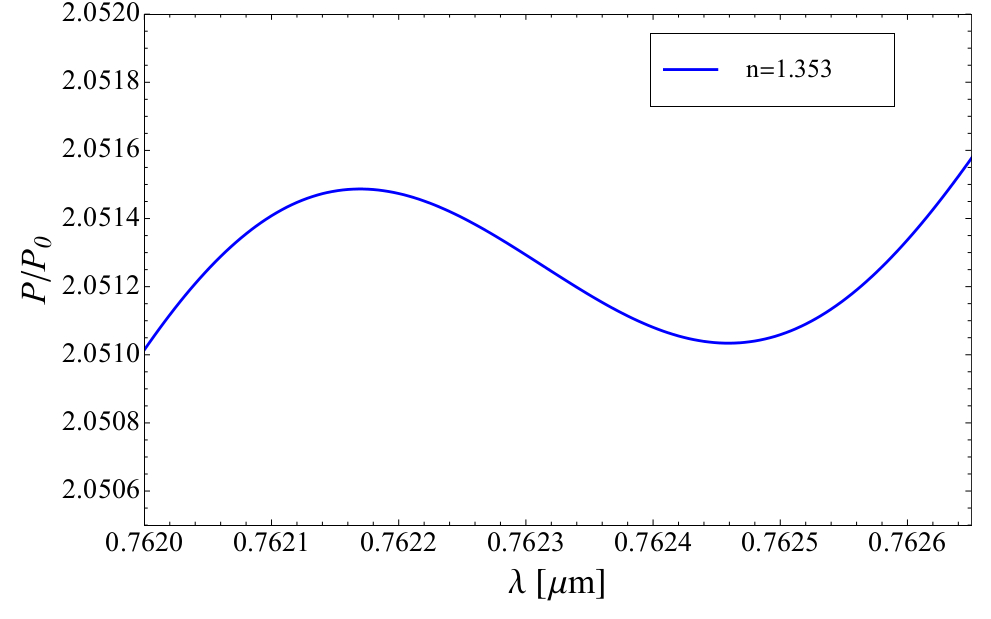}\\
\vspace{-5mm}
\hspace{12mm}\mbox{(b)}\\
\end{center}
\vspace{-5mm}\caption[The limiting \textit{zona} refractive indices 
for mode detection are explored for bovine embryos.]
{Spectra corresponding to the limiting \textit{zona} refractive index 
values for mode 
detection are estimated from the multilayer model with a surrounding 
medium of water. The lower and upper diameters measured from the bovine embryos 
of Section~\ref{subsec:st} are  
(a) $D = 124.4$ $\mu$m corresponding to a minimum index of 
$1.359$, and (b) $D = 168.2$ $\mu$m corresponding to a minimum index of $1.353$.
}
\label{fig:nlim}
\figrule
\end{figure}

Considering now a broad range of diameters, the limiting 
refractive index values of the outer layer, $\mathrm{n}_2$, 
may also be calculated, as a counterpart to the limiting diameters 
estimated in Fig.~\ref{fig:lim}. Using the range of diameters measured in 
Section~\ref{subsec:st}, $D = 146.3 \pm 21.9$ $\mu$m, the resulting 
signals simulated using the model are shown in Fig.~\ref{fig:nlim}.
Note that, as with Fig.~\ref{fig:lim}(a), the relatively small 
fluctuation in the emitter power is a consequence of the diameter and 
index set at the limit of detectability, however, the purpose here is 
to establish this extreme limit. 
For the smaller diameter bound, $D=124.4$ $\mu$m, the minimum refractive 
index for a resolvable mode in the multilayer model is $1.359$, 
while for the larger bound, $D=168.2$ $\mu$m, the minimum index is $1.353$. 
Note that minimum refractive index required to measure WGMs experimentally 
in an imperfect multilayer structure, such as an embryo, will be larger than 
these values obtained from simulation. 
It is clear that the behaviour of the resonance is more sensitive 
to changes in the refractive index near the boundary of the resonator 
than for changes in the diameter. Thus, the broad range of diameters expected 
for an embryo ($124.4$ to $168.2$ $\mu$m) result in a relatively 
constrained range of minimum required index values ($1.359$ to $1.353$), 
respectively. 

With these modelling predictions in mind, attention will now be turned to the 

\newpage
\noindent 
experimental results, beginning with an intermediary test case in which 
each of the selection criteria of Chapter~\ref{chpt:cel} can be examined, 
before continuing to the passive and active interrogation techniques as 
applied to embryos. 

\section{\cdb Experimental results}
\label{sec:exres}

The presentation of the experimental results follows the same order as 
Section~\ref{sec:meth}. First, the prism coupler method is used in the case of 
a silica glass microsphere that is known to support WGMs, and each of the 
selection criteria is examined. The successful tests in both water and 
handling medium 
narrow the challenges of producing modes within a biological cell to a few key 
possibilities. The microsphere is then replaced with a bovine embryo. Both the 
prism and taper coupling methods are applied in order to assess the scattering 
and absorption behaviour of the embryo.  Finally, the active interrogation 
methods are applied in order to explore effects that could not have been 
achieved with a tunable DFB laser alone. It is discovered that while passive 
interrogation leads to the measurement of modes close to the lower detection 
limit in the $Q$-factors, the inability of free space excitation of 
an active medium to excite modes along a preferred orientation renders 
the modes undetectable without further treatment of the surface effects. 
Therefore, a combination of multiple measurement techniques is required 
for active interrogation of embryos, such as the addition of a fibre taper 
to excite the active layer and collect the emitted radiation. 
A systematic study of the 
WGM peaks detected using the prism coupler method also leads to 
estimates of the geometric parameters of the 
embryo, including the refractive index of the \textit{zona pellucida},  
using the methodology developed in Section~\ref{sec:nondest}. 
These values can then be compared to the geometric proportions 
estimated from confocal microscopy in order to assess the accuracy of the 
WGM measurements.

\subsection{\cdb Test case: the silica microsphere}
\label{subsec:sil}

Recall that biological cells exhibit number of physical characteristics, each 
of which can have an impact on their ability to sustain WGMs. The relative 
effect of each of these characteristics on the resonance behaviour is 
non-trivial, and a thorough discussion and enumeration of these 
features can be found in Section~\ref{sec:selec}. As a way forward in removing 
the ambiguity of the competing effects of these physical characteristics, 
silica glass resonators are manufactured with diameters comparable 
to those of the bovine embryos. 
These resonators are known in advance to support WGMs, and thus the effect of 
a range of selection criteria can be tested one by one. 

Silica glass ($\mathrm{n}=1.46$ for $\lambda\approx 700$ nm) 
is chosen as the material that is closest in refractive index to the 
suggested values in the literature 
for glycoproteins of the type that comprise the embryo 
\textit{zona pellucida}, 
and are manufactured to diameters within the most conservative specifications 
of embryo diameter, $D=110$ $\mu$m 
(\textbf{Criterion~1}). 
Each microresonator is formed as an elongated spheroid to provide scope in 
testing the sphericity (\textbf{Criterion~2}), 
by altering the contact angle at the interface of the prism coupler or taper. 
The surface properties of the resonator (\textbf{Criteria~3} and \textbf{4}) 
represent 
the richest source of scientific investigation in this topic, 
particularly due to their intimate connection with the composition and chemical 
properties of the surrounding media (\textbf{Criterion~5}), 
which can affect the physical attributes of the \textit{zona} region in 
embryos. 
This will be investigated by exploring the behaviour of different media, 
particularly with respect to their salt content, 
and the effect of crystallisation on the generation of WGMs. 
The spot size and phase-matching condition 
of the coupling method (\textbf{Criterion~6}) 
is also refined in this test scenario, and the coupling to WGMs in 
imperfect resonators represents a crucial step in 
addressing the vision of the project 
as outlined in the Prologue. Finally, the scattering and absorption profiles 
of the silica microspheres and bovine embryos 
are directly compared (\textbf{Criterion~7}), and insights are drawn with 
regard to the outstanding challenges in the realisation of a biological 
resonator. 

\begin{figure}
\begin{center}
\begin{minipage}{0.49\linewidth}
\includegraphics[width=0.49\hsize]{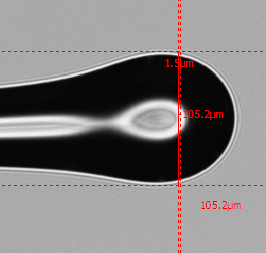}
\includegraphics[width=0.49\hsize]{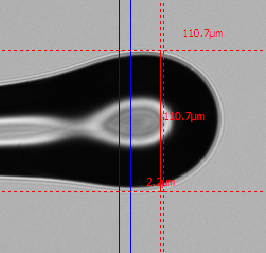}\\
\includegraphics[width=0.49\hsize]{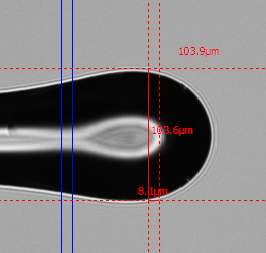}
\includegraphics[width=0.49\hsize]{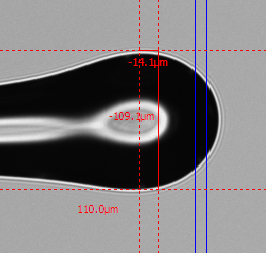}
\end{minipage}
\begin{minipage}{0.49\linewidth}
\includegraphics[width=0.99\hsize]{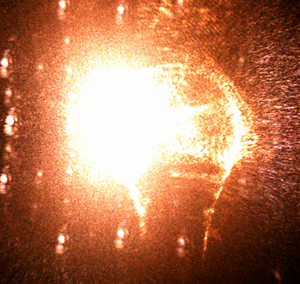}
\end{minipage}\\
\vspace{2mm}
\hspace{0cm} (a)\hspace{6.5cm} (b)
\end{center}
\vspace{-5mm}\caption[Silica glass microspheres are fabricated and tested.]
{Silica glass microspheres are fabricated such that they exhibit similar 
outer diameters to the embryos. 
They are constructed for the purpose 
of immersing in the embryo handling medium to test its effect on the WGM 
spectra. (a) Four similar-sized 
elongated microspheres are compared, with diameters of $105$, $111$, $104$ and 
$109$ $\mu$m, respectively. (b) 
Scattered light from a test silica microsphere exhibiting WGMs during 
experiment.}
\label{fig:sanalog}
\figrule
\end{figure}

\begin{figure}
\begin{center}
\includegraphics[width=0.8\hsize]{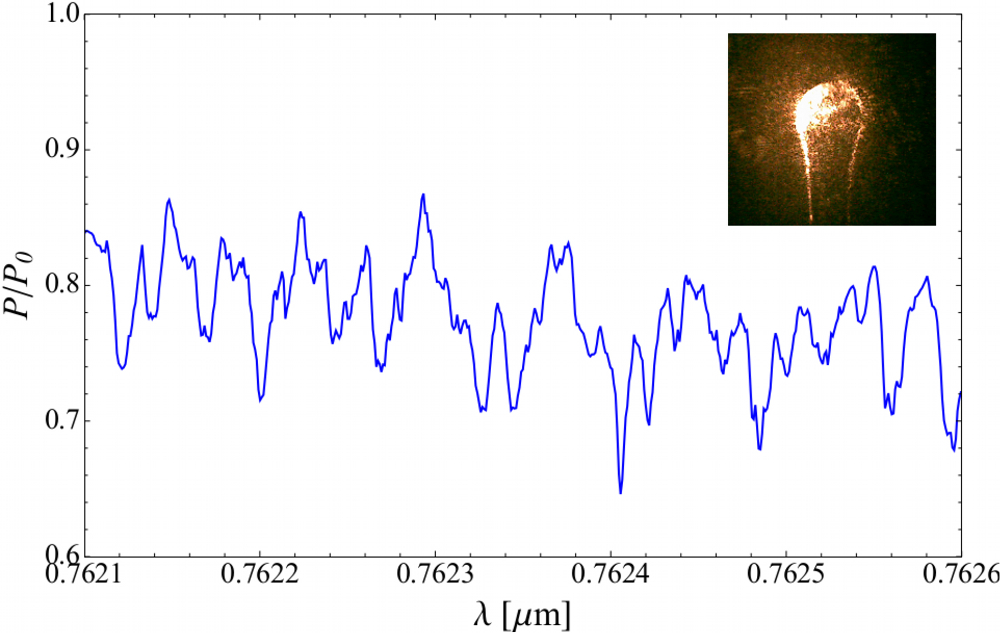}
\end{center}
\vspace{-5mm}\caption[A WGM signal obtained from the transmitted light from 
a silica glass microsphere
using the prism coupler method.]{A signal containing WGMs 
is obtained from a silica 
glass microsphere ($D=111$ $\mu$m), with a coupling efficiency of 
approximately $30$\%. 
The transmitted light is collected using the prism coupler method, with a 
surrounding medium of air. Note that the results 
have been clarified by removing noise. 
\emph{Inset:} an image 
of the scattered radiation of the microsphere.} 
\label{fig:prisspherecoarse}
\figrule
\end{figure}

Images of the microspheres are shown in Fig.~\ref{fig:sanalog}(a), 
comparing the diameters of each. 
An image of the scattered radiation from a silica resonator supporting WGMs 
excited using the prism coupler method 
is shown in Fig.~\ref{fig:sanalog}(b). 
Selecting the microsphere with a diameter of $111$ $\mu$m, the signal  
obtained from the scattered light 
is shown in Fig.~\ref{fig:prisspherecoarse}, 
scanning over a wavelength range of $762.1$ to $762.6$ nm, or roughly 
half the anticipated FSR. The surrounding medium, in this case, is air, 
thus removing any possible contaminating effects introduced 
by embryo handling media (the effect of the presence of media will be 
studied presently). In this case, the coupling efficiency is 
approximately $30$\%. 
The coupling condition is coarsely adjusted, and 
it is found that an angle of $30.84^\circ$ leads to 
$Q$-factors of $7.86\times10^4$. 
Further refinement of the angle leads
to an experimental upper limit of $Q \sim 10^{5-6}$ 
 \cite{Gorodetsky:99,NOTO20074466}. 
The polariser, shown in Fig.~\ref{fig:prissetup}, filters 
any erroneous polarisations emanating from the laser so that the incident 
beam couples selectively to certain modes only. 
Note that the results 
have been clarified by removing part of the fluctuation noise 
in the same manner as in Fig.~\ref{fig:fit} of Chapter~\ref{chpt:bub}. 

\begin{figure}
\begin{center}
\includegraphics[width=0.8\hsize]{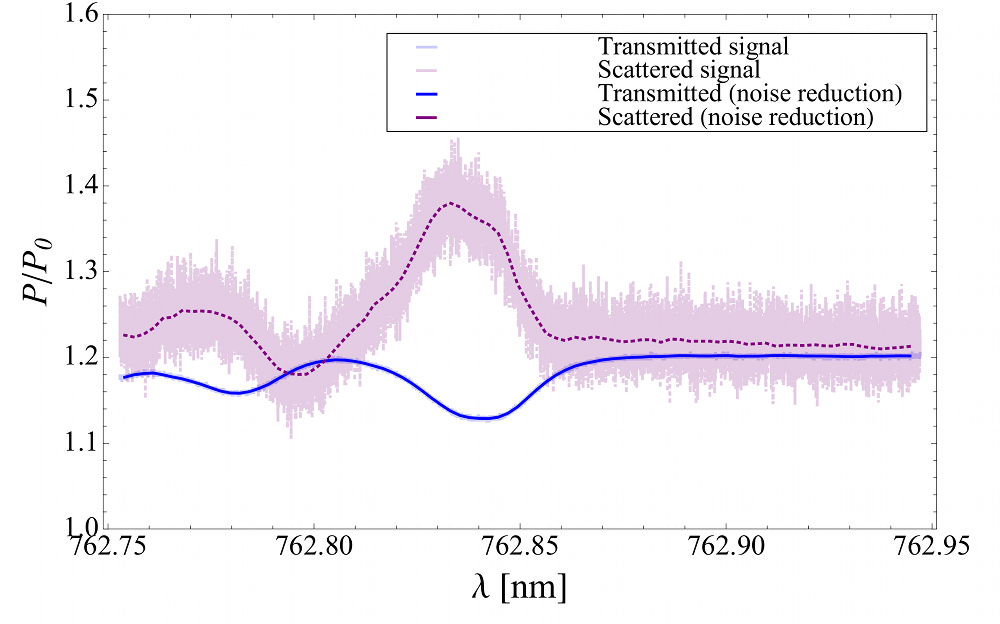}\\
\vspace{-3mm}
\hspace{2mm}\mbox{(a)}\\
\includegraphics[width=0.8\hsize]{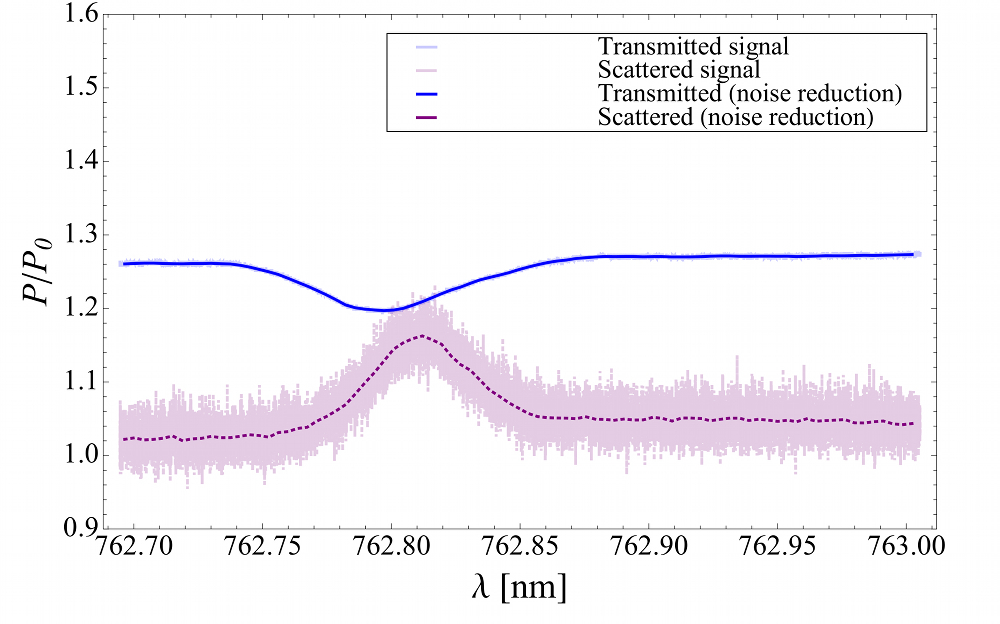}\\
\vspace{-3mm}
\hspace{2mm}\mbox{(b)}\\
\end{center}
\vspace{-5mm}\caption[Zoomed-in signal for a silica microsphere 
in water.] {Zoomed-in signal for a silica microsphere 
($D=111$ $\mu$m), normalised to the surrounding medium of water 
(a small power offset is introduced to move the signals away from zero). 
The wavelength 
range of the DFB laser is centred on $762.85$ nm, and the scattered radiation 
is collected with a $\times 40$ magnification objective. 
(a) An angle of incidence $\theta_i=27.38^\circ$ is chosen -- close to 
the critical angle for the fundamental modes of a silica 
microsphere at $\lambda=760$ nm, 
indicated in Table~\ref{tab:RIprism}. The $Q$-factor of the prominent mode in 
the scattered spectrum is $2.20\times10^4$. (b) The angle of incidence is then 
adjusted to $21.69^\circ$ in order to probe a different mode. 
The $Q$-factor is $1.74\times10^4$, reflecting the fact that the angle has 
been deviated from the critical angle. 
The results have been clarified by removing noise. 
}
\label{fig:prisspherewater}
\figrule
\end{figure}

The silica microsphere is now immersed in Milli-Q$^{\text{\textregistered}}$ 
water, and the experiment described above is repeated. 
A critical study of the WGM behaviour is then conducted by examining zoomed-in 
plots of a few example modes, shown in Fig.~\ref{fig:prisspherewater}.\footnote
{The detection of WGMs in this experiment was confirmed independently 
by Dr. Wenle Weng, The University of Adelaide.}
Note that the presence of WGMs is indicated definitively by the fact that each 
peak in the scattered spectrum corresponds exactly to a dip in the transmitted 
spectrum at the same wavelength. That is, as an increase in the power is 
measured from the scattered light, there is a 
corresponding reduction in the transmitted light owing to the presence of a 
resonator. Since these fluctuations are confined within a narrow wavelength 
region, they indicate the resonant behaviour associated with WGMs, as described 
in Section~\ref{sec:nut}. Note that the noise in the scattered signal 
is much larger than for the transmitted signal since less scattered 
light reaches the detector. To remove part of the noise, the wavelength 
resolution has been smoothed in the same way as in Fig.~\ref{fig:fit}. 

In Fig.~\ref{fig:prisspherewater}(a), the angle of incidence, $\theta_i$ as 
shown in Fig.~\ref{fig:pris}(a), is chosen to be $27.38^\circ$, near 
the critical angle for coupling to the fundamental modes of a silica 
microsphere, with a  
wavelength range scanned by the DFB laser of $200$ pm, centred on 
$762.85$ nm. 
By altering the angle of incidence, different modes may be selected using the 
prism coupler. In Fig.~\ref{fig:prisspherewater}(b), an angle of $21.69^\circ$ 
is selected, which allows the incoming beam to couple to modes that extend 
deeper within the resonator. The $Q$-factor of the 
mode centred at $762.81$ nm in the scattered spectrum is $1.74\times10^4$, 
reflecting the fact that the angle has been deviated 
from the critical angle.  
The mode selectivity of the prism coupler method gives rise to a high 
sensitivity of the coupling condition on the angle of incidence, requiring 
extremely precise determination of both the refractive index and angle in 
order to select a given mode.

\subsubsection{\cdb Osmolaric contamination and surface properties}

Recalling \textbf{Criterion~5} from Section~\ref{sec:selec}, 
the total concentration of soluble particles in the medium surrounding 
the resonator and coupler apparatus can affect the detectability of WGMs. 
In order to determine whether the surrounding environment used 
in the case of real-life embryos has an adverse effect on the clear 
measurement of modes, the microsphere is immersed in several different forms 
of media. 

In the first case, a droplet of PBS media, defined in Appendix~\ref{chpt:app4}, 
is placed over the resonator. Because of the high salt content of this type of 
medium, no WGMs are observed across a range of angles of incidence. 
Furthermore, after the removal of the droplet, and the subsequent 
decontamination of the prism surface, the silica resonator is replaced at the 
interface. Once again, no WGMs are detected, regardless of the 
incoming beam angle. The reason for this mode loss is that residual 
salt from any evaporated PBS media, which is heated by the radiation introduced 
from the DFB laser,  
crystallises and encrusts the resonator, affecting the 
surface properties, and drastically reducing the $Q$-factors of the modes. 
This is a crucial point in the investigation into the limiting 
factors that apply to biological resonators, 
offering guidance with the direction of scientific inquiry that represents the 
next key challenge to be addressed on this topic. While the surface properties 
undoubtedly have a profound impact on the ability of a prospective resonator to 
sustain WGMs, any methods that can assist 
in the smoothness, reduction of contamination, or removal of extraneous 
detritus can significantly improve its resonance properties. Techniques for 
smoothing or annealing the outer surface of the \textit{zona pellucida} 
of an embryo are discussed in more detail in Chapter~\ref{chpt:fdi}.

The next case to be considered for the surrounding environment of the 
microsphere is an equal-parts mix of MOPS+BSA handling medium (see 
Appendix~\ref{chpt:app4}) and Milli-Q$^{\text{\textregistered}}$ water. The 
experiment is then repeated. For this medium, a much lower salt 
content leads to improved resonance properties. 
The reduced osmotic pressure serves 
to increase the outer diameter of the embryo by a small fraction, leading to an 
improved $Q$-factor, as described under \textbf{Criterion~1} in 
Chapter~\ref{chpt:cel}.
The dilution of the handling medium will prove useful in the next 
section for the case of real-life embryos.  

\begin{figure}[H]
\begin{center}
\includegraphics[width=0.8\hsize]{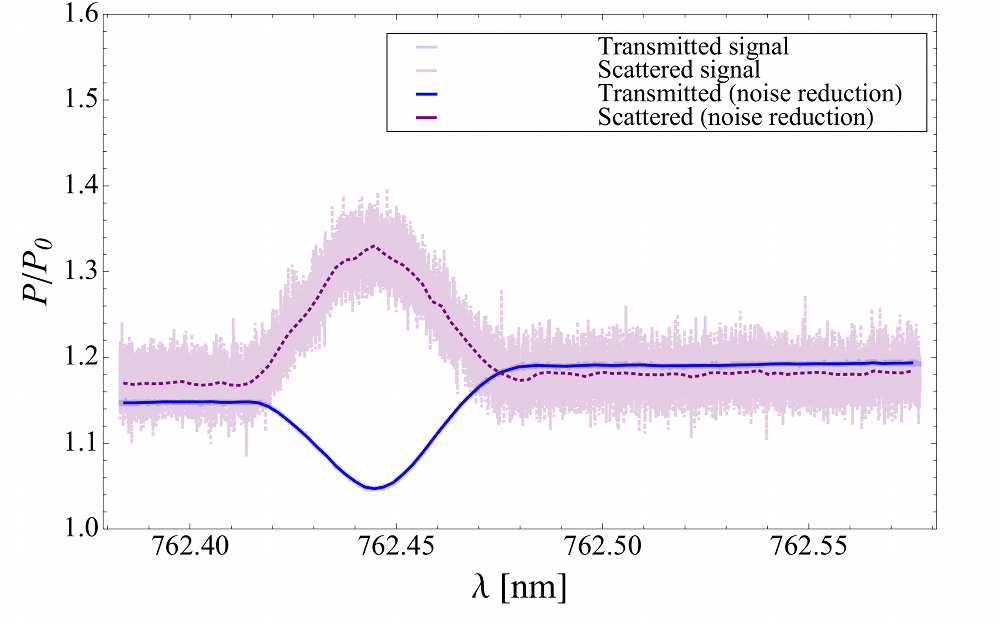}\\
\vspace{-5mm}
\hspace{4mm}\mbox{(a)}\\
\includegraphics[width=0.8\hsize]{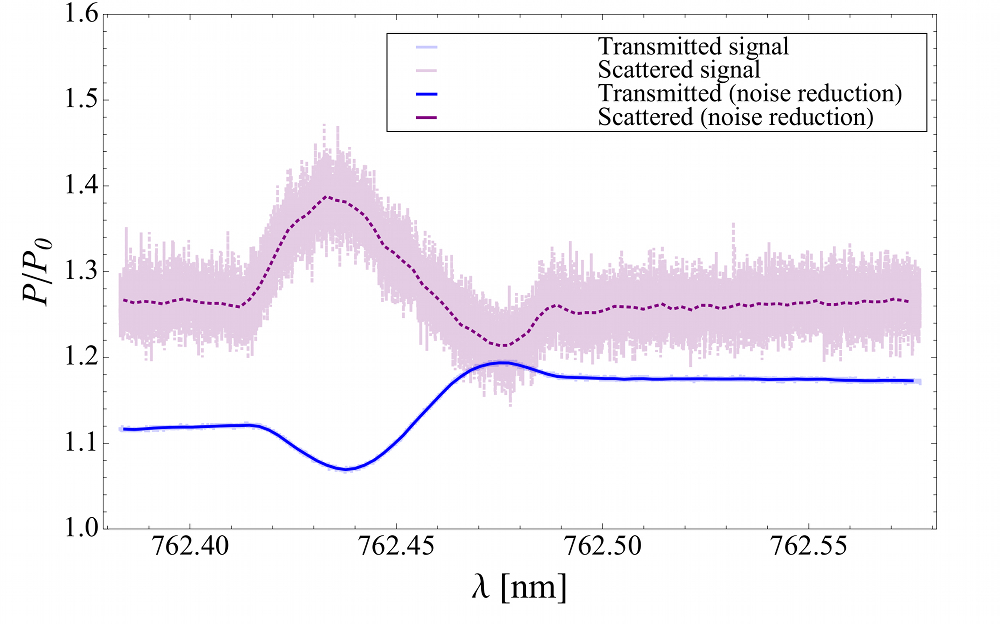}\\
\vspace{-5mm}
\hspace{4mm}\mbox{(b)}\\
\end{center}
\vspace{-5mm}\caption[Signal for a microsphere in an equal MOPS+BSA/water 
solution.]{Zoomed-in signal 
for a silica microsphere ($D=111$ $\mu$m) placed 
in an equal-part mixture of Milli-Q$^{\text{\textregistered}}$ water and MOPS+BSA 
handling medium. 
(a) The polariser of Fig.~\ref{fig:prissetup} is set to couple only to a single 
polarisation. (b) By adjusting the orientation of the polariser, different 
relative mode couplings may be selected. A secondary mode near a 
wavelength of $762.475$ nm 
is present. A small power offset is introduced to move the spectrum away from 
zero. The results have been smoothed to remove residual noise.}
\label{fig:prisspheremed}
\figrule
\end{figure}

WGM spectra measured in this scenario are shown in 
Fig.~\ref{fig:prisspheremed}. 
One key feature of these spectra is that very little degradation of the 
$Q$-factor is apparent. In Fig.~\ref{fig:prisspheremed}(a), the polariser 
shown in Fig.~\ref{fig:prissetup} is orientated so that a single polarisation 
is selected, as well as suppressing competing modes. In 
Fig.~\ref{fig:prisspheremed}(b), an alternative orientation of the polariser is 
selected, revealing a less prominent mode near the wavelength of $762.475$ nm. 
This test serves to illustrate an important point -- the selection of modes 
using the prism coupler method is greatly aided by the ability to remove 
competing modes of different polarisations, which are capable of overlapping, 
increasing the effective value of the $Q$-factor, and thus making the detection 
of clear modes more difficult. 
With these powerful tools in mind, an investigation of the resonance properties 
of embryos is now conducted.

\subsection{\cdb Passive interrogation of modes in embryos}
\label{subsec:pass}

\subsubsection{\cdb Prism coupler results}

Using the setup illustrated in Fig.~\ref{fig:prissetup}, 
fixed, denuded embryos are prepared and mounted in the same manner as the 
silica microsphere, making particular use of the suction holding-pipette to 
guide the cell precisely into position, as explained in Section~\ref{sec:samp}. 
Each embryo is placed in a small droplet of MOPS+BSA/water mixture 

\begin{figure}
\begin{center}
\includegraphics[height=0.45\hsize]{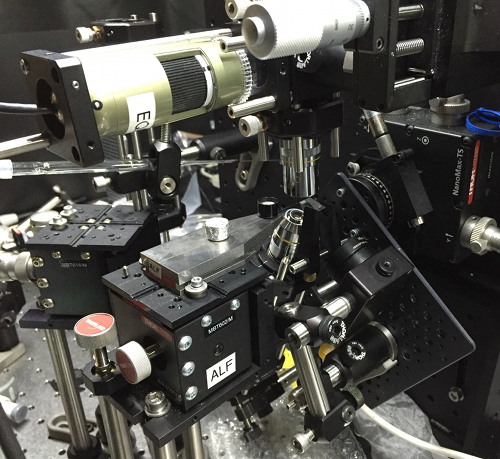}
\includegraphics[height=0.45\hsize]{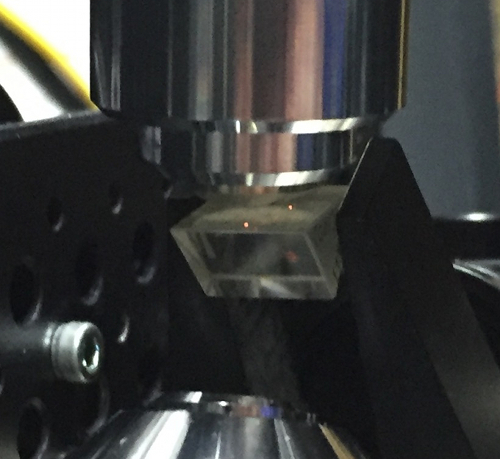}\\
\hspace{0cm} (a)\hspace{6.5cm} (b)
\end{center}
\vspace{-5mm}\caption[The prism coupler is altered to accommodate mounting 
embryos.]{The prism coupler setup is configured to accommodate the mounting of 
embryos within a droplet of medium/water solution. 
(a) A zoomed-out view showing 
the suction holding-pipette (\textit{centre left}), a $10$ mm length rutile 
prism with focusing and scattered light collection objectives, a $v$-groove 
mount for a microsphere at the end of a fibre tip (\textit{centre}), 
a detector for the transmitted light (\textit{centre right}), and a camera 
(\textit{centre top}). (b) A $\times 10$ magnification objective 
(\textit{bottom}) focuses light onto the top surface of the prism, causing a 
mounted embryo to glow (\textit{centre}). A $\times 40$ magnification objective 
collects the scattered radiation (\textit{top}), while the transmitted 
light exits the reverse side of the prism, collected by the detector behind.
}
\label{fig:priscell}
\figrule
\end{figure}

\newpage
\noindent at the edge 
of the rutile prism to ensure the minimum spot size is achieved. 
Figure~\ref{fig:priscell}(a) shows a zoomed-out view of the prism setup, which 
includes the suction holding-pipette for positioning the embryo within the 
droplet of the solution so that it exactly aligns with the laser spot on the 
surface of the prism. In Fig.~\ref{fig:priscell}(b), the glow of the embryo is 
apparent, indicating that real-life embryos do indeed scatter a portion of the 
incident radiation at approximately $760$ nm, and that not all the light is 
absorbed. This addresses \textbf{Criterion~7}, in that the relative 
magnitude of the absorption and scattering behaviour of embryos can 
potentially be measured using this technique. 

\begin{figure}
\begin{center}
\includegraphics[height=4.83cm]{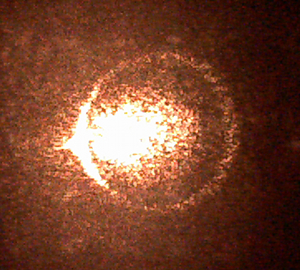}
\includegraphics[height=4.83cm]{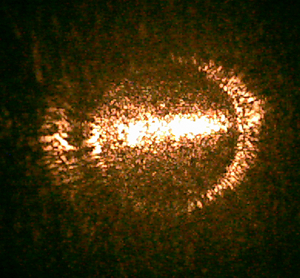}
\includegraphics[height=4.83cm]{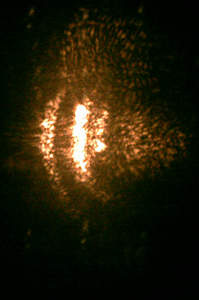}\\
\includegraphics[height=10.5cm]{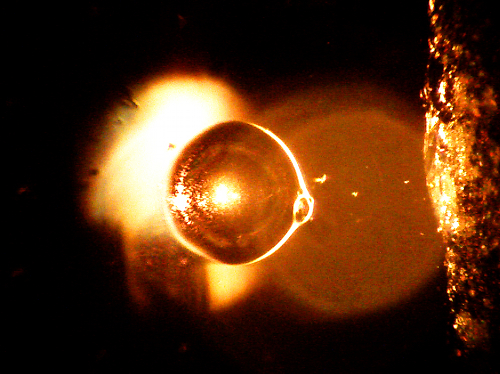}
\end{center}
\vspace{-5mm}\caption[Images of an embryo trapping laser light using the prism 
coupler.]{Top-down camera images of an embryo within a droplet of medium/water 
solution, mounted on the surface of a rutile prism. 
The images show a corona around the surface of the 
embryo. A zoomed-in inspection of the embryo 
indicates that the scattered light is 
emanating from the \textit{zona pellucida} (\textit{top 
images}). The structural components of the embryo, including the 
\textit{zona pellucida} and cytoplasm regions are clearly visible 
 (\textit{bottom image}).}
\label{fig:prispic}
\figrule
\end{figure}

A visual inspection of the embryo via the mounted camera
 is shown in Fig.~\ref{fig:prispic}, where the light 
appears as a ring or corona around the embryo. 
While this observation does not of itself indicate the presence of WGMs, 
nor that the radiation is entirely 
constrained by the \textit{zona pellucida} region, 
it is useful for distinguishing the 
individual structural components of the embryo.  
The \textit{zona pellucida} can be seen as a faint ring 
around the central globe of the cytoplasm. By continuously deviating the laser 
spot position, a reflection from the inside of the top  
boundary of the \textit{zona} is seen moving in the opposite 
direction. The droplet of solution is indicated by the sharp boundary 
(identified by the small air bubble on the right hand side). 
By comparing the effect with a silica microsphere 
resonator in the same solution, the measured scattering effect does not 
originate from residual salt within the media. 

\begin{figure}
\begin{center}
\includegraphics[width=0.85\hsize]{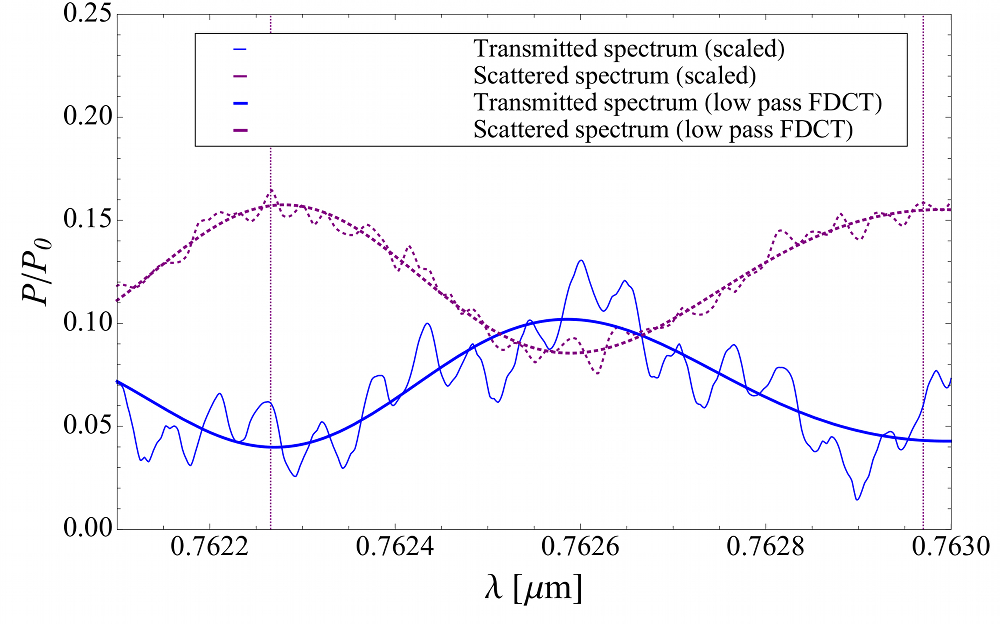}
\end{center}
\vspace{-5mm}\caption[Scattered and transmitted WGM spectra from an embryo.]
{WGM spectra for both scattered and transmitted radiation from an embryo 
(\textit{thin} lines), 
corresponding to an incidence of $\theta_i=19.95^\circ$, and a wavelength range 
of $762$ to $763$ nm. The high frequency noise oscillations in the spectra 
are then removed using a low-pass Fourier DCT (\textit{thick} lines). 
Vertical dotted \color{purple} \textit{purple} \cbl lines 
indicate the positions of the broad composite peaks measured from the 
experiment, which correspond to an FSR of $0.704$ nm. The measured quantities 
$\theta_i$ and FSR can be used to estimate the refractive index of the 
\textit{zona pellucida}, and the diameter, respectively, leading to values of 
$\mathrm{n}_Z = 1.532$, and $D=171.5$ $\mu$m. Note that the transmitted signal 
is two orders of magnitude greater than the scattered signal, and thus the 
spectra have been scaled for comparison of the mode positions.}
\label{fig:prisembryo}
\figrule
\end{figure}

An investigation of the scattered and transmitted spectra may now be 
carried out, as shown in Fig.~\ref{fig:prisembryo}. 
The angle of incidence used to isolate the clearest signal is $19.95^\circ$, 
with a DFB laser sweeping a range of wavelengths from $762$ to $763$ nm. The 
polariser is adjusted in order to clarify any underlying modes. 
Note that the magnitude of the collected 
transmitted power is two orders of magnitude greater than the scattered signal. 
The spectra have therefore been scale-adjusted in order to compare 
the positions of the broad envelopes in the spectra. 
While both signals contain high frequency noise, 
which appears as oscillations (particularly in the transmitted spectrum),  
this can be treated by resampling the data, 
applying a low-pass Fourier discrete cosine transform (DCT). 
It is clear from 
Fig.~\ref{fig:prisembryo} that there is a broad peak in the scattered signal, 
with a corresponding dip in the transmitted signal. This observation, taken 
together with the photographic information, is strongly indicative of the 
presence of modes. 
In fact, this is the first observation of WGMs sustained within an embryo. 
A range of other tests will now be carried out before a definitive answer is 
established.  

The spectrum may be used to estimate the physical parameters of the 
embryo. If the estimates lie within the bounds of the physical properties 
of the embryos, as established in Section~\ref{sec:oo}, this lends 
credence to the supposed generation of WGMs in cells. 
The angle of incidence for which the best set of spectra was achieved may be 
used to obtain a value for the refractive index of the material in which the 
radiation propagates, the \textit{zona pellucida}. Using 
Eq.~(\ref{eq:prisang}), a value of $\mathrm{n}_Z = 1.532$ is estimated. While 
this value is slightly larger than the upper estimate considered in 
Chapter~\ref{chpt:cel}, it is worth noting that Ref.~\cite{Voros2004553} 
indicates that certain protein structures may exhibit refractive indices as 
high as $1.55$. In addition, the value of $\mathrm{n}_Z$ may be 
increased relative to unfertilised embryos. This can occur due to 
the greater density of networked glycoproteins in the 
\textit{zona pellucida} immediately after fertilisation.  
Such an effect is present 
in the presumptive zygotes chosen for these experiments. This phase of 
embryo development also leads to a degree of \textit{zona} hardening 
\cite{Papi2010}, in conjunction with the increase in stiffness contributed by 
the fixing procedure of Appendix~\ref{sec:pfa}. As a result of these 
factors affecting the surface properties of the embryo, the measured 
$Q$-factor from the spectrum is $2.54\times10^3$. %Q:2.5407E3 
This value agrees well with the modelling predictions of 
Section~\ref{sec:modpred} that the $Q$-factor is expected to be of the 
order of $10^3$, including the effects derived from the physical 
parameters, described in the selection criteria of Section~\ref{sec:selec}. 

The spectra in Fig.~\ref{fig:prisembryo} may also be used to estimate the FSR, 
in the same manner reported in Section~\ref{sec:case}. The approximate peak 
positions in 
the scattered spectrum are shown as vertical dotted lines. Measuring an FSR of 
$0.704$ nm, an approximate value for the outer diameter may be determined using 
Eq.~(\ref{eqn:FSR}), as $D=171.5$ $\mu$m. 
Note that the FSR must be equal to or larger than the width of the 
envelope in the spectra to satisfy the Rayleigh criterion 
\cite{doi:10.1080/14786447908639684}. A full-width at 
half-maximum value less than a third of the FSR indicates that the 
envelopes are well-spaced within the spectrum. 
While this outer diameter lies within the normal range of diameters for bovine 
oocytes as reported in Ref.~\cite{Bo:13}, $150$ to $190$ $\mu$m 
with a variation of $23.5$\%, it is larger than the mean value of the sample 
of embryos used in the experiment, $146.3 \pm 21.9$ $\mu$m, by $17\%$ -- 
outside of one standard deviation. 
According to \textbf{Criterion~1}, a larger diameter embryo is more 
likely to exhibit modes than those with smaller diameters, for 
a fixed refractive index, $\mathrm{n}_Z$. Therefore, it is reasonable 
that a large number of experimental measurements should be required 
in order to obtain an embryo with a diameter this size. 
This, in part, explains why the generation of modes within embryos 
is not readily apparent, requiring multiple experimental techniques 
to answer the research question outlined in Section~\ref{sec:form}. 

The larger diameter can also be explained by the fact that a diluted 
mixture of handling medium is used to minimise any detrimental 
effects on the acquisition of the spectrum caused by residual salts or stray 
detritus in the solution. 
It has been reported that a decrease in oocyte volume of up to $40$\% can 
occur during hypo-osmolaric processes that occur naturally post-ovulation 
\cite{doi:10.1093/humupd/dmp045}, 
while the linear Boyle van't Hoff relationship 
between cell volume and osmolarity \cite{Mullen2007281} indicates 
that the radii of bovine embryos can be deformed up to 
$50$\% when placed under osmolar stress \cite{doi:10.1093/molehr/gam027}. 
Thus, the reduction in the osmolarity, 
\textbf{Criterion 5}, applies a uniform negative pressure on the \textit{zona} 
wall causing significant expansion of the surface of the embryo.

\subsubsection{\cdb Taper results}

\begin{figure}
\begin{center}
\includegraphics[height=0.45\hsize]{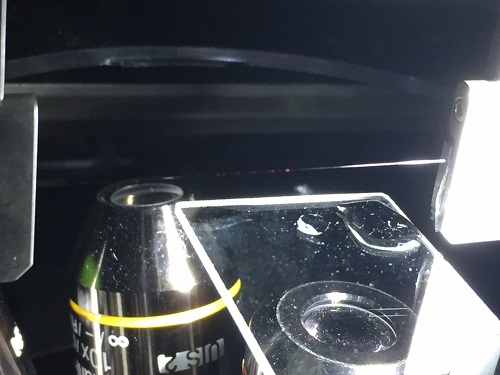}
\includegraphics[height=0.45\hsize]{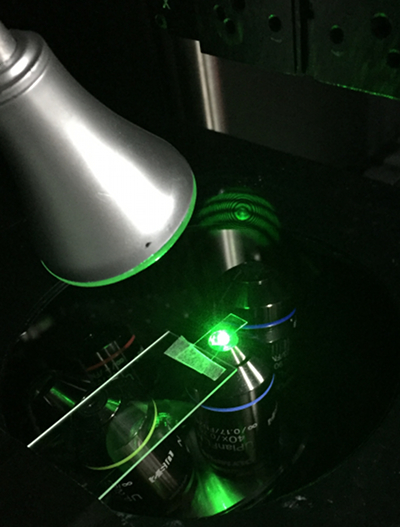}
\end{center}
\vspace{-5mm}\caption[Image of a fibre taper suspended above an embryo in 
solution.]
{(a) A fibre taper (\textit{centre}) suspended between the metal struts of a 
fork-shaped brace (\textit{centre left/right}) may be lowered onto an embryo, 
placed on a glass coverslip (\textit{centre/right bottom}). The droplet of 
solution containing the embryo must be placed over the lowered fibre taper to 
prevent breakage due to surface tension. The taper position may then be 
adjusted in order to optimise the coupling condition. (b) The taper 
method can be used in combination with free space coupling. 
The narrow sliver-shaped glass coverslip reduces the risk of abrasion 
of the taper when lowered in place.}
\label{fig:tapercell}
\figrule
\end{figure}

The taper coupler method potentially offers some advantages over the use of a 
prism, particularly the fact that it is able to constrain the orientation of 
the modes excited to within a single plane, leading to improved $Q$-factors and 
coupling efficiencies \cite{Riesen:15a,Knight:97,Dong:08}. 
An image of the experimental setup immediately prior to exposure of the taper 
to an embryo is shown in Fig.~\ref{fig:tapercell}(a). 
The additional space afforded by the use of a broad coverslip, rather than a 
restrictively-sized prism, allows a number of experimental configurations to be 
incorporated, such as the refractive index sensitivity measurement apparatus, 
discussed in Chapter~\ref{chpt:fdi}. 
However, the fabrication of a fibre taper, described in 
Section~\ref{subsec:fib}, 
is a precarious process, and the fragility of the taper leads to a number of 
drawbacks. First, a new taper must be manufactured immediately prior 
to each experiment. While fixed embryos need simply to be kept in a handling 
medium, 
the use of living cells in future experiments may present timing difficulties 
in carrying out measurements, since their survival rates drop severely after 
$30$ minutes without proper incubation. Second, the waist diameter required 
of a 
taper to achieve coupling to WGMs in resonators 
of the order of $150$ $\mu$m in diameter is approximately $3$ to $5$ $\mu$m 
\cite{PhysRevLett.85.74}. One study found that waist diameters of 
$1$ $\mu$m were required to obtain 
suitable coupling efficiency to smaller diameter ($15$ $\mu$m) 
high-index polystyrene microspheres \cite{Riesen:15a}. While 
this is feasible through the use of a Vytran, the large extent of the 
evanescent field within a droplet of solution exposes 
the taper to contamination from a range of biological debris, dust, and small 
particles that are difficult to filter out entirely. This effect can severely 
compromise the transmission of the taper while adding non-specific background, 
making the detection of a WGM signal difficult \cite{s150408968}. 

During the setup procedure, it is important to ensure that the taper is 
lowered close to the glass coverslip prior to the introduction of the droplet 
of solution containing the embryo. This prevents the fragile 
taper from experiencing upward pressure from the surface tension of the 
droplet, which may cause physical damage to the taper. 
A droplet of media of approximately $20$ $\mu$L is then added to the coverslip. 
Particularly in the case of 
larger volume pipettes, one must ensure that the nozzle velocity is 
sufficiently low to avoid breakage of the taper. 
Abrasion of the taper against the glass coverslip can be reduced by 
using a customised narrow sliver-shaped glass coverslip 
that does not extend beyond 
the tapered region, shown in Fig.~\ref{fig:tapercell}(b). 

\begin{figure}
\begin{center}
\includegraphics[height=0.35\hsize]{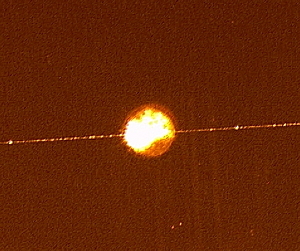}
\includegraphics[height=0.35\hsize]{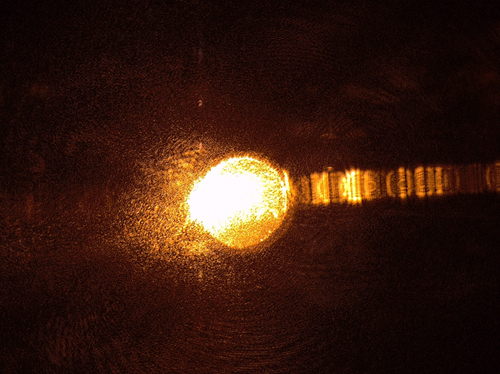}\\
\hspace{-0.5cm} (a)\hspace{5.7cm} (b)
\end{center}
\vspace{-5mm}\caption[Images of an embryo exposed to a fibre taper.]{
Images of an embryo inside a droplet of media/water solution, placed on a 
glass coverslip, and placed in the vicinity of a fibre taper. (a) The taper 
appears as a thin line placed on top of the embryo. The embryo scatters light, 
affecting the transmission of the taper. WGMs should appear as dips in the 
transmitted spectrum. (b) Focusing on the embryo instead of the taper, the 
scattered light forms a corona, similar to that of the prism 
coupler method; however, a clear WGM signal is not detectable in this case.}
\label{fig:tapermeas}
\figrule
\end{figure}

Images of the taper coupling method applied to an embryo are shown in 
Fig.~\ref{fig:tapermeas}. While there appears to be a glow from the light 
scattered by the embryo, it is unclear whether this effect 
is due to the resonance behaviour observed using the prism coupler method. 
In order to investigate whether any underlying modes have been detected, the 
transmitted spectrum is measured several times, each time detaching the taper 
from the embryo and replacing it so that the evanescent 
field is in the vicinity of the \textit{zona pellucida}. 

\begin{figure}
\begin{center}
\includegraphics[width=0.8\hsize]{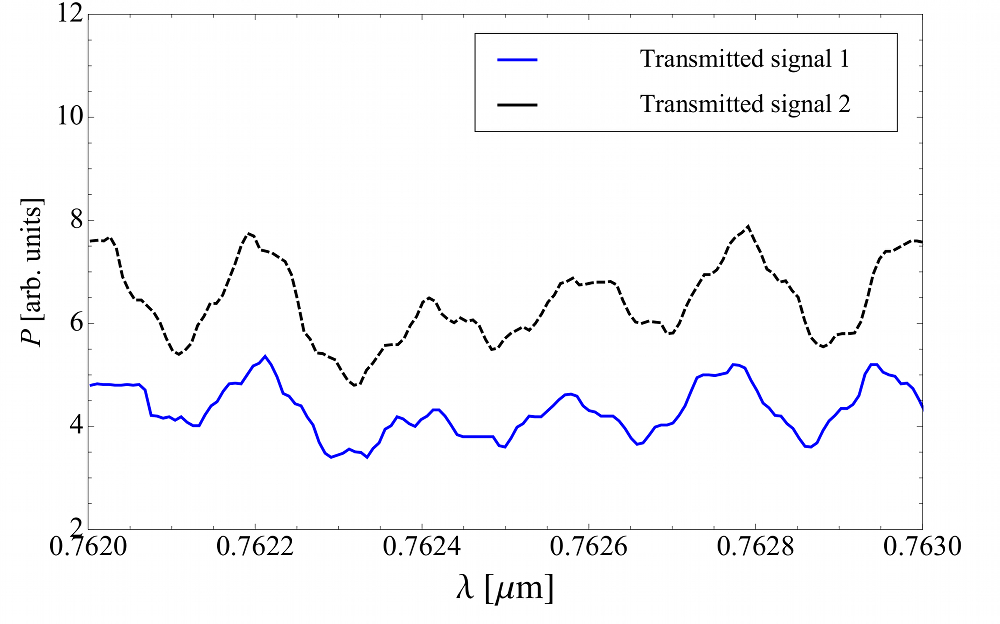}
\end{center}
\vspace{-5mm}\caption[Transmitted signals from an embryo using the fibre taper 
method.]{
Examples of transmitted power signals obtained from the taper coupling 
method. The signals show a similar pattern, while the embryo was detached and 
reattached in between each measurement. This indicates that the fluctuations 
are not dominated by WGMs, but high levels of noise. 
Such an effect is due to the small waist 
diameter of approximately $1$ $\mu$m required for phase-matching, 
which experiences loss through the evanescent field, and 
sensitivity to contaminants in the surrounding medium.}
\label{fig:taperdat}
\figrule
\end{figure}

In carrying out the measurement in conjunction with the taper, 
the high level of noise makes identification of modes inconclusive. 
Examples of signals collected using passive taper excitation 
are shown in Fig.~\ref{fig:taperdat}. 
The high noise within these signals 
 is a consequence of the small diameter of the taper waist 
required for the phase-matching 
condition. Small waist diameters lead to a large evanescent field, 
resulting in high levels of loss, as well as increasing the 
sensitivity of the taper to contaminants in the surrounding medium, 
as described in Sections~\ref{subsec:surr} and \ref{subsec:fib}.  

\subsection{\cdb Active interrogation of modes in embryos}
\label{subsec:ac}

\subsubsection{\cdb Quantum dot results -- combining interrogation methods}

Results for the quantum dot-coated embryos are now discussed, 
with a view to providing independent corroborating evidence for the generation 
of WGMs within embryos. 
While the introduction of either amino or carboxyl group 
functionalised dots directly into the handling media of the 
embryos does not result in significant emitter density on the 
surface of the \textit{zona} of each embryo, the use 
of crosslinked polyelectrolyte layers, described in Section~\ref{subsec:poly}, 
is more successful. 

\begin{figure}
\begin{center}
\includegraphics[height=0.4\hsize]{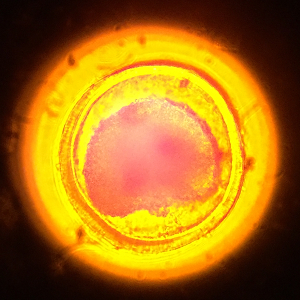}
\includegraphics[height=0.4\hsize]{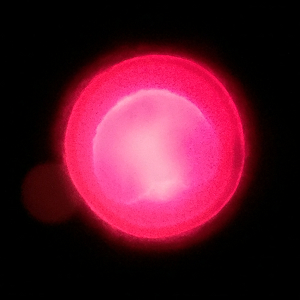}\\
\hspace{0cm} (a)\hspace{5.3cm} (b)
\end{center}
\vspace{-5mm}\caption[Quantum dot coated bovine embryo using 
polyelectrolyte cross-linking.]
{Bovine embryo with a quantum dot coating 
attached using the polyelectrolyte layering and cross-linking process 
described in Section~\ref{subsec:poly}. (a) The \textit{zona pellucida} 
and cytoplasm regions are visible under a while light source, 
with the emission from the cytoplasm excited 
using the $532$ nm CW laser clearly visible. 
(b) Removing the white light source, the emission from the quantum 
dots appears well distributed across the surface of the embryo.}
\label{fig:cellglow}
\figrule
\end{figure}

Following the procedure outlined in Section~\ref{subsec:poly}, 
amino group quantum dot coated embryos are prepared and excited 
with a $532$ nm wavelength CW laser. Images of the fluorescence 
emission from a sample embryo are shown in 
Fig.~\ref{fig:cellglow}. The \textit{zona} and cytoplasm regions 
are clearly visible under a white light source in Fig.~\ref{fig:cellglow}(a). 
When the white source is removed, in Fig.~\ref{fig:cellglow}(b), it is 
evident that the emitter density is well distributed over the surface 
of the embryo. 

The interrogation of the coated embryo can proceed in a number of ways. 
Free space excitation and collection is the simplest technique from a 
logistics standpoint, requiring no additional apparatus beyond 
that shown in the confocal microscope setup of Fig.~\ref{fig:exp}. 
However, the inability to select modes along a specific orientation 
limits the $Q$-factor and renders any underlying modes 
difficult to detect, as described in Section~\ref{sec:exres}. 
To mitigate this effect, a combination of techniques is employed 
to maximise the likelihood of measuring WGMs in the embryo. 
Specifically, a fibre taper may be used to collect the emitted radiation along 
a specific axis of revolution. 
The practical limitations of using fibre tapers for the collection 
of WGM signals explored in Section~\ref{subsec:pass} are ameliorated in this 
case, since the taper waist does not need to be perfectly phase matched with 
the WGMs within the resonator. In this case, tapers 
with diameters of $2$ to $3$ $\mu$m are employed in order to 
excite the quantum dots and/or collect the fluorescent radiation. 

\begin{figure}[H]
\begin{center}
\includegraphics[width=0.73\hsize]{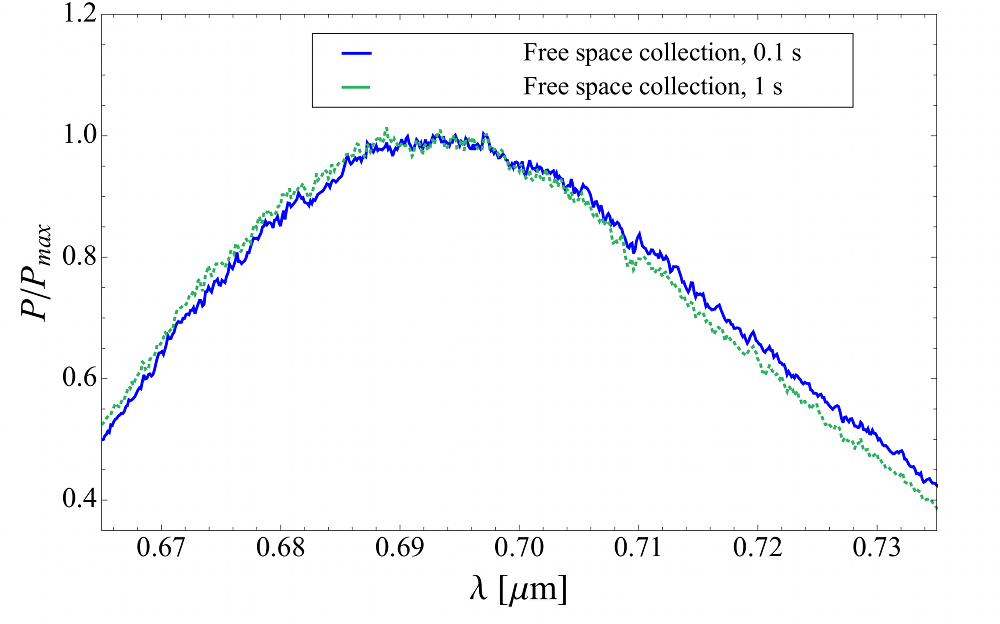}\\
\vspace{-5mm}
\hspace{4mm}\mbox{(a)}\\
\vspace{3mm}
\includegraphics[width=0.73\hsize]{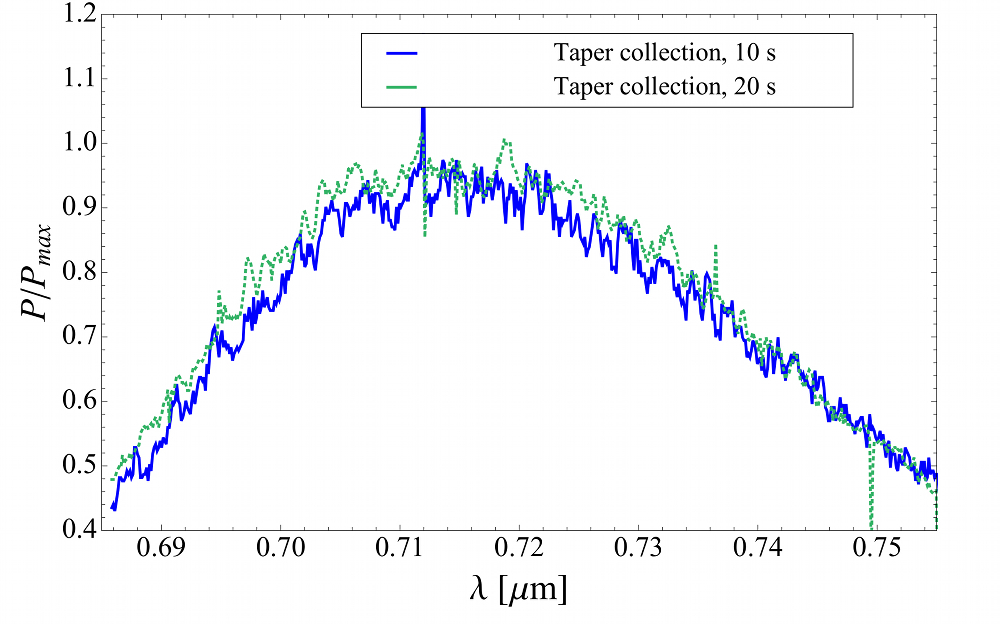}\\
\vspace{-5mm}
\hspace{4mm}\mbox{(b)}\\
\end{center}
\vspace{-9mm}\caption[Comparison of free space and taper collection 
for coated embryos.]{
Signals obtained 
 from the quantum dot coated embryo shown in 
Fig.~\ref{fig:cellglow}, prepared using the 
procedure explained in Section~\ref{subsec:poly}. 
The broad emission of the quantum dots allows a wider range of wavelengths to 
be probed. Several acquisition times are used in collecting the signal. 
(a) A $532$ nm CW laser is used to excite 
the quantum dots with free space collection of the emitted signal. 
(b) A taper coupler is used to collect the emitted signal. 
While a non-specific background is recorded due to the difficulty in 
phase-matching the taper to any WGMs within the embryo, there is 
a modulation of the envelope, which could indicate a WGM signal. 
Note that due to the reduced throughput of the taper,
the acquisition times for the taper collection are greater than those 
for free space collection. 
}
\label{fig:qdotcomb}
\figrule
\end{figure}

\begin{figure}[H]
\begin{center}
\includegraphics[width=0.73\hsize]{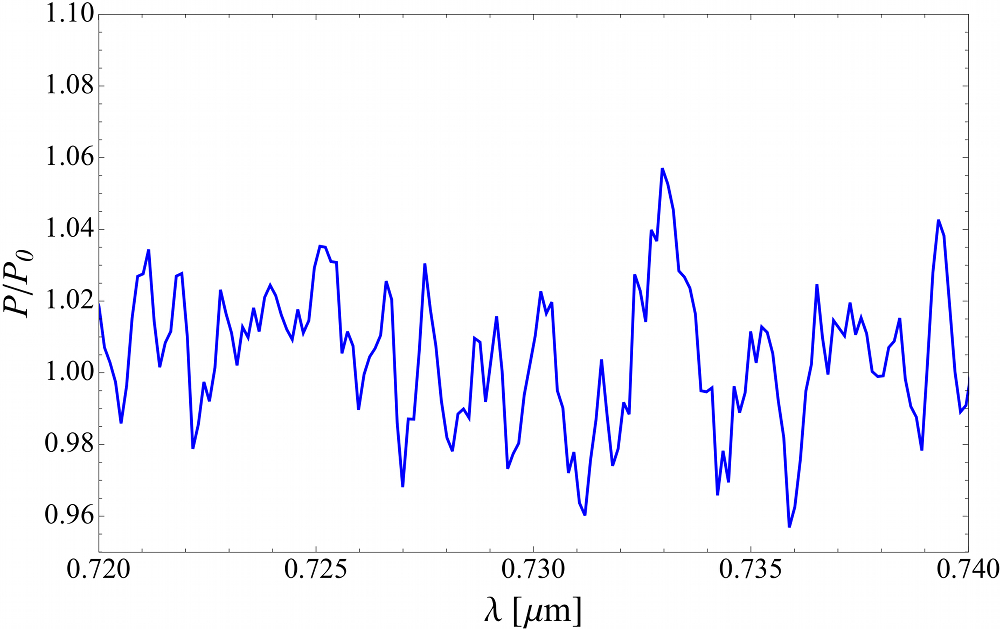}\\
\vspace{-3mm}
\hspace{2mm}\mbox{(a)}\\
\vspace{3mm}
\includegraphics[width=0.73\hsize]{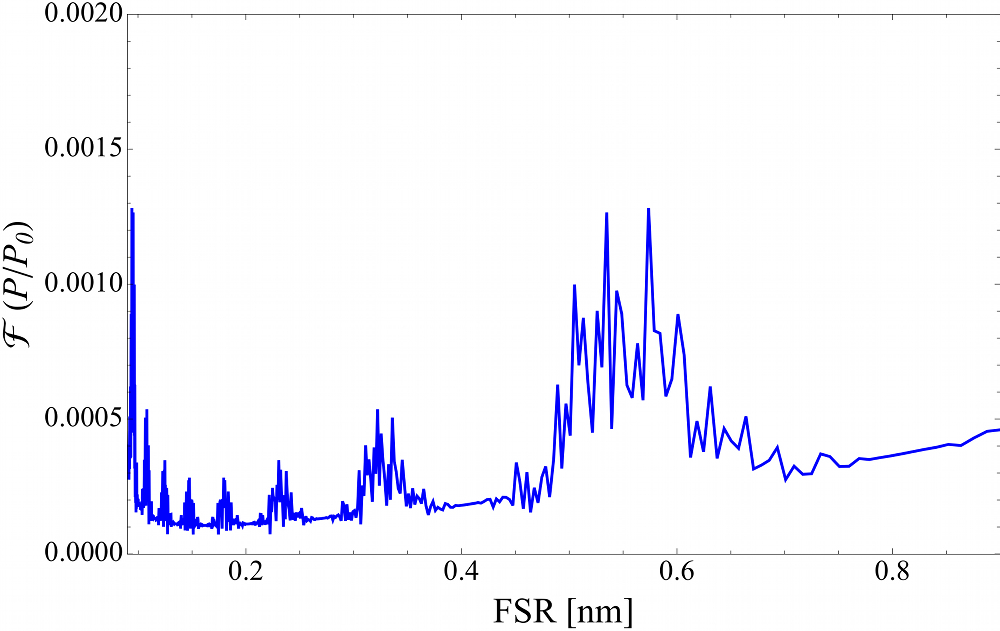}\\
\vspace{-3mm}
\hspace{2.5mm}\mbox{(b)}\\
\end{center}
\vspace{-9mm}\caption[Background subtraction and signal processing for 
coated embryos.]{
Signal analysis of the scattered light collected from the quantum 
dot coated embryo is carried out. 
(a) In order to reveal the underlying modulation of the scattered signal 
obtained from the quantum dot coated embryo, the residual background 
contribution from the quantum dots alone, $P_0$, is removed, 
resulting in a signal that represents 
the contribution purely from the presence of the embryo. 
(b) Fourier decomposition of the signals reveal a peak corresponding 
to an FSR of $0.55$ nm.
}
\label{fig:qdotsub}
\figrule
\end{figure}

A comparison of the signals obtained using free space or taper collection 
from the quantum dot coated embryo of 
Fig.~\ref{fig:cellglow} is shown in Fig.~\ref{fig:qdotcomb}. 
The broad emission of the quantum dots allows 
a wider range of wavelengths to be investigated compared to that 
of the passive interrogation results. 
The results are normalised to the maximum value of the power, $P_{\text{max}}$. 
In Fig.~\ref{fig:qdotcomb}(a), it is initially 
unclear whether the envelope,  corresponding to the scattered signal from 
the quantum dots, contains WGMs. 
By using the taper collection method, a portion of the background 
signal can be removed. The result is shown in Fig.~\ref{fig:qdotcomb}(b), 
noting that the plot has been re-normalised with respect 
to the new value for the maximum power recorded. 
The signal now exhibits a modulation atop the scattered signal, 
which may indicate the presence of modes. However, in order to clarify 
the result, a number of signal processing and experimentation techniques 
are used. 
First, the contribution to the signal from the quantum 
dots, in the absence of a structure, should be removed. While the 
signals shown in Fig.~\ref{fig:qdotcomb} have already been normalised to 
$P_{\text{max}}$, the background signal, $P_0$, must also divide the power 
spectrum to isolate the portion of the signal derived purely from the presence 
of the embryo. 
The result of this treatment of the taper collection results 
is shown in Fig.~\ref{fig:qdotsub}(a). 
Note that, for clarity, the wavelength axis has been scaled to 
$0.72$ to $0.74$ $\mu$m, in which the anticipated FSR of an embryo 
from the prism experiment, $0.704$ nm, can be clearly resolved when present. 
While this signal shows 
modulation of the main envelope, it is 
unclear whether there are embedded modes. 
A Fourier decomposition can determine whether 
a specific repeating pattern embedded in the data dominates. 
The result of this treatment of the taper collected signal with 
$20$ s acquisition time is displayed in Fig.~\ref{fig:qdotsub}(b). 

\newpage
The dominant peak in this Fourier spectrum is around $0.55$ nm, 
corresponding to a diameter of approximately $200$ $\mu$m, 
through use of Eq.~(\ref{eqn:FSR}). While this value lies outside the 
range of embryo diameters determined 
from Ref.~\cite{Bo:13}, and obtained from the 
embryos studied in Section~\ref{subsec:pass}, the peak is quite broad, 
indicating that a repeating pattern corresponding to WGMs 
is not easily determined. This may be a result of the effect 
noted in Sections~\ref{sec:spec}, \ref{sec:case}, and particularly in 
Section~\ref{sec:demo}, where nearby overlapping modes form composite 
peaks that alter the structure of the spectrum. 

\begin{figure}[H]
\begin{center}
\hspace{1mm}\includegraphics[width=0.78\hsize]
{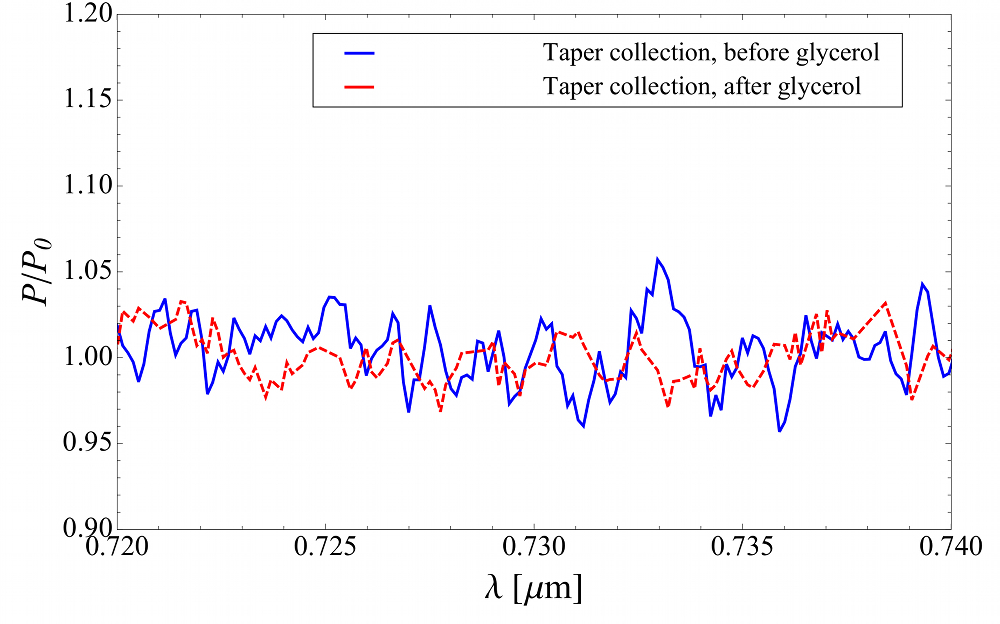}\\
\vspace{-4mm}
\hspace{4mm}\mbox{(a)}\\
\vspace{3mm}
\includegraphics[width=0.73\hsize]{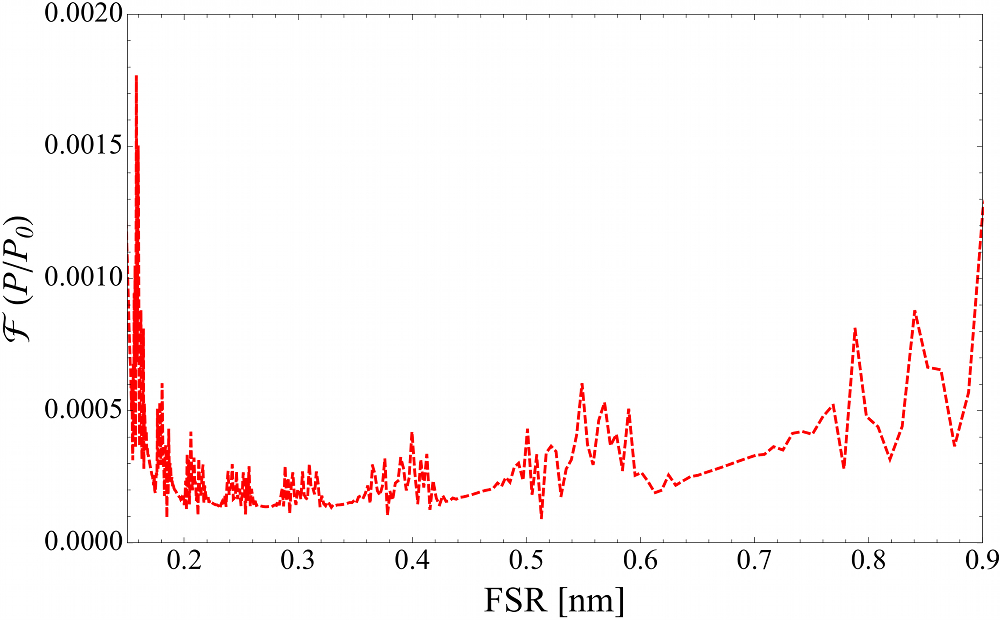}\\
\vspace{-3mm}
\hspace{4mm}\mbox{(b)}\\
\end{center}
\vspace{-9mm}\caption[Quenching of WGMs in an embryo by adding a droplet 
of glycerol.]{
Signals obtained from the quantum dot coated embryo shown in 
Fig.~\ref{fig:cellglow}. 
(a) The signal collected from a fibre taper is compared to the result obtained 
from adding droplet of $50$\% glycerol 
solution to the media, thus quenching all modes. The modulation 
within the scattered signal is subsequently reduced. 
(b) Fourier decomposition of scattered signal after the 
addition of the glycerol droplet. There is a reduction in the peak 
at an FSR of $0.55$ nm. 
}
\label{fig:qdotsig}
\figrule
\end{figure}

Next, the presence of modes in the signal shown in Fig.~\ref{fig:qdotsub}(a) 
can be tested experimentally by altering the refractive index of the 
surrounding medium. This is achieved by adding a droplet of glycerol solution 
to the surrounding medium of the embryo. Since glycerol is both soluble 
and has a higher refractive index ($1.467$) than that of water 
and handling media ($1.330$), any WGMs present inside the embryo 
will become quenched by the decrease in the refractive index contrast, 
as simulated in Section~\ref{sec:modpred}. 
A comparison of the normalised 
signals collected with a $20$ second exposure time, 
averaged over $20$ acquisitions, is shown before and after the 
addition of a $50$\% glycerol solution in Fig.~\ref{fig:qdotsig}(a). 
This concentration of glycerol solution is sufficient to 
achieve the critical surrounding index beyond which no WGMs 
can be sustained. 
While there is a marginal decrease in the clarity of mode structure 
in the signal, the reduction is nevertheless fairly subtle. 
An analysis of the Fourier spectrum after the addition of the glycerol 
solution, displayed in Fig.~\ref{fig:qdotsig}(b), shows that the 
main peak near the FSR value of $0.55$ nm is much less than that 
of Fig.~\ref{fig:qdotsub}(b). This indicates that the repeating 
pattern embedded within the signal is less prominent after glycerol 
is added. 

While the analysis of the results obtained through active interrogation 
of embryos is not as definitive as those obtained from the prism method, 
a number of other methods may be utilised in order to improve the 
distinctiveness of the measured modes. Examples such as the use of 
chemical or enzyme-based annealing of the \textit{zona pellucida} 
are anticipated to reduce the detrimental effect of surface roughness 
on the $Q$-factors of the modes, particularly the fundamental modes 
\cite{Rahachou:03}. 
These 

\newpage 
\noindent future directions are outlined more comprehensively in 
Chapter~\ref{chpt:fdi}. 

\vspace{-2mm}
\section{\cdb Conclusions from the experiments}
\label{sec:fin}

In this chapter, the measurement of WGMs within bovine embryos 
was sought, and a positive result was recorded. 

Using passive mode interrogation, modes present within the 
scattered and transmitted signals led to an estimate of the 
\textit{zona pellucida} refractive index of $\mathrm{n}_Z = 1.532$, 
an FSR of $0.704$ nm, and a diameter of $171.5$ $\mu$m.  
While the value of the diameter was consistent 
with the values expected from the literature 
\cite{Bo:13}, it was larger than the mean oocyte diameters measured 
prior to dilution of the handling medium, $146.31 \pm 0.22$ $\mu$m. 
While the dilution of the handling medium explains the 
increase in cell volume, due to the Boyle van't Hoff relationship 
between cell volume and osmolarity \cite{Mullen2007281}, a large 
number of trial experiments were required to obtain an embryo 
of optimal size and sphercity.  
The $Q$-factor of $2.54\times10^3$ was also consistent with the 
expected values from both modelling and experiment. 

In the context of the detection of WGMs in porcine adipocyte cells 
reported in the literature \cite{Humar2015}, a critical evaluation  
is necessary. 
A careful investigation of the generation of WGMs in a range of cells 
considered in this thesis, under ideal conditions, demonstrates that 
the challenge inherent in the detection of a clear signal. Successful 
detection is highly dependent on the interrogation method used, and the 
chemical preparation and handling of the cells prior to experimentation. 

\begin{table*}[t]
\begin{mdframed}[backgroundcolor=boxcol,hidealllines=true]
{
\begin{center}
\begin{mdframed}[backgroundcolor=boxcold,hidealllines=true]
\textbf{\color{white} \large Summary of evidence for WGM identification 
\cbl}\vspace{1mm}
\end{mdframed}
\end{center}
\cdb
\vspace{3mm}
\textbf{a)} Using the prism coupler, 
the scattered signal indicated the presence of a peak  
coinciding with a dip measured in the transmitted signal. 
This behaviour is expected from the quantisation condition of WGMs, 
introduced in Section~\ref{sec:nut}.
\vspace{2mm}\\ 
\textbf{b)} The $Q$-factor was measured as $2.54\times10^3$, 
which is consistent with the prediction of $10^3$ 
obtained from the modelling results and selection criteria.\vspace{2mm}\\ 
\textbf{c)} The FSR was measured as $0.704$ nm, and the 
refractive index as $\mathrm{n}_Z = 1.532$, obtained from the angle of 
incidence. Thus the embryo diameter was estimated as $D=171.5$ $\mu$m, 
which is consistent with the diameter values recorded in Ref~\cite{Bo:13}. 
\vspace{2mm}\\ 
\textbf{d)} Using a fluorescent coating of quantum dots with 
collection of the scattered signal using a fibre taper, a modulation 
of the main envelope indicated the presence of modes. \vspace{2mm}\\
\textbf{e)} A Fourier 
decomposition of the signal revealed a peak corresponding to an 
FSR of $0.55$, and an estimated diameter of $200$ $\mu$m; however, 
this lies just outside the bounds determined from Ref~\cite{Bo:13} 
and Section~\ref{subsec:pass}. \vspace{2mm}\\
\textbf{f)} The addition of a droplet of glycerol solution reduces 
the refractive index contrast of the embryo with its surrounding medium. 
A comparison of the Fourier spectra before and after the addition of 
glycerol shows a slight reduction in the dominant peak. 
\cbl 
\vspace{0mm}
}
\end{mdframed}
\end{table*}

In active interrogation of embryos, a fibre taper was required 
for collection of the scattered light  
in order to resolve the modulation of the background signal, 
potentially corresponding to WGMs. 
The presence of a background signal is a result of the imperfect 
phase-matching of the embryo with the taper waist, 
with diameters of $2$ to $3$ $\mu$m. 
To investigate the presence of WGMs, a range of signal processing 
and experimental 
techniques were employed. First, the signal was normalised to the 
residual background envelope. This served to isolate the contribution from the 
presence of the embryo. Second, a Fourier decomposition of the 
signal revealed a peak corresponding to an FSR of $0.55$ nm, 
corresponding to embryos of diameter $200$ $\mu$m -- a value that 
lies outside the bounds determined from the literature, 
and from observation of the embryos used for this thesis. 
However, the addition of a droplet of glycerol solution, reducing 
the refractive index contrast of the embryo with its surrounding medium, 
led to the reduction of the dominant peak at $0.55$ nm in the Fourier 
spectrum, indicating the quenching of modes. 

The results derived from the model, the experiments and the analysis 
integrate to form a prediction that WGMs can be generated within 
a biological cell, and this effect was observed within the bovine embryo. 
The development of embryos as viable biological resonators can 
advance in a range of directions from this point onwards. 
In the next chapter, key techniques are explained, which are able 
to address and refine the limiting factors, including the surface properties, 
and a number of innovative directions are proposed for future 
research endeavours in the field. 

\chapter{\cdb Future Directions: Towards a Biolaser Sensor}
\label{chpt:fdi}

The investigation of biological resonators represents the chief aim of this 
thesis, along with an understanding 
of the physical characteristics that underpin the ability of 
biological cells to exhibit WGMs. 

Refining the cell preparation methods, and introducing innovative 
techniques to facilitate the detection of WGMs, 
would serve to pave the way towards reliable cell resonator-based 
technologies in the future, including the ability to identify 
and track the behaviour of WGMs in the context of external macromolecules, 
or changes to the refractive index contrast of a cell. 
Such technologies are particularly attractive 
because resonances that are sustained by the cell itself are 
extremely sensitive to its internal features, as well as the surrounding 
environment. This means that, in principle, further development 
of WGM detection within a cell could provide vital information on the internal 
structure or status of the cell, as a complementary approach 
to the other label-free detection technologies 
\cite{doi:10.1167/iovs.03-0628,Freudiger1857,ELPS:ELPS200406070,
4014191,Kita:11,doi:10.1021/ac0513459,BIOT:BIOT200800316} mentioned in 
Chapter~\ref{chpt:intro}. 
Examples of this include the ability to detect macromolecules in the 
vicinity of the cell; the cell can be introduced to known biomolecules of 
interest, such as viruses or proteins, and the characteristic shift in the 
WGM positions due to changes in the surrounding refractive index can be 
assessed \cite{Vollmer:08}. 

In the specific case of embryo resonators, such innovations can have broad 
implications on health sciences, where the ability to avoid immune 
responses in an organism by using biological sensing modalities presents 
an attractive feature for technological development in the near future. 

The following sections include a few examples of the methods 
that can be used to improve the performance of biological 
cell resonators, as well as summarising key directions of future 
research that may facilitate the construction of a fully-fledged 
cell-based modality for sensing applications.

\section{\cdb Glycerol-based index sensitivity measurement}
\label{sec:RI}

The measurement of the refractive index sensitivity of a biological 
cell resonator represents  the next step in being able to identify 
modes and characterise the WGM shift in the presence of 
an external agent affecting the surrounding medium. 
While predictions of the sensitivity of an embryo 
from the multilayer model of Chapter~\ref{chpt:mod} 
are presented in Section~\ref{sec:modpred}, 
a comparable measurement of the sensitivity can be performed 
using the following method. 

A fluid with a higher refractive index than the medium surrounding the embryo, 
such as glycerol, can be circulated around the embryo during WGM measurement. 
The resultant change in the surrounding index, $\mathrm{n}_3$, can be estimated 
based on the relative refractive index difference between the two fluids 
and their relative molecular weights. 
This change in the index can be detected as a shift in the WGM peaks 
of the spectrum, and thus the sensitivity can be calculated 
using the definition, 
$\mathcal{S}\equiv \mathrm{d}\lambda/\mathrm{d}\mathrm{n}_3$ 
\cite{Reynolds:15}. 

\begin{figure}
\begin{center}
\includegraphics[width=0.4\hsize]{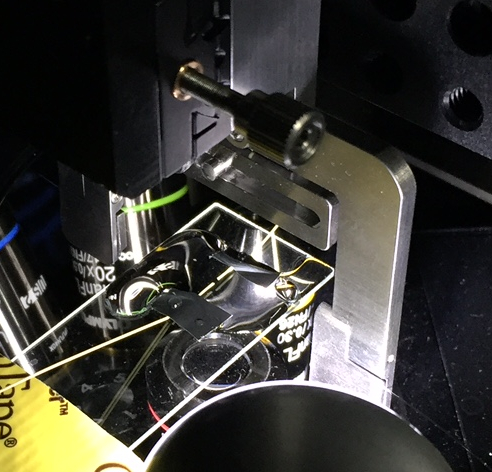}
\includegraphics[width=0.4\hsize]{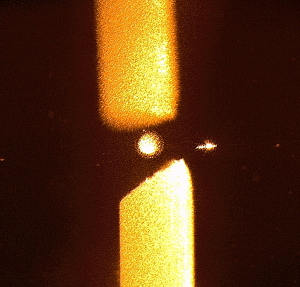}\\
\hspace{0.1cm} (a)\hspace{5.2cm} (b)
\end{center}
\vspace{-5mm}\caption[The refractive index sensitivity measurement apparatus 
is shown.]{The measurement of the refractive index sensitivity 
of a biological cell is performed by introducing two narrow 
tubes placed opposite one another, to expel and simultaneously 
to collect a high index fluid, such as a glycerol solution. (a) 
Two yellow IDEX $1572$ tubes are shown fixed to a glass coverslip 
mounted on a confocal microscope, using the setup illustrated 
in Fig.~\ref{fig:taper}. 
(b) A microscope image of a bovine embryo (\textit{centre}) placed 
between the two tubes (\textit{top} and \textit{bottom}) inside a droplet 
of handling media, 
which instigates a flow of high index fluid around the embryo. 
The excitation method in this case is a fibre taper (a bright spot 
through contamination loss is shown \textit{centre right}). 
}
\label{fig:RIsetup}
\figrule
\end{figure}

By using a series of different dilutions of glycerol 
($\mathrm{n}=1.47$) in water ($\mathrm{n}=1.33$), 
a range of refractive indices can be trialled, by adopting a 
tube-rigging system, shown in Fig.~\ref{fig:RIsetup}. 
Recalling the discussion in Section~\ref{subsec:pass}, the 
introduction of index sensitivity measuring apparatus requires 
physical space not easily afforded in the compact 
prism coupler setup described in Section~\ref{subsec:pris}. 
The inverted confocal microscope setup of Fig.~\ref{fig:exp} provides 
sufficient room to employ the glycerol flow method, shown in 
Fig.~\ref{fig:RIsetup}(a). However, 
one potential drawback is that the commencement of any flow 
of fluid causes the embryo to shift position, thus making consistent WGM 
measurements 
extremely challenging without more sophisticated optical 
compensation mechanisms. An alternative for handling this effect is to 
fix the embryo in place. 
This can be carried out conveniently 
in the case of the fibre taper coupling method. 
It is found that the taper itself acts as a brace, preventing the embryo from 
being dislodged from its initial position, as shown in 
Fig.~\ref{fig:RIsetup}(b). While the taper remains intact during this 
procedure, it is not known in what way this pressure applied from the embryo 
affects the behaviour of the resonances, and 
this would require further exploration. 

A specific concentration of glycerol solution can be chosen 
from a range of values shown in Table~\ref{tab:glycerol}, 
and placed in one of two SGE syringes, $250-500$ $\mu$L in volume, 
each connected to a length of yellow IDEX $1572$ tubing 
with an inner diameter of approximately $300$ $\mu$m. 
The tubes can then be placed opposite each another, leaving 
$4$ to $6$ nL of fluid between them in which the embryo will be situated. 
The tube ends and embryo are then situated within a droplet of handling media. 
The upper tube expels the glycerol solution, while the lower 
tube collects the expelled fluid at the same rate, so as to avoid 
contamination of the droplet of media. 
Since the flow between the tube ends must be kept low, for example, 
less than $10$ $\mu$L per hour to avoid dislodging the embryo, the diffusion 
rate of the glycerol solution in media plays an important role in assessing 
when the flow rate is consistent across the embryo. When this condition is 
met, the shift in the WGM peaks reaches its converged value. 

\begin{table*}
\caption[The refractive index shift expected for a range of 
concentrations of glycerol solution.]{For a given concentration of 
glycerol solution flowing around a resonator, the refractive index 
is shifted above the base value for water, $\mathrm{n}_3=1.33$. 
This is obtained for a linear correlation between the dilution factor 
and the refractive index for the relative molecular weights of glycerol 
and water. The density of glycerol is $1.26$ g/mL.}
\vspace{-6pt}
  \newcommand\T{\rule{0pt}{2.8ex}}
  \newcommand\B{\rule[-1.4ex]{0pt}{0pt}}
  \begin{center}
    \begin{tabular}{ccc}
      \hline
      \hline
      \T\B            
       glycerol concentration \quad \qquad & \qquad \quad refractive 
index shift (above $\mathrm{n}_3=1.33$)  \\ 
      \hline     
      $20$\% & \qquad\qquad\qquad $0.03528$ \\
      $10$\% & \qquad\qquad\qquad $0.01764$ \\
      $5$\% &  \qquad\qquad\qquad $0.00882$ \\
      $2$\% &  \qquad\qquad\qquad $0.00353$ \\
      \hline
    \end{tabular}    
  \end{center}
  \label{tab:glycerol}
\end{table*}

\section{\cdb \textit{Zona pellucida} annealing procedures} 
\label{sec:anneal}

One of the principal discoveries in Chapter~\ref{chpt:wgm} is the dramatic 
influence of the surface roughness on the ability of a resonator to 
sustain modes. While the bovine embryo is able to sustain modes at the 
limit of detectability, there is scope for improving the 
$Q$-factors of the modes by applying a specialised range of annealing 
procedures selected for these kinds of cells. 

The annealing procedures can be divided into two groups: chemical-based 
erosion of the \textit{zona pellucida} \cite{Jones06,Yano2007}, 
and enzyme-based 
thinning methods \cite{Fong,Frank}. 
The motivation for the use of these techniques in reproductive 
biology is to induce qualitative changes in the \textit{zona} region in order 
to facilitate hatching of the blastocyst \cite{Jones06,Yano2007},  
and to collect cells for biopsies for preimplantation 
genetic screening \cite{Geber2011}, for which competing 
methods such as partial \textit{zona} (mechanical) dissection 
and the use of a diode laser are used for assisted hatching 
\cite{doi:10.1093/humrep/17.5.1239,Geber2011}. 
An investigation of annealing with respect to cell resonators 
should focus on the relative smoothness of the cells after treatment. 
One particular advantage of the \textit{zona} thinning techniques 
over the total mechanical removal of the \textit{zona} itself, apart 
from avoiding  
exposure of the embryo to chemical, mechanical and bacterial risks, 
is that the chemical and enzyme-based annealing can partially thin 
the \textit{zona} uniformly \cite{Yano2007}. Thus, the innermost layer 
of the \textit{zona} is retained, offering control over the thickness of the 
remaining \textit{zona}, as well as the $Q$-factor, 
sensitivity and FSR of any modes sustained within, as described in 
Chapters~\ref{chpt:bub} and \ref{chpt:mod}. 

Of the chemical annealing methods, the use of an artificial 
interstitial fluid known as Acidified Tyrode's Solution (ATS)
has shown success in thinning the \textit{zona}. (The recipe for ATS 
is reproduced for reference in Appendix~\ref{chpt:app4}.) 
An embryo is attached to a holding pipette, with a secondary micropipette 
containing the ATS to induce an even flow over the embryo through expulsion 
for a limited duration \cite{Yano2007,Geber2011}. 
The solution dissolves the outer layer of the \textit{zona}, 
with the inner layer remaining, which is both more compact and 
resilient after treatment \cite{Yano2007}. Thus, the exposed inner layer 
of the \textit{zona} has the potential to exhibit superior surface 
properties to the untreated \textit{zona}, improving the quality 
of the modes. 

The enzyme annealing method uses pronase to achieve similar 
results in removing the \textit{zona} through digestion \cite{Fong,Frank}. 
By using a low concentration of pronase ($10$ IU/mL) diluted 
with G-$2^{\text{\texttrademark}}$ media,\footnote{Vitrolife, 
V{\"a}stra Fr{\"o}lunda, V{\"a}stra G{\"o}talands L{\"a}n, Sweden, 
\textit{G-}$2^{\text{\texttrademark}}$ \textit{media}, see
\href{http://www.vitrolife.com/en/Products/G-SeriesTM-media/G-2/}
{\cdb http://www.vitrolife.com/en/Products/G-SeriesTM-media/G-2/}} 
an initial stretching and softening of the \textit{zona} 
can be achieved, and the rate of digestion can be controlled 
\cite{doi:10.1093/humrep/17.5.1239}. 

While both these methods of annealing are capable of exposing 
the inner layer of the \textit{zona} through partial removal 
of the glycoproteins, ATS is usually preferred in clinical practice, 
due to the fact that 
it allows the embryo to survive during parthenogenic activation in the absence 
of calcium \cite{clone}. 
For the purposes of sustaining modes, both these methods are expected to be 
comparable with each other.

\section{\cdb Use of lasing to enhance detection of resonances} 

The autofluorescence properties of cells, and the introduction 
of artificial fluorescent coatings have been studied throughout 
this thesis; however, the possibility of constructing a biological cell 
\emph{laser} has not been explored in the context of self-supported WGMs, 
such as those within an embryo. 
While the use of resonator technology \emph{within} biological cells has 
been demonstrated in the literature, 
either by implanting microspheres \textit{in vivo} \cite{Himmelhaus2009418}, 
or exploiting natural oil cavities within certain animal cells \cite{Humar2015},
the capability of biological cells to exhibit lasing unassisted has not yet 
been fully realised. 

The potential of biological cells to support lasing modes was first studied 
in Ref.~\cite{Gather2011}, where GFP 
was used to excite modes, facilitated by placing the cell in a mirrored cavity. 
Achieving lasing within a resonator can potentially enhance the $Q$-factors 
of the modes \cite{vanderMolen:06,C5LC00670H,Francois:16,
LPOR:LPOR201600265}, as well as lower the detection limit for 
sensing applications \cite{Francois:15a}. While this is an attractive prospect, 
biological cells are significantly more fragile than most artificial 
resonators, and thus the lasing threshold may lie beyond the inherent 
damage thresholds of the cells. 
Since the lasing threshold takes the form $A\times V_{\text{eff}}/Q^2$ 
for an effective mode volume $V_{\text{eff}}$ \cite{Spillane2002}, 
where the gain coefficient $A$ is highly dependent on the distribution 
of the fluorescent medium, high density of fluorophores can lead to 
self-quenching \cite{ADFM:ADFM200801583}. 
This indicates that there is an optimal fluorophore density for achieving 
a low lasing threshold. 
However, the low $Q$-factors of the embryos 
studied in Chapter~\ref{chpt:wgm} ($Q=2.54\times10^3$) 
lead to a strong opposing effect whereby the lasing threshold 
is increased significantly, due to the fact that $A\propto 1/Q^2$. 
The damage threshold, on the other hand, is likely to be less than
that of polystyrene microspheres -- 
$300$ mJ/cm$^2$ using a pump wavelength of $532$ nm 
\cite{B201308H,Francois:16a}. One recent study reports 
the damage threshold for nucleolus-like bodies in mammalian oocytes 
as less than $3\times 10^{11}$ W/cm$^2$ using a $30$ fs pulse 
($<9$ mJ/cm$^2$) \cite{Shakhov2016}. 

As a result, the realisation of a biological cell laser will require 
careful management of the lasing and damage thresholds. 
A solution for pursuing
 this particular direction, however, may lie in the field 
of genetic engineering.

\section{\cdb Genetic modification}

While the genetic modification of cells has been discussed in 
Chapter~\ref{chpt:cel} in the case of yeast, specific isolation 
of genes that control fluorescence represents an important 
sub-branch of genetics that has seen much use in sensing 
applications since the discovery of GFP within jellyfish of the species 
\textit{Aequorea Victoria} \cite{doi:10.1021/bi00610a004}. 
While recent studies of cells that incorporate GFP 
have shown an ability to sustain WGMs with artificial assistance 
from a pre-built mirrored cavity \cite{Gather2011} mentioned above, 
the genetic modification of the cell itself to express GFP on its own 
may alleviate the practical difficulties in introducing an artificial 
fluorophore explored in Section~\ref{subsec:fl}. This avoids toxicity issues 
altogether, as well as the challenges in attaching or otherwise 
incorporating fluorescent material into a cell. 

One such method for the modification of a cell to express fluorescent 
proteins is \emph{clustered regularly interspaced short palindromic repeats}, 
or CRISPR \cite{Horvath167}. While CRISPR has been used 
to carry out gene tagging \cite{Lackner2015}, and to insert genetic sequences into 
embryonic stem cells in mice \cite{Kimura2015}, 
and while oocytes have been used to improve the gene editing 
methodology behind CRISPR \cite{Xie2016}, its use as an efficient 
method for the introduction of fluorescent proteins into oocytes remains to 
be fully explored.

\section{\cdb Living resonators}

The interrogation of living matter represents the culmination 
of the interrogation methods presented herein, since the ability to 
report on the status of a living cell has the potential to 
impact health sciences by leading to new diagnostic tools, 
aiding in the early identification of diseases and pathologies that 
cannot easily be studied using more conventional techniques. 
However, the use of living cells presents the researcher with a number 
of additional complications to address. 

The first complication is simply a logistical one -- the life-time of 
many living cells, particularly embryos, is limited to a window of 
approximately $15$ minutes or less, without a thermal mount or 
carefully prepared oxygen/carbon dioxide control system to maintain 
the required environment provided by a standard incubator. 
Compounding this limitation with the additional treatments of the cells to 
incorporate fluorophores or quantum dots, and the fabrication of fibre tapers, 
 the future development of these techniques 
as applied to living cells is initially a daunting proposition. 
However, these challenges can, in principle, be readily addressed with 
proper engineering to facilitate the preparation, mounting 
and interrogation of cells within this limited window of time. 
As an additional direction, the use of genetically 
engineered cells, which require no additional \textit{in vitro} processes 
to be carried out post-incubation, 
presents an enticing opportunity for future development. 

An additional feature in the context of living cells is the fact that they
 are in a state of continual 
growth, and in the case of embryos, progressing through 
the stages of maturation at a rate of a stage per day or faster. 
The notion of using embryos in the \emph{presumptive zygotes} phase 
can no longer be treated as a static quality of the cell. 
As the cells mature, their surface properties inevitably change 
\cite{Papi2010,Yanez2016}. This presents an opportunity to gain 
an understanding of \emph{how} the WGMs change within a living cell. 
If WGMs can be sustained in one phase of development while they are 
unable to be sustained in other stages, the generation of modes may 
be used as an indicator to gauge the stage of development of the embryo, 
their health, and their responses to the surrounding media, including 
neighbouring macromolecules contained therein.

\section{\cdb Cells as sensors}

In this chapter, a variety of methods for extending the work 
of this thesis on generating WGMs within biological cells has been 
presented. While the immediate focus of the next challenges in this 
area of study is on the surface properties of the cells, 
annealing procedures for smoothing the outer layers of embryos, 
improving the $Q$-factors through lasing, and improving the fluorescent 
signal via genetic modification, the long-term goal remains the 
development of a sensing modality for cells. 

Since cell biology represents a crucial area of study in pathology, 
diagnosis and medical technology, steps toward a novel sensing 
technology capable of exploring 
new aspects of cells is of paramount importance. 
Whispering gallery modes represent one such research direction in the 
development of sensing technologies, not only because of their high sensitivity 
to the internal and external environment of the cell, but because it is 
the cell itself that is capable of supporting the modes. 
Without the requirement of 
labelled proteins using specialised fluorophore markers, the light 
can be generated on site -- within the cell itself, and the natural geometry 
of the cell provides a reporting modality for its current status. 
Using the cells themselves to report on their own status, 
without the addition of complex machinery or probes that must 
enter the cell, avoids the risks of contamination, immune response or physical 
damage. In summary, the generation of whispering gallery modes within 
a biological cell represents a conceptually elegant paradigm for the 
future realisation of autonomous biological cell sensors. 

\chapter{\cdb Conclusion}
\label{chpt:con}

\section{\cdb Summary}

In this thesis, the ability to sustain 
whispering gallery modes in biological cells was explored, 
by considering the conditions under which resonances 
can take place in an optical cavity. 
In the first chapters of the thesis, 
the properties of the emitted energy spectra from 
resonators were examined, and the behaviour of the 
wavelength positions of the modes, the free spectra range 
and the quality factors were investigated in order to 
interpret the critical conditions beyond which no 
modes can be sustained. 

This understanding was applied to single and multilayer 
resonator architectures, and the principles of Mie 
scattering theory were extended to develop an efficient, 
general model for multilayer microspheres in Chapter~\ref{chpt:mod}. 
The unified nature of the multilayer model permitted its 
application in a number of settings, including a range of 
mode excitation strategies typically used in the field of
 biosensing that have resisted the development of a 
comprehensive model up to the present 
time. One such method in particular, active 
mode interrogation involving a uniform layer of fluorescent 
material, was able to be simulated for the first time 
in \emph{multilayer} resonators, and the spectral properties 
were examined. 

The development of sophisticated models for understanding the 
properties of resonators was accomplished in order to target the 
main research objective -- to discover whether whispering gallery 
modes can be sustained within a biological cell. 
A detailed study of the physical parameters associated with resonators, 
with cells as a particular example, was conducted to 
distil the critical selection criteria required of a resonator, 
in Chapter~\ref{chpt:cel}. 
In this study, the limit of the detectability of WGMs within imperfect 
resonators was investigated. 
Criteria for the cases where modes can no longer be sustained or detected 
were developed, as well as 
the mechanisms for the deterioration of the spectrum. 
These criteria were applied in a range of examples of biological 
cells to demonstrate how they can be used in practice. 

After a comprehensive analysis of the features inherent in the 
range of biological resonator candidates considered, the 
bovine embryo was selected as the most viable choice for further 
study and experimental treatment. 
The models, methods, understanding of the spectra and experimental 
techniques explored in this thesis were brought to bear in 
combination in order to analyse the resonance properties of embryos 
in Chapter~\ref{chpt:wgm}. 
%The culmination of this work in Section~\ref{sec:den} 
This work demonstrated 
that, through the measurement of scattered and transmitted spectra, as well 
as visual confirmation, 
 whispering gallery modes can be sustained within a biological 
cell, in this case, the embryo.

\section{\cdb Methodological evaluation and final analysis}

An analysis of the techniques and findings of the thesis are as follows. 

While Chapters~\ref{chpt:sph} and \ref{chpt:bub} developed 
the Finite Difference Time Domain method, employing state-of-the-art 
supercomputing resources to examine the transient behaviour of the 
electromagnetic fields, the principal focus has been on understanding 
the properties of the whispering gallery mode spectra. 
By carefully examining the radiation collection time and the grid resolution, 
converged spectra derived from this general computational framework were 
matched to the results of Mie scattering theory, providing a robust 
and comprehensive calculation methodology for future research endeavours 
into novel resonator architecture. These methods are, in essence, 
exploratory technologies, which require significant computational resources 
in practice. 

With the understanding gained through the computational methods, the 
research aim could be further refined, focusing only on the conditions 
of mode loss for a multilayer resonator. 
The development of the multilayer model in Chapter~\ref{chpt:mod}, 
while valuable in its own right as an efficient and general tool, 
was used principally to address this aim. The extraction 
of the free spectral range, quality factors and index sensitivity from 
a simulation, using carefully chosen input parameters to align 
closely with those of a candidate cell, yielded a robust prediction 
to which experimental results could be compared. 

Since the modelling predictions do not incorporate all the physical 
parameters and defects of a real-life cell, an understanding 
of which physical parameters impact the generation of 
whispering gallery modes, and their relative impact on the quality 
of the modes, was necessary to establish prior to measurement. 
As a result of this evaluation, 
the studies conducted in Chapter~\ref{chpt:cel} clearly illustrate 
how the selection criteria are applied in practice, as well as 
determine the most likely viable candidate cell to be used in 
the remainder of this project. 

Extensive characterisation of the physical properties of the 
bovine embryo represented an important step in understanding 
how the modes might be realised, and which geometric 
and material parameters could serve to hamper the detection of 
modes. Studies of the shape, diameter, refractive index, 
surface properties, scattering and absorption profiles and 
the effect of the osmolarity of their surrounding handling media 
were conducted in order to isolate the most crucial physical feature 
to be addressed by way of improving the clarity of the modes in the future. 
It was found that the selectivity of the prism coupler 
method resulted in the clearest detection of modes in the experiments 
conducted in this project. The use of the polariser was a vital tool 
for the clarification of the modes, by suppressing nearby competing 
modes of differing polarisations, thereby allowing a mode structure 
to be detected. Furthermore, the surface properties of the resonator 
were identified as the principal physical attribute leading to 
reduced quality factors, 
and thus highest priority to be addressed in the future for embryo-based 
resonator technology. 

A number of experimental methods were employed in order 
to measure the underlying modes within the structure,  
in Sections~\ref{subsec:pass} and \ref{subsec:ac}.  
It was found 
that the free spectral range and quality factors of the resonances 
obtained from the prism coupler method 
were consistent with the predictions from the modelling, and the 
selection criteria-based feasibility study. 
In the case of active interrogation using quantum dot coated 
embryos, a modulation of the fluorescent envelope was detected, 
and taper coupling was employed to remove a portion of the background 
radiation, as an attempt to resolve the underlying modes. 
A Fourier signal analysis 
of the measured radiation revealed a resonance corresponding to 
a resonator diameter lying outside the bounds determined in 
Section~\ref{subsec:pass}, placing tension upon the clear identification 
of modes. The broad peak in the Fourier 
decomposition indicated that a repeating pattern corresponding to WGMs 
was not easily determined in this experiment. 
While further development in measurement techniques will no doubt elucidate 
more properties of the detected signal modulation, 
the structure of the Fourier analysis is anticipated to originate from 
broad, composite peaks within the emitted power spectrum that exhibit 
significant overlap with each other. 
Finally, the behaviour of the spectrum was recorded while a droplet of 
glycerol solution was added to the surrounding medium of the embryo, 
as a method for increasing the refractive index of the medium, 
and thus quenching any WGMs that may be present. A marginal decrease in the 
clarity of the mode structure was noted; however, this result taken 
alone on its own merits was inconclusive. 

A complete summary of the evidence for each experiment, 
discussed in Section~\ref{sec:fin}, 
presents a compelling picture of resonances within an embryo. 
While the detection of clear modes posed many technical challenges 
as outlined in Chapters~\ref{chpt:cel} and \ref{chpt:wgm}, each 
experimental method revealed unique advantages that shed light on 
different aspects of the viability of biological cells as resonators. 
While it is clear that further refinement and investigation will 
be necessary for the effective utilisation of the definitive proof-of-concept 
contained in this thesis, it is anticipated that these research endeavours 
will become a vital asset for the future realisation of autonomous biological 
cell sensors.

Thus, I shall conclude with the following summary statement: 

\subsection{\cdb Concluding statement}

The understanding of the spectral behaviour of resonators, 
together with the predictions from sophisticated models and 
carefully-chosen selection criteria, have been applied to a 
biological cell -- the bovine embryo, leading to the detection 
of whispering gallery modes sustained within the cell. 
This is an original contribution of this thesis.

\titleformat{\chapter}[display]
  {\vspace{5cm}\flushright\fontseries{b}\fontsize{80}{100}\selectfont}
  {\fontseries{b}\fontsize{100}{130}\selectfont\textcolor{numcol}\thechapter}
  {0pt}
  {\Huge\bfseries}[]
\chapter*{\cdb \decofourright\,\, Epilogue}
\label{chpt:epi}

\addcontentsline{toc}{chapter}{Epilogue}
\renewcommand{\chaptermark}[1]{\markboth{\textbf{\chaptername #1}}{}}
\chaptermark{\cdb \decofourright\,\, Epilogue}
\renewcommand{\chaptermark}[1]{\markboth{\textbf{\chaptername\ \thechapter. \,\, #1}}{}}

\textit{``In a thought experiment the scientist puts nature to the test 
in his mind's eye. The thought experiment combines the visual and abstract 
mathematical modes of thinking in a manner that can permit the scientist 
to ``see'' the deep structure in a problem situation.''} \vspace{5mm}\\
\indent
A. I. Miller, ``Imagery and Intuition in Creative Scientific Thinking: 
Albert Einstein's Invention of the Special Theory of Relativity,'' in 
\textit{Creative People at Work, Twelve Cognitive Case Studies}, 
(D. B. Wallace and H. E. Gruber, eds.), p.~172 (Oxford University, 1989 Ed.) 
\cite{Miller}.\\

\section*{\cdb Novel sensing technologies and beyond}
\addcontentsline{toc}{section}{\cdb Novel sensing technologies and beyond}

The journey laid out in this thesis has traversed across multiple 
disciplines, and incorporated many different aspects of science 
that play an important role in contemporary research. 
In the quotation above, 
Einstein's scientific approach successfully 
integrated experimental data from the 
laboratory and conceptual frameworks in order to explore natural processes. 
Recalling the motivation for the vision described in the Prologue, 
``to have a cell act as a resonator is to have a new window into its internal 
structure and its immediate environment'', 
the realisation of this vision opens the door into new directions 
in biology, chemistry and physics. 

While the discovery of embryos as novel resonators is a compelling 
result, and the development of biosensing in a way that 
circumvents the need for cumbersome interrogation methods is a promising 
possibility for the future, 
laying the groundwork for new technologies 
represents only one aspect of the value of integrative research. 
Beyond facilitating the development of novel utilities, science 
involves the exploration of the mysteries of nature to gain 
a new understanding of the world. 
In this context, the goal of a scientific endeavour is not merely 
to predict, nor simply to measure or analyse, but to take these aspects 
together in order to synthesise a coherent picture that provides new meaning 
and insights into the whole. 

In this thesis, the understanding gained in how 
whispering gallery modes behave in novel resonator architectures 
prompted the development of a unified multilayer model. 
The physical limits for non-ideal resonators were 
distilled into selection criteria, using the predictions afforded by 
the model. The predictions, the measurement of 
the modes within a biological cell, and the analysis of the results, 
together form an integrated picture of resonances within a cell, 
and the discovery of purely biological resonators. 
\\

\textit{``If we microscope inward, we see molecules underlying our physical 
structure. Go even more micro, and we see atoms comprising the molecules. 
An even more fine-grained focus gets us to the now accepted reality 
that the vast majority of an atom itself is empty space\ldots\, 
we can see that even what we think of as the physical nature of the world, 
the world of things comprised of mass, is actually made up of very dense 
energy.''} \vspace{5mm}\\
\indent
D. J. Siegel, 
\textit{Mind, a journey to the heart of being human}, 
p.~318 (W. W. Norton, 2017) 
\cite{siegel2017mind}.\\

\appendix
\titleformat{\chapter}[display]
  {\vspace{-1.5cm}\flushright\fontseries{b}\fontsize{80}{100}\selectfont}
  {\fontseries{b}\fontsize{100}{130}\selectfont\textcolor{numcol}\thechapter}
  {0pt}
  {\Huge\bfseries}[]
\chapter{\cdb Vector Spherical Harmonics}
\label{chpt:app1}

In this appendix, the notation and properties of the vector spherical harmonics 
used in Chapter~\ref{chpt:mod} and Appendix~\ref{chpt:app2} are explained.

\section{\cdb Notation}

Considering the spherical symmetry of the resonators used in 
Chapter~\ref{chpt:mod}, 
the use vector spherical harmonics (VSH) is important for the construction of 
a general model from first principles. 
Different conventions exist for the definitions of the VSH in the
literature, for example, in atomic physics, or electrodynamics (see 
Refs.~\cite{BohrenHuffman} and \cite{Blatt78}). 
For this work, the definitions given by Barrera \cite{Barrera85} are modified 
to the following form 
\begin{align}
\label{C001a}
\mathbf{Y}_{lm}&=Y_{lm}(\theta,\phi)\mathbf{\hat{r}},\\
\bm{\Psi}_{lm}    &=(\frac{1}{\mathrm{i}\,\sqrt{l(l+1)}})r\bm{\nabla}Y_{lm}
(\theta,\phi),\\
\bm{\Phi}_{lm}(\theta,\phi) &=
(\frac{1}{\mathrm{i}\,\sqrt{l(l+1)}})\mathbf{r}\times\bm{\nabla}Y_{lm}
(\theta,\phi),
\label{C001c}
\end{align}
where $Y_{lm}(\theta,\phi)$ are standard scalar spherical harmonics 
\cite{Jackson89}, and $l$ and $m$ 
are the azimuthal and polar quantum numbers, respectively. 

These definitions of the VSH can be related to the other forms that 
appear in the literature. Key examples are as follows

\begin{align}
\mathbf{X}_{lm}(\theta,\phi)  &  =(\frac{1}{\mathrm{i}})(\frac{1}{\sqrt{l(l+1)}
})\mathbf{r}\times\bm{\nabla}Y_{lm}(\theta,\phi)\text{  from 
Ref.~\cite{Liang:04,Jackson89},}\label{B01}\\
\mathbf{Y}_{llm}(\theta,\phi)  &  =(\frac{1}{\mathrm{i}})(\frac{1}{\sqrt{l(l+1)}
})\mathbf{r}\times\bm{\nabla}Y_{lm}(\theta,\phi)\text{  from Eq.~(5.9.14) of 
Ref.~\cite{edmonds1996angular},}\label{B02}\\
\mathbf{Y}_{L}^{(m)}  &  =(\frac{1}{\mathrm{i}})(\frac{1}{\sqrt{l(l+1)}
})\mathbf{r}\times\bm{\nabla}Y_{lm}(\theta,\phi),\text{ \ }\mathbf{Y}
_{L}^{(e)}=-(\frac{1}{\mathrm{i}})(\frac{1}{\sqrt{l(l+1)}})r\bm{\nabla}
Y_{lm}(\theta,\phi),\nonumber\\
&  
(\frac{1}{\mathrm{i}})Y_{lm}(\theta,\phi)\mathbf{\hat{r}=-Y}_{L}^{(o)}\text{  from 
Ref.~\cite{Moroz05}}. \label{B03}
\end{align}
Comparing these equations with the orthonormal functions used in this work, 
Eqs.~(\ref{C001a}) through (\ref{C001c}), the following relations hold 
\begin{align}
\mathbf{Y}_{llm} &=\mathbf{X}_{lm}=\mathbf{Y}_{L}^{(m)}=\bm{\Phi
}_{lm},\label{B04}\\
\qquad\mathbf{Y}_{L}^{(e)}    &=-\bm{\Psi}_{lm},\\
\qquad(\frac{1}{\mathrm{i}})\mathbf{Y}_{L}^{(o)}    &=\mathbf{Y}_{lm}.  
\end{align}

\section{\cdb Properties}

Several salient properties of the VSH are now summarised.  
First, the VSH are \\ orthogonal 
\begin{equation}
\mathbf{Y}_{lm}\cdot\bm{\Psi}_{lm}=\bm{\Phi}_{lm}%
\cdot\bm{\Psi}_{lm}=\bm{\Phi}_{lm}\cdot\mathbf{Y}_{lm}=0. 
\label{A01}%
\end{equation}
Second, they are orthonormal in Hilbert space
\begin{align}
\int\mathbf{Y}_{lm}\cdot\mathbf{Y}_{l^{\prime}m^{\prime}}^{\ast}\mathrm{d}\Omega &
=\int\bm{\Psi}_{lm}\cdot\bm{\Psi}_{l^{\prime}m^{\prime}}^{\ast}\mathrm{d}\Omega
=\int\bm{\Phi}_{lm}\cdot\bm{\Phi}_{l^{\prime}m^{\prime}}^{\ast}d\Omega
=\delta_{ll^{\prime}}\delta_{mm^{\prime}%
},\nonumber\label{A02}\\
\int\mathbf{Y}_{lm}\cdot\bm{\Psi}_{l^{\prime}m^{\prime}}^{\ast}\mathrm{d}\Omega
&  =\int\bm{\Psi}_{lm}\cdot\bm{\Phi}_{l^{\prime}m^{\prime}%
}^{\ast}\mathrm{d}\Omega=\int\mathbf{Y}_{lm}\cdot\bm{\Phi}_{l^{\prime}m^{\prime}}^{\ast}
\mathrm{d}\Omega=0.
\end{align}

The following summation rules apply to dyadic products of the VSH
\begin{align}
\label{A04a}
\sum_{m=-l}^{m=l}\bm{\Phi}_{lm}\bm{\Phi}_{lm}^{\ast}  &
=\sum_{m=-l}^{m=l}\bm{\Psi}_{lm}\bm{\Psi}_{lm}^{\ast}%
=\frac{2l+1}{8\pi}(\mathbf{\hat{e}}_{\theta}\mathbf{\hat{e}}_{\theta}+\mathbf
{\hat{e}}_{\phi
}\mathbf{\hat{e}}_{\phi}),\\
\sum_{m=-l}^{m=l}\mathbf{Y}_{lm}\mathbf{Y}_{lm}^{\ast
}&=\frac{2l+1}{4\pi}(\mathbf{\hat{e}}_{r}\mathbf{\hat{e}}_{r}),\\
\sum_{m=-l}^{m=l}\bm{\Phi}_{lm}\mathbf{Y}_{lm}^{\ast}  &  =\sum
_{m=-l}^{m=l}\bm{\Phi}_{lm}\bm{\Psi}_{lm}^{\ast}=\sum
_{m=-l}^{m=l}\mathbf{Y}_{lm}\bm{\Psi}_{lm}^{\ast}=0,
\label{A04c}
\end{align}
where $\mathbf{\hat{e}}$ represents unit vectors in the basis of the 
coordinate space. 

In addition, the following vector properties are used explicitly in the  
derivation of the model described in Chapter~\ref{chpt:mod}, as shown in 
Appendix~\ref{chpt:app2}
\begin{align}
\bm{\nabla}\times\left(  f(r)\mathbf{Y}_{lm}\right)   &  =\frac{1}%
{r}f(r)\bm{\Phi}_{lm},\\
\bm{\nabla}\times\left(
f(r)\bm{\Psi}_{lm}\right)  &=\left(  \frac{\mathrm{d}f}{\mathrm{d}r}+\frac{1}%
{r}f(r)\right)  \bm{\Phi}_{lm},\label{A03}\\
\bm{\nabla}\times\left(  f(r)\bm{\Phi}_{lm}\right)   &
=-\frac{\sqrt{l(l+1)}}{ir}f\mathbf{Y}_{lm}-\left(  \frac{\mathrm{d}f}
{\mathrm{d}r}+\frac{1}%
{r}f\right)  \bm{\Psi}_{lm}.
\end{align}

\addtocontents{toc}{\protect\newpage}
\chapter{\cdb Multilayer Examples}
\label{chpt:app2}

The general formulae presented for the multilayer model in 
Chapter~\ref{chpt:mod} 
can be adapted to consider a number of specific scenarios, particularly in 
terms of the number of dipole sources, their orientation, or 
if a uniform layer of sources is used. This appendix presents the more detailed 
derivations of these formulae, which could not be presented in 
Chapter~\ref{chpt:mod} without compromising the flow of the narrative 
structure of the thesis.

\section{\cdb Scattered power in the outermost region}

The first example involves the derivation of the general formula for the total 
radiated power of the system. First, the behaviour of the electric and magnetic 
fields in the far field must be considered. The total transverse parts of the 
fields in Eq.~(\ref{R016}) in the outermost region $N+1$, and for 
$r\gg r_j^{\prime}$,  are given by
\begin{equation}
\mathbf{EH}_{N+1}^{\text{T}}(r)=M_{N+1}(r)\mathbf{A}_{N+1}+\theta
(r-r_{N+1}^{\prime})M_{N+1}(r)\mathbf{a}_{N+1H}(r_{N+1}^{\prime}).
\end{equation}
In the limit of $r\rightarrow\infty$, this equation leads to the following 
forms for the scattered fields
\begin{align}
\mathbf{E}_{sc}  &= -\sum_{l,m}\left(  \frac{\mathrm{i}}{\mathrm{n}_{N+1}}\right)
\frac{1}{k_{N+1}r}\chi_{l}^{^{\prime}}(k_{N+1}r)\bm{\Psi}_{lm}%
(\theta,\phi)[B_{N+1}+a_{N+1EH}(r_{N+1}^{\prime}%
)]\nonumber\label{R027}\\
&+\sum_{l,m}\bm{\Phi}_{lm}(\theta,\phi)\frac{1}{k_{N+1}r}\chi
_{l}(k_{N+1}r)[D_{N+1}+a_{N+1MH}(r_{N+1}^{\prime})],\\
\mathbf{H}_{sc}&=\sum_{l,m}(\frac{1}{\mu_{_{N+1}}})\bm{\Phi}%
_{lm}(\theta,\phi)\frac{1}{k_{N+1}r}\chi_{l}(k_{N+1}r)[B_{N+1}+a_{N+1EH}%
(r_{N+1}^{\prime})]\nonumber\label{R028}\\
&+\sum_{l,m}(\frac{\mathrm{i}\mathrm{n}_{N+1}}{\mu_{_{N+1}}})\frac{1}{k_{N+1}r}\chi
_{l}^{^{\prime}}(k_{N+1}r)\bm{\Psi}_{lm}[D_{N+1}+a_{N+1MH}%
(r_{N+1}^{\prime})].
\end{align}
In the above equations, the property $j_{l}(kr)\rightarrow0$ as $r\rightarrow
\infty$ has been used. The total scattered power through a sphere of radius $r$ 
can then be calculated:  
\begin{align}
P_{\text{total}}  &  =r^{2}\int\mathbf{S}_{sc}.\mathbf{\hat{r}}d\Omega
=\frac{c}{8\pi\mu_{N+1}}r^{2}\int(\mathbf{E}_{sc}\times\mu_{N+1}%
\mathbf{H}_{sc}^{\ast}).\mathbf{\hat{r}}d\Omega\nonumber\\
&  =\frac{c}{8\pi\mu_{N+1}}\sum_{l,m}\Big(\frac{-\mathrm{i}}{\mathrm{n}_{N+1}
k_{N+1}^{2}%
}\Big)\chi_{l}^{^{\prime}}k_{N+1}r\,\chi_{l}^{\ast}(k_{N+1}r)\left\vert
B_{N+1}+a_{N+1EH}(r_{N+1}^{\prime})\right\vert ^{2}\nonumber\\
&  + \mathrm{i}\frac{\mathrm{n}_{N+1}}{k_{N+1}^{2}}\chi_{l}(k_{N+1}r)\chi_{l}%
^{\ast^{\prime}}(k_{N+1}r)\left\vert D_{N+1}+a_{N+1MH}(r%
_{N+1}^{\prime})\right\vert ^{2},
\end{align}
where the orthonormal properties of $\bm{\Psi}_{lm}(\theta,\phi)$ and 
$\bm{\Phi}_{lm}(\theta,\phi)$ functions given in
Eq.~(\ref{A01}) through (\ref{A03}) have been used. Note that in the limit of 
$r\rightarrow
\infty$, it can be shown that $\chi_{l}(z)= zh_{l}^{(1)}(z)\rightarrow
\mathrm{i}^{-l-1}\exp({\mathrm{i} z})$ and $\chi_{l}^{^{\prime}}(z)\rightarrow
\mathrm{i}^{-l}\exp({\mathrm{i} z})$ \cite{Moroz05}, and hence
\begin{align}
P_{\text{total}}&=\frac{c}{8\pi}\sqrt{\frac{\epsilon_{N+1}}{\mu_{N+1}}}\frac
{1}{k_{N+1}^{2}}\times\nonumber\\
&\sum_{l,m}[\Big(\frac{1}{\mathrm{n}_{N+1}^{2}}\Big)\left\vert B_{N+1}%
+a_{N+1EH}(r_{N+1}^{\prime})\right\vert ^{2}+\left\vert
D_{N+1}+a_{N+1MH}(r_{N+1}^{\prime})\right\vert ^{2}]. \label{R0300}%
\end{align}
The coefficients $B_{N+1}$, $D_{N+1}$, $a_{N+1EH} 
(r_{N+1}^{\prime})$ and $a_{N+1MH}(r_{N+1}^{\prime})$ are known,  
based on Eqs.~(\ref{R023}), (\ref{R0241}), (\ref{R013}), and
(\ref{R014}) respectively. Note that if there is no dipole in the outermost
region, then $a_{N+1EH}(r_{N+1}^{\prime})=a_{N+1MH}(r_{N+1}^{\prime})=0$.

\section{\cdb Single dipole embedded in a layer}

In this section, the scenario is selected where there is only one dipole in the
layer $j$, for $j=1,\ldots, N+1$. This is useful when comparing simulations 
with results from embedded nanoparticles \cite{Kaufman2012,Ilchenko:13,
Tao21062016}. In this case, according to \textbf{Scenario 2} of 
Section~\ref{sec:rec}, the following 
field coefficients may be obtained using the transfer matrix approach
\begin{align}
A_{1}  &=-\frac{(-S_{21}S_{12}+S_{11}S_{22})}{S_{22}}\mathcal{D}_{1}, \\
\quad C_{1}  &=\frac{(S_{43}S_{34}-S_{33}S_{44})}{S_{44}}\mathcal{D}_{3},\\
B_{N+1}  &=\mathcal{D}_{2}+\frac{S_{21}}{S_{22}}\mathcal{D}_{1},\\ 
D_{N+1}  &=\mathcal{D}_{4}+\frac{S_{43}}{S_{44}}\mathcal{D}_{3}.
\label{R031} 
\end{align}
In addition, the dipole vector introduced in Chapter~\ref{chpt:mod} takes the 
following form
\begin{align}
\mathbfcal{D}  &  =T(N+1,j)(1-\delta_{j,N+1})\mathbf{a}_{jH}-T(N+1,j)(1-\delta
_{j,1})\mathbf{a}_{jL}\nonumber\\
&  =\left(
\begin{array}
[c]{c}%
T_{12}^{j}(1-\delta_{j,N+1})a_{jEH}-T_{11}^{j}(1-\delta_{j,1})a_{jEL}\\
T_{22}^{j}(1-\delta_{j,N+1})a_{jEH}-T_{21}^{j}(1-\delta_{j,1})a_{jEL}\\
T_{34}^{j}(1-\delta_{j,N+1})a_{jMH}-T_{33}^{j}(1-\delta_{j,1})a_{jML}\\
T_{44}^{j}(1-\delta_{j,N+1})a_{jMH}-T_{43}^{j}(1-\delta_{j,1})a_{jML}%
\end{array}
\right).\label{R0311}
\end{align}
Thus, the coefficients required for the general formula in Eq.~(\ref{R0300}) 
can be obtained for this specific case of a single dipole 
in a layer. For the contributions to the TM modes, it is found that
\begin{align}
B_{N+1}+a_{N+1EH}(r_{N+1}^{\prime})&=
(T_{22}^{j}+\frac{S_{21}}{S_{22}}T_{12}^{j})(1-\delta_{j,N+1})a_{jEH}(r_{j}^{\prime})
\nonumber\\
&-(T_{21}^{j}+\frac{S_{21}}{S_{22}}T_{11}^{j})(1-\delta_{j,1})a_{jEL}(r_{j}^{\prime})+
\delta_{j,N+1}a_{jEH}(r_{j}^{\prime})\nonumber\\
&  =\alpha_{l}a_{jEH}(r_{j}^{\prime})-\beta_{l}a_{jEL}(r_{j}^{\prime}),\label{R032}
\end{align}
where $\alpha_{l}\equiv(T_{22}^{j}+\frac{S_{21}}{S_{22}}T_{12}^{j})(1-\delta_{j,N+1})+
\delta_{j,N+1}$ 
and $\beta_{l}\equiv(T_{21}^{j}+\frac{S_{21}}{S_{22}}T_{11}^{j})(1-\delta_{j,1})$. 
Similarly, for the contributions to the TE modes one may write
\begin{align}
 D_{N+1}+a_{N+1MH}(r_{N+1}^{\prime})&=
 (T_{44}^{j}+\frac{S_{43}}{S_{44}}T_{34}^{j})(1-\delta_{j,N+1}%
)a_{jMH}(r_{j}^{\prime})\nonumber\\
&-(T_{43}^{j}+\frac{S_{43}}{S_{44}}T_{33}%
^{j})(1-\delta_{j,1})a_{jML}(r_{j}^{\prime})+\delta_{j,N+1}%
a_{jMH}(r_{j}^{\prime})\nonumber\\
&  =\gamma_{l}a_{jMH}(r_{j}^{\prime})-\zeta_{l}a_{jML}(r_{j}^{\prime}),\label{R0321}
\end{align}
where $\gamma_{l}\equiv(T_{44}^{j}+\frac{S_{43}}{S_{44}}T_{34}^{j})(1-\delta_{j,N+1})+
\delta_{j,N+1}$
and $\zeta_{l}\equiv(T_{43}^{j}+\frac{S_{43}}{S_{44}}T_{33}^{j})(1-\delta_{j,1})$. 
Equations~(\ref{R032}) and
(\ref{R0321}) are general, and can be applied to a dipole in any layer $j$, 
including the outermost layer $N+1$ or the innermost layer $1$.

By considering the forms of the dipole coefficients that appear in 
Section~\ref{subsec:trans},
 the field coefficients $B_{N+1}+a_{N+1EH}(r_{N+1}^{\prime})$ and 
$D_{N+1}+a_{N+1MH}(r_{N+1}^{\prime})$, which appear in
the total scattered power of Eq.~(\ref{R0300}), may be found explicitly
\begin{align}
B_{N+1}+a_{N+1EH}(r_{N+1}^{\prime})  &  =4\pi k_{j}^{2}\sqrt
{\frac{\mu_{j}}{\epsilon_{j}}}\mathbf{P}\cdot\bm{\nabla}^{\prime
}_j\mathbf{\times\{[}\alpha_{l}j_{l}(k_{j}r_{j}^{\prime})-\beta_{l}h_{l}%
^{(1)}(k_{j}r_{j}^{\prime})]\bm{\Phi}_{lm}^{\ast}(\theta_{j}^{\prime
},\phi_{j}^{\prime}\mathbf{)\}},\label{R034}\\
D_{N+1}+a_{N+1MH}(r_{N+1}^{\prime})  &  =4\pi\mathrm{i} k_{j}^{3}\frac
{1}{\epsilon_{j}}[\gamma_{l}j_{l}(k_{j}r_{j}^{\prime})-\zeta_{l}h_{l}%
^{(1)}(k_{j}r_{j}^{\prime})]\mathbf{P}\cdot\bm{\Phi}_{lm}^{\ast}%
(\theta_{j}^{\prime},\phi_{j}^{\prime}). \label{R035}%
\end{align}
Using the properties of orthonormal functions $\mathbf{Y}_{lm}$,
$\bm{\Psi}_{lm}$, and $\bm{\Phi}_{lm}(\theta,\phi)$ as stated in 
Appendix~\ref{chpt:app1}, 
in conjunction with Eq.~(\ref{A03}), 
the coefficients may be written in the following forms
\begin{align}
B_{N+1}+a_{N+1EH}(r_{N+1}^{\prime})&=
\mathbf{P}\cdot[f_{l}(r_j^{\prime})\mathbf{Y}_{lm}+g_{l}(r_j^{\prime})\bm{\Psi}_{lm}],\\
D_{N+1}+a_{N+1MH}(r_{N+1}^{\prime})&=\mathbf{P}\cdotp_{l}(r_j^{\prime})\bm{\Phi}_{lm}^{\ast}, 
\end{align}
where $f_{l}$, $g_{l}$ and $p_{l}$ are functions
of $r_j^{\prime}$. Hence,  
the properties of the dyadic products of $\mathbf{Y}_{lm}$, $\bm{\Psi
}_{lm}$, and $\bm{\Phi}_{lm}$ from Eqs.~(\ref{A04a}) through (\ref{A04c}) may be 
used to perform the summation over $m$ and simplify the field coefficients 
that appear in the general formulae of Eq.~(\ref{R0300}) 
\begin{align}
&  \sum_{l,m}\left\vert B_{N+1}+a_{N+1EH}(r_{N+1}^{\prime
})\right\vert ^{2} =\nonumber\label{R036}\\
&  16\pi^{2}k_{j}^{6}(\frac{\mu_{j}}{\epsilon_{j}})\sum_{l}\Bigg\{\frac{2l+1}{4\pi
}l(l+1)\frac{\left\vert \mathbf{\mathbf{[}}\alpha_{l}j_{l}(k_{j}r_{j}^{\prime
})-\beta_{l}h_{l}^{(1)}(k_{j}r_{j}^{\prime})\mathbf{]}\right\vert ^{2}}%
{k_{j}^{2}r^{\prime2}}\left\vert \mathbf{P}_{r}\right\vert ^{2}\nonumber\\
&  +\frac{2l+1}{8\pi}\frac{\left\vert \{\alpha_{l}\frac{\mathrm{d}}%
{\mathrm{d}r_{j}^{\prime}}[r_{j}^{\prime}j_{l}(k_{j}r_{j}^{\prime})]-\beta
_{l}\frac{\mathrm{d}}{\mathrm{d}r_{j}^{\prime}}[r_{j}^{\prime}h_{l}(k_{j}%
r_{j}^{\prime})]\}\right\vert ^{2}}{k_{j}^{2}r_{j}^{\prime2}}(\left\vert
\mathbf{P}_{\theta}\right\vert ^{2}+\left\vert \mathbf{P}_{\phi}\right\vert
^{2})\Bigg\},\\
&  \sum_{l,m}\left\vert D_{N+1}+a_{N+1MH}(r_{N+1}^{\prime
})\right\vert ^{2} =\nonumber\label{R037}\\
&  16\pi^{2}k_{j}^{6}(\frac{1}{\epsilon_{j}^{2}})\sum_{l}\frac{2l+1}{8\pi
}\left\vert [\gamma_{lm}j_{l}(k_{j}r_{j}^{\prime})-\zeta_{lm}h_{l}^{(1)}%
(k_{j}r_{j}^{\prime})]\right\vert ^{2}(\left\vert \mathbf{P}_{\theta
}\right\vert ^{2}+\left\vert \mathbf{P}_{\phi}\right\vert ^{2}).
\end{align}
Note that the quantities $\mathbf{P}_{r},$ $\mathbf{P}_{\theta},$ and 
$\mathbf{P}_{\phi}$ are the polar components of the polarisation vector 
$\mathbf{P}$. Based on the above
equations, the total scattered power from a sphere can be expressed 
as the sum of the emitted powers due to the normal and transverse components 
of $\mathbf{P}$
\begin{align}
P_{\text{total}}    &=P_{\bot}+P_{\Vert}=\frac{c}{2}\sqrt{\frac{\epsilon_{N+1}}%
{\mu_{N+1}}}\frac{k_{j}^{4}\mathrm{n}_{j}^{2}}{\mathrm{n}_{N+1}^{2}}\frac{1}
{\epsilon_{j}^{2}%
}\times\nonumber\\
&\sum_{l}(2l+1)\Bigg\{(\frac{\mathrm{n}_{j}^{2}}{\mathrm{n}_{N+1}^{2}})l(l+1)
\frac{\left\vert
\mathbf{\mathbf{[}}\alpha_{l}j_{l}(k_{j}r_{j}^{\prime})-\beta_{l}h_{l}%
^{(1)}(k_{j}r_{j}^{\prime})\mathbf{]}\right\vert ^{2}}{k_{j}^{2}r_{j}%
^{\prime2}}\left\vert \mathbf{P}_{r}\right\vert ^{2}\nonumber\\
&+\!\Bigg[\!\Big(\frac{\mathrm{n}_{j}^{2}}{\mathrm{n}_{N+1}^{2}}\Big)
\frac{\left\vert \{\alpha_{l}%
\frac{\mathrm{d}}{\mathrm{d}r_{j}^{\prime}}[r_{j}^{\prime}j_{l}(k_{j}%
r_{j}^{\prime})]-\beta_{l}\frac{\mathrm{d}}{\mathrm{d}r_{j}^{\prime}}%
[r_{j}^{\prime}h_{l}(k_{j}r_{j}^{\prime})]\}\right\vert ^{2}}{k_{j}^{2}%
r_{j}^{\prime2}}\!\nonumber\\
&+\!\left\vert [\gamma_{l}j_{l}(k_{j}r_{j}^{\prime})-\zeta
_{l}h_{l}^{(1)}(k_{j}r_{j}^{\prime})]\right\vert ^{2}\!\Bigg]
\!\left[  \frac{\left\vert \mathbf{P}_{\theta}\right\vert
^{2}\!+\!\left\vert \mathbf{P}_{\phi}\right\vert ^{2}}{2}\right]\Bigg\}.
\label{R038}
\end{align}

One can normalise the powers $P_{\bot}$ and $P_{\Vert}$ to powers
radiated by a dipole in a bulk material with material properties 
$(\mathrm{n}_{j},\epsilon_{j},\mu_{j})$,
using $P_{\bot}^{0}=ck_{j}^{4}\left\vert \mathbf{P}_{r}\right\vert ^{2}/
(3\epsilon_{j}\mathrm{n}_{j})$ and $P_{\Vert}%
^{0}=ck_{j}^{4}(\left\vert \mathbf{P}_{\theta}\right\vert
^{2}+\left\vert \mathbf{P}_{\phi}\right\vert ^{2})/(3\epsilon_{j}\mathrm{n}_{j})$ 
to obtain
\begin{align}
&\frac{P_{\bot}}{P_{\bot}^{0}}    =\frac{1}{2}\sqrt{\frac{\epsilon_{N+1}}%
{\mu_{N+1}}}\frac{\mathrm{n}_{j}^{2}}{\mathrm{n}_{N+1}^{2}}\frac{3\mathrm{n}_{j}}
{\epsilon_{j}}\sum
_{l}(\frac{\mathrm{n}_{j}^{2}}{\mathrm{n}_{N+1}^{2}})(2l+1)l(l+1)\frac{\left\vert
\mathbf{\mathbf{[}}\alpha_{l}j_{l}(k_{j}r_{j}^{\prime})-\beta_{l}h_{l}%
^{(1)}(k_{j}r_{j}^{\prime})\mathbf{]}\right\vert ^{2}}{k_{j}^{2}r_{j}%
^{\prime2}},\label{R039}\\
\frac{P_{\Vert}}{P_{\Vert}^{0}}    &= \frac{1}{4}\sqrt{\frac{\epsilon_{N+1}%
}{\mu_{N+1}}}\frac{\mathrm{n}_{j}^{2}}{\mathrm{n}_{N+1}^{2}}\frac{3\mathrm{n}_{j}}
{\epsilon_{j}} 
\sum_{l}(2l+1)\Bigg\{\Bigg[\Big(\frac{\mathrm{n}_{j}^{2}}{\mathrm{n}_{N+1}^{2}}
\Big)\frac{\left\vert
\{\alpha_{l}\frac{\mathrm{d}}{\mathrm{d}r_{j}^{\prime}}[r_{j}^{\prime}%
j_{l}(k_{j}r_{j}^{\prime})]-\beta_{l}\frac{\mathrm{d}}{\mathrm{d}r_{j}%
^{\prime}}[r_{j}^{\prime}h_{l}(k_{j}r_{j}^{\prime})]\}\right\vert ^{2}}%
{k_{j}^{2}r_{j}^{\prime2}}\nonumber\\
  &+\left\vert [\gamma_{l}j_{l}(k_{j}r_{j}^{\prime})-\zeta_{l}h_{l}%
^{(1)}(k_{j}r_{j}^{\prime})]\right\vert ^{2}\Bigg]\Bigg\}.\label{R040}
\end{align}
These two equations exactly match Eqs.~(\ref{Pbot}) and (\ref{Pvert}) in 
Chapter~\ref{chpt:mod}.

\section{\cdb Deriving an active layer case}
\label{sec:act}

Consider a multilayer structure where one of the
layers, $j$, consists of active material. In this context, a uniform 
distribution of randomly-oriented dipoles, with density 
$\rho(r_j^{\prime})=1,$ is introduced into that layer. As a result, one must 
integrate Eqs.~(\ref{R039}) and (\ref{R040}) with respect
to the variable $r_j^{\prime}$, which is located within the layer $j$. Since the 
dipoles are randomly oriented, and the orientations of the constituent 
dipoles are averaged over the distribution, 
$\left\langle P_{\text{total}}/P^{0}\right\rangle =
\frac{1}{3}\left\langle P_{\bot}/P_{\bot}^{0}\right\rangle 
+\frac{2}{3}\left\langle P_{\Vert}/P_{\Vert}^{0}\right\rangle$,
one may calculate the normal and transverse components of the emitted power 
independently, even in the case of an active layer. In the case of the normal 
component, one finds
\begin{align}
\left\langle \frac{P_{\bot}}{P_{\bot}^{0}}\right\rangle  &  =\frac{1}{2}%
\sqrt{\frac{\epsilon_{N+1}}{\mu_{N+1}}}\frac{\mathrm{n}_{j}^{2}}
{\mathrm{n}_{N+1}^{2}}\!%
\frac{3\mathrm{n}_{j}}{\epsilon_{j}}\times\nonumber\\
&\sum_{l}(\frac{\mathrm{n}_{j}^{2}}{\mathrm{n}_{N+1}^{2}%
})(2l+1)l(l+1)\!\!\int_{j}\frac{\left\vert \mathbf{\mathbf{[}}\alpha_{l}%
j_{l}(k_{j}r_{j}^{\prime})-\beta_{l}h_{l}^{(1)}(k_{j}r_{j}^{\prime}%
)\mathbf{]}\right\vert ^{2}}{k_{j}^{2}r_{j}^{\prime2}}\mathrm{d}^{3}r_{j}^{\prime}
\Big{/}\!\!\!\int_{j}\mathrm{d}^{3}r_{j}^{\prime},\label{R041}
\end{align}
which can be simplified to a summation over a one-dimensional integral over the 
shell region that forms the $j^{\mathrm{th}}$ layer
\begin{align}
\left\langle \frac{P_{\bot}}{P_{\bot}^{0}}\right\rangle &=\frac{1}{2}\sqrt{
\frac{\epsilon_{N+1}}{\mu_{N+1}}}\frac{\mathrm{n}_{j}^{2}}%
{\mathrm{n}_{N+1}^{2}}\frac{3\mathrm{n}_{j}}{k_{j}^{2}\epsilon_{j}V_{j\text{shell}}}
4\pi\sum_{l}(\frac{\mathrm{n}_{j}^{2}}{\mathrm{n}_{N+1}^{2}})l(l+1)I_{l}^{(1)},\\
\text{where}\,\,I_{l}^{(1)}  &\equiv(2l+1)\int_{j}\left\vert \mathbf{\mathbf{[}}
\alpha_{l}j_{l}(k_{j}r_{j}^{\prime})-\beta_{l}h_{l}^{(1)}(k_{j}r_{j}^{\prime}
)\mathbf{]}\right\vert ^{2}\mathrm{d}r_{j}^{\prime}.
\end{align}
As explained in Chapter~\ref{chpt:mod}, the volume of the shell, 
$V_{j\text{shell}}=4\pi(r_{j}^{\prime2} - r_{j-1}^{\prime2})$, 
is defined so that if $j=1$, then $r_0$ is taken to be zero, since the relevant 
volume is simply a sphere bounded by the innermost radius. 

In the case of the transverse component of the power, one may similarly 
express the result as a summation over a set of one-dimensional integrals, in 
this case, $I_{l}^{(2)}$ and $I_{l}^{(3)}$
\begin{align}
\left\langle \frac{P_{\Vert}}{P_{\Vert}^{0}}\right\rangle  &  =\frac{1}%
{4}\sqrt{\frac{\epsilon_{N+1}}{\mu_{N+1}}}\frac{\mathrm{n}_{j}^{2}}
{\mathrm{n}_{N+1}^{2}}%
\frac{3\mathrm{n}_{j}}{k_{j}^{2}\epsilon_{j}}(\frac{\mathrm{n}_{j}^{2}}
{\mathrm{n}_{N+1}^{2}})4\pi
\sum_{l}(2l+1)\Bigg\{\Bigg[\int_{j}(\frac{\mathrm{n}_{j}^{2}}
{\mathrm{n}_{N+1}^{2}})\Big\vert
\{\alpha_{l}\frac{\mathrm{d}}{\mathrm{d}r_{j}^{\prime}}[r_{j}^{\prime}%
j_{l}(k_{j}r_{j}^{\prime})]\nonumber\\
&-\beta_{l}\frac{\mathrm{d}}{\mathrm{d}r_{j}%
^{\prime}}[r_{j}^{\prime}h_{l}(k_{j}r_{j}^{\prime})]\}\Big\vert
^{2}
  +k_{j}^{2}r_{j}^{\prime2}\left\vert [\gamma_{l}j_{l}(k_{j}r_{j}^{\prime
})-\zeta_{l}h_{l}^{(1)}(k_{j}r_{j}^{\prime})]\right\vert ^{2}\mathrm{d}r_{j}^{\prime
}/\int_{j}\mathrm{d}^{3}r_{j}^{\prime}\Bigg]\Bigg\},\label{R042}\\
&  =\frac{1}{4}\sqrt{\frac{\epsilon_{N+1}}{\mu_{N+1}}}\frac{\mathrm{n}_{j}^{2}}%
{\mathrm{n}_{N+1}^{2}}\frac{3\mathrm{n}_{j}}{k_{j}^{2}\epsilon_{j}V_{j\text{shell}}}
4\pi\sum_{l}(\frac{\mathrm{n}_{j}^{2}}{\mathrm{n}_{N+1}^{2}})I_{l}^{(2)}+I_{l}^{(3)},\\
I_{l}^{(2)}  &  =(2l+1)\int_{j\text{shell}}\left\vert \{\alpha_{l}%
\frac{\mathrm{d}}{\mathrm{d}r_{j}^{\prime}}[r_{j}^{\prime}j_{l}(k_{j}%
r_{j}^{\prime})]-\beta_{l}\frac{\mathrm{d}}{\mathrm{d}r_{j}^{\prime}}%
[r_{j}^{\prime}h_{l}(k_{j}r_{j}^{\prime})]\}\right\vert ^{2}\mathrm{d}r_{j}^{\prime},\\
I_{l}^{(3)}  &  =(2l+1)\int_{j\text{shell}}k_{j}^{2}r_{j}^{\prime2}\left\vert
[\gamma_{l}j_{l}(k_{j}r_{j}^{\prime})-\zeta_{l}h_{l}^{(1)}(k_{j}r_{j}^{\prime
})]\right\vert ^{2}\mathrm{d}r_{j}^{\prime}.
\end{align}
The total emitted  power then takes the form
\begin{align}
\left\langle \frac{P_{\text{total}}}{P^{0}}\right\rangle  &  =\frac{1}{3}
\left\langle\frac{P_{\bot}}{P_{\bot}^{0}}\right\rangle +\frac{2}{3}\left\langle
\frac{P_{\Vert}}{P_{\Vert}^{0}}\right\rangle \nonumber\\
&  =\frac{1}{2}\sqrt{\frac{\epsilon_{N+1}}{\mu_{N+1}}}\frac{\mathrm{n}_{j}^{2}}%
{\mathrm{n}_{N+1}^{2}}\frac{\mathrm{n}_{j}}{k_{j}^{2}\epsilon_{j}V_{j\text{shell}}}
4\pi\sum_{l}\left[  (\frac{\mathrm{n}_{j}^{2}}{\mathrm{n}_{N+1}^{2}})l(l+1)
I_{l}^{(1)}+(\frac{\mathrm{n}_{j}^{2}}{\mathrm{n}_{N+1}^{2}})I_{l}^{(2)}+I_{l}^{(3)}
\right].
\label{R043}
\end{align}
While this result exactly matches Eq.~(\ref{Ptot}), the computation of the 
integrals may be carried out prior to numerical calculation, using the 
properties of the Bessel and Hankel functions as follows. 

Let the following 
functional form be defined, which relates the integral over the Bessel 
and/or Hankel functions to a sequence of terms and their derivatives 
\begin{equation}
\Psi\lbrack p_{l}q_{l}](z)=\int p_{l}(z)q_{l}(z)\mathrm{d}z
=\text{constant+}\frac{1}
{2}\Big\{[z-\frac{l(l+1)}{z}]p_{l}q_{l}-\frac{1}{2}(p_{l}q_{l}^{\prime}
+p_{l}^{\prime}q_{l})+zp_{l}^{\prime}q_{l}^{\prime}\Big\}. \label{R044}
\end{equation}
Here, $p_{l}(z)$ and $q_{l}(z)$ can be any of the Riccati-Bessel or 
Riccati-Hankel functions, $\psi(x)=xj_{l}(x)$ or $\chi(x)=xh_{l}^{(1)}(x)$. 
One may then expand the integrals appearing in Eq.~(\ref{R043}), beginning 
with the contributions to the TM modes
\begin{align}
\left[  l(l+1)I_{l}^{(1)}+I_{l}^{(2)}\right]   &=\frac{1}{k_{j}}\Big\{
%
%T1
|\alpha_{l}|^{2}(l+1)\big(\Psi\lbrack|\psi_{l-1}|^2](k_{j}r_{j})-\Psi\lbrack|
\psi_{l-1}|^2(k_{j}r_{j-1})\big)\nonumber\\
%
%T2
 &+|\alpha_{l}|^{2}l\big(\Psi\lbrack|\psi_{l+1}|^2](k_{j}r_{j})-\Psi\lbrack
|\psi_{l+1}|^2](k_{j}r_{j-1})\big)\nonumber\\
%
%T3
&  +|\beta_{l}|^{2}(l+1)\big(\Psi\lbrack|\chi_{l-1}|^2](k_{j}r_{j})-\Psi\lbrack
|\chi_{l-1}|^2](k_{j}r_{j-1})\big)\nonumber\\
%
%T4
&  +|\beta_{l}|^{2}l\big(\Psi\lbrack|\chi_{l+1}|^2](k_{j}r_{j})-\Psi\lbrack
|\chi_{l+1}|^2](k_{j}r_{j-1})\big)\nonumber\\
%
%T5
&  -(l+1)\Big(\alpha_{l}\beta_{l}^*\big(\Psi\lbrack\psi_{l-1}\chi_{l-1}^*](k_{j}%
r_{j})-\Psi\lbrack\psi_{l-1}\chi_{l-1}^*\big)(k_{j}r_{j-1})]\nonumber\\
%
%T5A
  &+\alpha_{l}^*\beta_{l}\big(\Psi\lbrack\psi_{l-1}^*\chi_{l-1}](k_{j}%
r_{j})-\Psi\lbrack\psi_{l-1}^*\chi_{l-1}](k_{j}r_{j-1})\big)\Big)\nonumber\\
%
%T6
&  -l\Big(\alpha_{l}\beta_{l}^*\big(\Psi\lbrack\psi_{l+1}\chi_{l+1}^*](k_{j}r_{j}%
)-\Psi\lbrack\psi_{l+1}\chi_{l+1}^*](k_{j}r_{j-1})\big)\nonumber\\
%
%T6A
  &+\alpha_{l}^*\beta_{l}\big(\Psi\lbrack\psi_{l+1}^*\chi_{l+1}](k_{j}r_{j}%
)-\Psi\lbrack\psi_{l+1}^*\chi_{l+1}](k_{j}r_{j-1})\big)\Big)
\Big\},
\label{R045}
\end{align}
where $r_{j}$ and $r_{j-1}$ are the radii of the upper and lower interfaces
of the region $j$, respectively. 
The contributions to the TE modes may be calculated in a similar fashion 
\begin{align}
I_{l}^{(3)} &=(2l+1)\int_{j\text{shell}}k_{j}^{2}r_{j}^{\prime2}\left\vert
[\gamma_{l}j_{l}(k_{j}r_{j}^{\prime})-\zeta_{l}h_{l}^{(1)}(k_{j}r_{j}^{\prime
})]\right\vert ^{2}\mathrm{d}r_{j}^{\prime}\\
&=
\frac{1}{k_{j}}(2l+1)\Big\{
%
%T7
|\gamma_{l}|^{2}\big(\Psi\lbrack|\psi_{l}|^2%
](k_{j}r_{j})-\Psi\lbrack|\psi_{l}|^{2}](k_{j}r_{j-1})\big)\nonumber\\
%
%T8
&  +|\zeta_{l}|^{2}\big(\Psi\lbrack|\chi_{l}|^{2}](k_{j}r_{j})-\Psi\lbrack|\chi_{l}|%
^{2}](k_{j}r_{j-1})\big)\nonumber\\
%
%T9
&  -\Big(\gamma_{l}\zeta_{l}^*\big(\Psi\lbrack\psi_{l}\chi_{l}^*](k_{j}r_{j})-\Psi
\lbrack\psi_{l}\chi_{l}^*](k_{j}r_{j-1})\big)\nonumber\\
%
%T9A
&  +\gamma_{l}^*\zeta_{l}\big(\Psi\lbrack\psi_{l}^*\chi_{l}](k_{j}r_{j})-\Psi
\lbrack\psi_{l}^*\chi_{l}](k_{j}r_{j-1})\big)\Big)
\Big\}\label{R046}.
\end{align}
With $\left[  l(l+1)I_{l}^{(1)}+I_{l}^{(2)}\right]  $ and $I_{l}^{(3)}$ now known,
they may be substituted into  Eq.~(\ref{R043}) in order to calculate the 
total averaged power for an uniform distribution of dipoles in one of the 
layers of a multilayer microsphere. This result is computed numerically, 
and the results are presented in Chapter~\ref{chpt:mod} and 
Appendix~\ref{chpt:app3}.

\chapter{\cdb Transfer Matrix Method Verification}
\label{chpt:app3}

The active multilayer microsphere model introduced in Chapter~\ref{chpt:mod} 
has been shown to provide powerful functionality and guidance 
in examining the properties of whispering gallery modes within resonators. 

While the highlights of the results are explained in the body of the thesis, 
it is important, for due diligence, to outline exactly how the multilayer model 
has been verified with extant models in the literature. 
This is achieved by selecting the special cases that correspond to each of the 
models mentioned throughout the thesis, and described in this appendix. 
Since the multilayer model is more general than these 
previous models, prompting its development as a powerful tool for examining a 
number of resonator designs not easily studied thus far, the comparisons 
between it and extant models predominantly involves a simplification of the 
multilayer model, followed by a 
mathematical check to ensure that identical results are produced.

\section{\cdb Chew model for an unlayered microsphere}
\label{sec:chew}

Consider the Chew model, from which one is able to calculate 
the emitted power of an unlayered microsphere in a dielectric medium 
\cite{PhysRevA.13.396,Chew:87a,Chew:88}. The first 
scenario of interest is the case where a dipole source is placed in the 
outermost region. 

\subsection{\cdb Single dipole in the outermost region}

Assume that a dipole source is located in the outermost region, 
which corresponds to the choice $j=N+1=2$, in the multilayer model. 
Then, the vector $\mathbfcal{D}$ defined in Eq.~(\ref{R020}) simplifies to 
\begin{equation}
\mathbfcal{D}=-T^{2}(2,2)\mathbf{a}_{2L}=-I_{4\times4}\,\mathbf{a}_{2L}. 
\label{AC06}%
\end{equation}
Using Eq.~(\ref{R032}) and (\ref{R0321}) from Appendix~\ref{chpt:app2}, one 
can then find the field coefficients 
$B_{2}+a_{2EH}(r_j^{\prime})$ and $D_{2}+a_{2MH}(r_j^{\prime})$ in this scenario, 
which take the forms
\begin{align}
B_{2}+a_{2EH}(r_{2}^{\prime})  &  =\alpha_{l}a_{2EH}-\beta_{l}%
a_{2EL};\text{ where }\alpha_{l}=1\text{ and }\beta_{l}=(T_{21}^{2}%
+\frac{S_{21}}{S_{22}}T_{11}^{2})=\frac{S_{21}}{S_{22}},\label{AC072}\\
D_{2}+a_{2MH}(r_{2}^{\prime})  &  =\gamma_{l}a_{2MH}-\zeta_{l}%
a_{2ML};\text{ where }\gamma_{l}=1\text{ and }\zeta_{l}=(T_{43}+\frac{S_{43}%
}{S_{44}}T_{33})=\frac{S_{43}}{S_{44}}. \label{AC082}%
\end{align}
Having identified the coefficients $\alpha_{l}$, $\beta_{l}$, $\gamma_{l}$ and 
$\zeta_{l}$, the normal and transverse components of the emitted power simplify 
to
\begin{align}
\frac{P_{\bot}}{P_{\bot}^{0}}  &  =\frac{3}{2}\sum_{l}(2l+1)l(l+1)\frac
{\left\vert \mathbf{\mathbf{[}}j_{l}(k_{2}r_{2}^{\prime})+\frac{S_{21}}%
{S_{22}}h_{l}^{(1)}(k_{2}r_{2}^{\prime})\mathbf{]}\right\vert ^{2}}{k_{2}%
^{2}r_{2}^{\prime2}},\label{AC09}\\
\frac{P_{\Vert}}{P_{\Vert}^{0}}  &  =\frac{3}{4}\sum_{l}(2l+1)\Big\{  \Big[
\frac{\left\vert \{\frac{\mathrm{d}}{\mathrm{d}r_{2}^{\prime}}[r_{2}^{\prime
}j_{l}(k_{2}r_{2}^{\prime})]+\frac{S_{21}}{S_{22}}\frac{\mathrm{d}}%
{\mathrm{d}r_{2}^{\prime}}[r_{2}^{\prime}h_{l}^{(1)}(k_{2}r_{2}^{\prime
})]\}\right\vert ^{2}}{k_{2}^{2}r_{2}^{\prime2}}\nonumber\\
&+\left\vert [j_{l}(k_{2}%
r_{2}^{\prime})+\frac{S_{43}}{S_{44}}h_{l}^{(1)}(k_{2}r_{2}^{\prime
})]\right\vert ^{2}\Big]  \Big\}  . \label{AC010}%
\end{align}
Since $S$ is defined as $T^{-1}$, it follows that $S_{21}/S_{22}%
=-T_{21}/T_{11}$ and $S_{43}/S_{44}=-T_{43}/T_{33}$. 
In addition, the scattering matrix takes the form 
$T(2,1)=M_{2}^{-1}(r_{1})M_{1}(r_{1})$, and hence the continuity condition 
of Section~\ref{subsec:trans} can be used to find the ratios
\begin{align}
\frac{S_{21}}{S_{22}}  &  =-\frac{[\mathrm{n}_{1}\mu_{2}\psi_{l}^{\prime}(k_{2}%
r_{1})\psi_{l}(k_{1}r_{1})-\mathrm{n}_{2}\mu_{1}\psi_{l}(k_{2}r_{1})\psi_{l}^{\prime
}(k_{1}r_{1})]}{[\mathrm{n}_{2}\mu_{1}\psi_{l}^{\prime}(k_{1}r_{1})\chi_{l}(k_{2}%
r_{1})-\mathrm{n}_{1}\mu_{2}\psi_{l}(k_{1}r_{1})\chi_{l}^{\prime}(k_{2}r_{1}%
)]}\label{AC011}\\
&  =\frac{[(\epsilon_{2}/\mathrm{n}_{2})\psi_{l}(k_{2}r_{1})\psi_{l}^{\prime}%
(k_{1}r_{1})-(\epsilon_{1}/\mathrm{n}_{1})\psi_{l}^{\prime}(k_{2}r_{1})\psi
_{l}(k_{1}r_{1})]}{[(\epsilon_{2}/\mathrm{n}_{2})\psi_{l}^{\prime}(k_{1}r_{1}%
)\chi_{l}(k_{2}r_{1})-(\epsilon_{1}/\mathrm{n}_{1})\psi_{l}(k_{1}r_{1})\chi
_{l}^{\prime}(k_{2}r_{1})]},\\
\frac{S_{43}}{S_{44}}  &  =-\frac{\mathrm{n}_{2}\mu_{1}\psi_{l}^{\prime}(k_{2}%
r_{1})\psi_{l}(k_{1}r_{1})-\mathrm{n}_{1}\mu_{2}\psi_{l}(k_{2}r_{1})\psi_{l}^{\prime
}(k_{1}r_{1})}{\mathrm{n}_{1}\mu_{2}\psi_{l}^{\prime}(k_{1}r_{1})\chi_{l}(k_{2}%
r_{1})-\mathrm{n}_{2}\mu_{1}\psi_{l}(k_{1}r_{1})\chi_{l}^{\prime}(k_{2}r_{1}%
)}\label{AC012}\\
&  =\frac{(\epsilon_{1}/\mathrm{n}_{1})\psi_{l}(k_{2}r_{1})\psi_{l}^{\prime}%
(k_{1}r_{1})-(\epsilon_{2}/\mathrm{n}_{2})\psi_{l}^{\prime}(k_{2}r_{1})\psi
_{l}(k_{1}r_{1})}{(\epsilon_{1}/\mathrm{n}_{1})\psi_{l}^{\prime}(k_{1}r_{1})\chi
_{l}(k_{2}r_{1})-(\epsilon_{2}/\mathrm{n}_{2})\psi_{l}(k_{1}r_{1})\chi_{l}^{\prime
}(k_{2}r_{1})}.%
\end{align}
Having found the coefficients $S_{21}/S_{22}$ and $S_{43}/S_{44}$, 
one may directly calculate $P_{\bot}/P_{\bot}^{0}$ and $P_{\Vert}/P_{\Vert}^{0}$. 
It is found that Eqs.~(\ref{AC011}) and (\ref{AC012}) are identical to
 Eqs.~(6) and (7) of Ref.~\cite{Chew:87a}.

\subsection{\cdb Single dipole in the innermost region}

Now assume a dipole is in the innermost region, then $j=1$, 
and from Eqs.~(\ref{R013}), (\ref{R0201}) and (\ref{R020}) it is found that
\begin{equation}
\mathbfcal{D}=T(2,1)\mathbf{a}_{jH}=\left(  T_{12}a_{1EH},T_{22}a_{1EH}
,T_{34}a_{1MH},T_{44}a_{1MH}\right). 
\end{equation}
Using Eq.~(\ref{R032}) and
(\ref{R0321}), the coefficients $B_{2}+a_{2EH}(r_{1}^{\prime})$ and
$D_{2}+a_{2MH}(r_{1}^{\prime})$ can be found
\begin{align}
B_{N+1}+a_{N+1EH}(r_{2}^{\prime})  &  =(T_{22}+\frac{S_{21}}{S_{22}%
}T_{12})a_{jEH}\label{AC013}\\
&  =\alpha_{l}a_{_{1}EH}-\beta_{l}a_{1EL};\text{ where }\alpha_{l}%
=(T_{22}+\frac{S_{21}}{S_{22}}T_{12})\text{ and }\beta_{l}=0,
\end{align}
and similarly
\begin{align}
D_{N+1}+a_{N+1MH}(r_{2}^{\prime})  &  =(T_{44}+\frac{S_{43}}{S_{44}%
}T_{34})a_{1MH}\label{AC014}\\
&  =\gamma_{l}a_{1MH}-\zeta_{l}a_{1ML};\text{ where }\gamma_{l}=(T_{44}%
+\frac{S_{43}}{S_{44}}T_{34})\text{ and }\zeta_{l}=0.
\end{align}

Now the forms of the scattering matrix, $T$, and its inverse, $S$, 
can be specified. 
According to Eq.~(\ref{R020}), $\ S=T^{-1}(2,1)$, and hence $(S_{22}T_{22}%
+S_{21}T_{12})=(S_{44}T_{44}+S_{43}T_{34})=1$. As a result, the following 
simple forms for the coefficients can be found, $\alpha
_{l}=1/S_{22}$ and $\gamma_{l}=1/S_{44}$. Furthermore, since
$S=T^{-1}(2,1)$, $S_{22}=\det(T^{\text{TM}})^{-1}%
T_{11}$ and $S_{44}=\det(T^{\text{TE}})^{-1}T_{33}$, and hence Eqs.~(\ref{R0191}) 
and (\ref{R0192}) can be used to find the matrix element corresponding 
to the TM modes
\begin{align}
S_{22}  &  =[\det(M_{2})^{-1}\det(M_{1})]^{-1}\!\!\frac{\mathrm{i}}{\mathrm{n}_{2}
\mu_{1}}\frac{k_{2}}{k_{1}}\frac{\mathrm{n}_{2}}{\mathrm{n}_{1}}[\mathrm{n}_{2}
\mu_{1}\psi_{l}^{\prime}%
(k_{1}r_{1})\chi_{l}(k_{2}r_{1})-\mathrm{n}_{1}\mu_{2}\psi_{l}(k_{1}r_{1})\chi
_{l}^{\prime}(k_{2}r_{1})],\text{ }\label{AC015}\\
&  =-\mathrm{i}\frac{k_{1}\mathrm{n}_{1}}{\epsilon_{2}\epsilon_{1}}\sqrt{
\frac{\epsilon
_{2}}{\mu_{2}}}\frac{\mathrm{n}_{1}}{\mathrm{n}_{2}}[\epsilon_{1}r_{1}j_{l}
(k_{1}r_{1})\chi^{\prime
}(k_{2}r_{1})-\epsilon_{2}\psi_{l}^{\prime}(k_{1}r_{1})rh_{l}(k_{2}r_{1})].
\end{align}
In the same manner, 
the matrix element corresponding to the TE modes takes the form
\begin{align}
S_{44}  &  =[\det(M_{2})^{-1}\det(M_{1})]^{-1}\frac{\mathrm{i}}{\mathrm{n}_{2}
\mu_{1}}\frac{k_{2}}{k_{1}}[\mathrm{n}_{1}\mu_{2}\psi_{l}^{\prime}(k_{1}r_{1})\chi_{l}%
(k_{2}r_{1})-\mathrm{n}_{2}\mu_{1}\psi_{l}(k_{1}r_{1})\chi_{l}^{\prime}(k_{2}%
r_{1})],\label{AC016}\\
&  =\mathrm{i}(\frac{\mathrm{n}_{1}}{\mu_{2}})[\frac{\mathrm{n}_{2}}
{\epsilon_{2}}\psi_{l}^{\prime
}(k_{1}r_{1})\chi_{l}(k_{2}r_{1})-\frac{\mathrm{n}_{1}}{\epsilon_{1}}\psi_{l}%
(k_{1}r_{1})\chi_{l}^{\prime}(k_{2}r_{1})].
\end{align}
Now that the TE and TM contributions have been simplified from the 
general transfer matrix approach to a a form composed of Bessel and Hankel 
functions and their derivatives, 
the formulae for the normalised emitted powers ${P_{\bot}}/{P_{\bot}^{0}}$ 
and ${P_{\Vert}}/{P_{\Vert}^{0}}$ may be written
\begin{align}
\frac{P_{\bot}}{P_{\bot}^{0}}  &  =\frac{1}{2}\sqrt{\frac{\epsilon_{2}}%
{\mu_{2}}}\frac{3\mathrm{n}_{1}\epsilon_{1}}{(k_{1}r_{1})^{2}}\sum_{l}(2l+1)l(l+1)
\frac{\left\vert j_{l}(k_{1}r_{1}^{\prime})\right\vert ^{2}}{k_{1}^{2}r_{1}%
^{\prime2}\left\vert D_{l}\right\vert ^{2}}, \label{AC018} \\
\frac{P_{\Vert}}{P_{\Vert}^{0}}  &  =\frac{1}{4}\sqrt{\frac{\epsilon_{2}}%
{\mu_{2}}}\frac{3\mathrm{n}_{1}\epsilon_{1}}{(k_{1}r_{1})^{2}}\sum_{l}(2l+1)\left\{
\frac{\left\vert \frac{\mathrm{d}}{\mathrm{d}r_{1}^{\prime}}[r_j^{\prime}%
j_{l}(k_{1}r_{1}^{\prime})]\right\vert ^{2}}{k_{1}^{2}r_{1}^{\prime
2}\left\vert D_{l}\right\vert ^{2}}+\frac{\mu_{1}\mu_{2}}{\epsilon_{2}%
\epsilon_{1}}\frac{\left\vert j_{l}(k_{1}r_{1}^{\prime})\right\vert ^{2}%
}{\left\vert D_{l}^{\prime}\right\vert ^{2}}\right\}, \label{AC017} \\
\text{for\,\,}D_{l}&\equiv[\epsilon_{1}j_{l}(k_{1}r_{1})\chi_{l}^{^{\prime}}(k_{2}%
r_{1})-\epsilon_{1}\psi_{l}^{^{\prime}}(k_{1}r_{1})h_{l}(k_{2}r_{1})],
\text{\,\,and\,\,} D_{l}^{\prime}\equiv D_{l}(\epsilon\leftrightarrow\mu).
\label{AC020}%
\end{align}
These equations exactly match 
Eqs.~(1), (2) and (3) of Ref.~\cite{Chew:88}, respectively.

\subsection{\cdb Active inner regions}

\begin{figure}[H]
\hspace{-2mm}
\begin{center}
\includegraphics[width=0.8\hsize,angle=0]{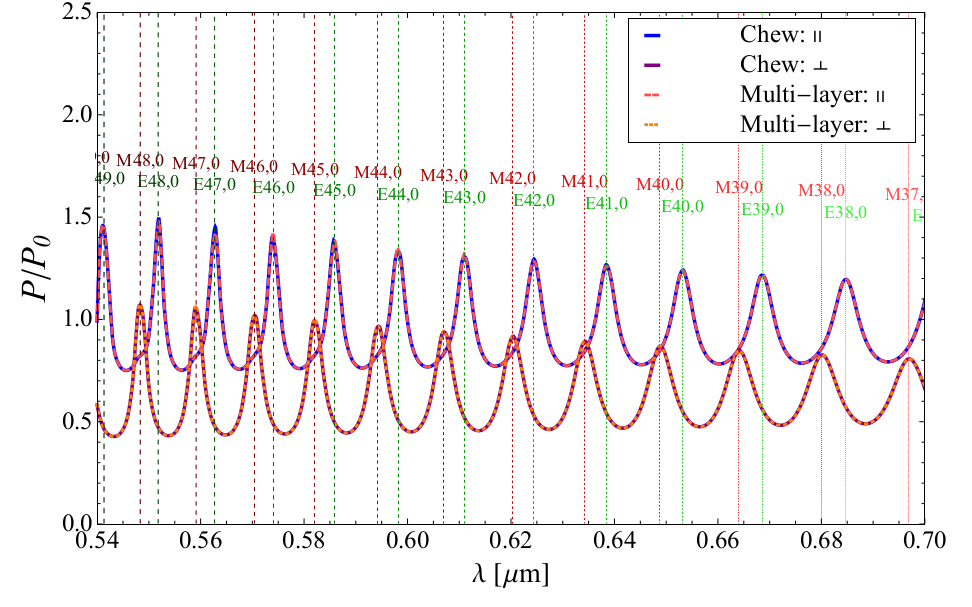}\\
\vspace{-3mm}
\hspace{3mm}\mbox{(a)}\\
\vspace{2mm}
\includegraphics[width=0.8\hsize,angle=0]{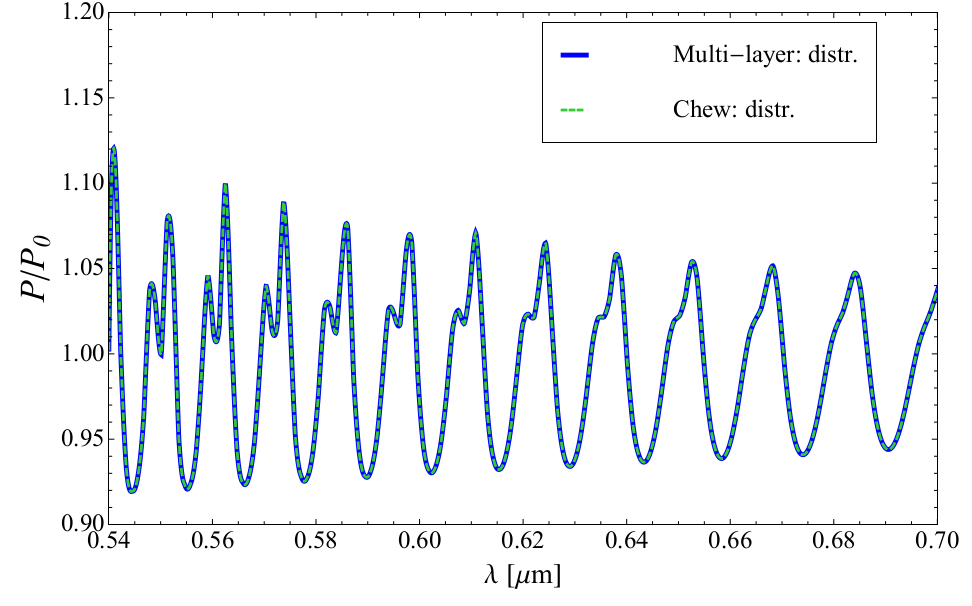}\\
\vspace{-3mm}
\hspace{5mm}\mbox{(b)}\\
\end{center}
\vspace{-2mm}\caption[The multilayer model convergences to the Chew model for 
$N=1$.]{{\protect{Demonstration that themultilayer model reproduces the 
microsphere model results for an example $D=6$ $\mu$m,
$\mathrm{n}_{1}=1.59$, and $\mathrm{n}_{2}=1.33$. (a) Spectra for single-dipole 
excitation in both tangential  and radial orientations.
(b) Spectra for a uniform distribution of dipoles are also shown.}}}
\label{fig:match}
\figrule
\end{figure}

Assume that the inner region is now filled with randomly oriented
dipoles. Then, Eqs.~(\ref{R043}), (\ref{R045}), and (\ref{R046}) may be 
used, 
together with $\alpha_{l}=\frac{1}{S_{22}}$, $\beta_{l}=0$, $\zeta=0$, and
$\gamma_{l}=\frac{1}{S_{44}}$, calculated in the previous section, to simplify
the form of the integral components of the TM modes
\begin{equation}
\frac{1}{(2l+1)}\left[  l(l+1)I_{l}^{(1)}+I_{l}^{(2)}\right]  =\frac
{\alpha_{l}^{2}}{(2l+1)k_{1}}\{(l+1)[\Psi\lbrack\psi_{l-1}^{2}](k_{1}%
r_{1})+l^{2}[\Psi\lbrack\psi_{l+1}^{2}](k_{1}r_{1})\}. \label{AC022}%
\end{equation}
Note that in the case of a microsphere, $r_0$ is set to zero, 
as well as all functionals of the form
$\Psi\lbrack\cdot](k_{1}r_{0})$ from Eq.~(\ref{R044}). 
Similarly, for the TE modes
\begin{equation}
I_{l}^{(3)}=\frac{1}{k_{1}}(2l+1)\gamma_{l}^{2}\Psi\lbrack\psi_{l}^{2}%
](k_{1}r_{1}). \label{AC023}%
\end{equation}
Possessing known 
forms for $\frac{1}{(2l+1)}\left[  l(l+1)I_{l}^{(1)}+I_{l}^{(2)}\right]$ and
$I_{l}^{(3)}$, and substituting into the equation 
\begin{equation}
\left\langle \frac{P_{\text{total}}}{P^{0}}\right\rangle 
  =\frac{3}{2}\sqrt{\frac{\epsilon_{2}}{\mu_{2}}}\frac{\mathrm{n}_{1}^{2}}
{\mathrm{n}_{2}^{2}%
}\frac{\mathrm{n}_{1}}{k_{1}^{2}\epsilon_{1}r_{1}^{3}}\sum_{l}\left[  l(l+1)I_{l}%
^{(1)}+I_{l}^{(2)}+I_{l}^{(3)}\right], 
\label{AC024}
\end{equation}
one may check to see if this result is consistent with the Chew model 
through numerical analysis. 
A numerical comparison of both the single dipole of 
both normal and transverse polarisations, and a uniform layer of dipoles, 
is shown in Fig.~\ref{fig:match}. 
The multilayer model for a single layer matches the 
microsphere case for a vanishingly small layer coating, or
 a vanishingly small internal sphere size. 
Furthermore, it is found that both of these limits converge to within 
numerical precision.

\section{\cdb Johnson model for an unlayered microsphere}
\label{sec:john}

Recall the Johnson model \cite{Johnson:93} for the mode positions of a 
microsphere. Setting $N=1$, and then using Eq.~(\ref{R0201}), the scattering 
matrix takes the form
\begin{equation}
T(2,1)=M_{2}^{-1}(r_{1})M_{1}(r_{1}). 
\label{AC01}
\end{equation}
The elements of the scattering matrix can then be extracted
\begin{align}
T_{11}  &  =\frac{\mathrm{i}}{\mathrm{n}_{2}\mu_{1}}G_{11}=\frac{\mathrm{i}}
{\mathrm{n}_{2}\mu_{1}}\frac{k_{2}%
}{k_{1}}\frac{\mathrm{n}_{2}}{\mathrm{n}_{1}}[\mathrm{n}_{2}\mu_{1}\psi_{l}^{\prime}
(k_{1}r_{1})\chi
_{l}(k_{2}r_{1})-\mathrm{n}_{1}\mu_{2}\psi_{l}(k_{1}r_{1})\chi_{l}^{\prime}(k_{2}%
r_{1})],\label{AC02}\\
T_{33}  &  =\frac{\mathrm{i}}{\mathrm{n}_{2}\mu_{1}}\frac{k_{2}}{k_{1}}G_{33}=
\frac{\mathrm{i}}{\mathrm{n}_{2}%
\mu_{1}}\frac{k_{2}}{k_{1}}[\mathrm{n}_{1}\mu_{2}\psi_{l}^{\prime}(k_{1}r_{1})\chi
_{l}(k_{2}r_{1})-\mathrm{n}_{2}\mu_{1}\psi_{l}(k_{1}r_{1})\chi_{l}^{\prime}(k_{2}%
r_{1})]. \label{AC03}%
\end{align}
By making the assumption that $\mu_1=\mu_2$, which is a required condition in 
the derivation of the Johnson model, 
the TM and TE resonance conditions, $T_{11}=0$ and $T_{33}=0$, 
lead to the following characteristic equations
\begin{align}
\mathrm{n}_{2}\frac{\psi_{l}^{^{\prime}}(k_{1}r_{1})}{\psi_{l}(k_{1}r_{1})}  &
=\mathrm{n}_{1}\frac{\chi_{l}^{^{\prime}}(k_{2}r_{1})}{\chi_{l}(k_{2}r_{1})}\text{ for TM
resonances,}\label{AC04}\\
\mathrm{n}_{1}\frac{\psi_{l}^{^{\prime}}(k_{1}r_{1})}{\psi_{l}(k_{1}r_{1})}  &
=\mathrm{n}_{2}\frac{\chi_{l}^{^{\prime}}(k_{2}r_{1})}{\chi_{l}(k_{2}r_{1})}\text{ for TE
resonances.} \label{AC05}%
\end{align}

\newpage
\noindent Equations.~(\ref{AC04}) and (\ref{AC05}) exactly match Eqs.~(19) and 
(13) in \cite{Teraoka:06a}, respectively.  
Furthermore, these equations are consistent 
with the formulae for the Mie scattering coefficients as stated in  
Eq.~(4.53) of Ref.~\cite{BohrenHuffman}. 
It should be noted that in Eq.~(33) of Ref.~\cite{Moroz05} 
the labels of the TE and TM modes are interchanged.

\section{\cdb Teraoka-Arnold model for a microsphere with a single layer}
\label{sec:TA}

In the case of a microsphere coated by a single layer, $N=2$, and 
the following form for the scattering matrix, $T(3,1)$, is found 
\begin{equation}
T(3,1)=M_{3}^{-1}(r_{2})M_{2}(r_{2})M_{2}^{-1}(r_{1})M_{1}(r_{1})=\frac{\mathrm{i}k_{3}
}{\mathrm{n}_{2}\mu_{2}}\frac{\mathrm{i}k_{2}}{\mathrm{n}_{1}\mu_{1}}G(3,2)G(2,1). 
\label{AC025}
\end{equation}
One can find the TM and TE resonances of this coated microsphere
 by setting $T_{11}=0$ and $T_{33}=0$, respectively
\begin{align}
T_{11}  &  =\frac{\mathrm{i}k_{3}}
{\mathrm{n}_{2}\mu_{2}}\frac{\mathrm{i}k_{2}}{\mathrm{n}_{1}\mu_{1}}%
[G_{11}(3,2)G_{11}(2,1)+G_{12}(3,2)G_{21}(2,1)]=0,\label{AC026}\\
T_{33}  &  =\frac{\mathrm{i}k_{3}}
{\mathrm{n}_{2}\mu_{2}}\frac{\mathrm{i}k_{2}}{\mathrm{n}_{1}\mu_{1}}%
[G_{33}(3,2)G_{33}(2,1)+G_{34}(3,2)G_{43}(2,1)]=0. \label{AC027}%
\end{align}
This results in the following characteristic equations for the resonance 
positions 
\begin{align}
\frac{\mu_{2}\mathrm{n}_{3}\chi_{l}(k_{3}r_{2})}{\mathrm{n}_{2}\mu_{3}\chi_{l}^{\prime}%
(k_{3}r_{2})}&=\frac{\frac{C}{D}\psi_{l}(k_{2}r_{2})+\chi_{l}(k_{2}r_{2}%
)}{\frac{C}{D}\psi_{l}^{\prime}(k_{2}r_{2})+\chi_{l}^{\prime}(k_{2}r_{2}%
)}\text{ (TM), 
} \label{AC028} \\
\frac{\mathrm{n}_{2}\mu_{3}\chi_{l}(k_{3}r_{2})}{\mathrm{n}_{3}\mu_{2}\chi_{l}^{\prime}%
(k_{3}r_{2})}&=\frac{\frac{E}{F}\psi_{l}(k_{2}r_{2})+\chi_{l}(k_{2}r_{2}%
)}{\frac{E}{F}\psi_{l}^{\prime}(k_{2}r_{2})+\chi_{l}^{\prime}(k_{2}r_{2}%
)}\text{ \ (TE), 
} \label{AC029}
\end{align}
where
\begin{align}
\frac{C}{D}  &  =\frac{[\mathrm{n}_{2}\mu_{1}\psi_{l}^{\prime}(k_{1}r_{1})\chi
_{l}(k_{2}r_{1})-\mathrm{n}_{1}\mu_{2}\psi_{l}(k_{1}r_{1})\chi_{l}^{\prime}(k_{2}
r_{1})]}{[\mathrm{n}_{1}\mu_{2}\psi_{l}^{\prime}(k_{2}r_{1})\psi_{l}(k_{1}r_{1}
)-\mathrm{n}_{2}\mu_{1}\psi_{l}(k_{2}r_{1})\psi_{l}^{\prime}(k_{1}r_{1})]}
\label{AC030}\\
\frac{E}{F}  &  =\frac{[\mathrm{n}_{1}\mu_{2}\psi_{l}^{\prime}(k_{1}r_{1})\chi
_{l}(k_{2}r_{1})-\mathrm{n}_{2}\mu_{1}\psi_{l}(k_{1}r_{1})\chi_{l}^{\prime}(k_{2}
r_{1})]}{[\mathrm{n}_{2}\mu_{1}\psi_{l}^{\prime}(k_{2}r_{1})\psi_{l}(k_{1}r_{1}
)-\mathrm{n}_{1}\mu_{2}\psi_{l}(k_{2}r_{1})\psi_{l}^{\prime}(k_{1}r_{1})]}.
\label{AC031}
\end{align}
The characteristic equations in Eqs.~(\ref{AC028}) and (\ref{AC029}) 
exactly match Eq.~(7) in \cite{Teraoka:07} and Eq.~(10) in \cite{Teraoka:06}, 
respectively. 
Furthermore, a consistency check is carried out by comparing the mode positions 
of the spectrum and the behaviour of the positions as a function of layer 
thickness, for both the multilayer model and the Teraoka-Arnold model. 

\begin{figure}[H]
\begin{center}
\includegraphics[width=0.8\hsize,angle=0]{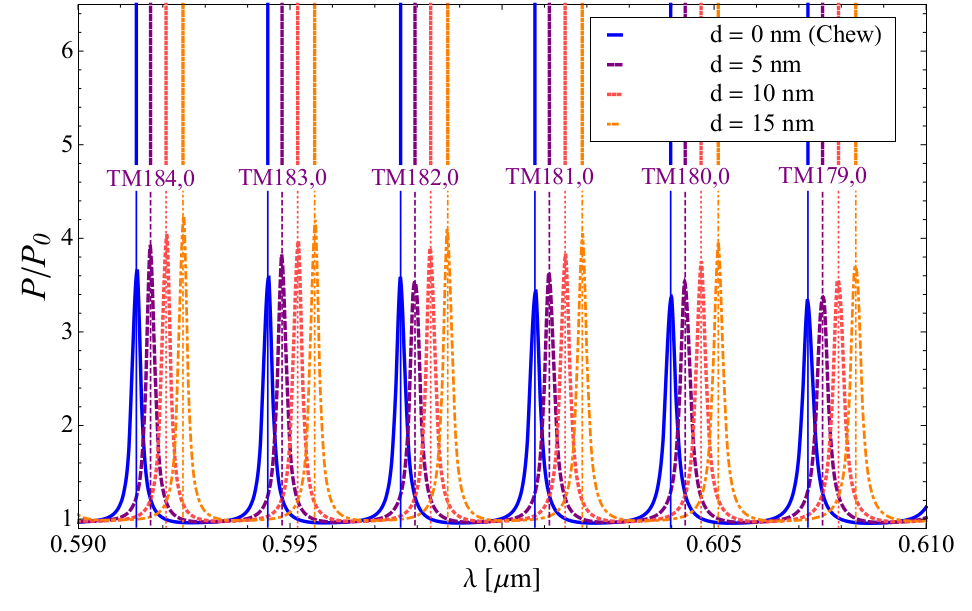}\\
\vspace{-5mm}
\hspace{3mm}\mbox{(a)}\\
\vspace{3mm}
\includegraphics[width=0.8\hsize,angle=0]{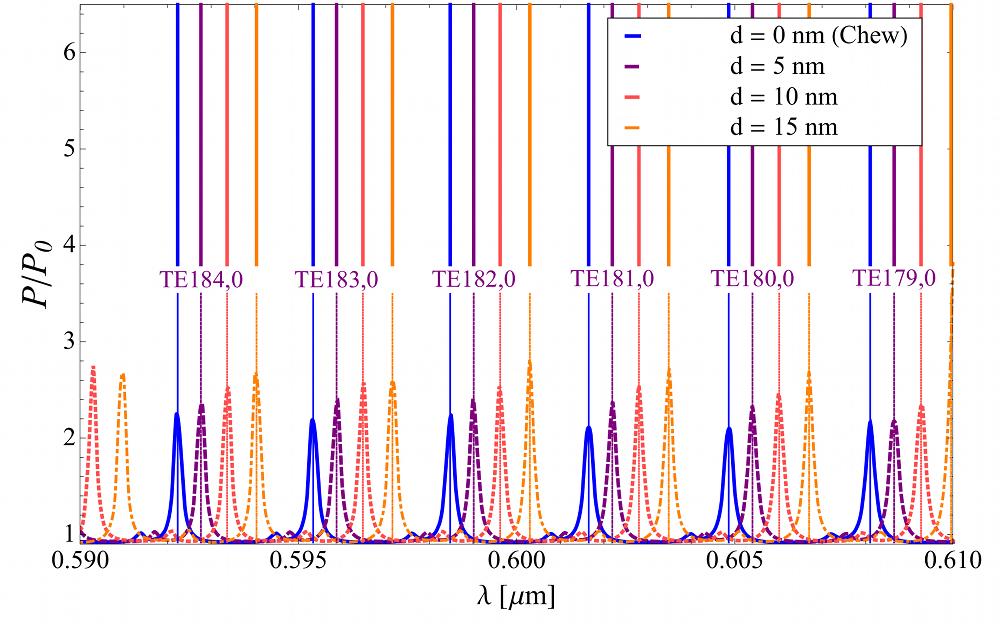}\\
\vspace{-5mm}
\hspace{1.5mm}\mbox{(b)}\\
\vspace{-8mm}
\end{center}
\caption[The multilayer model is verified by the Teraoka-Arnold model.]{{\protect{Spectra obtained from
a simulated silica microsphere, with dispersion, coated with a high refractive 
index layer ($\mathrm{n}_{2}=1.7$) with an outer diameter of $25$ $\mu$m, 
surrounded by water ($n_3=1.33$). A single electric dipole is oriented (a) 
in the radial direction and (b) in the tangential direction. The 
\textit{thin} vertical lines indicate the mode positions obtained from 
the multilayer model of Chapter~\ref{chpt:mod}, whereas the \textit{thick} 
vertical lines indicate the positions obtained from the 
Teraoka-Arnold model \cite{Teraoka:06,Teraoka:07}.}}}
\label{fig:arnold}
\figrule
\end{figure}

Consider an example of a silica microsphere, with dispersion included through 
the Sellmeier equation \cite{Malitson:65}, coated with a single high refractive 
index layer of the same value as in Ref.~\cite{Teraoka:06}, $n_{2}=1.7$. 
The microsphere is surrounded by water, $n_{3}=1.33$, and the thickness of the 
coating $d$ is changed from $5$ nm to $15$ nm. 
Figure~\ref{fig:arnold} shows the results obtained from the multilayer model 
for an electric dipole placed just outside the surface, with normal and 
transverse orientations considered separately. 
A range of wavelengths $590$ to $610$ nm is simulated, and the outer diameter 
is kept fixed at $25$ $\mu$m.  
The \textit{thin} vertical lines marking the position of the resonances, 
with their corresponding mode numbers and labels, are obtained from the 
structure resonance positions of Eqs.~(\ref{R0261}) and (\ref{R0262}) for a 
two-layer microsphere, setting $N=2$ in the general multilayer formalism. 
The \textit{thick} vertical lines indicate the resonance positions 
obtained from the Teraoka-Arnold model \cite{Teraoka:06,Teraoka:07}, which 
agree exactly. Note that in the case of the parallel excitation in 
Fig.~\ref{fig:arnold}(b), there is a small additional 
contribution from the TM mode, as expected from Eq.~(\ref{R040}). 
It is found that in both models there is a systematic shift in the 
prominent WGM peaks towards higher wavelengths as the thickness of the layer is 
increased. The free spectral range, however, remains largely unchanged over 
this range of wavelength values. In the limit $d\rightarrow 0$, the results 
match the simple case of the Chew model \cite{Chew:87a} as anticipated.

\newpage
\section{\cdb Yariv model of scattering for concentric resonators}

In the model that has been developed for spherical
concentric `onion' resonators   
\cite{Xu:03,Xu:04,Liang:04}, denoted the Yariv model, 
the transfer matrix approach is adopted, and therefore a comparison with the 
multilayer model introduced in Chapter~\ref{chpt:mod} becomes more 
straightforward. 
A crucial difference between the Yariv model and the other models discussed 
in this thesis is that, in this case, the modes are excited via an 
incident wave only, and the scattering cross section is considered, rather than 
the emitted power. Since the incident field takes the form \cite{Liang:04}
\begin{equation}
\mathbf{E}_{\text{inc}} = \mathbf{\hat{e}}\,e^{\mathrm{i} k r} 
= \sum_{l=1}^\infty \mathrm{i}^l \frac{\sqrt{4\pi (2l+1)}}{2}\Big[j_l(k r) 
(\mathbf{X}_{l,+1}+\mathbf{X}_{l,-1}) + \frac{1}{k}\bm{\nabla}\times j_l(k r) 
(\mathbf{X}_{l,+1}-\mathbf{X}_{l,-1})\Big], 
\end{equation}
it is clear that the quantum number $m$ takes only the values $\pm 1$. 
Therefore, there is no summation, as was performed 
in the case of Eqs.~(\ref{R036}) and 
(\ref{R037}). The form of the field coefficients in each layer, however, is 
consistent with the multilayer model, as can be seen in Eqs.~(7a) and (7b) of 
Ref.~\cite{Liang:04}
\begin{align}
\left[\begin{array}[c]{c}\mathbf{H}\\\mathbf{E}\end{array}\right]
&= 
\left[
\begin{array}
[c]{cc}
j_l(k_j r)\mathbf{X}_{lm}  & h_l^{(1)}(k_j r)\mathbf{X}_{lm} \\
Z_j \frac{\mathrm{i}}{k_j}\bm{\nabla}\times j_l(k_j r)\mathbf{X}_{lm} & 
Z_j \frac{\mathrm{i}}{k_j}\bm{\nabla}\times h_l^{(1)}(k_j r)\mathbf{X}_{lm}
\end{array}
\right]
\cdot
\left[\begin{array}[c]{c}A_j\\B_j\end{array}\right] \text{ (TM) },
\\
\left[\begin{array}[c]{c}\mathbf{E}\\\mathbf{H}\end{array}\right]
&= 
\left[
\begin{array}
[c]{cc}
j_l(k_j r)\mathbf{X}_{lm}  & h_l^{(1)}(k_j r)\mathbf{X}_{lm} \\
\frac{-\mathrm{i}}{Z_j k_j}\bm{\nabla}\times j_l(k_j r)\mathbf{X}_{lm} & 
\frac{-\mathrm{i}}{Z_j k_j}\bm{\nabla}\times h_l^{(1)}(k_j r)\mathbf{X}_{lm}
\end{array}
\right]
\cdot
\left[\begin{array}[c]{c}C_j\\D_j\end{array}\right] \text{ (TE) },
\end{align}
where $Z_j\equiv\sqrt{\mu_{N+1}/(\epsilon_{N+1} \epsilon_j)}$ is an impedance 
function, and $k = \sqrt{\epsilon_j}\omega/c$. 
By recognising that the definition of the transverse VSH, 
$\mathbf{X}_{lm}$ in Refs.~\cite{Liang:04,Jackson89}, is equivalent to 
the function $\bm{\Phi}_{lm}$ in Eq.~(\ref{B04}), 
the relations among the field coefficients and the values of the fields 
may be compared to Eqs~(\ref{R002}) and (\ref{R003}) in the multilayer model, 
and the forms are found to be consistent with each other.

\section{\cdb Algorithm scaling behaviour}
\label{sec:scal}

The precise implementation of the algorithm for the calculation of the emitted 
power is presented in the code online (see Appendix~\ref{sec:code}). 
In this version of the code, the summation, $\sum
_{l=1}^{\infty}$, is calculated to an upper bound, $l_{max}$, determined by a
prescribed tolerance $\tau$, so that
\begin{equation}
\frac{|(P/P^{0})_{l=l_{max+1}}-(P/P^{0})_{l=l_{max}}|}{(P/P^{0})_{l=l_{max+1}%
}} < \tau.
\end{equation}
This prescription is sufficient so long as the behavior of $P/P^{0}$ is
convergent, which is the case unless unstable parameter regions are selected 
\cite{Yang:03}. At each value of $l>1$, the spherical Bessel and Hankel 
functions are calculated using their recursion relations 
\cite{abramowitz1964handbook}, and function calls are minimised within 
the algorithm in order to
improve the efficiency of the calculation.

\begin{figure}[t]
\begin{center}
\includegraphics[width=0.65\hsize,angle=0]{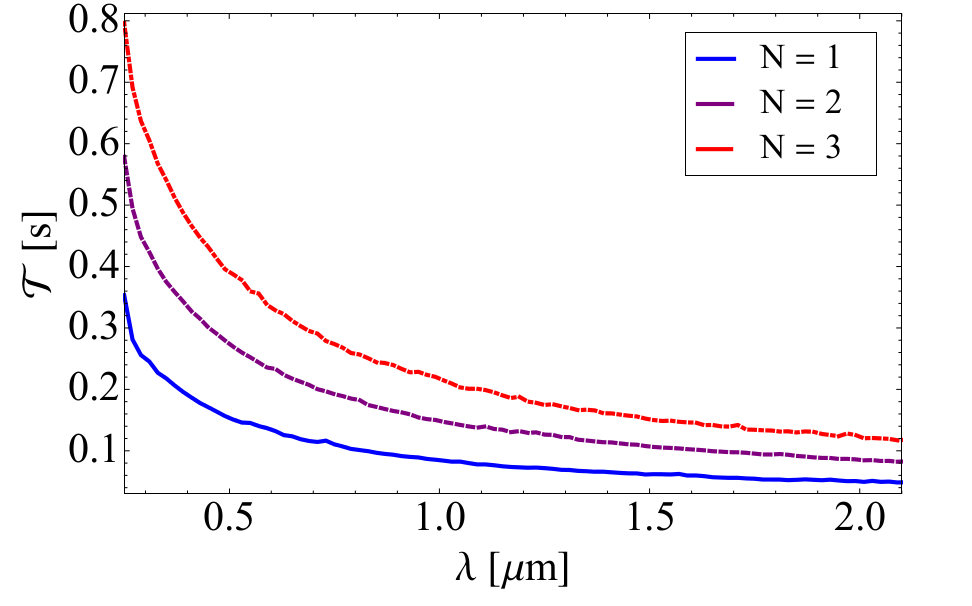}\\
\vspace{-3mm}
\hspace{3mm}\mbox{(a)}\\
\includegraphics[width=0.65\hsize,angle=0]{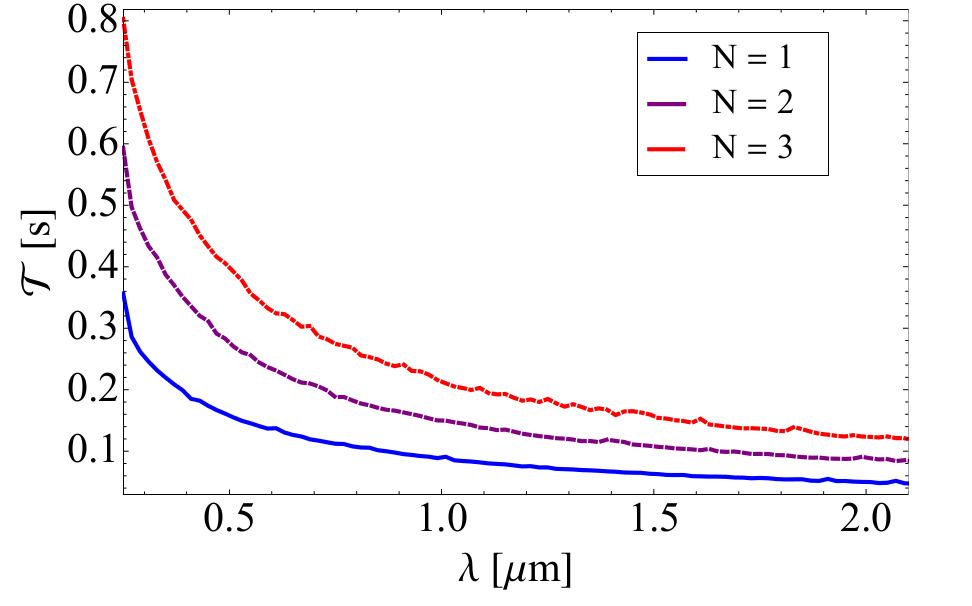}\\
\vspace{-3mm}
\hspace{3mm}\mbox{(b)}\\
\includegraphics[width=0.65\hsize,angle=0]{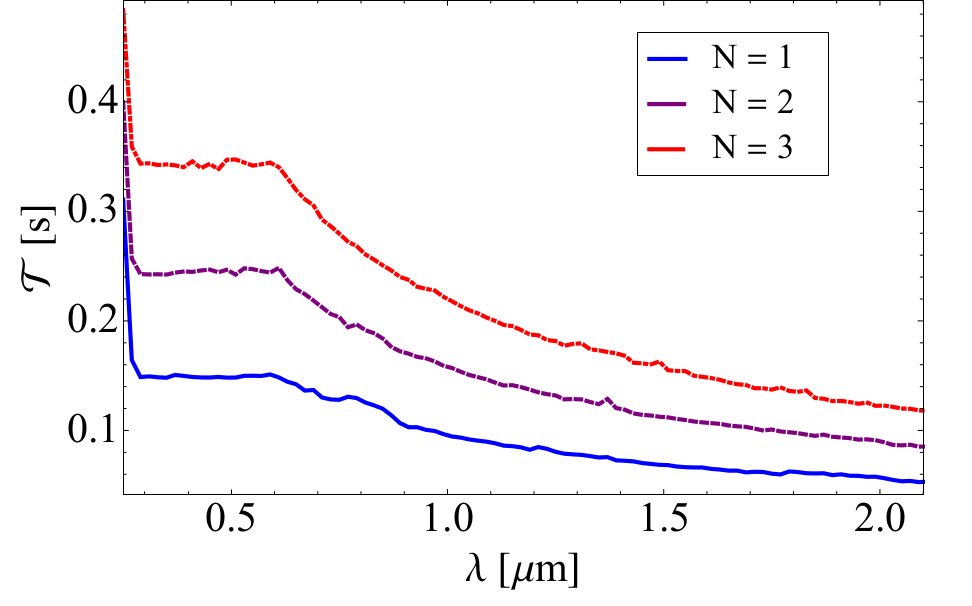}\\
\vspace{-3mm}
\hspace{3mm}\mbox{(c)}\\
\end{center}
\vspace{-6mm}\caption[The execution times of the multilayer model are examined.]
{{\protect{The execution time 
$\mathcal{T}$ of the formulae $P/P^{0}(\lambda)$ as a function of wavelength. 
The results are shown for numbers of layers $N=1$, $2$ and $3$. The results 
for a single-dipole excitation oriented (a) in the tangential direction and 
(b) in the radial direction are similar in magnitude. (c) The results of a 
single uniform distribution of dipoles within the center of the sphere begin to
plateau as the wavelength becomes small. The tolerance selected is 
$\tau=1\times10^{12}$.}}}
\label{fig:time}
\figrule
\end{figure}

Examples of the scaling behavior of the execution times ($\mathcal{T}$) for the 
functions $P/P^{0}(\lambda)$ with respect to wavelength, for numbers of layers 
$N=1$, $2$ and $3$, are shown in Fig.~\ref{fig:time} 
for a fixed outer diameter of $25$ $\mu$m. The results are fairly
insensitive to the layer thickness, allowing Fig.~\ref{fig:time} to be an 
accurate
measure of the execution time for a given number of layers and prescribed
tolerance, $\tau$. Furthermore, 
it is found that the implementation of the recursion
relations results in an improvement of approximately one order of magnitude
in the execution time compared to function-call based methods. 
This represents a significant improvement in the performance efficiency 
of the algorithm for the calculation of the WGM spectrum from a 
multilayer resonator. 

\chapter{\cdb Chemical Compositions of Media}
\label{chpt:app4}

\section{\cdb Standard non-selective yeast media (YEPD) \cbl}

Yeast Extract Peptone Dextrose (YEPD) agar promotes the growth 
of \textit{Saccharomyces cerevisiae}. For more information, see the 
\href{http://www.himedialabs.com/intl/en/products/Molecular-Biology/Molecular-Biology-Growth-Media-S-Cerevisiae-and-S-Pombe/YPD-YEPD-Growth-Agar-G038}
{\cb HiMedia product documentation}. 

The proportions of the ingredients in solid or powder form are as follows. 

\newpage
\begin{mdframed}[backgroundcolor=boxcol,hidealllines=true]
{
\begin{center}
\begin{mdframed}[backgroundcolor=boxcold,hidealllines=true]
\textbf{\color{white} \large YEPD Recipe \cbl}
\end{mdframed}
\end{center}
\vspace{-2mm}
$\bullet$ yeast extract -- $15\%$\\
$\bullet$ agar -- $23\%$\\
$\bullet$ peptone -- $31\%$\\
$\bullet$ dextrose -- $31\%$\\
\noindent\textbf{\cdb Method:\cbl} A YEPD solution of concentration 
approximately $65$ mg/mL is made with Milli-Q$^{\text{\textregistered}}$ water and 
placed in an autoclave at $121^\circ$C for $15$ minutes for sterilisation. The 
resultant pH at $25^\circ$C is $6.5\pm0.2$. 
}
\end{mdframed}

\section{\cdb MLA algal media \cbl}

A variety of algae of the family \textit{Volvocaceae}, 
including \textit{Eudorina}-\textit{Pandorina}, may be cultivated in MLA media 
\cite{Bolch1996}. For more information, see the 
\href{http://www.marine.csiro.au/microalgae/methods/Media%20CMARC%20recipes.htm#MLA}
{\cb CSIRO media recipes\cbl}.\\

\begin{mdframed}[backgroundcolor=boxcol,hidealllines=true]
{
\begin{center}
\begin{mdframed}[backgroundcolor=boxcold,hidealllines=true]
\textbf{\color{white} \large MLA Recipe \cbl}
\end{mdframed}
\end{center}
\vspace{-2mm}
\noindent\textbf{\cdb Stock solutions 
(per litre Milli-Q}$^{\text{\textregistered}}$\textbf{\cdb):}\cbl\\
$\bullet$ MgSO$_4\cdot7$H$_2$O\, -- \,$49.4$ g\\
$\bullet$ NaNO$_3$\, -- \,$85.0$ g\\
$\bullet$ K$_2$HPO$_4$\, -- \,$6.96$ g\\
$\bullet$ H$_3$BO$_3$\, -- \,$2.47$ g\\
$\bullet$ H$_2$SeO$_3$\, -- \,$1.29$ mg\\
$\bullet$ NaHCO$_3$\, -- \,$16.9$ g\\
$\bullet$ CaCl$_2\cdot2$H$_2$O\, -- \,$29.4$ g\\
\noindent\textbf{\cdb Vitamins (per 100 mL Milli-Q}$^{\text{\textregistered}}$
\textbf{\cdb):}\cbl\\
$\bullet$ Biotin \, -- \,$0.05$ mL primary stock\\
$\bullet$ B$12$ \, -- \,$0.05$ mL primary stock\\
$\bullet$ Thiamine HCl \, -- \,$10.0$ mg \\
\newpage
\indent\textbf{\cdb Vitamin primary stocks 
(per 100 mL Milli-Q}$^{\text{\textregistered}}$\textbf{\cdb):}\cbl\\
\indent$\bullet$ Biotin primary stock \, -- \,$10$ mg\\
\indent$\bullet$ B$12$ primary stock \, -- \,$10$ mg\\
\noindent\textbf{\cdb Micronutrients 
(per 800 mL Milli-Q}$^{\text{\textregistered}}$\textbf{\cdb):}\cbl\\
$\bullet$ Na$_2$EDTA \, -- \,$4.36$ g (dissolve on low heat)\\
$\bullet$ FeCl$_3\cdot6$H$_2$O \, -- \,$1.58$ g\\
$\bullet$ NaHCO$_3$ \, -- \,$0.60$ g \\
$\bullet$ MnCl$_2\cdot4$H$_2$O \, -- \,$0.36$ g\\
$\bullet$ Micronutrient primary stocks \, -- \,$10$ mL each \\
\indent\textbf{\cdb Micronutrient primary stocks 
(per litre Milli-Q}$^{\text{\textregistered}}$\textbf{\cdb):}\cbl\\
\indent$\bullet$ CuSO$_4\cdot5$H$_2$O\, -- \,$1.0$ g\\
\indent$\bullet$ ZnSO$_4\cdot7$H$_2$O\, -- \,$2.2$ g\\
\indent$\bullet$ CoCl$_2\cdot6$H$_2$O\, -- \,$1.0$ g\\
\indent$\bullet$ Na$_2$MoO$_4\cdot2$H$_2$O\, -- \,$0.6$ g\\
\noindent\textbf{\cdb Method:\cbl} Autoclave the distilled water to sterilise. 
Make up micronutrient stock to $1$ litre. Store all stock solutions in a 
refrigerator at $4^\circ$C. 
Then, prepare MLA $\times$ 40 concentrated nutrient mixture to $250$ mL.\\
\noindent\textbf{\cdb MLA $\times$ 40 concentrated nutrient mixture 
(per $130$ mL Milli-Q}$^{\text{\textregistered}}$\textbf{\cdb):}\cbl\\
$\bullet$ MgSO$_4\cdot7$H$_2$O stock \, -- \,$10$ mL\\
$\bullet$ NaNO$_3$ stock \, -- \,$20$ mL\\
$\bullet$ K$_2$HPO$_4$ stock \, -- \,$50$ mL\\
$\bullet$ H$_3$BO$_3$ stock \, -- \,$10$ mL\\
$\bullet$ H$_2$SeO$_3$ stock \, -- \,$10$ mL\\
$\bullet$ Vitamin stock \, -- \,$10$ mL\\
$\bullet$ Micronutrient stock \, -- \,$10$ mL\\
Filter-sterilise using $0.22$ mm filter into a sterile $250$ mL Schott bottle. 
To make $1$ litre of MLA, add $964$ mL of Milli-Q$^{\text{\textregistered}}$ water 
into a sterile $1$ litre Schott bottle, add $25$ mL of MLA $\times 40$ 
nutrients, $10$ mL of NaHCO$_3$ stock, $1$ mL of sterile CaCl$_2\cdot2$H$_2$O 
stock and mix well. 
}
\end{mdframed}

\section{\cdb \textit{Eudorina}-\textit{Pandorina} geometric analysis \cbl}

\begin{figure}[h]
\begin{center}
\includegraphics[width=0.8\hsize]{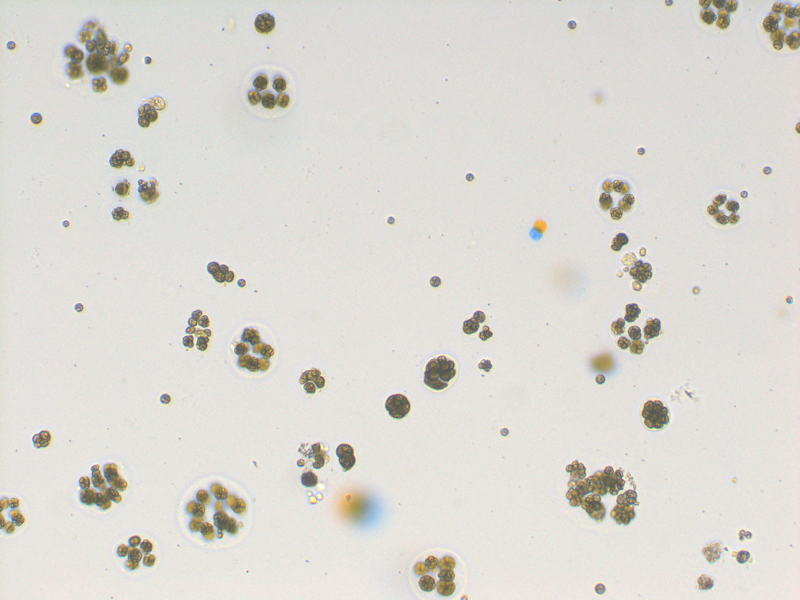}
\end{center}
\caption[Image of \textit{Eudorina}-\textit{Pandorina} algae - assessing 
geometric properties.]{An image of algae of the 
\textit{Eudorina}-\textit{Pandorina} genus used for a statistical 
assessment of their geometric properties. 
\textit{Image:} 
produced with the assistance of Mr. Steven Amos, 
The University of Adelaide.}
\label{fig:eugeo}
\figrule
\end{figure}

Once algae of the \textit{Eudorina}-\textit{Pandorina} genus have been 
cultivated in sufficient quantity, an analysis on their geometric 
characteristics is carried out. 
An example of a sample of the solution containing the 
\textit{Eudorina}-\textit{Pandorina} algae suspended 
in MLA media is shown in Fig.~\ref{fig:eugeo}. 
This image is used as a representative sample for extracting the geometric 
parameters of the identifiable particles. The results of the analysis are 
shown in Table~\ref{tab:eugeo}.

\vspace{21mm}

\newpage
\vspace{5mm}
\begin{table}[t]
\includepdf[pages=1]{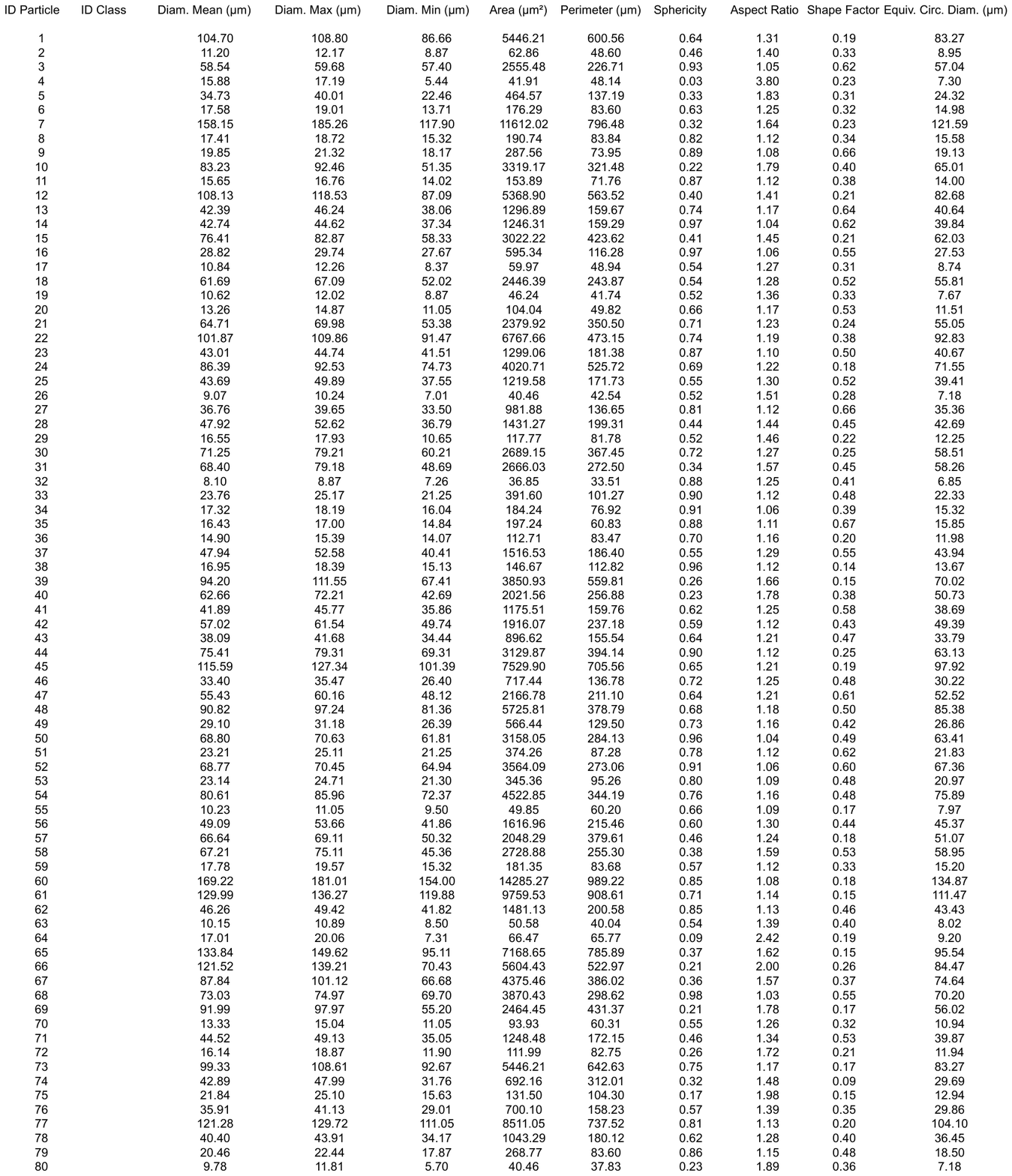}
\vspace{17.1cm}
\caption[An analysis of the geometric properties of a sample of 
\textit{Eudorina}-\textit{Pandorina} algae.]{An analysis of the geometric 
properties of the  particles identified in the 
\textit{Eudorina}-\textit{Pandorina} algae image shown in Fig.~\ref{fig:eugeo}.}
\label{tab:eugeo}
\end{table}

%\newpage
\section{\cdb Paraffin formaldehyde (PFA) oocyte fixing method \cbl}
\label{sec:pfa}
\vspace{-10mm}
One of the most common ways of \textit{fixing} oocytes and embryos, which 
essentially entails placing the cells into a permanently storable state and 
preventing all metabolic activity, is that of washing them in paraffin 
formaldehyde (PFA). Typically, the cells are placed in PFA of a concentration 
of between $2\%$ and $4\%$ for approximately $30$ minutes at 
$4^\circ$C, and then placed in one of the following handling media.

\subsection{\cdb Phosphate-buffered saline (PBS) media \cbl}
\label{sec:zona}

The handling and storage of mammalian oocytes and embryos can involve a number 
of different procedures and media. In this thesis, two methods in particular 
are investigated. The \textit{zona} salt solution, or PBS, represents the first 
of these media, and the method for fixing 
these cells is as follows. For more information, see 
\href{http://www.protocolsonline.com/recipes/phosphate-buffered-saline-pbs/}
{\cb Protocols Online\cbl}. \\

\begin{mdframed}[backgroundcolor=boxcol,hidealllines=true]
{
\begin{center}
\begin{mdframed}[backgroundcolor=boxcold,hidealllines=true]
\textbf{\color{white} \large PBS Recipe \cbl}
\end{mdframed}
\end{center}
\vspace{-2mm}
$\bullet$ NaCl -- $8$ g \\
$\bullet$ KCl -- $0.2$ g \\
$\bullet$ Na$_2$HPO$_4$ -- $1.44$ g \\
$\bullet$ KH$_2$PO$_4$ -- $0.24$ g \\
\noindent\textbf{\cdb Method:\cbl} $1$ litre of $1\times$ PBS media 
is prepared with $800$ mL of Milli-Q$^{\text{\textregistered}}$ water, and 
sequentially adding the ingredients above. 
After this has been achieved, add HCl to the mixture so the pH of the solution 
has decreased to $7.4$ at $25^\circ$C. 
Sterilise using an autoclave for $20$ min at $121^\circ$C and store at 
$25^\circ$C. 
}
\end{mdframed}

\subsection{\cdb MOPS-buffered wash and bovine serum albumin (BSA) \\ 
handling media \cbl}
\label{sec:mops}

The second choice of media for storing mammalian oocytes and embryos 
considered in this thesis 
is that of a solution composed of (3-(N-morpholino)propanesulfonic 
acid (MOPS)-buffered wash\footnote{Cook Medical Australia, $95$ 
Brandl St, Brisbane Technology Park, Eight 
Mile Plains, QLD $4113$, Australia, \textit{Research Vitro Wash}, Catalogue 
Number: K-RVWA-$50$.} 
and bovine albumin serum (BSA). 
This medium has a lower salt content than PBS, and its recipe is as follows. \\

\begin{mdframed}[backgroundcolor=boxcol,hidealllines=true]
{
\begin{center}
\begin{mdframed}[backgroundcolor=boxcold,hidealllines=true]
\textbf{\color{white} \large MOPS+BSA Recipe \cbl}
\end{mdframed}
\end{center}
\vspace{-2mm}
$\bullet$ MOPS-buffered Research Vitro Wash$^1$ -- $280$ mOsm \\
$\bullet$ BSA -- $4$ mg/mL \\
\noindent\textbf{\cdb Method:\cbl} A MOPS-buffered solution of wash medium is 
diluted to an osmolarity of $280$ mOsm. BSA is added to a concentration is $4$ 
mg/mL. The final pH should be in the range $7.3$ to $7.5$. 
Live culturing must be handled at $39^\circ$C. However, this medium may also be 
used for fixed cells and stored in a refrigerator at $4^\circ$C. 
}
\end{mdframed}

\newpage
\section{\cdb Chemical annealing - Acidified Tyrode's Solution \cbl}
\label{sec:ats}
\vspace{-1.1cm}
In Section~\ref{sec:anneal} of Chapter~\ref{chpt:fdi}, the method 
for thinning or \emph{annealing} the \textit{zona pellucida} region 
of embryos is described. This can be achieved using other chemical-based 
or enzyme-based solutions, exposed to the outer layer of the \textit{zona} 
for a limited time. The predominant chemical-based annealing procedure 
uses Acidified Tyrode's Solution (ATS), whose recipe is as follows \cite{tyrodes}. 
For more information, see the 
\href{http://www.origio.com/products/acidified-tyrodes-solution/}
{\cb Origio product datasheet\cbl}.\\

\begin{mdframed}[backgroundcolor=boxcol,hidealllines=true]
{
\begin{center}
\begin{mdframed}[backgroundcolor=boxcold,hidealllines=true]
\textbf{\color{white} \large ATS Recipe \cbl}
\end{mdframed}
\end{center}
\vspace{-2mm}
$\bullet$ NaCl -- $137$ mM \\
$\bullet$ KCl -- $2.70$ mM \\
$\bullet$ MgCl$_2$ -- $1.00$ mM \\
$\bullet$ CaCl$_2$ -- $1.80$ mM \\
$\bullet$ Na$_2$HPO$_4$ -- $0.20$ mM \\
$\bullet$ NaHCO$_3$ -- $12.0$ mM \\
$\bullet$ D-glucose -- $5.50$ mM 
}
\end{mdframed}

\chapter{\cdb Lists of Publications}
\label{chpt:pub}

The following lists of published journal articles, conference papers and 
open-source code were produced 
during the time of candidature, based on work submitted to the degree of 
Doctor of Philosophy. For more information, visit my academic website: 
\href{http://drjonathanmmhallfrsa.wordpress.com}
{\cb http://drjonathanmmhallfrsa.wordpress.com\cbl}.

\vspace{-3mm}
\section{\cdb Peer-reviewed published journal articles}
\label{sec:jou}

\vspace{-4mm}
\subsubsection{\cdb Lead author:}
\vspace{-2mm}

\noindent\textbf{1.} ``Method for predicting whispering gallery mode spectra of 
spherical microresonators''; \textbf{J. M. M. Hall}, 
S. Afshar V., M. R. Henderson, A. Fran\c{c}ois, T. Reynolds, N. Riesen, 
T. M. Monro; \textbf{Optics Express, Vol. 23, Issue 8}, 
pp. 9924-9937 (2015) -- in-text Ref.~\cite{Hall:15}; \\
DOI: \href{http://dx.doi.org/10.1364/OE.23.009924}
{\cb http://dx.doi.org/10.1364/OE.23.009924\cbl}, \\
web: \href{http://www.opticsinfobase.org/oe/abstract.cfm?uri=oe-23-8-9924}
{\cb http://www.opticsinfobase.org/oe/abstract.cfm?uri=oe-23-8-9924 \cbl}. 
\\

\noindent\textbf{2.} ``Determining the geometric parameters of microbubble 
resonators from their spectra''; \textbf{J. M. M. Hall}, 
A. Fran\c{c}ois, S. Afshar V., N. Riesen, M. R. Henderson, T. Reynolds, 
T. M. Monro; 
\textbf{Journal of the Optical Society of America B  34}, pp. 2699-2706 (2017) 
-- in-text Ref.~\cite{Hall:17}; \\
DOI: \href{http://doi.org/10.1364/JOSAB.34.002699}
{\cb http://doi.org/10.1364/JOSAB.34.002699\cbl}, \\
web: \href{https://www.osapublishing.org/josab/abstract.cfm?uri=josab-34-1-44}
{\cb https://www.osapublishing.org/josab/abstract.cfm?uri=josab-34-1-44\cbl}.
\\

\noindent\textbf{3.} ``Unified theory of whispering gallery multilayer 
microspheres with single dipole or active layer sources''; 
\textbf{J. M. M. Hall}, T. Reynolds, M. R. Henderson, N. Riesen, T. M. Monro, 
S. Afshar, V.; \textbf{Optics Express, Vol. 25, Issue 6}, 
pp. 6192-6214 (2017) -- in-text Ref.~\cite{Hall:17a}; \\ 
DOI: \href{http://doi.org/10.1364/OE.25.006192}
{\cb http://doi.org/10.1364/OE.25.006192\cbl}, \\
web: \href{http://www.osapublishing.org/oe/abstract.cfm?uri=oe-25-6-6192}
{\cb http://www.osapublishing.org/oe/abstract.cfm?uri=oe-25-6-6192\cbl}. 

\vspace{-4mm}
\subsubsection{\cdb Co-author:}
\vspace{-2mm}

\noindent\textbf{4.} ``Optimization of whispering gallery resonator design for 
biosensing applications''; T. Reynolds, M. R. Henderson, 
A. Fran\c{c}ois, N. Riesen, \textbf{J. M. M. Hall}, S. Afshar V., S. J. 
Nicholls, 
T. M. Monro; \textbf{Optics Express, Vol. 23, Issue 13}, 
pp. 17067-17076 (2015) -- in-text Ref.~\cite{Reynolds:15}; \\ 
DOI: \href{http://dx.doi.org/10.1364/OE.23.017067}
{\cb http://dx.doi.org/10.1364/OE.23.017067\cbl}, \\
web: \href{http://www.osapublishing.org/oe/abstract.cfm?uri=oe-23-13-17067}
{\cb http://www.osapublishing.org/oe/abstract.cfm?uri=oe-23-13-17067\cbl}.
\\

\noindent\textbf{5.} ``Combining whispering gallery mode lasers and 
microstructured optical fibers: limitations, applications and 
perspectives for in-vivo biosensing”; A. Fran\c{c}ois, T. Reynolds, N. Riesen, 
\textbf{J. M. M. Hall}, M. R. Henderson, E. Zhao, S. Afshar V. 
and T. M. Monro; \textbf{MRS Advances, 2059-8521}, pp. 1-12 (2016) -- 
in-text Ref.~\cite{Francois:16a}; \\
DOI: \href{http://dx.doi.org/10.1557/adv.2016.342}
{\cb http://dx.doi.org/10.1557/adv.2016.342\cbl}, \\
web: \href{http://journals.cambridge.org/action/displayAbstract?fromPage=online&aid=10321592&fileId=S205985211600342X}
{\cb http://journals.cambridge.org/action/displayAbstract?fromPage=online\\
\&aid=10321592\&fileId=S205985211600342X\cbl}.
\\

\noindent\textbf{6.} ``Fluorescent and lasing whispering gallery mode 
microresonators for sensing applications''; T. Reynolds, N. Riesen, 
A. Meldrum, X. Fan, \textbf{J. M. M. Hall}, T. M. Monro and A. Fran\c{c}ois; 
\textbf{Laser \& Photonics Reviews}, 1600265, 
-- in-text Ref.~\cite{LPOR:LPOR201600265}; \\
DOI: \href{http://dx.doi.org/10.1002/lpor.201600265}
{\cb http://dx.doi.org/10.1002/lpor.201600265\cbl},\\
web: \href{http://onlinelibrary.wiley.com/wol1/doi/10.1002/lpor.201600265/full}
{\cb http://onlinelibrary.wiley.com/wol1/doi/10.1002/lpor.201600265/full\cbl}.

\section{\cdb Published conference proceedings}

\noindent\textbf{7.} ``Predicting the whispering gallery mode spectra of 
microresonators''; \textbf{J. M. M. Hall}, S. Afshar V., 
M. R. Henderson, A. Fran\c{c}ois, T. Reynolds, N. Riesen, T. M. Monro; 
\textbf{Proc. SPIE 9343, Laser Resonators, Microresonators, and 
Beam Control XVII}, 93431Y (2015) 
-- in-text Ref.~\cite{doi:10.1117/12.2078526}; \\
DOI: \href{http://dx.doi.org/10.1117/12.2078526}
{\cb http://dx.doi.org/10.1117/12.2078526\cbl}, \\
web: \href{http://proceedings.spiedigitallibrary.org/proceeding.aspx?articleid=2194939}
{\cb http://proceedings.spiedigitallibrary.org/proceeding.aspx?articleid=2194939\cbl}.

\section{\cdb Press releases}

\noindent\textbf{8.} ``Shining new light on the body''; 
A. Greentree and \textbf{J. M. M. Hall}; 
\textbf{National Computational Infrastructure (NCI)}, 
press release,  \\ 
web: 
\href{http://nci.org.au/research/shining-new-light-on-the-body/}
{\cb 
http://nci.org.au/research/shining-new-light-on-the-body/}. \\

\noindent\textbf{9.} ``WGM Laser-Tipped Fiber for Biomedical Applications''; 
A. Fran\c{c}ois, S. Afshar V., T. M. Monro, T. Reynolds, 
N. Riesen, \textbf{J. M. M. Hall} and M. R. Henderson; 
\textbf{Optics \& Photonics News}, Year in Optics 2015, Nano-optics,  \\ 
web: 
\href{http://www.osa-opn.org/home/articles/volume_26/december_2015/extras/wgm_laser-tipped_fiber_for_biomedical_applications/#.VmDGrPl94fQ}
{\cb 
http://www.osa-opn.org/home/articles/volume\_26/december\_2015/extras/wgm\_laser-tipped\_fiber\_for\_biomedical\_applications/\#.VmDGrPl94fQ\cbl}.

\section{\cdb Code produced as part of this thesis}
\label{sec:code}

The following directories of original code, produced for this thesis, 
have been placed online for public availability. 

\boxed{\noindent\textbf{1. FDTD-based toolkit for MEEP:}}\\
\href{https://sourceforge.net/projects/npps/files/FDTD_WGM_Simulator}
{\cb https://sourceforge.net/projects/npps/files/FDTD\_WGM\_Simulator\cbl} --- 
associated with Refs.~\cite{Hall:15,Hall:17}.

\boxed{\noindent\textbf{2. Active multilayer microsphere WGM simulator for 
MATLAB:}}\\
\href{http://www.photonicsimulation.net}{\cb http://www.photonicsimulation.net
\cbl} --- associated with Ref.~\cite{Hall:17a}.

\section{\cdb Conferences and workshops organised}

In organising the following conferences and workshops, I functioned 
as the Chair and Founder of the IPAS Science Network (ISN) in the 
Institute for Photonics and Advanced Sensing (IPAS), 
Chair of the International Conference on Optics, Atoms \& Laser Applications 
(IONS-KOALA) 2014 Organising Committee, 
Inaugural Elected Representative of the Faculty of Sciences 
on the Alumni Council of The University of Adelaide, 
Secretary of the Optical Society of America (OSA) and 
International Society for Optical Engineering (SPIE) 
Adelaide Chapters, 
Listed Expert of the Australian Science Media Centre (AusSMC) and 
Committee Member of the SA Branch of the Australian Institute of Physics 
(AIP) and the Australian Research Council (ARC) 
Centre for Nanoscale BioPhotonics (CNBP) Early Career Researcher (ECR) 
Committee. In chronological order: 

\noindent$\bullet$ OSA Scientific 
Photography Competition 2013, The University of Adelaide, SA, 
Australia. 

\noindent$\bullet$ IONS-KOALA 2014 Welcome Reception and Opening Address, 
The University of Adelaide, SA, Australia. 

\noindent$\bullet$ IONS-KOALA 2014 Industry Workshop, representative 
of the OSA and SPIE, The University of Adelaide, SA, Australia. 

\noindent$\bullet$ IONS-KOALA 2014 Banquet Presentation Day and 
New Horizons in Science Grant Award Ceremony, McLaren Vale, SA, Australia 

\noindent$\bullet$ IONS-KOALA 2014 Scientific Seminars, Session Chair,\\ 
The University of Adelaide, SA, Australia.

\noindent$\bullet$ OSA / IONS-KOALA Co-sponsored Scientific Photography 
Competition 2014, The University of Adelaide, SA, Australia. 

\noindent$\bullet$ IONS-KOALA 2014 Closing Address and Prize Ceremony,\\
The University of Adelaide, SA, Australia. 

\noindent$\bullet$ IPAS Research Presentation Event 2015, Convenor as Chair 
of the ISN,\\ The University of Adelaide, SA, Australia. 

\noindent$\bullet$ Alumni Council Convenor for Alumni Ambassadors and 
Class Champions 2015, Engagement Branch, The University of Adelaide, SA, 
Australia.

\begin{figure}[H]
\begin{center}
\includegraphics[width=0.9\hsize]{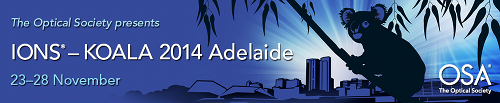}
\end{center}
\end{figure}

\section{\cdb Conferences and workshops participated}

\noindent$\bullet$  2014 International Conference on Nanoscience and 
Nanotechnology (ICONN) / 23$^{\text{rd}}$ Australian Conference on Microscopy and 
Microanalysis (ACMM), The University of Adelaide, SA, Australia.

\noindent$\bullet$ CNBP Inaugural Workshop 2014, The University of Adelaide, 
SA, Australia.

\noindent$\bullet$ Australian Nanotechnology Network (ANN) ECR Workshop 2014, 
University of Technology Sydney (UTS), NSW, Australia: 
\textit{Developing a method for predicting the whispering gallery mode 
spectrum of micro-resonators}.

\noindent$\bullet$ CNBP Workshop 2014, Macquarie University, 
NSW, Australia: \textit{Living resonators: cells as sensors}.

\noindent$\bullet$ Royal Institution of Australia (RiAus), 
ECR Grant Writing Workshop, Science Exchange Centre, SA, Australia.

\noindent$\bullet$ CNBP Annual Retreat 2014, Port Lincoln, SA, Australia:\\
\textit{Towards living cell sensors}.

\noindent$\bullet$ IONS-KOALA 2014, The University of Adelaide, SA, 
Australia:\\ \textit{Resonator sensors}. 

\noindent$\bullet$ 5$^{\text{th}}$ Mount Lofty Workshop on 
Frontier Technologies for Nervous System Function \& Repair 2014, Mount Lofty, 
SA: Contribution to seminar \textit{Biophotonics: living resonators}, 
\textit{via} Prof. Andre Luiten. 

\noindent$\bullet$ University of Adelaide Research Symposium 2014, 
\textit{Cells as Lasers}. 

\noindent$\bullet$ IPAS Research Presentation Event, 
The University of Adelaide, SA, Australia: 
\textit{Toward cell-lasers in biophotonics}. 

\noindent$\bullet$ SPIE Photonics West, 
LASE (Laser Technology and Industrial Laser Conference), 
Laser Resonators, Microresonators and Beam Control XVII, 
Moscone Center, San Francisco, CA, USA 2015: 
\textit{Predicting the whispering gallery mode spectra of microresonators}. 

\noindent$\bullet$ CNBP ECR Presentation Series 2015, The University of 
Adelaide, SA, Australia:
\textit{Living cells: sensors of the future?}

\noindent$\bullet$ ETHOS Australia Pty Ltd Workshop, 
``Industry engaged research: Seven principles guiding an enduring 
relationship'', Adelaide, SA, Australia.

\noindent$\bullet$ Royal Melbourne Institute of Technology (RMIT) CNBP 
Launch, Melbourne, VIC, Australia.

\noindent$\bullet$ Commonwealth Scientific and Industrial Research 
Organisation (CSIRO) and CNBP Partnership Launch, Melbourne, VIC, Australia. 

\noindent$\bullet$ IPAS Scientific Management Committee, 
Strategic Planning Meeting 2015, The University of 
Adelaide, SA, Australia.

\noindent$\bullet$ The Australian and New Zealand Conference on Optics and 
Photonics (ANZCOP) 2015, The University of Adelaide, SA, Australia: 
\textit{Multilayer resonator sensitivity analysis}.

\noindent$\bullet$ Australian Conference on Optics, Lasers and Spectroscopy 
(ACOLS) 2015, The University of Adelaide, SA, Australia: 
\textit{Microbubble resonator simulation using FDTD}.

\noindent$\bullet$ 40$^{\text{th}}$ Australian Conference on Optical Fibre 
Technology (ACOFT) 2015, The University of Adelaide, 
SA: Contribution to seminar \textit{Combining whispering gallery mode lasers 
and microstructured optical fibres for in-vivo biosensing applications}, 
\textit{via} Dr. Alexandre Fran\c{c}ois.

\noindent$\bullet$ Innovation Voucher Program Workshop 2016, 
Flinders University, Tonsley, SA, Australia.

\noindent$\bullet$ CNBP Annual Retreat 2015, Lake Macquarie, NSW, Australia: 
\textit{Biolaser}.

\noindent$\bullet$ Contribution to seminar 
\textit{Unlocking the secrets within, using light - from wine to embryos}, 
\textit{via} Prof. Tanya Monro, The Shine Dome, Acton, ACT, Australia.

\noindent$\bullet$ Robinson Research Institute in reproductive health, 
group research presentation, CNBP, SA, Australia.

\noindent$\bullet$ ACOFT 2016, The Australian National University (ANU), 
Canberra, ACT, Australia: Contribution to seminar 
\textit{A Unified Model for Active Multilayer Microsphere Resonators}, 
\textit{via} Assoc. Prof. Shahraam Afshar V.

\noindent$\bullet$ SPIE BioPhotonics Australasia 2016, 
Adelaide Convention Centre, SA, Australia. 

\noindent$\bullet$ CNBP Annual Retreat 2016, Victor Harbor, SA, Australia: 
\textit{Towards an Embryo Laser.} 

\noindent$\bullet$ CSIRO \textit{ON Prime} Pre-Accelerator Program 2016, 
Australian Technology Park, Sydney, NSW, Australia.

\noindent$\bullet$ Australian eChallenge Venture Showcase 2016, 
Adelaide Convention Centre, SA, Australia.

\section{\cdb Journals - acting as reviewer}

\noindent$\bullet$ \textbf{\cdb Applied Optics (OSA)\cbl} 

\noindent$\bullet$ \textbf{\cdb The Journal of the Optical Society of 
America A (OSA)
\cbl} 

\noindent$\bullet$ \textbf{\cdb Optics Express (OSA)\cbl} 

\noindent$\bullet$ \textbf{\cdb IEEE Photonics Technology Letters \cbl} 

\noindent$\bullet$ \textbf{\cdb Physical Review Letters, 
American Physical Society (APS)\cbl} 

\noindent$\bullet$ \textbf{\cdb Physical Review A, C and D (APS)\cbl}

\section{\cdb Grants and funding}

\noindent$\bullet$ ARC Laureate Scholarship 2014-2017.

\noindent$\bullet$ ANN Invited Presenter Travel Grant, UTS, 2014.

\noindent$\bullet$ ISN Funding 2014-2016 - \$10,000 p.a.

\noindent$\bullet$ IONS-KOALA 2014 Institutional and Industry 
Sponsorship - \$56,000 + in-kind support: \\ \textbf{\cdb Gold sponsor\cbl}: 
OSA; \\
\textbf{\cdb Silver sponsors\cbl}: 
SPIE, The ARC Centre of Excellence in Ultrahigh 
Bandwidth Devices for Optical Systems (CUDOS - Universities of 
Sydney, Macquarie, Monash, Swinburne, RMIT, UTS and ANU), 
The Centre for Micro-Photonics: Swinburne University of Technology, 
MQ Photonics: Macquarie University, Lastek Pty Ltd (Industry), 
IPAS, CNBP and The Australian Optical Society (AOS);
 \\ 
\textbf{\cdb Bronze sponsors\cbl}: 
The AIP National Executive, Maptek (Industry), 
Coherent Scientific (Industry), Grey Innovation Technology 
Commercialisation (Industry), The Photon Factory: University of Auckland  
(New Zealand), Edmund Optics (Industry), NewSpec (Industry), Griffith 
University, The ARC T-ray Facility, School of Electrical \& Electronic 
Engineering (The University of Adelaide) and The Department of State 
Development (DSD) (formerly Department for Manufacturing, Innovation, Trade, 
Resources and Energy (DMITRE)), Government of South Australia. \\
\textbf{\cdb Auxilliary sponsors\cbl}:
Photon Scientific (Industry), Melbourne
School of Physical Sciences (formerly 
School of Chemistry and Physics, The University of Adelaide);\\
\textbf{\cdb Travel grant sponsors\cbl}: 
OSA, Lastek Pty Ltd, IPAS and The New Zealand Institute of Physics. 

\noindent$\bullet$ IPAS Pilot Project Scheme 2016 - \$5,000.

\section{\cdb Recognition and presentations}

\noindent$\bullet$ Election to the  Alumni Council of The University of 
Adelaide as Representative for the Faculty of Sciences 2014-2015.

\noindent$\bullet$ Australian Academy of Science EMCR Forum Nomination 2014.

\noindent$\bullet$ Australian Academy of Science Young Tall Poppy Award 
Nomination 2015.

\noindent$\bullet$ IPAS Research Presentation Awards 2016 - Honourable 
Mention.

\noindent$\bullet$ Australian eChallenge 2016 -  Research Commercialisation 
1$^{\text{st}}$ Prize - \$10,000.

\noindent$\bullet$ Australian eChallenge 2016 - Medical Innovations 
1$^{\text{st}}$ Prize - \$10,000.

\titleformat{\chapter}[display]
  {\vspace{5cm}\flushright\fontseries{b}\fontsize{80}{100}\selectfont}
  {\fontseries{b}\fontsize{100}{130}\selectfont\textcolor{numcol}\thechapter}
  {0pt}
  {\Huge\bfseries}[]
\renewcommand{\bibname}{\cdb \textbf{Bibliography}}
\addcontentsline{toc}{chapter}{Bibliography}

\bibliographystyle{osajnl}
\bibliography{references}

\end{document}